\documentclass[twocolumn]{aastex631}

\usepackage{makecell}
\usepackage{subfigure}

\newcommand{\Bbarolo}{{$^{\rm 3D}$\textsc{Barolo}}}

\begin{document}
\title{Constraining Quasar Feedback from Analysis of the Hydrostatic Equilibrium of the Molecular Gas in Their Host Galaxies}
\author[0000-0001-7232-5355]{Qinyue Fei}
\affiliation{Department of Astronomy, School of Physics, Peking University, Beijing 100871, P. R. China}
\affiliation{Kavli Institute for Astronomy and Astrophysics, Peking University, Beijing 100871, P. R. China}
\affiliation{Kavli Institute for the Physics and Mathematics of the Universe (Kavli IPMU, WPI), UTIAS, Tokyo Institutes for Advanced Study, University of Tokyo, Chiba, 277-8583, Japan}

\author[0000-0003-4956-5742]{Ran Wang}
\correspondingauthor{Ran Wang}
\affiliation{Department of Astronomy, School of Physics, Peking University, Beijing 100871, P. R. China}
\affiliation{Kavli Institute for Astronomy and Astrophysics, Peking University, Beijing 100871, P. R. China}

\author[0000-0002-8136-8127]{Juan Molina}
\affiliation{Instituto de F\'isica y Astronom\'ia, Universidad de Valpara\'iso, Avda. Gran Breta\~na 1111, Valpara\'iso, Chile}

\author[0000-0001-6947-5846]{Luis C. Ho}
\affiliation{Department of Astronomy, School of Physics, Peking University, Beijing 100871, P. R. China}
\affiliation{Kavli Institute for Astronomy and Astrophysics, Peking University, Beijing 100871, P. R. China}

\author[0000-0002-4569-9009]{Jinyi Shangguan}
\affiliation{Max-Planck-Institut f\"{u}r Extraterrestrische Physik (MPE), Giessenbachstr., D-85748 Garching, Germany}

\author[0000-0002-8686-8737]{Franz E. Bauer}
\affil{Instituto de Astrof{\'{\i}}sica and Centro de Astroingenier{\'{\i}}a, Facultad de F{\'{i}}sica, Pontificia Universidad Cat{\'{o}}lica de Chile, Casilla 306, Santiago 22, Chile}
\affiliation{Millennium Institute of Astrophysics (MAS), Nuncio Monse{\~{n}}or S{\'{o}}tero Sanz 100, Providencia, Santiago, Chile} \affiliation{Space Science Institute, 4750 Walnut Street, Suite 205, Boulder, Colorado 80301}

\author[0000-0001-7568-6412]{Ezequiel Treister}
\affil{Instituto de Astrof{\'{\i}}sica and Centro de Astroingenier{\'{\i}}a, Facultad de F{\'{i}}sica, Pontificia Universidad Cat{\'{o}}lica de Chile, Casilla 306, Santiago 22, Chile}

\begin{abstract}
    We investigate the kinematics and dynamics of the molecular and ionized gas in the host galaxies of three Palomar-Green quasars at low redshifts, benefiting from the archival millimeter-wave interferometric and optical integral field unit data. We study the kinematics of both cold molecular and hot ionized gas by analyzing the CO and H$\alpha$ data cubes, and construct the mass distributions of our sample through gas dynamics, utilizing a priori knowledge regarding the galaxy light distribution. We find no systematic offset between the stellar mass derived from our dynamical method and that from the broad-band photometry and mass-to-light ratio, suggesting the consistency of both methods. We then study the kinetic pressure and the weight of the interstellar medium using our dynamical mass model. By studying the relationship between kinetic pressure and gravitational pressure of the quasar host galaxies, we find an equivalence in the hydrostatic equilibrium states of ISM in the quasar host galaxies, similar to the result of gas equilibrium in normal star-forming galaxies, suggesting minimal quasar feedback. Regarding non-circular motion as indicative of quasar-driven outflows, we observe an exceptionally low coupling efficiency between molecular gas outflow and AGN bolometric luminosities. These results demonstrate the marginal influence of the central engine on the properties of cold molecular gas in quasar host galaxies. 
\end{abstract}

\section{Introduction}
The empirical scaling relationships between supermassive black holes (SMBHs) and their host galaxies have been established for several decades \citep{Magorrian+1998, Ferrarese+2000, Gebhardt+2000}, suggesting a strong correlation between these two components in their early evolutionary stages \citep{Kormendy&Ho2013, Heckman+2014}. Nonetheless, the underlying physical mechanisms governing this coevolution remain a topic of ongoing debate. Energy feedback originating from active galactic nuclei (AGNs) is believed to play an important role in modulating galaxy evolution \citep{Fabian2012}. In specific instances where SMBHs are accreting materials with high efficiency, they can initiate the production of high luminosity levels within the AGN. The resultant radiative or mechanical energy fluxes can then initiate pronounced outflows, capable of expelling cold gas from the galaxy's confines \citep{Silk&Rees1998, Zubovas+2012}. 

Quasars, being the most luminous AGNs, emerge as particularly fitting subjects to study the potential implications of AGN feedback. Studies focused on the properties of host galaxies of low-$z$ quasars provided a good opportunity to understand how BHs affect their host galaxies, as proposed by both observations and simulations \citep{Sanders+1988, Hopkins+2008, Treister+2010}. ``Quasar/radiation mode'' feedback has been widely investigated with the large population of quasars \citep{Feruglio+2010, Veilleux+2013, Tombesi+2015, Rupke+2011}. This mechanism is hypothesized to exert a pivotal role within the widely accepted evolutionary paradigm for AGNs, particularly those arising from gas-rich, major merger-driven scenarios \citep{Sanders+1988}, as the systems transition from an initial, dust-obscured state to their ultimate unobscured quasar phase \citep{Hopkins+2008}. Leveraging the advances achieved over several decades in the realm of multi-band instrumentation, manifestations of AGN-driven outflows have been observed with notable regularity \citep{Forster-Schreiber+2014, Perna+2015, Woo+2016, Nesvadba+2008}, encompassing the realms of cold molecular gas \citep{Rupke+2011, Sturm+2011, Cicone+2014, Garcia-Burillo+2014, Feruglio+2020}, neutral gas \citep{Shapley+2003, Martin2005, Sugahara+2017, Gatkine+2019}, and ionized gas \citep{Nesvadba+2008, Bischetti+2017}. However, the contributions of these AGN feedback activities to the subsequent evolution of the host galaxies are still under debate \citep{Woo+2017}. 

On one hand, observations reveal that gas within quasar-host galaxies can be removed effectively by AGN feedback \citep{Fiore+2017, Fluetsch+2019}, concomitant with substantial suppression of star formation rates (SFRs) \citep{Ho+2003, Alatalo+2015, Ellison+2016, Leslie+2016}. However, recent studies focused on the most powerful AGNs at low redshifts suggest an absence of robust gas outflows \citep{Shangguan+2020a, Shangguan+2020b, Molina+2022}, challenge our understanding of the coevolutionary relationship between SMBHs and galaxies. This result suggests our understanding of the AGN-host galaxy coevolution is still very limited. Therefore it is necessary to further investigate the properties of the cold gas in more quasar host galaxies, as well as the potential impact of AGN feedback on the cold gas.


In previous work, we analyzed the gas dynamics in one of the Palomar-Green (PG) quasars \citep{Boroson&Green1992}, I\,Zw\,1. The kinetic pressure of the molecular gas in the nuclear disk is in equilibrium with the gravitational pressure perpendicular to the galactic disk plane, thereby implying a limited impact of AGN feedback within that particular galaxy \citep{Fei+2023}. It is crucial to study whether the impact of AGN feedback is uniform in a larger sample. Moreover, it is necessary to compare the properties of cold molecular gas within AGN host galaxies and those in normal star-forming galaxies. To address these inquiries, we have broadened the analysis of the cold gas kinematics to encompass additional objects from the PG quasar sample. We study the gas dynamics between the PG quasar hosts and compare it to a control sample of normal star-forming galaxies.


This paper is organized as follows: Section \ref{sec2: Sample} introduces the quasar and star-forming galaxy sample. The analysis method is described in Section \ref{sec3: methods}. Section \ref{sec4: gas dynamics} introduces the mass models of galaxies using gas dynamics. Section \ref{sec5: discussions} presents the main results and discussions. And Section \ref{sec6: summary} is the summary and conclusion. This work adopts following parameters for a $\Lambda$CDM cosmology: $\Omega_m=0.308,\Omega_\Lambda = 0.692$, and $H_0=67.8\,{\rm km\,s^{-1}\,Mpc^{-1}}$ \citep{Planck+2016}.

\section{Sample and Data reduction}
\label{sec2: Sample}
\begin{figure}
    \centering
    \includegraphics[width=\linewidth]{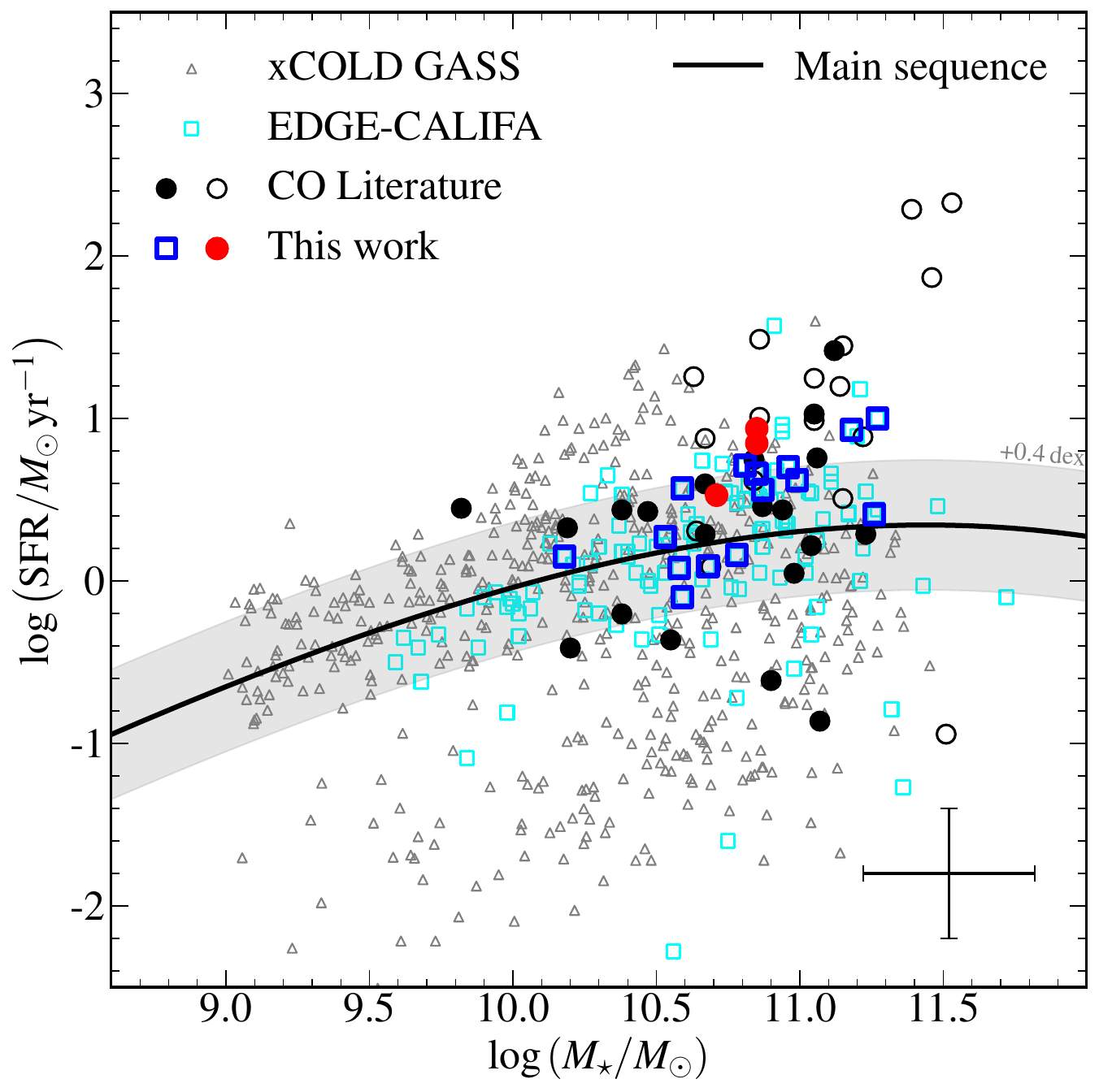}
    \caption{The stellar mass versus SFR for both quasar host galaxies and inactive galaxies. Black and white circles denote quasars with CO detections \citep{Evans+2006, Shangguan+2020a}, and red points indicate three targets with spatially-resolved CO observation \citep{Molina+2021, Molina+2023} that are used in this work. The EDGE-CALIFA data are represented by light-blue and dark-blue squares, with the latter corresponding to a subsample chosen for comparison with the quasar hosts. Additionally, we include the xCOLD GASS sample \citep{Saintonge+2017} as gray triangles. The solid line on the plot represents the main sequence of star-forming galaxies at $z\sim 0$ \cite{Saintonge+2016}, while the gray shaded band illustrates its $\pm0.4\,$dex uncertainties. The typical uncertainties associated with the host galaxy measurements are depicted by the error bar in the bottom right corner of the plot.}
    \label{Fig1: sample selection}
\end{figure}

\begin{deluxetable*}{ccccccc}
    \tablenum{1}
    \tablecaption{Galaxy Properties\label{Tab1: sample}}
    \tablewidth{0pt}
    \tablehead{
    \colhead{Name} & \colhead{R.A.} & \colhead{Decl.} & \colhead{$z$} & \colhead{$D_L$} & \colhead{$\log M_\star$} & \colhead{SFR}\\
    \colhead{} & \colhead{(J2000.0)} & \colhead{(J2000.0)} & \colhead{} & \colhead{(Mpc)} & \colhead{$\left(M_\odot\right)$} & \colhead{$\left(M_\odot\,\rm yr^{-1}\right)$}
    }
    \decimalcolnumbers
    \startdata
    PG\,0923$+$129 & 09:26:03.29 & $+$12:44:03.6 & 0.029 & 131.2 & 10.71 & 3.4\\
    PG\,1126$-$041 & 11:29:16.66 & $-$04:24:07.6 & 0.060 & 277.5 & 10.85 & 8.7\\
    PG\,2130$+$099 & 21:32:27.81 & $+$10:08:19.5 & 0.063 & 292.3 & 10.85 & 7.1\\
    IC\,0944       & 13:51:30.87 & $+$14:05:31.2 & 0.023 & 100.8 & 11.26 & 2.6\\
    IC\,1199       & 16:10:34.35 & $+$10:02:24.3 & 0.016 & 68.3  & 10.78 & 1.4\\
    IC\,1683       & 01:22:39.00 & $+$34:26:13.1 & 0.016 & 69.7  & 10.76 & 3.5\\
    NGC\,0496      & 01:23:11.60 & $+$33:31:44.0 & 0.020 & 87.5  & 10.85 & 4.6\\
    NGC\,2906      & 09:32:06.22 & $+$08:26:29.7 & 0.007 & 37.7  & 10.59 & 0.8\\
    NGC\,3994      & 11:57:36.87 & $+$32:16:38.2 & 0.010 & 44.7  & 10.59 & 3.7\\
    NGC\,4047      & 12:02:50.68 & $+$48:38:10.3 & 0.011 & 49.1  & 10.87 & 3.6\\
    NGC\,4644      & 12:42:42.66 & $+$55:08:43.7 & 0.016 & 71.6  & 10.68 & 1.2\\
    NGC\,4711      & 12:48:45.87 & $+$35:19:57.7 & 0.013 & 58.8  & 10.58 & 1.2\\
    NGC\,5480      & 14:06:21.58 & $+$50:43:30.3 & 0.006 & 27.0  & 10.18 & 1.4\\
    NGC\,5980      & 15:41:30.40 & $+$15:47:15.3 & 0.014 & 59.4  & 10.81 & 5.1\\
    NGC\,6060      & 16:05:51.98 & $+$21:29:05.6 & 0.015 & 63.2  & 10.99 & 4.2\\
    NGC\,6301      & 17:08:32.74 & $+$42:20:20.3 & 0.027 & 121.4 & 11.18 & 8.3\\
    NGC\,6478      & 17:48:38.35 & $+$51:09:25.9 & 0.023 & 97.4  & 11.27 & 10.0\\
    UGC\,09067     & 14:10:45.46 & $+$15:12:33.1 & 0.026 & 114.5 & 10.96 & 5.0
    \enddata
    \tablecomments{(1) Name of targets; (2) Right ascension; (3) Declination; (4) Redshift; (5) Luminosity distance; (6) Stellar mass derived from \citep{Shangguan+2018,Bolatto+2017}. In order to convert from \citet{Salpeter1955} to the Kroupa-like IMF, we divide the stellar mass by 1.5 following \cite{Kennicutt+2012}; (7) Star formation rate. The typical uncertainties associated with the host galaxy measurements are $\sim0.1\,$dex for stellar mass and $\sim0.3\,$dex for SFR.}
\end{deluxetable*}

\subsection{Quasar sample}
\label{subsec2.1: quasar sample}
We have leveraged archival Atacama Large Millimeter/submillimeter Array (ALMA) and Multi Unit Spectroscopic Explorer (MUSE) observations to conduct spatially-resolved mapping at approximately $1\farcs0$ scales for three quasars with redshifts $z\lesssim 0.06$. These galaxies were extracted from a larger sample of 87 quasars belonging to the Palomar-Green survey \citep{Boroson&Green1992}. The PG quasars were selected based on their optical/ultraviolet colors, making them a representative sample of luminous, broad-line (type 1) AGNs. This sample was selected irrespective of dust or gas content. It is also one of the most extensively studied quasar samples, with a wealth of multi-wavelength data available for both the AGN and the host galaxy. These data covers a wide range of wavelengths, including X-ray \citep{Reeves&Turner2000, Bianchi+2009}, optical \citep{Boroson&Green1992, Ho&Kim2009}, mid-IR \citep{Shi+2014, Xie+2021}, far-IR \citep{Petric+2015, Shangguan+2018}, mm-wave \citep{Shangguan+2020a, Shangguan+2020b}, and radio \citep{Kellermann+1989, Kellermann+1994} wavelengths. This comprehensive dataset allows us to perform detailed spectral energy distribution (SED) modeling and accurately estimate the global SFR and gas content of the host galaxies \citep{Shangguan+2018}. We have access to Hubble Space Telescope (HST) imaging data at a resolution of approximately $0\farcs1$ in the optical and near-IR for a substantial fraction of the sample \citep{Kim+2008, Zhang+2016, Kim+2017, Zhao+2021}. This combination of high-resolution imaging and multi-wavelength data provides a powerful tool for studying the properties and dynamics of these AGN host galaxies in detail.

We constructed the sample with several selection criteria. We focused on galaxies with extended molecular and ionized gas distributions, requiring at least 4 beams along the half-major axis. This criterion ensures that we have sufficient spatial coverage to study the distribution and kinematics of gas components. Additionally, we selected galaxies without bars or close companions, as the presence of such features could significantly perturb the gravitational potential and result in different gas dynamics compared to galaxies without bars. The above-mentioned selection criteria allow the reliable application of dynamical analysis. Both the final sample and the initial sample are depicted in Figure \ref{Fig1: sample selection}. It is worth noting that the quasar host galaxies and inactive galaxies selected here span a similar range of stellar mass and SFR, with the same morphology, making the inactive galaxy sample an ideal reference for comparison purposes.

The ALMA observations carried out under program 2018.1.0006.S (PI: F. Bauer), have been formally documented in \cite{Molina+2021}. These observations are characterized by a spatial resolution ranging from approximately $0\farcs4$ to $1\farcs4$ and were conducted in Band 6 to detect the $^{12}$CO$(J=2-1)$ transition [with a rest frequency of $\nu_{\rm rest}=230.58\,\rm GHz$; hereafter CO(2--1)] as well as the underlying continuum emission. The ALMA data were combined with previous Atacama Compact Array (ACA) observations \citep{Shangguan+2020a}, and reduced with Common Astronomy Software Application (CASA; \citealt{McMullin+2007}), with channel width of $\sim11\,\rm km\,s^{-1}$.

The MUSE observations of six PG quasar host galaxies were previously presented in \cite{Molina+2022}, with a mean spectral resolution of $R\approx 3000$ or full width at half maximum (FWHM) of approximately $\sim$2.65\,\AA. The procedure to subtract the AGN emission from the host galaxy was presented in \cite{Molina+2022}. The observed data cubes are decomposed into AGN component, stellar component, and ionized gas component by fitting the spectra of each spaxel using the stellar emission template \citep{Cappellari2017}, the AGN template (e.g., \citealt{Greene&Ho2005}), and the Gaussian profiles for emission lines (see more details in \citealt{Molina+2022}). The H$\alpha$ data cube is then used in dynamical analysis.

\subsection{Inactive star-forming galaxies}
\label{subsec2.2: comparison sample}
We selected a subset of nearby inactive galaxies from the Calar Alto Legacy Integral Field Area (CALIFA; \citealt{Sanchez+2012, Sanchez+2014, Sanchez+2016}) and the Extragalactic Database for Galaxy Evolution (EDGE) survey \citep{Bolatto+2017} to compare them with the host galaxies of quasars. This selection was based on matching the stellar mass and star-formation rate ranges to ensure similar properties between the two samples. We notice that our quasars have larger bulge-to-disk ratio compared to the control sample, respectively. However, we note that this selection is not sensitive to our results. The EDGE-CALIFA sample is composed of galaxies with redshifts in the range of $0.015<z<0.03$, making it a representative collection of star-forming galaxies in the local universe \citep{Sanchez+2012, Walcher+2014, Bolatto+2017}. We note that the EDGE-CALIFA sample has relatively lower redshifts compared to the quasar sample, while the spatial resolution of the CO data is similar to that of the quasar sample, which is $\sim 1-1.5\,$kpc.

The EDGE-CALIFA survey \citep{Bolatto+2017} measured the $^{12}$CO($J=1-0$) transition [with a rest frequency of $\nu_{\rm rest}=115.27\,\rm GHz$; hereafter referred to as CO(1--0)] in 126 nearby galaxies with CARMA in the D and E configurations. Full details of the survey, data reduction, and masking techniques were presented in \cite{Bolatto+2017}, and we present a brief overview here. The EDGE sample is the largest sample of galaxies with spatially resolved CO, with a typical angular resolution of $\sim 4''.5$ (corresponding to $\sim1.5\,$kpc at the mean distance of the sample) and a spectral resolution of $\sim 20\,\rm km\,s^{-1}$. The CALIFA\footnote{\url{https://califa.caha.es/}} survey \citep{Sanchez+2012} is an optical integral field units (IFU) survey of local galaxies, which is representative of the general galaxy population. CALIFA provided two sets of data cubes with different spectral resolutions \citep{Sanchez+2016}, and we adopted the low-resolution data in our analysis, covering the H$\alpha$ emission line, with a spectral resolution of $\sim 160\,\rm km\,s^{-1}$ and a spatial resolution of $\sim2''.5$. We note that the spectral resolution of CALIFA data is too poor to provide a reliable measurement of the velocity dispersion of the ionized gas. A more reliable velocity dispersion should be obtained from the CALIFA high-resolution data \citep{Levy+2018}. However, the velocity dispersion of the H$\alpha$ line is only used in calculating the asymmetric drift correction and estimating the circular velocities. In addition, the velocity dispersion of the ionized gas is comparable to the velocity dispersion of H$\gamma$ line \citep{Levy+2018}, and is much smaller compared to the rotation velocities ($V/\sigma \gtrsim$ 5). Therefore it contributes a little to the circular velocities, i.e., the asymmetric drift correction is limited.

To characterize the emission line properties of the star-forming galaxies, we followed the analysis approach outlined in \cite{Molina+2022} to mitigate potential systematic uncertainties arising from variations in stellar population synthesis models. In brief, we employed the penalized pixel-fitting method (\texttt{pPXF}; \citealt{Cappellari&Emsellem2004, Cappellari2017}) on Voronoi binned data cubes \citep{Cappellari&Copin2003} with a signal-to-noise ratio (SNR) of 50 (for two galaxies, NGC\,0496 and NGC\,6301, the SNR is set to be 20). For each spaxel, the spectrum was modeled as a combination of a scaled \texttt{pPXF} best-fit stellar continuum model corresponding to the Voronoi cell spectrum representing the stellar component and Gaussian functions representing the nebular emission lines. At this step, we consider the instrumental resolution by adding the line spread function (LSF) line width in quadrature within the line model. The basic properties of the quasar host galaxies and the inactive star-forming galaxies are summarized in Table \ref{Tab1: sample}.

\begin{deluxetable*}{cccccc}
    \tablenum{2}
    \tablecaption{Fitting Parameters of CO Intensity Map\label{Tab2: int fitting}}
    \tablewidth{0pt}
    \tablehead{
        \colhead{Name} & \colhead{$I_e$} & \colhead{$R_e$} & \colhead{$n$} & \colhead{$b/a$} & \colhead{$\phi_{\rm phot}$} \\
        \colhead{} & \colhead{$\left(\rm Jy\,beam^{-1}\,km\,s^{-1}\right)$} & \colhead{$\left(''\right)$} & \colhead{} & \colhead{} & \colhead{$\left(^\circ \right)$}
    }
        \decimalcolnumbers
        \startdata
        PG\,0923$+$129 & ${1.11}_{{-{0.01}}}^{{+{0.01}}}$ & ${1.49}_{{-{0.01}}}^{{+{0.01}}}$ & ${0.51}_{{-{0.01}}}^{{+{0.01}}}$ & ${0.67}_{{-{0.01}}}^{{+{0.01}}}$ & ${59.36}_{{-{0.47}}}^{{+{1.40}}}$ \\
        PG\,1126$-$041 & ${0.80}_{{-{0.02}}}^{{+{0.02}}}$ & ${3.66}_{{-{0.04}}}^{{+{0.04}}}$ & ${1.00}_{{-{0.03}}}^{{+{0.03}}}$ & ${0.21}_{{-{0.01}}}^{{+{0.01}}}$ & ${154.84}_{{-{0.15}}}^{{+{0.15}}}$ \\
        PG\,2130$+$099 & ${1.68}_{{-{0.06}}}^{{+{0.05}}}$ & ${1.50}_{{-{0.02}}}^{{+{0.02}}}$ & ${2.39}_{{-{0.07}}}^{{+{0.08}}}$ & ${0.57}_{{-{0.01}}}^{{+{0.01}}}$ & ${50.09}_{{-{1.65}}}^{{+{0.81}}}$ \\
        IC\,0944 & ${1.92}_{{-{0.03}}}^{{+{0.03}}}$ & ${9.60}_{{-{0.02}}}^{{+{0.01}}}$ & ${3.84}_{{-{0.06}}}^{{+{0.06}}}$ & ${0.18}_{{-{0.00}}}^{{+{0.00}}}$ & ${109.25}_{{-{0.15}}}^{{+{0.15}}}$ \\
        IC\,1199 & ${0.55}_{{-{0.02}}}^{{+{0.02}}}$ & ${6.57}_{{-{0.01}}}^{{+{0.01}}}$ & ${1.88}_{{-{0.05}}}^{{+{0.06}}}$ & ${0.57}_{{-{0.02}}}^{{+{0.02}}}$ & ${166.45}_{{-{0.93}}}^{{+{0.89}}}$ \\
        IC\,1683 & ${2.98}_{{-{0.23}}}^{{+{0.27}}}$ & ${2.39}_{{-{0.11}}}^{{+{0.10}}}$ & ${4.60}_{{-{0.22}}}^{{+{0.22}}}$ & ${0.73}_{{-{0.01}}}^{{+{0.01}}}$ & ${155.78}_{{-{1.01}}}^{{+{1.01}}}$ \\
        NGC\,0496 & ${0.70}_{{-{0.08}}}^{{+{0.08}}}$ & ${6.18}_{{-{0.43}}}^{{+{0.49}}}$ & ${1.77}_{{-{0.11}}}^{{+{0.12}}}$ & ${0.67}_{{-{0.01}}}^{{+{0.01}}}$ & ${38.06}_{{-{1.41}}}^{{+{1.43}}}$ \\
        NGC\,2906 & ${2.54}_{{-{0.03}}}^{{+{0.03}}}$ & ${3.00}_{{-{0.03}}}^{{+{0.02}}}$ & ${0.22}_{{-{0.01}}}^{{+{0.01}}}$ & ${0.41}_{{-{0.00}}}^{{+{0.00}}}$ & ${92.01}_{{-{0.33}}}^{{+{0.32}}}$ \\
        NGC\,3994 & ${9.83}_{{-{0.09}}}^{{+{0.09}}}$ & ${1.69}_{{-{0.01}}}^{{+{0.01}}}$ & ${0.22}_{{-{0.01}}}^{{+{0.01}}}$ & ${0.40}_{{-{0.00}}}^{{+{0.00}}}$ & ${3.42}_{{-{0.21}}}^{{+{0.21}}}$ \\
        NGC\,4047 & ${5.36}_{{-{0.04}}}^{{+{0.03}}}$ & ${2.63}_{{-{0.01}}}^{{+{0.01}}}$ & ${0.55}_{{-{0.01}}}^{{+{0.01}}}$ & ${0.81}_{{-{0.00}}}^{{+{0.00}}}$ & ${108.79}_{{-{0.58}}}^{{+{0.58}}}$ \\
        NGC\,4644 & ${1.61}_{{-{0.02}}}^{{+{0.02}}}$ & ${4.14}_{{-{0.05}}}^{{+{0.05}}}$ & ${0.10}_{{-{0.00}}}^{{+{0.00}}}$ & ${0.50}_{{-{0.01}}}^{{+{0.01}}}$ & ${36.23}_{{-{0.66}}}^{{+{0.67}}}$ \\
        NGC\,4711 & ${1.28}_{{-{0.03}}}^{{+{0.03}}}$ & ${5.66}_{{-{0.05}}}^{{+{0.02}}}$ & ${0.32}_{{-{0.02}}}^{{+{0.02}}}$ & ${0.38}_{{-{0.01}}}^{{+{0.01}}}$ & ${40.77}_{{-{0.82}}}^{{+{0.83}}}$ \\
        NGC\,5480 & ${1.00}_{{-{0.01}}}^{{+{0.01}}}$ & ${2.67}_{{-{0.00}}}^{{+{0.00}}}$ & ${1.41}_{{-{0.02}}}^{{+{0.02}}}$ & ${0.67}_{{-{0.01}}}^{{+{0.01}}}$ & ${56.02}_{{-{0.81}}}^{{+{0.78}}}$ \\
        NGC\,5980 & ${3.19}_{{-{0.02}}}^{{+{0.02}}}$ & ${5.71}_{{-{0.01}}}^{{+{0.00}}}$ & ${1.14}_{{-{0.01}}}^{{+{0.01}}}$ & ${0.24}_{{-{0.00}}}^{{+{0.00}}}$ & ${9.27}_{{-{0.10}}}^{{+{0.10}}}$ \\
        NGC\,6060 & ${1.42}_{{-{0.00}}}^{{+{0.00}}}$ & ${6.81}_{{-{0.00}}}^{{+{0.00}}}$ & ${1.00}_{{-{0.00}}}^{{+{0.00}}}$ & ${0.44}_{{-{0.00}}}^{{+{0.00}}}$ & ${102.00}_{{-{0.00}}}^{{+{0.00}}}$ \\
        NGC\,6301 & ${1.06}_{{-{0.01}}}^{{+{0.01}}}$ & ${11.36}_{{-{0.04}}}^{{+{0.02}}}$ & ${0.10}_{{-{0.00}}}^{{+{0.00}}}$ & ${0.52}_{{-{0.01}}}^{{+{0.01}}}$ & ${113.26}_{{-{0.70}}}^{{+{0.70}}}$ \\
        NGC\,6478 & ${5.08}_{{-{0.03}}}^{{+{0.03}}}$ & ${7.00}_{{-{0.02}}}^{{+{0.02}}}$ & ${0.39}_{{-{0.01}}}^{{+{0.01}}}$ & ${0.36}_{{-{0.00}}}^{{+{0.00}}}$ & ${28.21}_{{-{0.12}}}^{{+{0.11}}}$ \\
        UGC\,09067 & ${4.15}_{{-{0.09}}}^{{+{0.09}}}$ & ${5.20}_{{-{0.05}}}^{{+{0.05}}}$ & ${0.54}_{{-{0.02}}}^{{+{0.02}}}$ & ${0.37}_{{-{0.00}}}^{{+{0.00}}}$ & ${9.59}_{{-{0.30}}}^{{+{0.31}}}$ \\
        \enddata
    \tablecomments{(1) Name of target; (2) Intensity at effective radius; (3) Effective radius in the unit of arcsec; (4) The S{\'e}rsic index; (5) The minor-to-major axis ratio; (6) The position angle of the major axis. North$=0^\circ$, east=$90^\circ$}
\end{deluxetable*}

\section{Analysis}
\label{sec3: methods}

\subsection{Distribution of the Molecular Gas}
\label{subsec3.1: gas distribution}


\figsetgrpstart
\figsetgrpnum{2.1}
\figsetgrptitle{The CO intensity map, best-fit S\'ersic model, and residual of PG\,0923$+$129}
\figsetplot{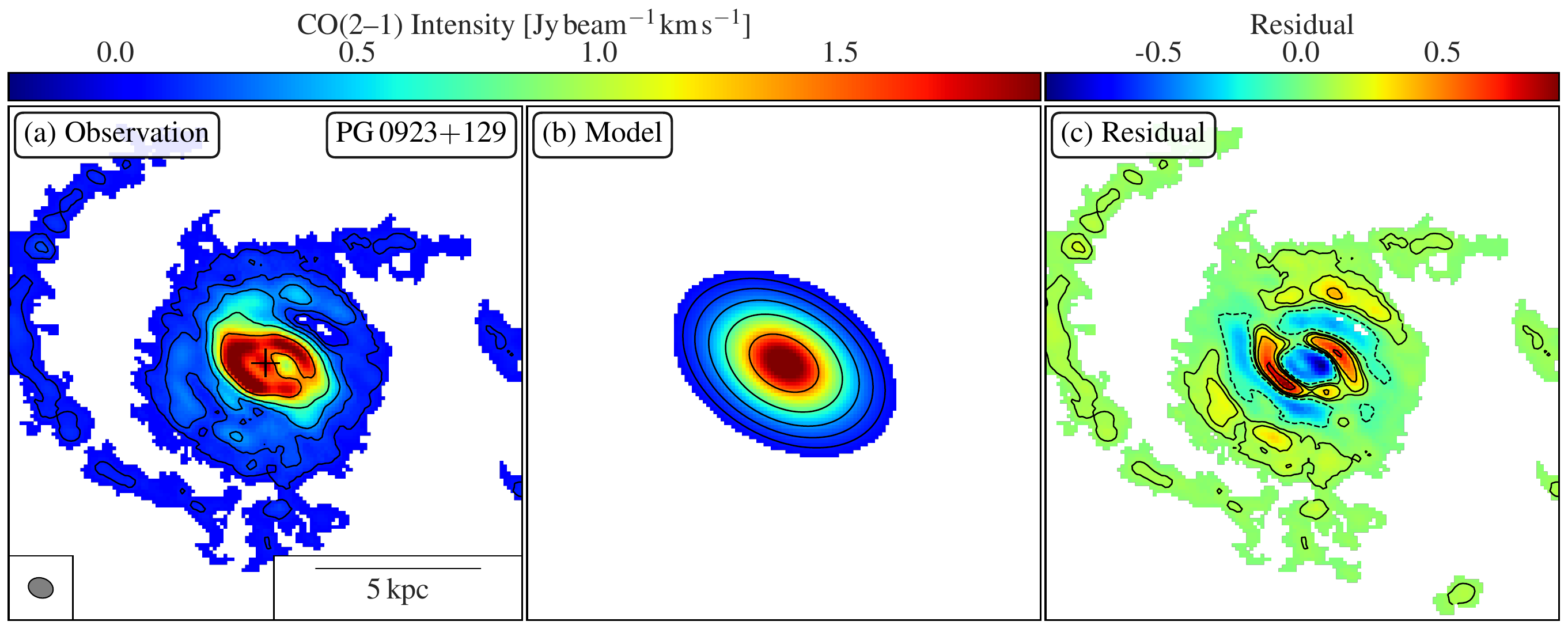}
\figsetgrpnote{The fitting result of a quasar host galaxy PG\,0923$+$129. Each column showcases the corresponding information for a specific galaxy.
The black contours overlaid on panels (a) and (b) depict the velocity-integrated intensity maps derived from the observed data
and the model, respectively. Panel (c) displays the residuals, highlighting the differences between the observed data and the
best-fit model. The North and East directions are indicated by the top and left orientations, respectively. The synthesized
beams are shown at the bottom-left corner of panel (a), and scale bars are included at the bottom-right corner.}
\figsetgrpend

\figsetgrpstart
\figsetgrpnum{2.2}
\figsetgrptitle{The CO intensity map, best-fit S\'ersic model, and residual of PG1126-041}
\figsetplot{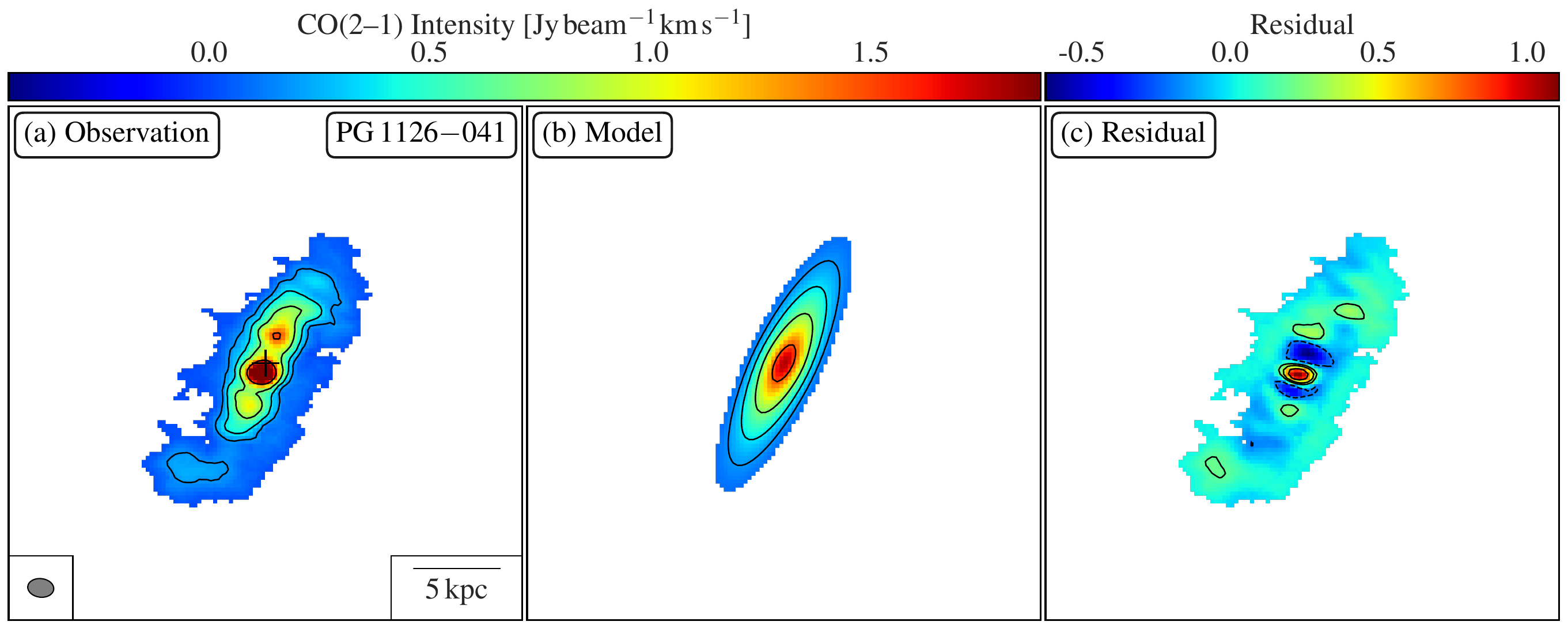}
\figsetgrpnote{The fitting result of a quasar host galaxy PG\,1126-041. Each column showcases the corresponding information for a specific galaxy.
The black contours overlaid on panels (a) and (b) depict the velocity-integrated intensity maps derived from the observed data
and the model, respectively. Panel (c) displays the residuals, highlighting the differences between the observed data and the
best-fit model. The North and East directions are indicated by the top and left orientations, respectively. The synthesized
beams are shown at the bottom-left corner of panel (a), and scale bars are included at the bottom-right corner.}
\figsetgrpend

\figsetgrpstart
\figsetgrpnum{2.3}
\figsetgrptitle{The CO intensity map, best-fit S\'ersic model, and residual of PG2130+099}
\figsetplot{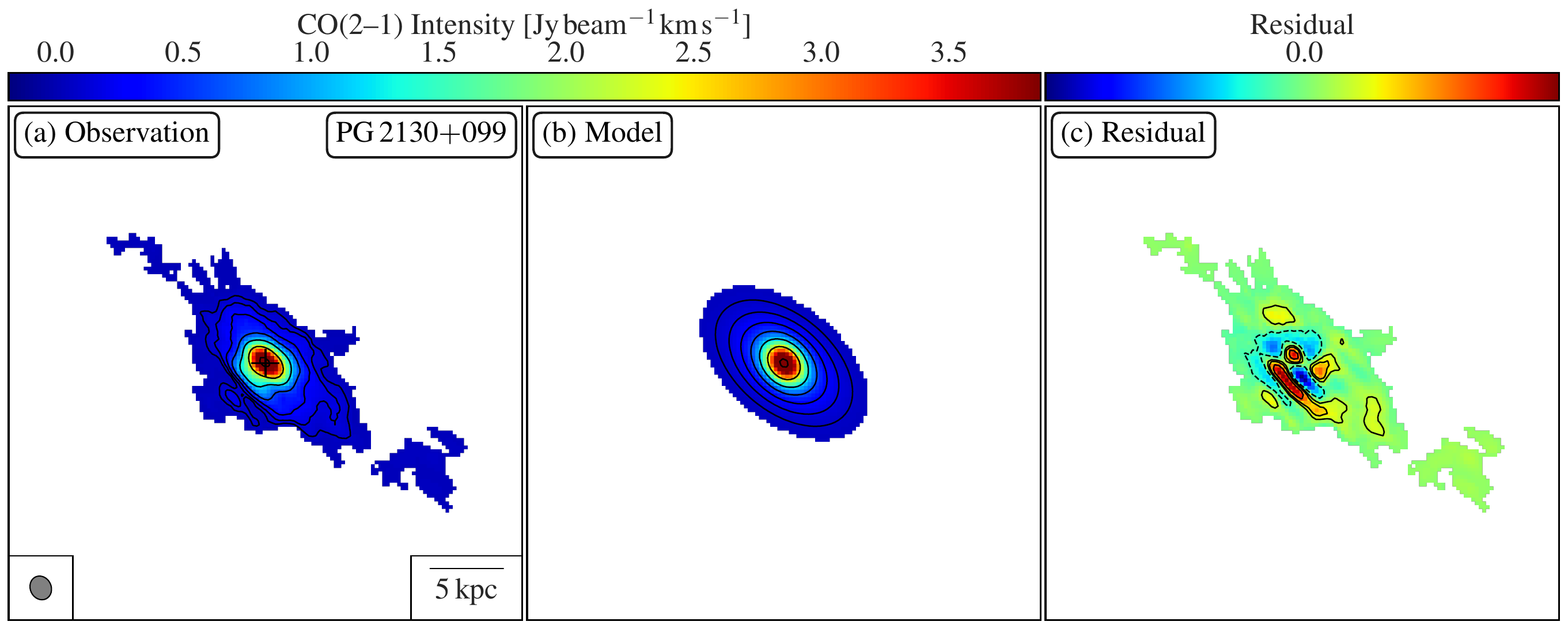}
\figsetgrpnote{The fitting result of a quasar host galaxy PG\,2130$+$099. Each column showcases the corresponding information for a specific galaxy.
The black contours overlaid on panels (a) and (b) depict the velocity-integrated intensity maps derived from the observed data
and the model, respectively. Panel (c) displays the residuals, highlighting the differences between the observed data and the
best-fit model. The North and East directions are indicated by the top and left orientations, respectively. The synthesized
beams are shown at the bottom-left corner of panel (a), and scale bars are included at the bottom-right corner.}
\figsetgrpend

\figsetgrpstart
\figsetgrpnum{2.4}
\figsetgrptitle{The CO intensity map, best-fit S\'ersic model, and residual of IC0944}
\figsetplot{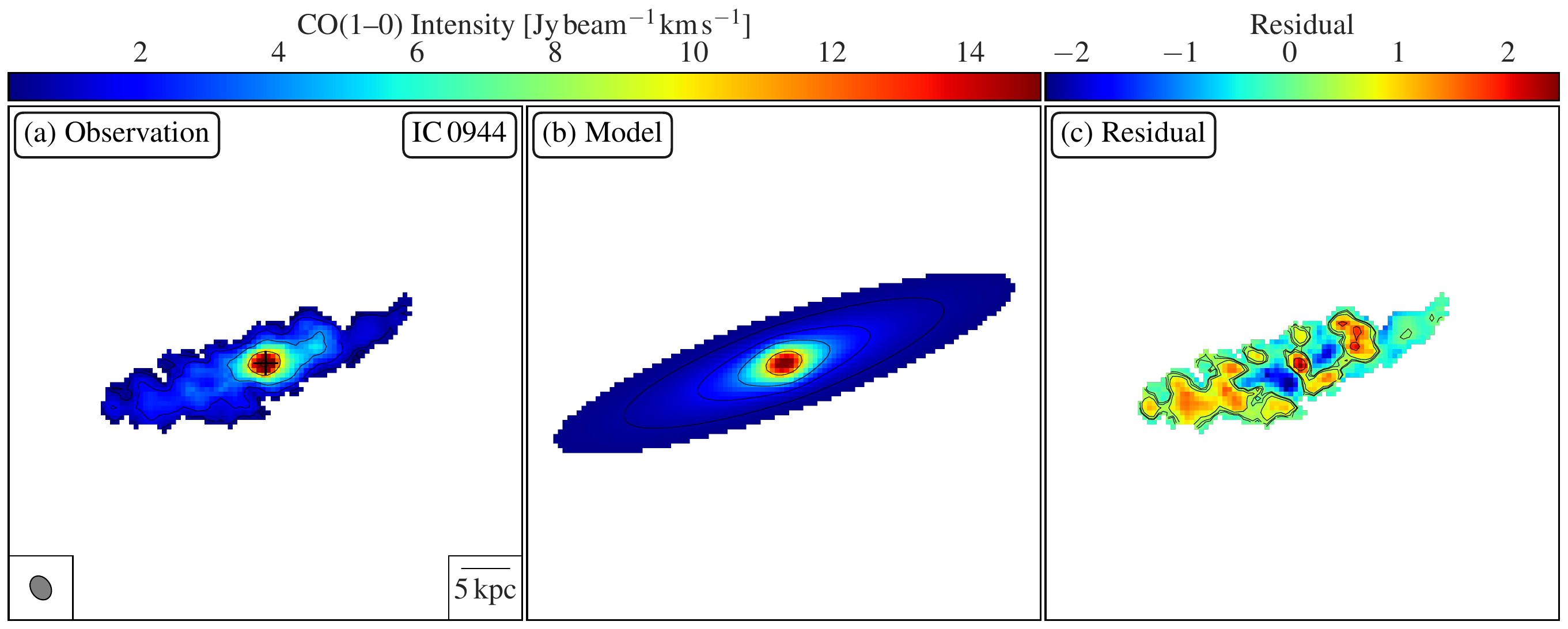}
\figsetgrpnote{The fitting result of an inactive star-forming galaxy IC0944. Each column showcases the corresponding information for a specific galaxy.
The black contours overlaid on panels (a) and (b) depict the velocity-integrated intensity maps derived from the observed data
and the model, respectively. Panel (c) displays the residuals, highlighting the differences between the observed data and the
best-fit model. The North and East directions are indicated by the top and left orientations, respectively. The synthesized
beams are shown at the bottom-left corner of panel (a), and scale bars are included at the bottom-right corner.}
\figsetgrpend

\figsetgrpstart
\figsetgrpnum{2.5}
\figsetgrptitle{The CO intensity map, best-fit S\'ersic model, and residual of IC1199}
\figsetplot{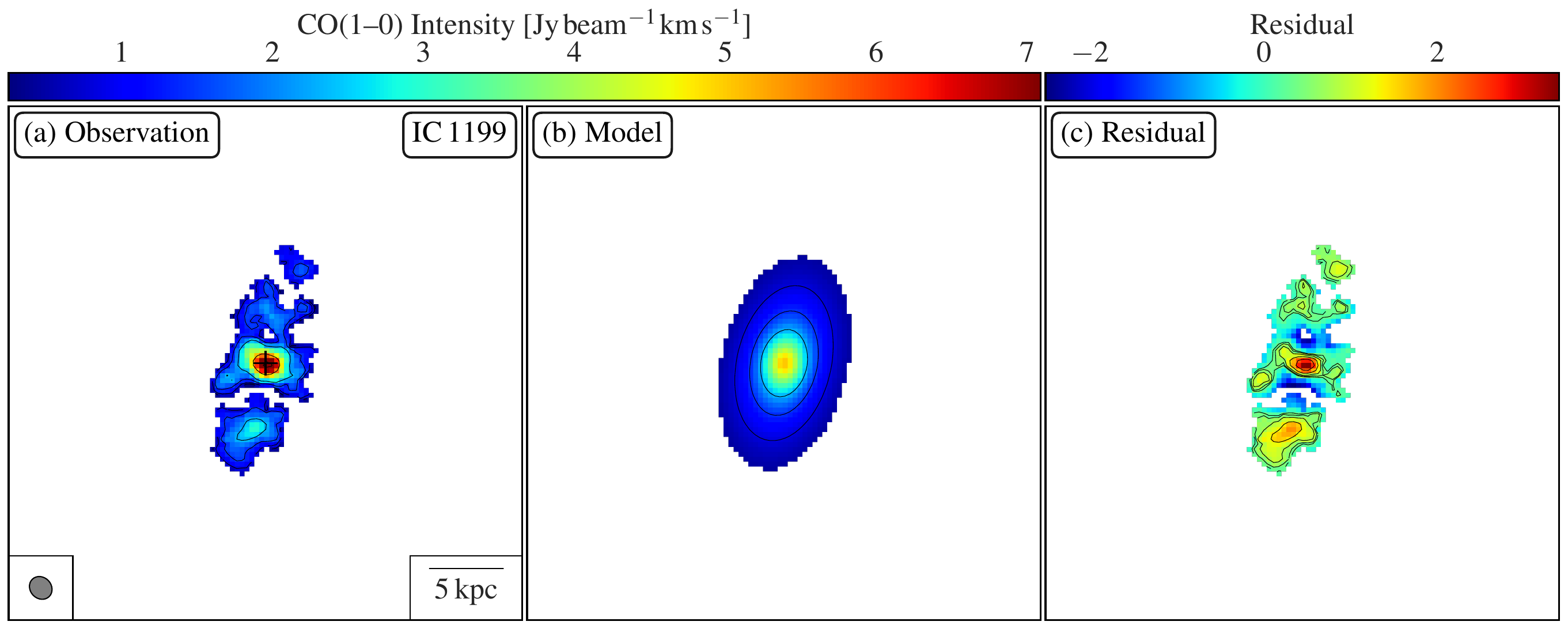}
\figsetgrpnote{The fitting result of an inactive star-forming galaxy IC1199. Each column showcases the corresponding information for a specific galaxy.
The black contours overlaid on panels (a) and (b) depict the velocity-integrated intensity maps derived from the observed data
and the model, respectively. Panel (c) displays the residuals, highlighting the differences between the observed data and the
best-fit model. The North and East directions are indicated by the top and left orientations, respectively. The synthesized
beams are shown at the bottom-left corner of panel (a), and scale bars are included at the bottom-right corner.}
\figsetgrpend

\figsetgrpstart
\figsetgrpnum{2.6}
\figsetgrptitle{The CO intensity map, best-fit S\'ersic model, and residual of IC1683}
\figsetplot{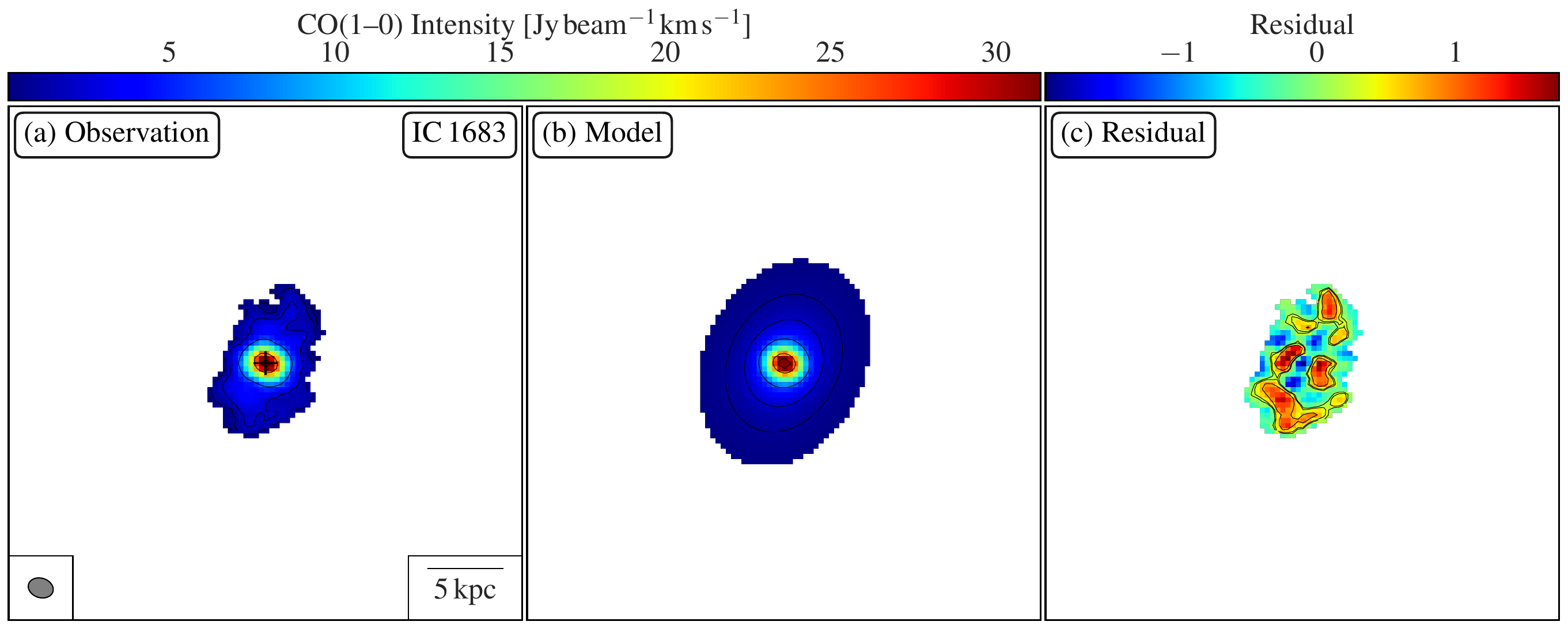}
\figsetgrpnote{The fitting result of an inactive star-forming galaxy IC1683. Each column showcases the corresponding information for a specific galaxy.
The black contours overlaid on panels (a) and (b) depict the velocity-integrated intensity maps derived from the observed data
and the model, respectively. Panel (c) displays the residuals, highlighting the differences between the observed data and the
best-fit model. The North and East directions are indicated by the top and left orientations, respectively. The synthesized
beams are shown at the bottom-left corner of panel (a), and scale bars are included at the bottom-right corner.}
\figsetgrpend

\figsetgrpstart
\figsetgrpnum{2.7}
\figsetgrptitle{The CO intensity map, best-fit S\'ersic model, and residual of NGC0496}
\figsetplot{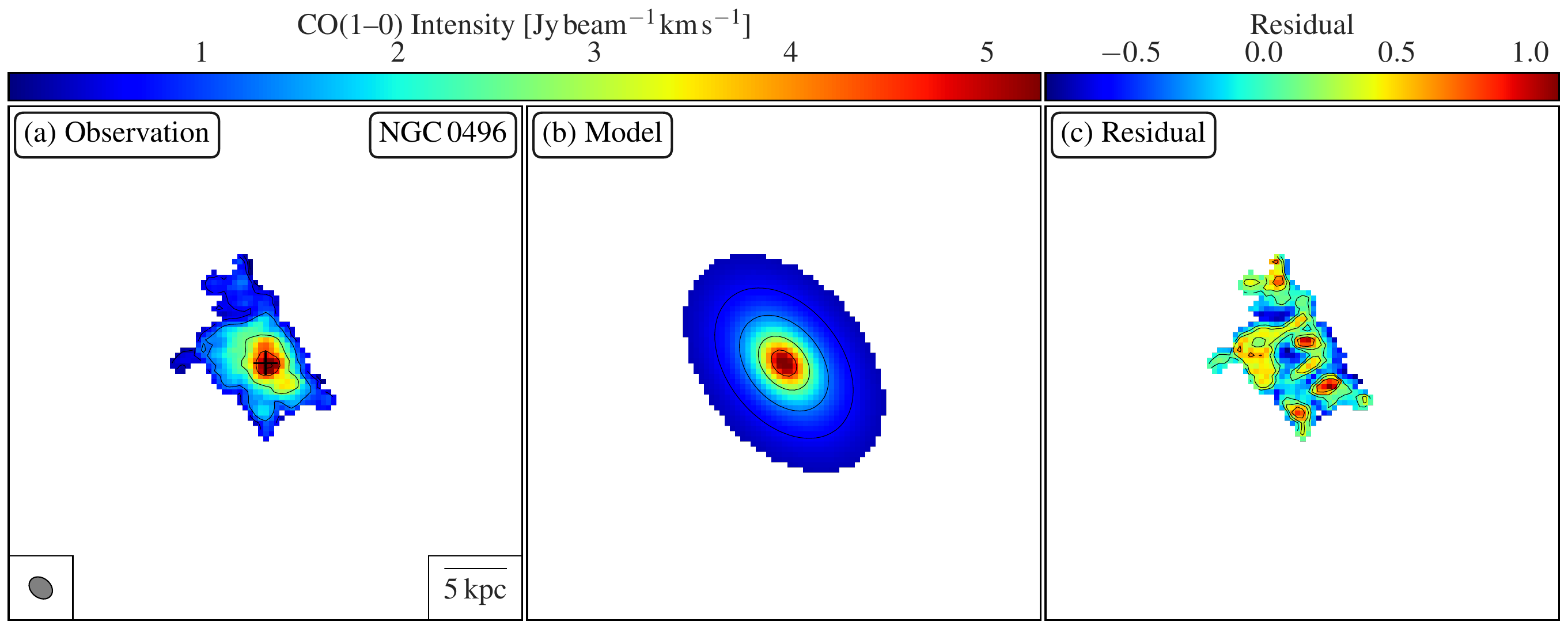}
\figsetgrpnote{The fitting result of an inactive star-forming galaxy NGC0496. Each column showcases the corresponding information for a specific galaxy.
The black contours overlaid on panels (a) and (b) depict the velocity-integrated intensity maps derived from the observed data
and the model, respectively. Panel (c) displays the residuals, highlighting the differences between the observed data and the
best-fit model. The North and East directions are indicated by the top and left orientations, respectively. The synthesized
beams are shown at the bottom-left corner of panel (a), and scale bars are included at the bottom-right corner.}
\figsetgrpend

\figsetgrpstart
\figsetgrpnum{2.8}
\figsetgrptitle{The CO intensity map, best-fit S\'ersic model, and residual of NGC2906}
\figsetplot{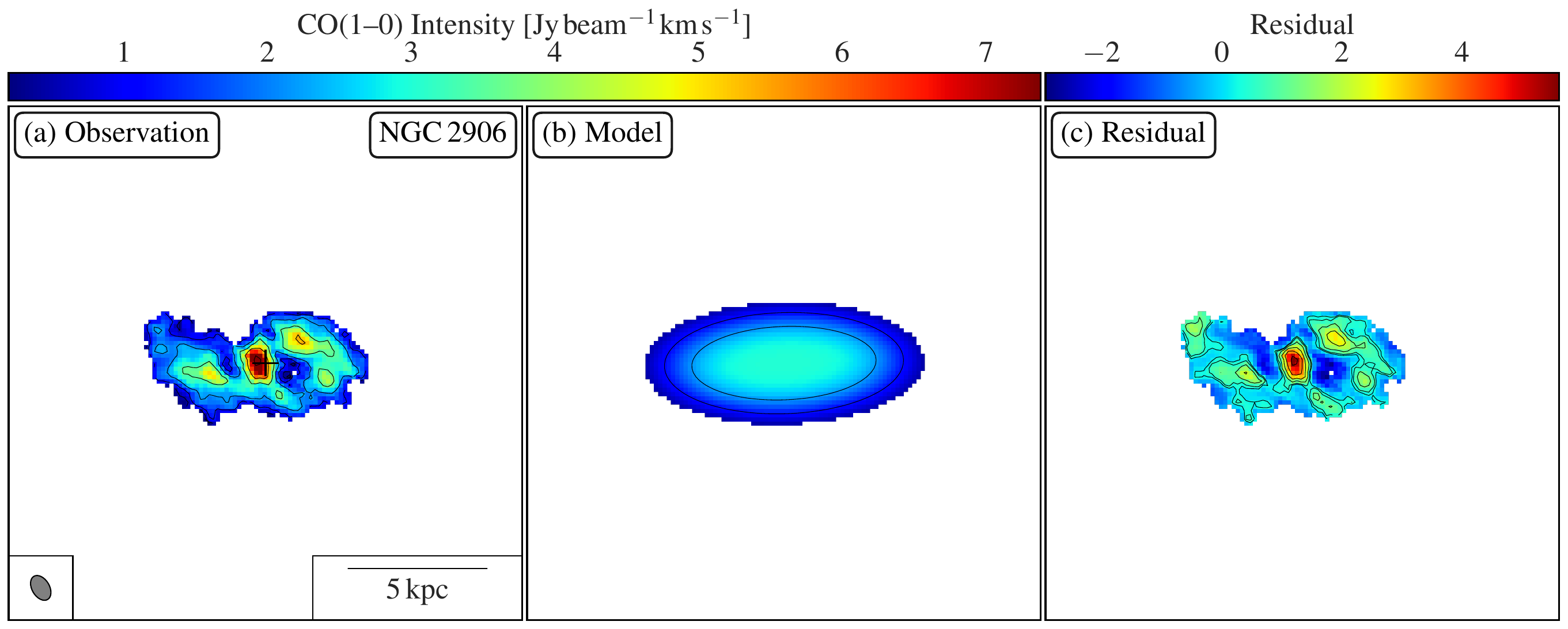}
\figsetgrpnote{The fitting result of an inactive star-forming galaxy NGC2906. Each column showcases the corresponding information for a specific galaxy.
The black contours overlaid on panels (a) and (b) depict the velocity-integrated intensity maps derived from the observed data
and the model, respectively. Panel (c) displays the residuals, highlighting the differences between the observed data and the
best-fit model. The North and East directions are indicated by the top and left orientations, respectively. The synthesized
beams are shown at the bottom-left corner of panel (a), and scale bars are included at the bottom-right corner.}
\figsetgrpend

\figsetgrpstart
\figsetgrpnum{2.9}
\figsetgrptitle{The CO intensity map, best-fit S\'ersic model, and residual of NGC3994}
\figsetplot{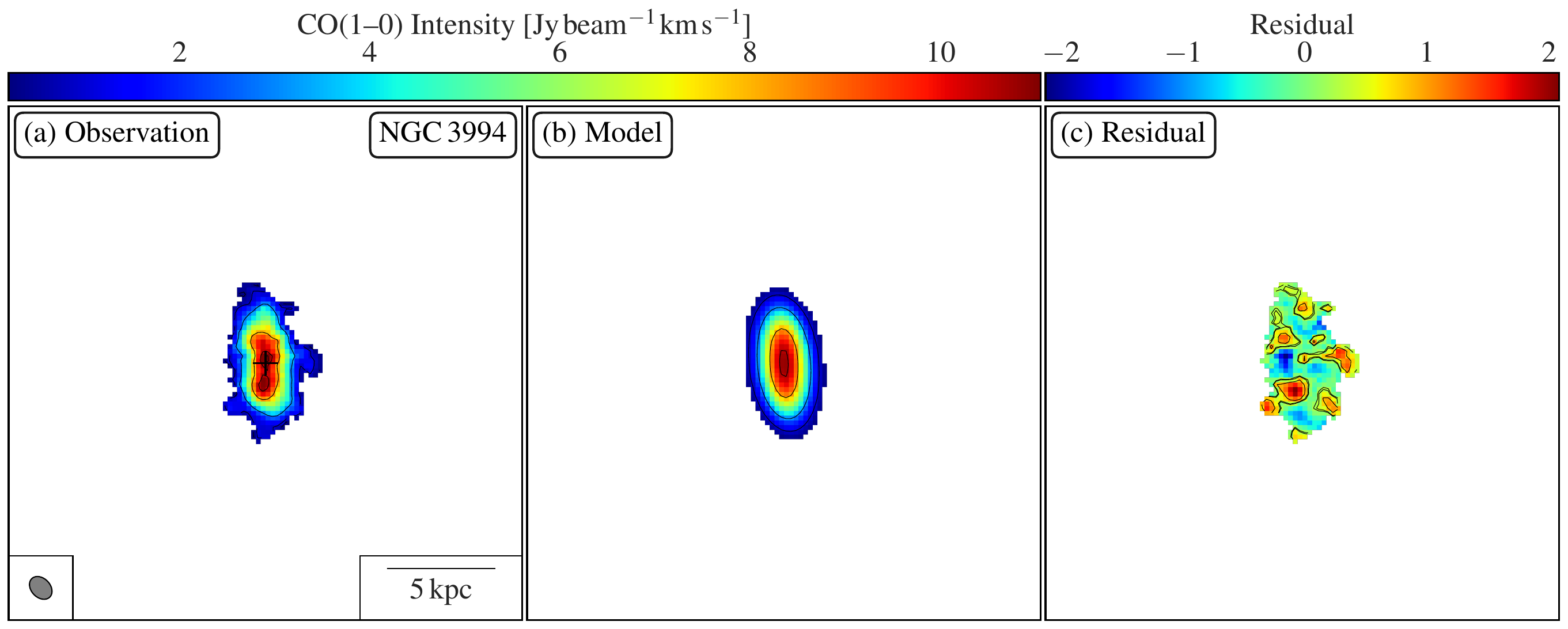}
\figsetgrpnote{The fitting result of an inactive star-forming galaxy NGC3994. Each column showcases the corresponding information for a specific galaxy.
The black contours overlaid on panels (a) and (b) depict the velocity-integrated intensity maps derived from the observed data
and the model, respectively. Panel (c) displays the residuals, highlighting the differences between the observed data and the
best-fit model. The North and East directions are indicated by the top and left orientations, respectively. The synthesized
beams are shown at the bottom-left corner of panel (a), and scale bars are included at the bottom-right corner.}
\figsetgrpend

\figsetgrpstart
\figsetgrpnum{2.10}
\figsetgrptitle{The CO intensity map, best-fit S\'ersic model, and residual of NGC4047}
\figsetplot{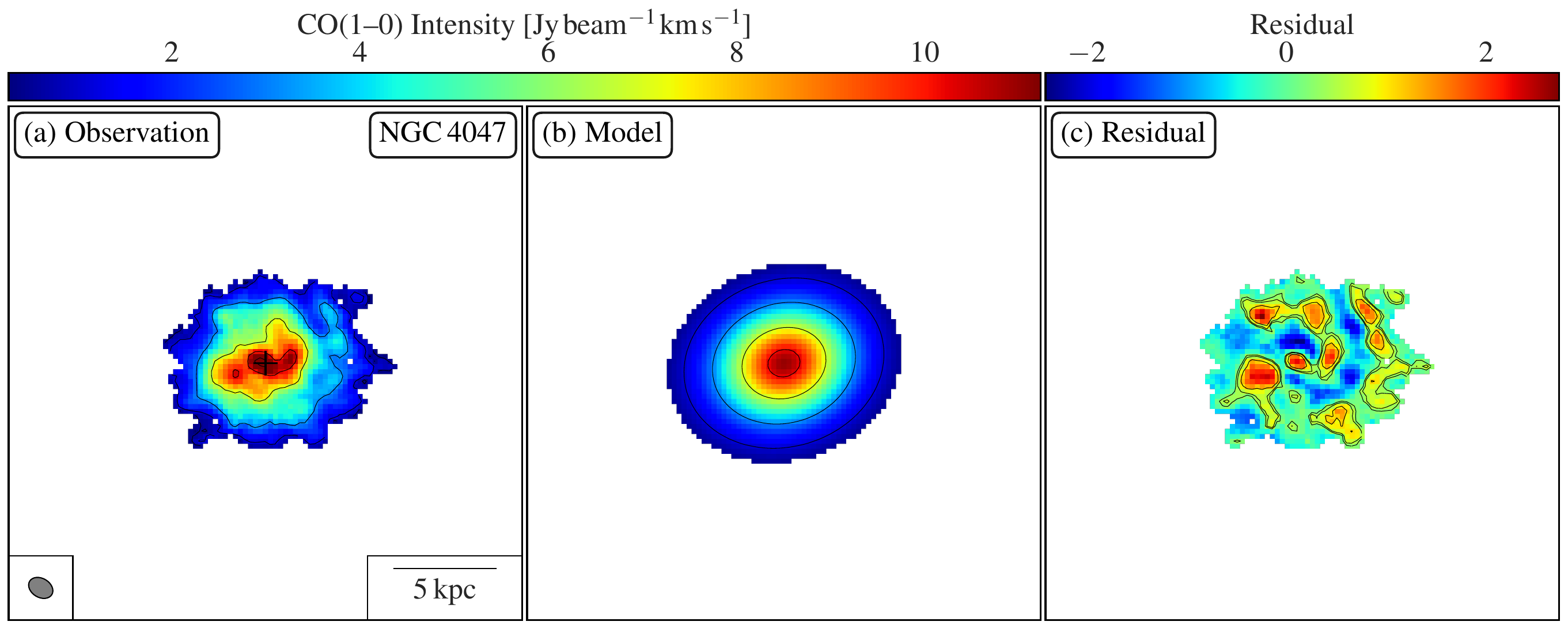}
\figsetgrpnote{The fitting result of an inactive star-forming galaxy NGC4047. Each column showcases the corresponding information for a specific galaxy.
The black contours overlaid on panels (a) and (b) depict the velocity-integrated intensity maps derived from the observed data
and the model, respectively. Panel (c) displays the residuals, highlighting the differences between the observed data and the
best-fit model. The North and East directions are indicated by the top and left orientations, respectively. The synthesized
beams are shown at the bottom-left corner of panel (a), and scale bars are included at the bottom-right corner.}
\figsetgrpend

\figsetgrpstart
\figsetgrpnum{2.11}
\figsetgrptitle{The CO intensity map, best-fit S\'ersic model, and residual of NGC4644}
\figsetplot{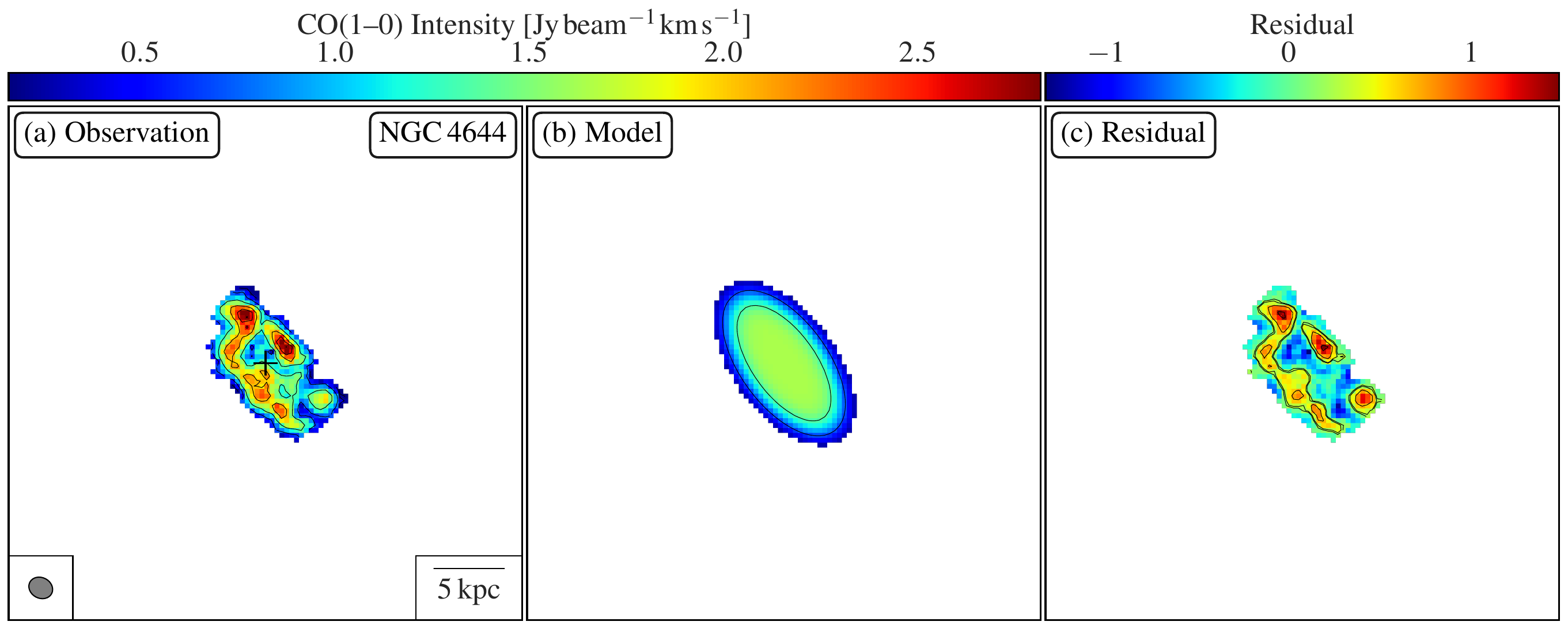}
\figsetgrpnote{The fitting result of an inactive star-forming galaxy NGC4644. Each column showcases the corresponding information for a specific galaxy.
The black contours overlaid on panels (a) and (b) depict the velocity-integrated intensity maps derived from the observed data
and the model, respectively. Panel (c) displays the residuals, highlighting the differences between the observed data and the
best-fit model. The North and East directions are indicated by the top and left orientations, respectively. The synthesized
beams are shown at the bottom-left corner of panel (a), and scale bars are included at the bottom-right corner.}
\figsetgrpend

\figsetgrpstart
\figsetgrpnum{2.12}
\figsetgrptitle{The CO intensity map, best-fit S\'ersic model, and residual of NGC4711}
\figsetplot{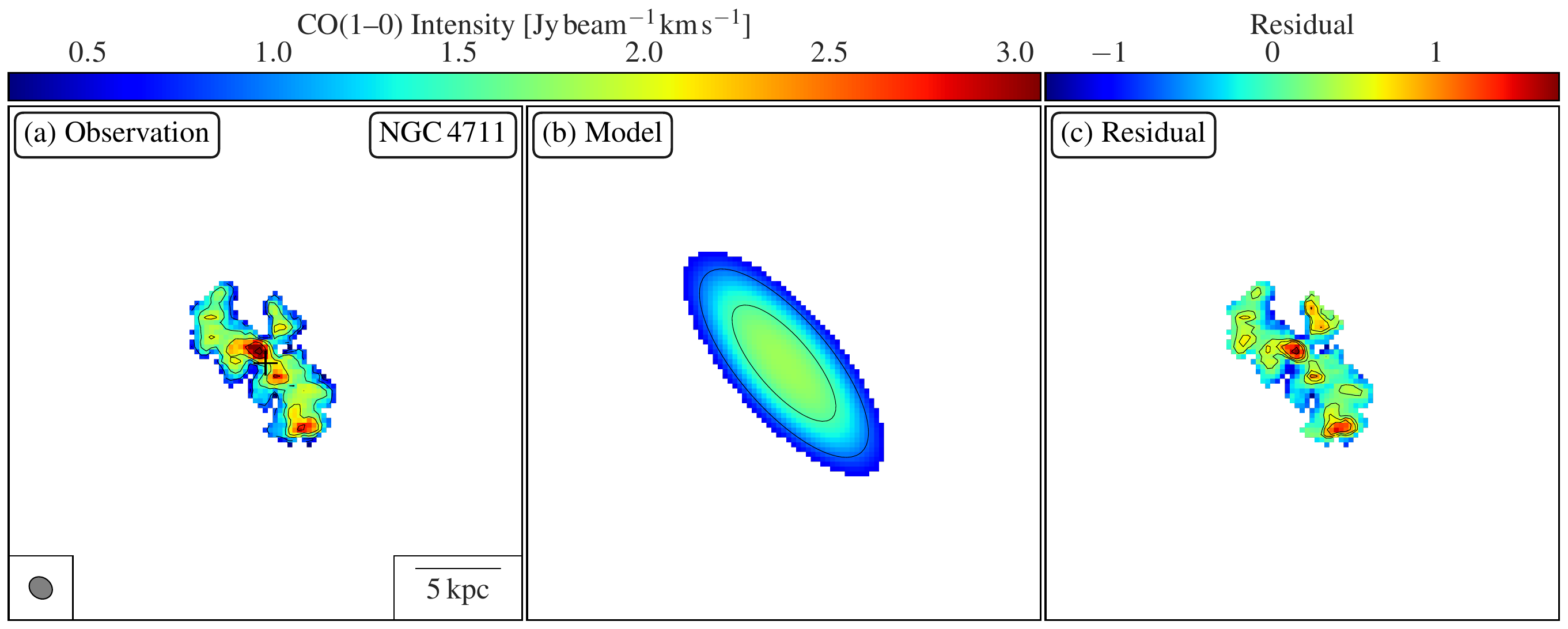}
\figsetgrpnote{The fitting result of an inactive star-forming galaxy NGC4711. Each column showcases the corresponding information for a specific galaxy.
The black contours overlaid on panels (a) and (b) depict the velocity-integrated intensity maps derived from the observed data
and the model, respectively. Panel (c) displays the residuals, highlighting the differences between the observed data and the
best-fit model. The North and East directions are indicated by the top and left orientations, respectively. The synthesized
beams are shown at the bottom-left corner of panel (a), and scale bars are included at the bottom-right corner.}
\figsetgrpend

\figsetgrpstart
\figsetgrpnum{2.13}
\figsetgrptitle{The CO intensity map, best-fit S\'ersic model, and residual of NGC5480}
\figsetplot{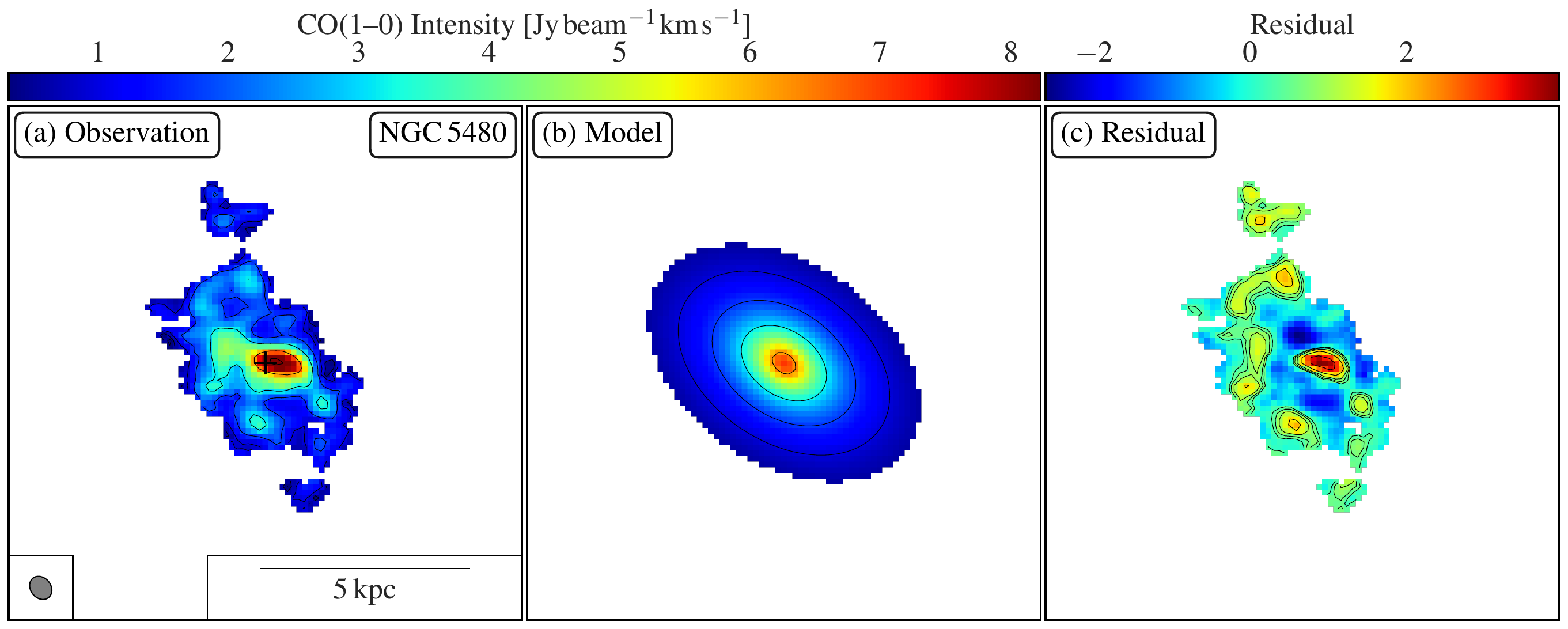}
\figsetgrpnote{The fitting result of an inactive star-forming galaxy NGC5480. Each column showcases the corresponding information for a specific galaxy.
The black contours overlaid on panels (a) and (b) depict the velocity-integrated intensity maps derived from the observed data
and the model, respectively. Panel (c) displays the residuals, highlighting the differences between the observed data and the
best-fit model. The North and East directions are indicated by the top and left orientations, respectively. The synthesized
beams are shown at the bottom-left corner of panel (a), and scale bars are included at the bottom-right corner.}
\figsetgrpend

\figsetgrpstart
\figsetgrpnum{2.14}
\figsetgrptitle{The CO intensity map, best-fit S\'ersic model, and residual of NGC5980}
\figsetplot{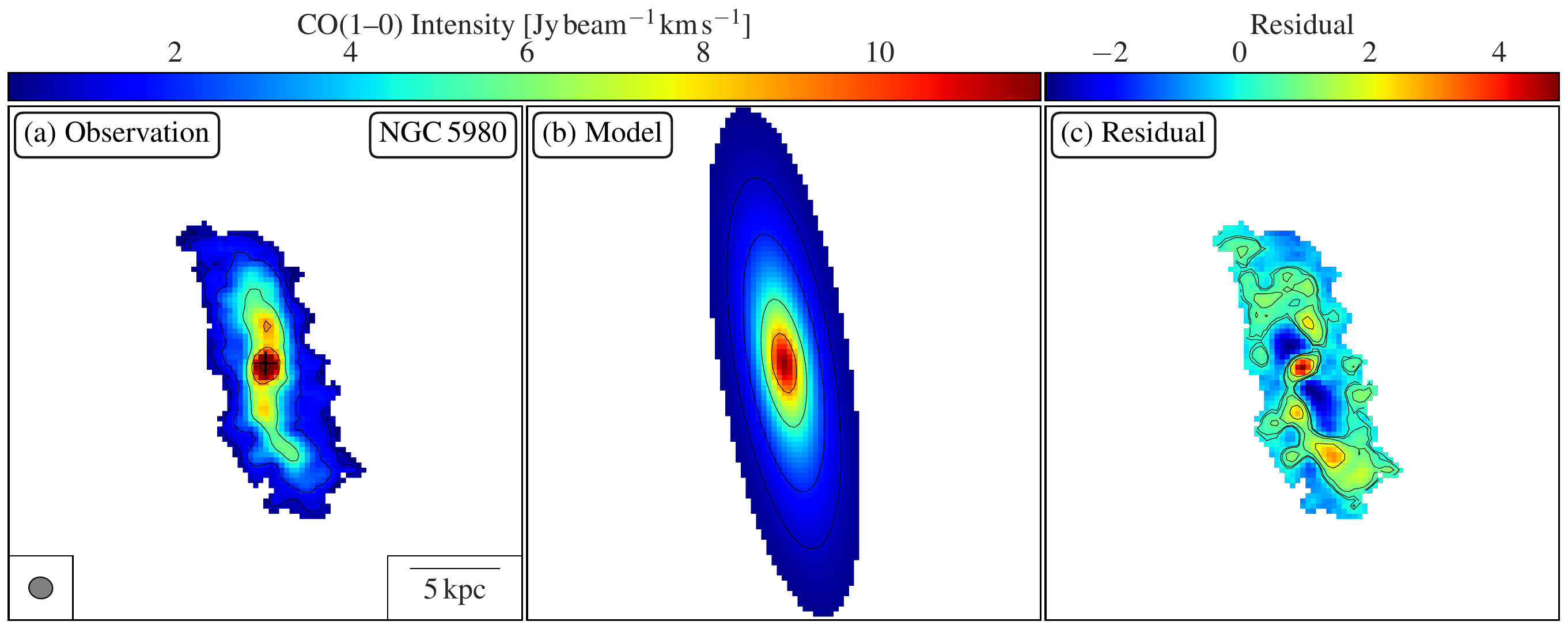}
\figsetgrpnote{The fitting result of an inactive star-forming galaxy NGC5980. Each column showcases the corresponding information for a specific galaxy.
The black contours overlaid on panels (a) and (b) depict the velocity-integrated intensity maps derived from the observed data
and the model, respectively. Panel (c) displays the residuals, highlighting the differences between the observed data and the
best-fit model. The North and East directions are indicated by the top and left orientations, respectively. The synthesized
beams are shown at the bottom-left corner of panel (a), and scale bars are included at the bottom-right corner.}
\figsetgrpend

\figsetgrpstart
\figsetgrpnum{2.15}
\figsetgrptitle{The CO intensity map, best-fit S\'ersic model, and residual of NGC6060}
\figsetplot{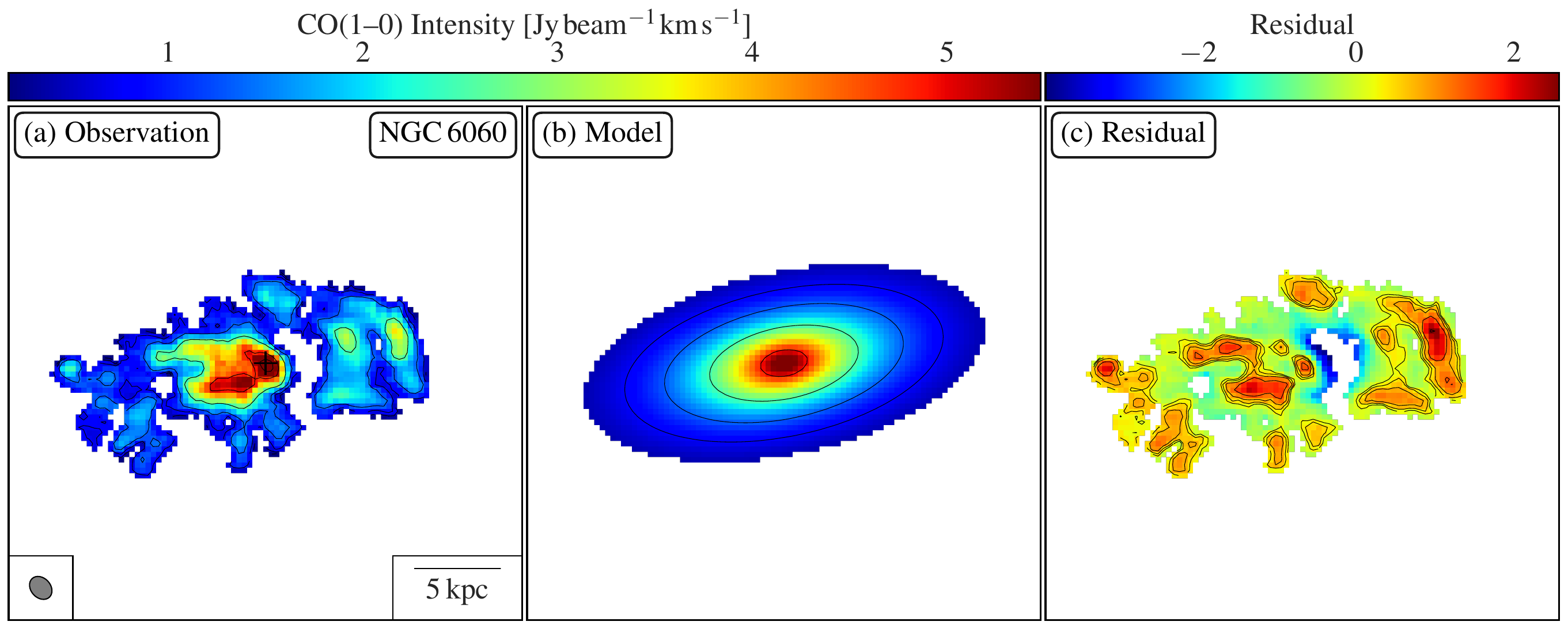}
\figsetgrpnote{The fitting result of an inactive star-forming galaxy NGC6060. Each column showcases the corresponding information for a specific galaxy.
The black contours overlaid on panels (a) and (b) depict the velocity-integrated intensity maps derived from the observed data
and the model, respectively. Panel (c) displays the residuals, highlighting the differences between the observed data and the
best-fit model. The North and East directions are indicated by the top and left orientations, respectively. The synthesized
beams are shown at the bottom-left corner of panel (a), and scale bars are included at the bottom-right corner.}
\figsetgrpend

\figsetgrpstart
\figsetgrpnum{2.16}
\figsetgrptitle{The CO intensity map, best-fit S\'ersic model, and residual of NGC6301}
\figsetplot{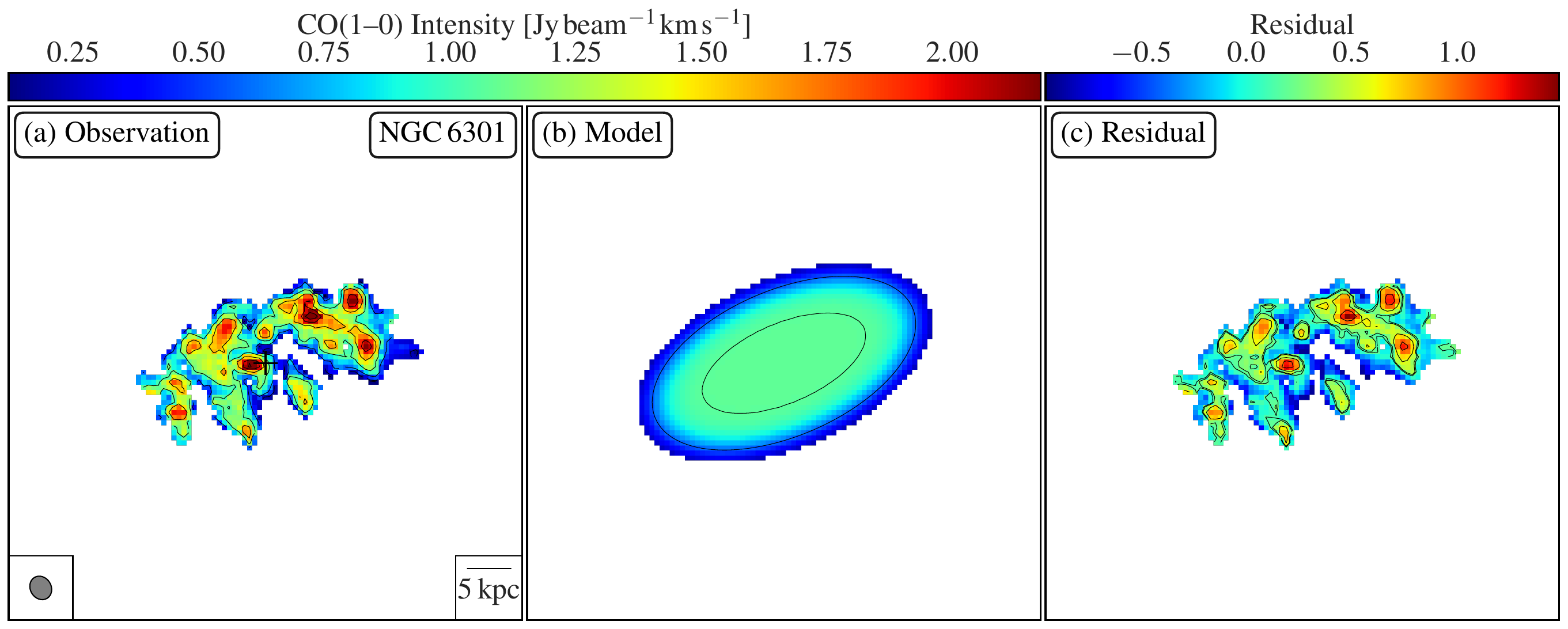}
\figsetgrpnote{The fitting result of an inactive star-forming galaxy NGC6301. Each column showcases the corresponding information for a specific galaxy.
The black contours overlaid on panels (a) and (b) depict the velocity-integrated intensity maps derived from the observed data
and the model, respectively. Panel (c) displays the residuals, highlighting the differences between the observed data and the
best-fit model. The North and East directions are indicated by the top and left orientations, respectively. The synthesized
beams are shown at the bottom-left corner of panel (a), and scale bars are included at the bottom-right corner.}
\figsetgrpend

\figsetgrpstart
\figsetgrpnum{2.17}
\figsetgrptitle{The CO intensity map, best-fit S\'ersic model, and residual of NGC6478}
\figsetplot{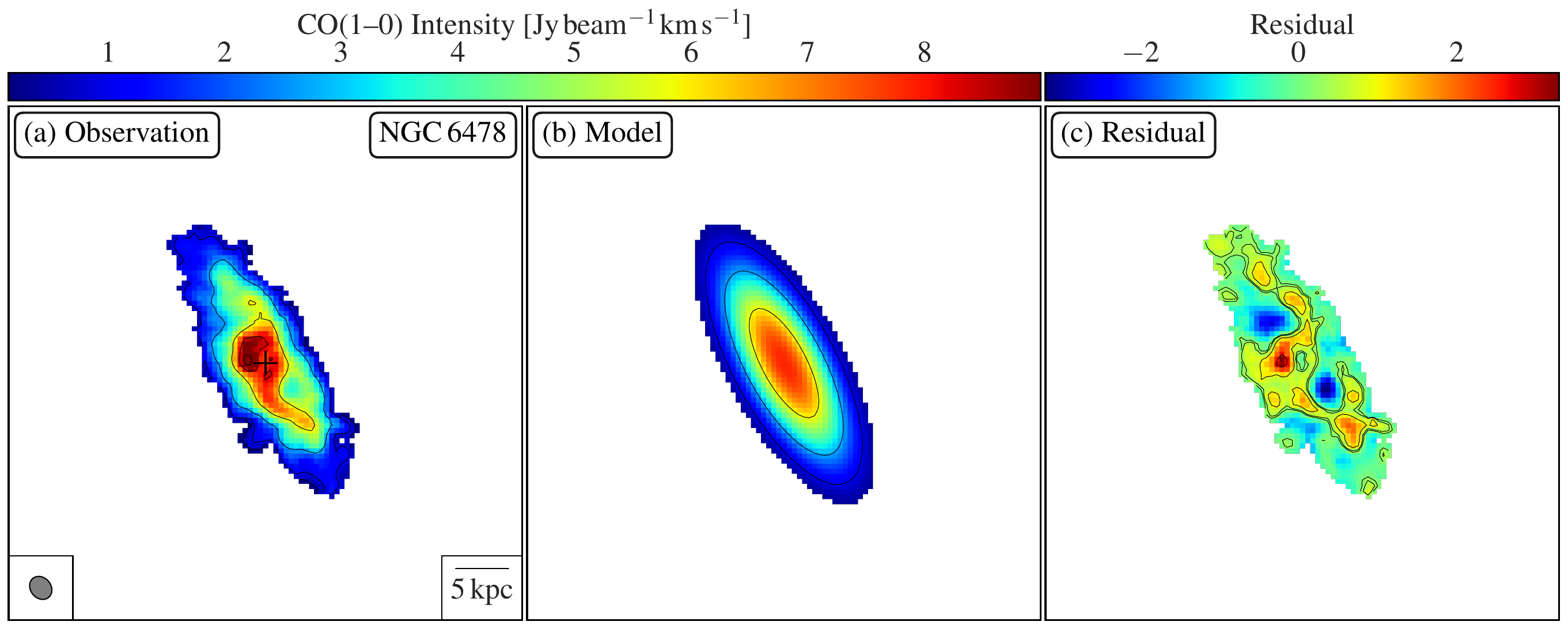}
\figsetgrpnote{The fitting result of an inactive star-forming galaxy NGC6478. Each column showcases the corresponding information for a specific galaxy.
The black contours overlaid on panels (a) and (b) depict the velocity-integrated intensity maps derived from the observed data
and the model, respectively. Panel (c) displays the residuals, highlighting the differences between the observed data and the
best-fit model. The North and East directions are indicated by the top and left orientations, respectively. The synthesized
beams are shown at the bottom-left corner of panel (a), and scale bars are included at the bottom-right corner.}
\figsetgrpend

\figsetgrpstart
\figsetgrpnum{2.18}
\figsetgrptitle{The CO intensity map, best-fit S\'ersic model, and residual of UGC09067}
\figsetplot{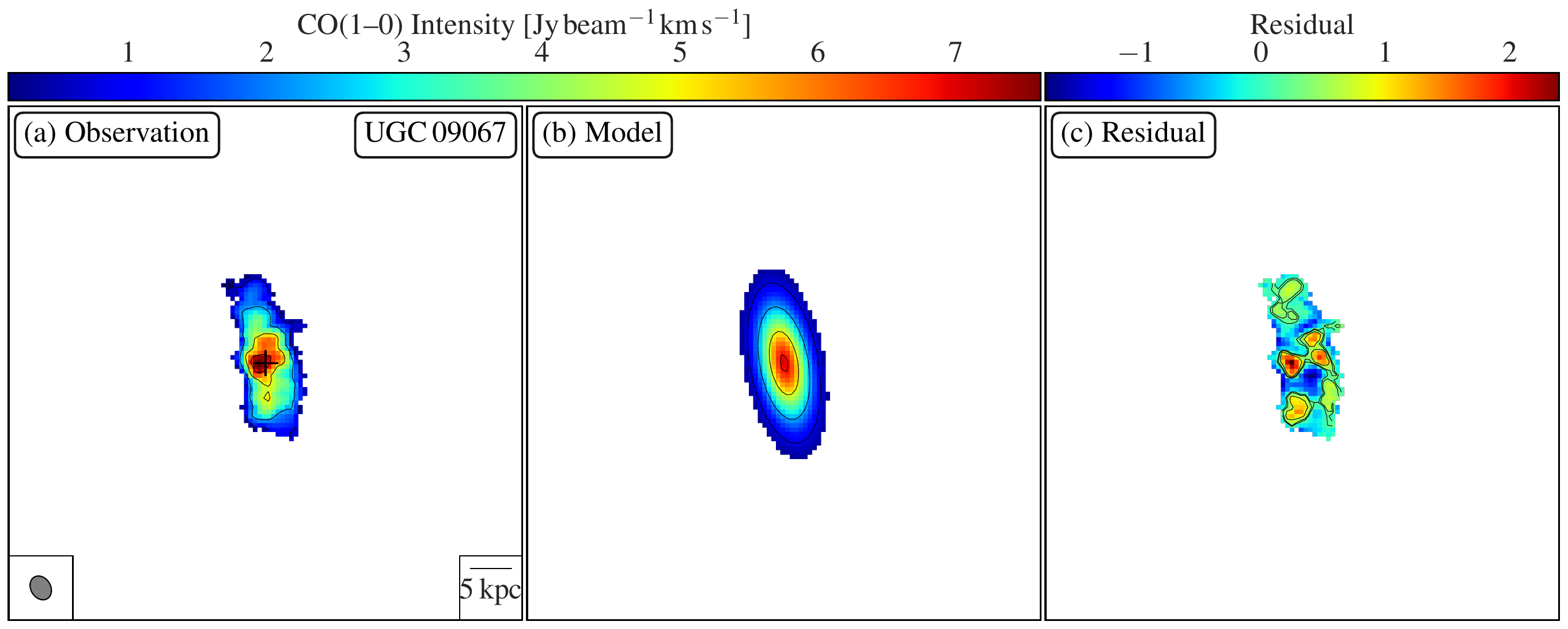}
\figsetgrpnote{The fitting result of an inactive star-forming galaxy UGC09067. Each column showcases the corresponding information for a specific galaxy.
The black contours overlaid on panels (a) and (b) depict the velocity-integrated intensity maps derived from the observed data
and the model, respectively. Panel (c) displays the residuals, highlighting the differences between the observed data and the
best-fit model. The North and East directions are indicated by the top and left orientations, respectively. The synthesized
beams are shown at the bottom-left corner of panel (a), and scale bars are included at the bottom-right corner.}
\figsetgrpend

\begin{figure*}
    \centering
    \includegraphics[width=0.72\linewidth]{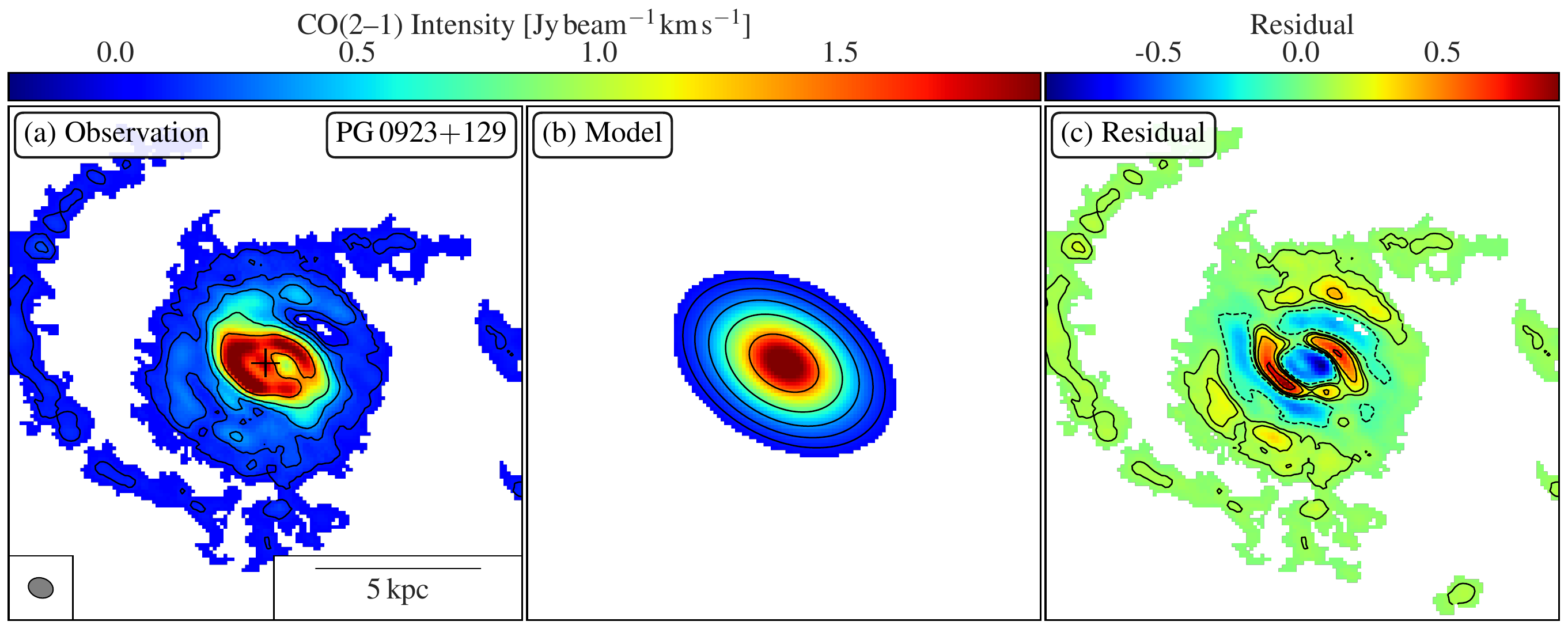}
    \includegraphics[width=0.72\linewidth]{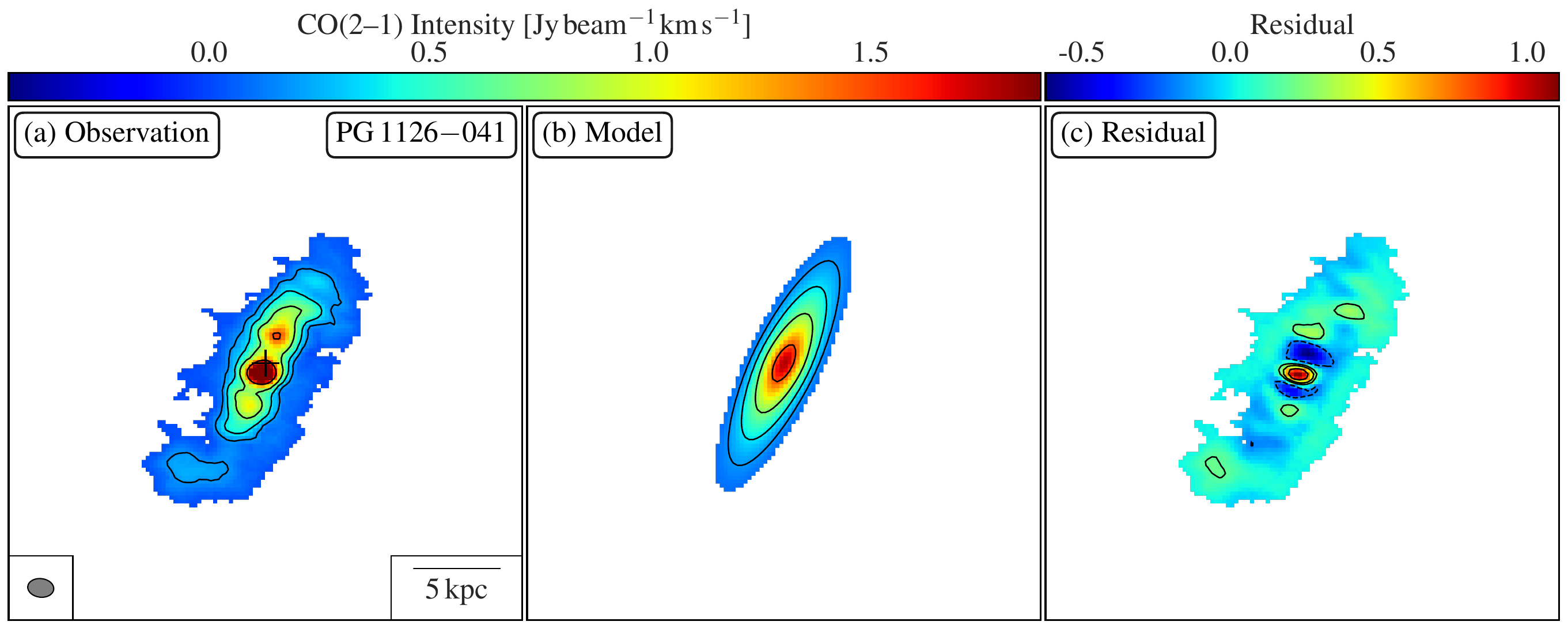}
    \includegraphics[width=0.72\linewidth]{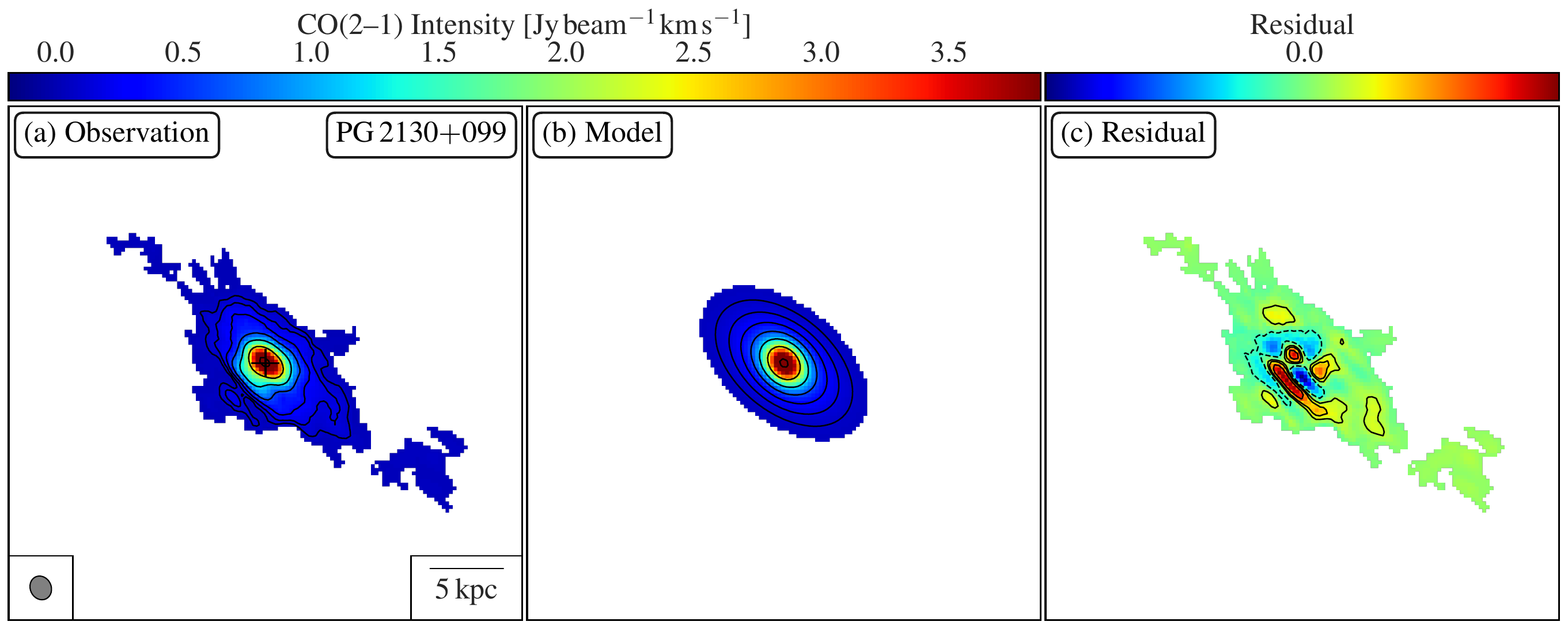}
    \includegraphics[width=0.72\linewidth]{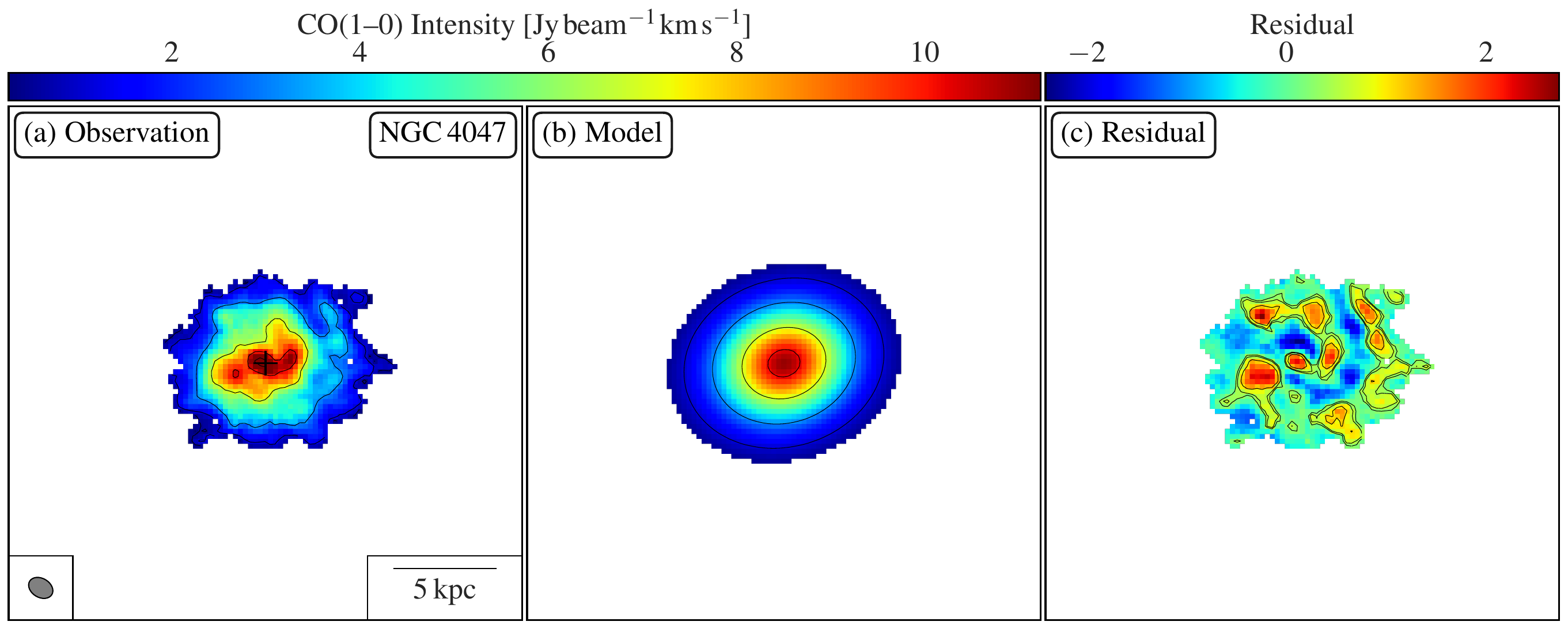}
    \caption{The velocity-integrated intensity maps, best-fit results, and residual maps are presented for quasar host galaxies and one of the inactive star-forming galaxies, NGC\,4047. Each column showcases the corresponding information for a specific galaxy. The black contours overlaid on {\it panels (a)} and {\it (b)} depict the velocity-integrated intensity maps derived from the observed data and the model, respectively. {\it Panel (c)} displays the residuals, highlighting the differences between the observed data and the best-fit model. The North and East directions are indicated by the top and left orientations, respectively. The synthesized beams are shown at the bottom-left corner of {\it panel (a)}, and scale bars are included at the bottom-right corner.}
    \label{Fig2: intensity maps}
\end{figure*}

To construct the intrinsic distribution of molecular gas in our sample, we followed the same analysis procedure described in \cite{Fei+2023}. We derived the CO(2–1) intensity maps for quasar host galaxies and CO(1–0) for inactive galaxies using a blanking mask generated with the Python package \texttt{maskmoment}. The mask was initiated at $\geq$5$\sigma$ peaks in the data cube and expanded to include the surrounding 2$\sigma$ contour. These "dilated-mask" moment maps, as described by \cite{Bolatto+2017}, effectively capture low-level signals while reducing noise.

To quantitatively describe the gas distribution, We modeled the CO(2–1) intensity maps for quasar hosts and CO(1–0) for inactive galaxies using a radial profile described by a \cite{Sersic1963} function, which defines surface brightness $I(R)$ based on parameters like the effective radius $R_e$, S{'e}rsic index $n$, and a coefficient $b_n$ that ensures $R_e$ corresponds to the half-light radius. The model was constructed using \texttt{Astropy} and convolved with the synthesized beam to produce the line-intensity map. This two-dimensional model includes six free parameters: $I_e$, $n$, $R_e$, the photometric position angle $\phi_{\rm phot}$, the minor-to-major axis ratio $b/a$, and the on-sky center location $(x_0,,y_0)$.

To find the best-fitting model, we used the Python package \texttt{emcee}, which employs the affine-invariant ensemble sampler for Markov Chain Monte Carlo (MCMC) sampling. The log-likelihood function was optimized, comparing the observed surface brightness at each pixel $z_i$ (with associated noise $\sigma_i$) to the model value $z_i^m$. We present the line-intensity map, best-fit model, and residual image for the three quasar host galaxies and an example inactive galaxy (NGC\,4047) in Figure \ref{Fig2: intensity maps}. The best-fit parameters for all objects are shown in Table \ref{Tab2: int fitting}.

We note for both quasar hosts and star-forming samples, the S\'ersic model hardly fully describes the CO intensity map, which is likely due to the clumpy- or ring-like structure of the molecular gas, e.g., PG\,0923+129, NGC\,4644 (see the fitting parameters in Table \ref{Tab2: int fitting} and the results in Figure \ref{Fig2: intensity maps}). However, we note that our dynamical model, in particular, the derived stellar mass, is not sensitive to this issue. In these cases, the gravitational potential is dominated by the other components rather than the gas, and the gas dynamics are not significantly affected by the gas distribution. Therefore, we do not exclude these galaxies from our analysis.

\subsection{Gas kinematics}
\label{subsec3.2: gas kinematics}
\begin{deluxetable*}{ccccc}
    \tablenum{3}
    \tablecaption{Kinematics information about galaxies\label{Tab3: kinematics}}
    \tablewidth{0pt}
    \tablehead{
        \colhead{Name} & \colhead{$\phi_{\rm mol}$} & \colhead{$\phi_{\rm ion}$} & \colhead{$i_{\rm mol}$} & \colhead{$i_{\rm ion}$} \\
        \colhead{} & \colhead{($^\circ$)} & \colhead{($^\circ$)} & \colhead{($^\circ$)} & \colhead{($^\circ$)} 
    }
    \decimalcolnumbers
    \startdata
    PG\,0923$+$129 & $259.56 \pm 3.92$ & $260.73 \pm 2.01$ & $42.44 \pm 0.71$ & $40.86 \pm 3.56$ \\
    PG\,1126$-$041 & $142.21 \pm 6.78$ & $145.93 \pm 3.60$ & $75.64 \pm 2.10$ & $67.02 \pm 0.81$ \\
    PG\,2130$+$099 & $229.21 \pm 8.47$ & $234.60 \pm 5.43$ & $68.52 \pm 0.70$ & $63.15 \pm 2.08$ \\
    IC\,0944 & $280.80 \pm 2.16$ & $283.57 \pm 7.27$ & $79.51 \pm 1.69$ & $78.68 \pm 0.82$ \\
    IC\,1199  & $162.22 \pm 9.18$ & $161.54 \pm 2.85$ & $68.46 \pm 1.58$ & $68.15 \pm 1.70$ \\
    IC\,1683  & $197.84 \pm 7.61$ & $202.14 \pm 5.52$ & $61.42 \pm 5.73$ & $56.89 \pm 2.43$ \\
    NGC\,0496 & $222.39 \pm 13.83$ & $217.55 \pm 3.03$ & $59.87 \pm 4.29$ & $58.79 \pm 1.22$ \\
    NGC\,2906 & $81.72 \pm 4.13$ & $80.82 \pm 3.15$ & $59.30 \pm 2.44$ & $59.77 \pm 1.02$ \\
    NGC\,3994 & $4.98 \pm 5.72$ & $6.85 \pm 1.74$ & $67.54 \pm 3.57$ & $60.14 \pm 1.58$ \\
    NGC\,4047 & $283.65 \pm 2.29$ & $287.69 \pm 3.54$ & $44.24 \pm 2.71$ & $43.74 \pm 1.75$ \\
    NGC\,4644 & $236.39 \pm 5.66$ & $232.16 \pm 3.31$ & $70.00 \pm 3.96$ & $72.92 \pm 1.92$ \\
    NGC\,4711 & $34.90 \pm 11.83$ & $41.36 \pm 4.16$ & $61.20 \pm 1.89$ & $60.63 \pm 1.75$ \\
    NGC\,5480 & $358.27 \pm 4.71$ & $12.60 \pm 2.91$ & $43.82 \pm 3.27$ & $46.51 \pm 2.28$ \\
    NGC\,5980 & $195.12 \pm 3.25$ & $194.91 \pm 4.49$ & $72.64 \pm 0.89$ & $68.22 \pm 2.21$ \\
    NGC\,6060 & $289.09 \pm 6.44$ & $277.59 \pm 3.82$ & $68.97 \pm 2.82$ & $65.93 \pm 1.70$ \\
    NGC\,6301 & $111.80 \pm 5.67$ & $109.60 \pm 3.93$ & $52.24 \pm 3.18$ & $54.00 \pm 1.74$ \\
    NGC\,6478 & $213.24 \pm 3.40$ & $214.30 \pm 2.50$ & $71.89 \pm 1.49$ & $73.58 \pm 1.80$ \\
    UGC\,09067& $196.23 \pm 3.99$ & $192.76 \pm 1.88$ & $60.61 \pm 0.64$ & $61.21 \pm 1.82$ \\
    \enddata
    \tablecomments{(1) Galaxy name; (2) Position angle of the major axis of the molecular gas kinematics; (3) Position angle of the major axis of the ionized gas kinematics; (4) Inclination angle of the molecular gas disk; (5) Inclination angle of the ionized gas disk.}    
\end{deluxetable*}



\begin{figure*}
    \centering
    \includegraphics[width=\linewidth]{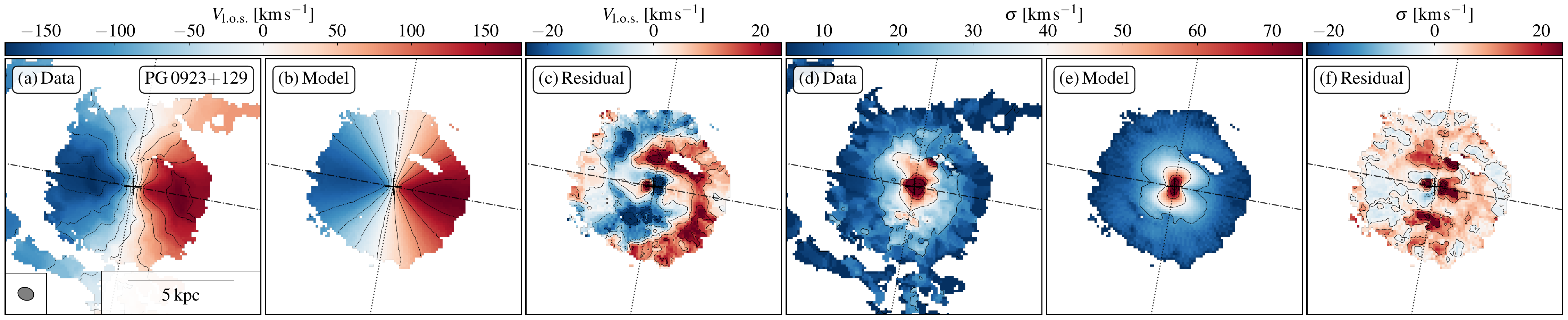}
    \includegraphics[width=\linewidth]{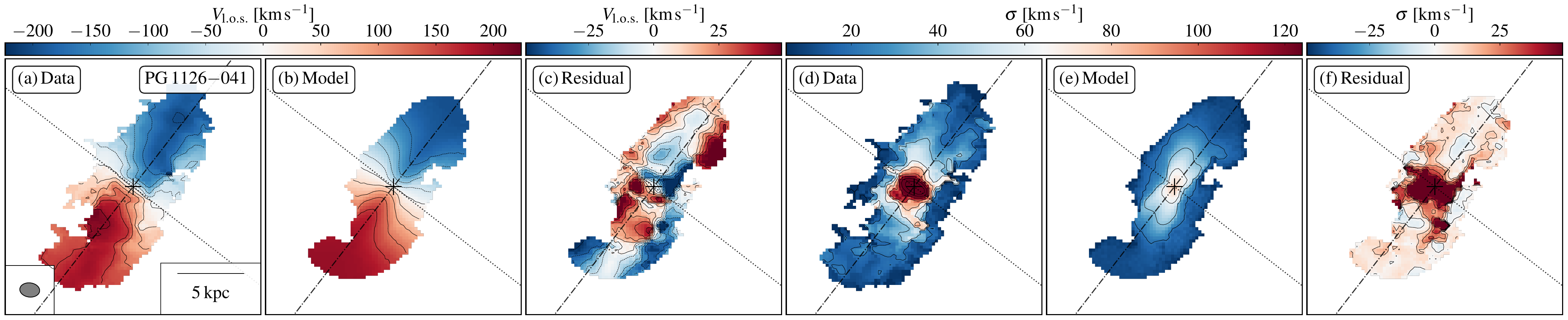}
    \includegraphics[width=\linewidth]{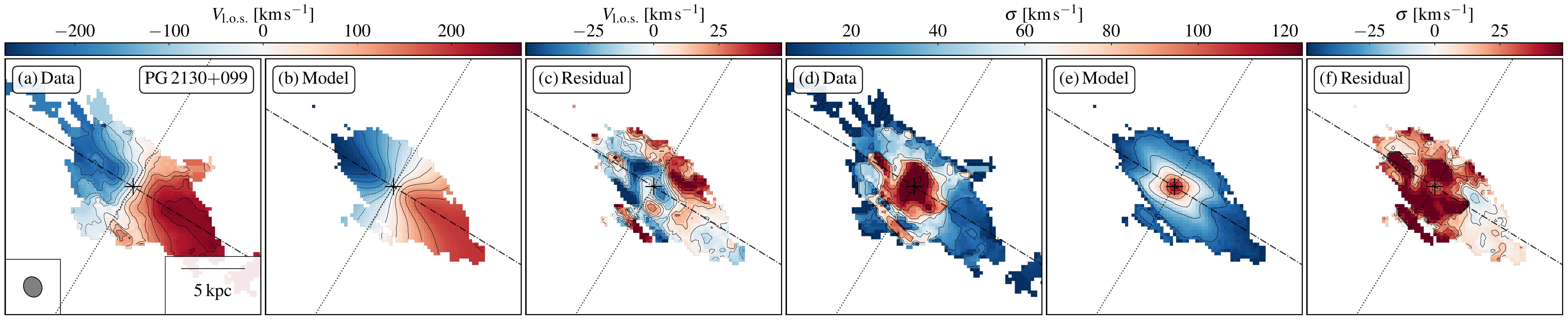}
    \includegraphics[width=\linewidth]{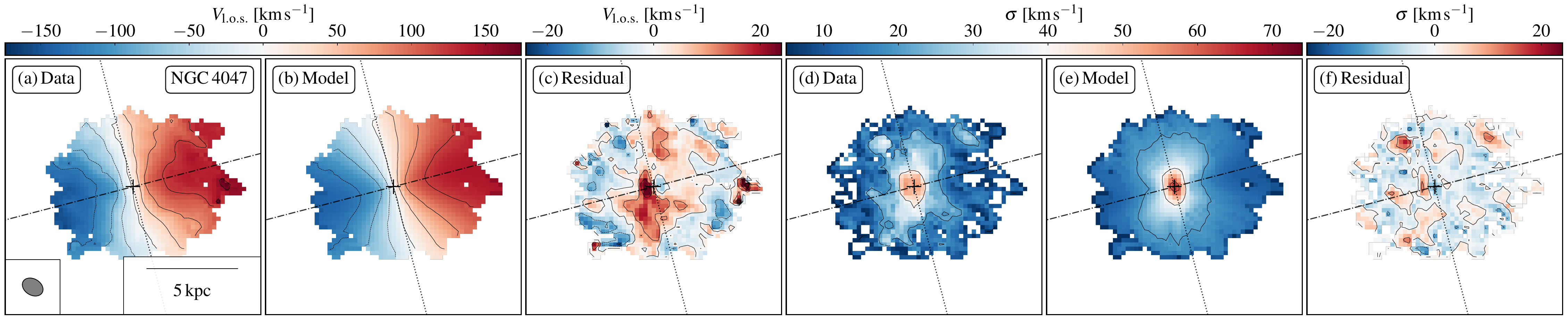}
    \caption{The \Bbarolo\  best-fit results for the CO(2--1) emission of quasar host galaxies and for the CO(1--0) emission of one inactive star-forming galaxy NGC\,4047. Panels (a) and (b) depict the line-of-sight (los) velocity maps and the corresponding best-fit models. Panel (c) illustrates the residual map, showing the differences between the observation and the best-fit model. Panels (d) and (e) exhibit the velocity dispersion maps, respectively. Panel (f) shows the residual map of the velocity dispersion.
    The contours in panels (a) and (b) are plotted starting from $-200\,{\rm km\,s^{-1}}$, with steps of $40\,{\rm km\,s^{-1}}$. The contours in panels (d) and (e) start from $0\,{\rm km\,s^{-1}}$, with steps of $20\,{\rm km\,s^{-1}}$. The residual maps' contours range from $-50\,{\rm km\,s^{-1}}$ to $50\,{\rm km\,s^{-1}}$, in steps of $10\,{\rm km\,s^{-1}}$. Each subpanel features a black cross, indicating the kinematic center. The dashed-dotted and dotted lines represent the major and minor kinematic axes, respectively. Synthesized beams are depicted at the bottom-left corner of each panel, and scale bars are provided at the bottom-right corner.
    }
    \label{Fig3: kinematics for CO}
\end{figure*}

\begin{figure*}
    \centering
    \includegraphics[width=\linewidth]{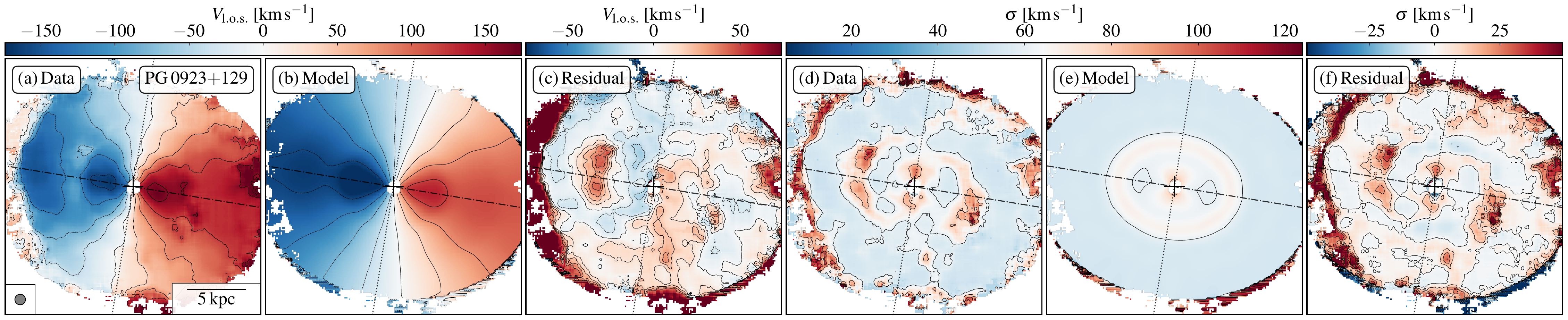}
    \includegraphics[width=\linewidth]{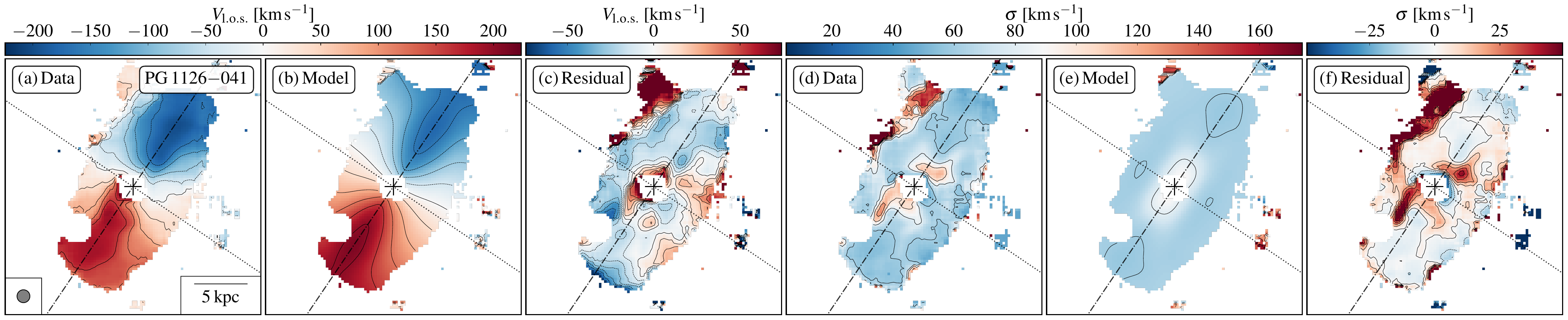}
    \includegraphics[width=\linewidth]{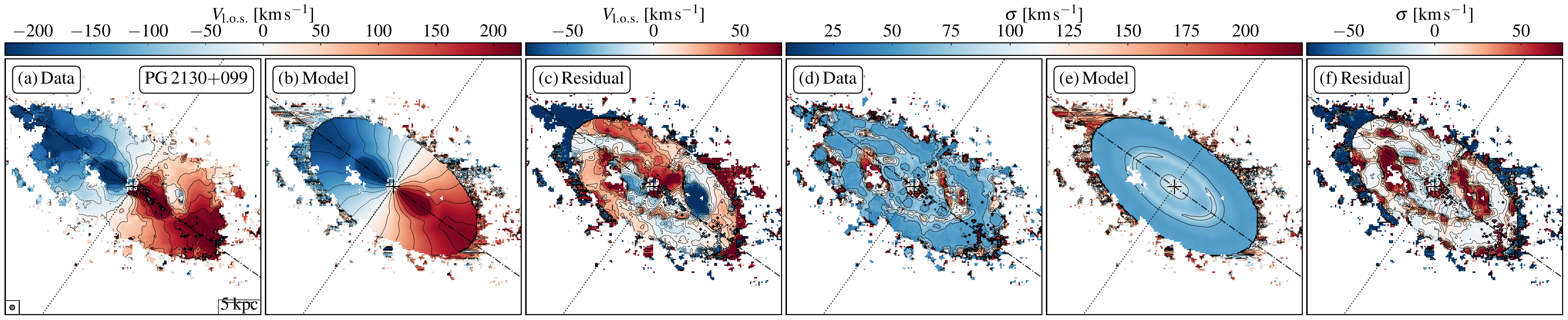}
    \includegraphics[width=\linewidth]{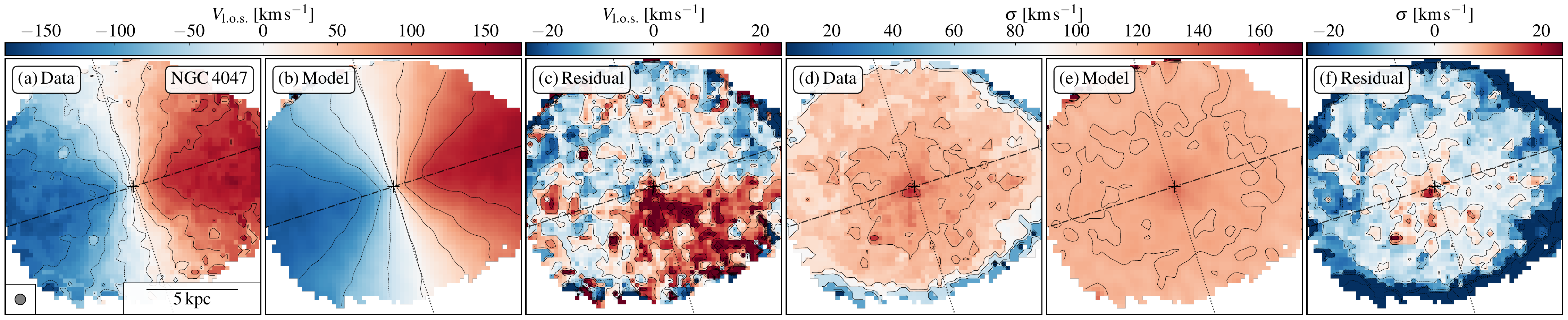}
    \caption{The \Bbarolo\  best-fit results for the H$\alpha$ emission of quasar hosts and of star-forming galaxy. The label of each subpanel, and the contours of each panel are the same as that in Figure \ref{Fig3: kinematics for CO}.}
    \label{Fig4: kinematics for Ha}
\end{figure*}

\begin{figure}
    \centering
    \includegraphics[width=\linewidth]{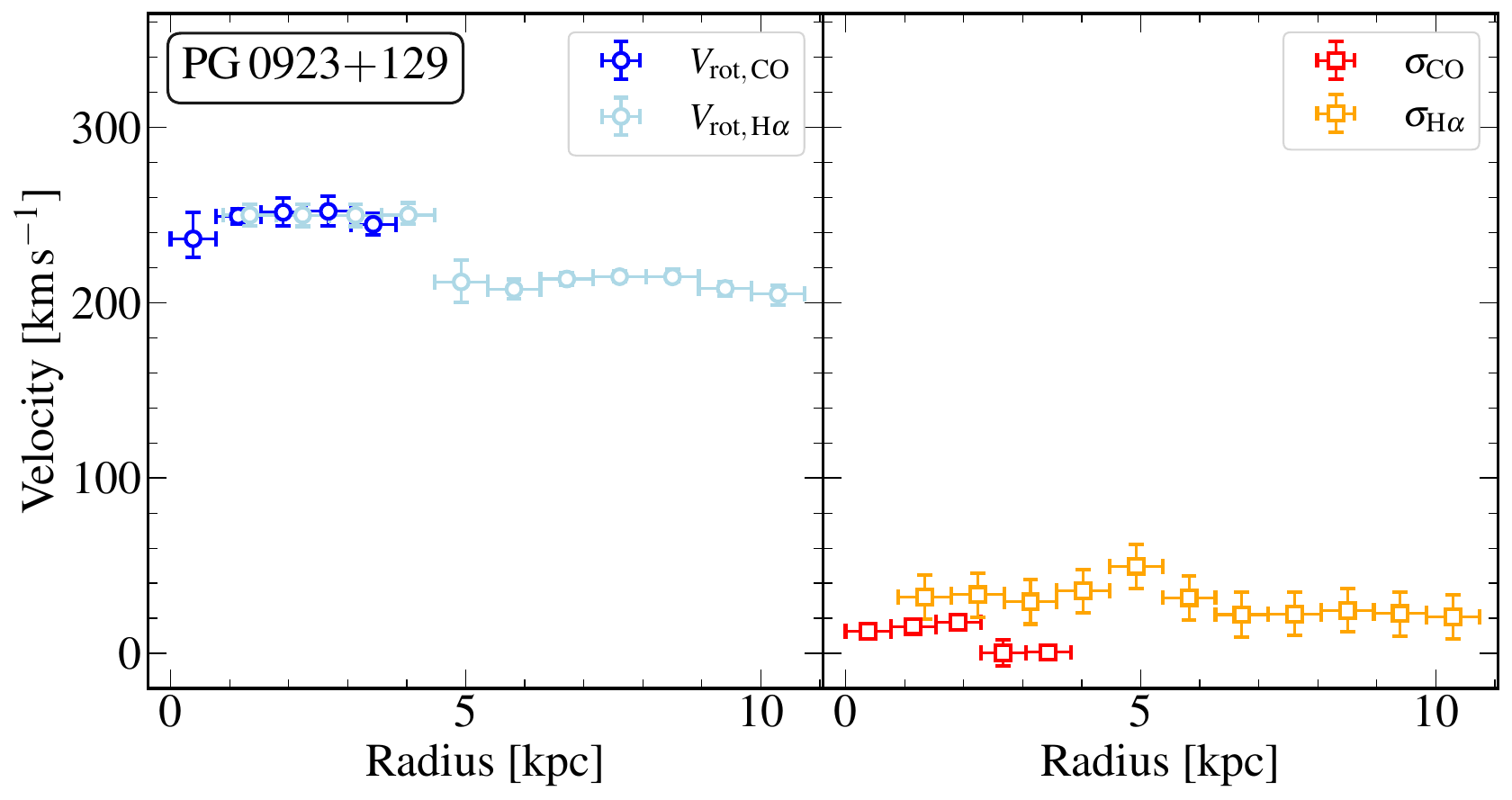}
    \includegraphics[width=\linewidth]{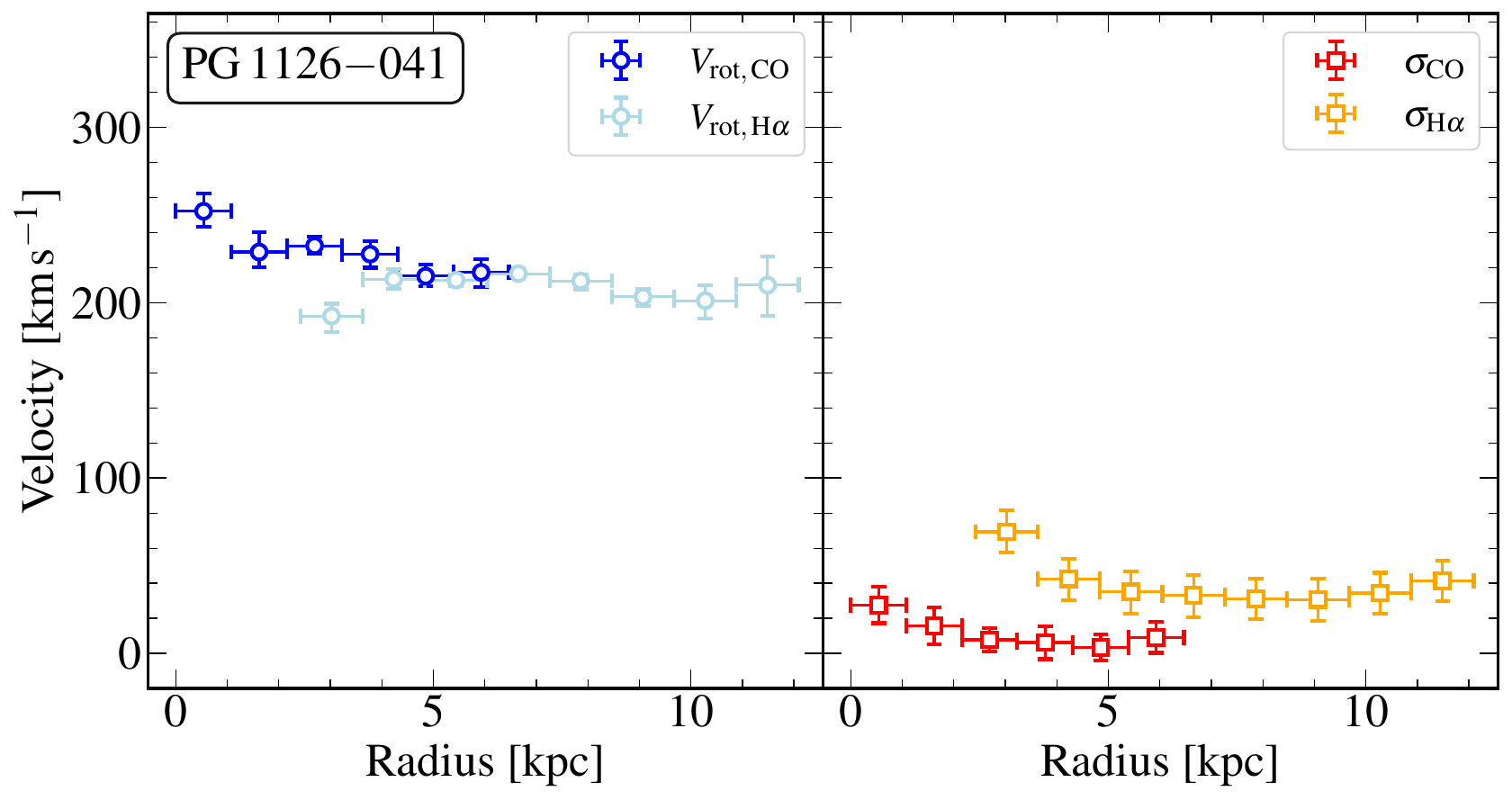}
    \includegraphics[width=\linewidth]{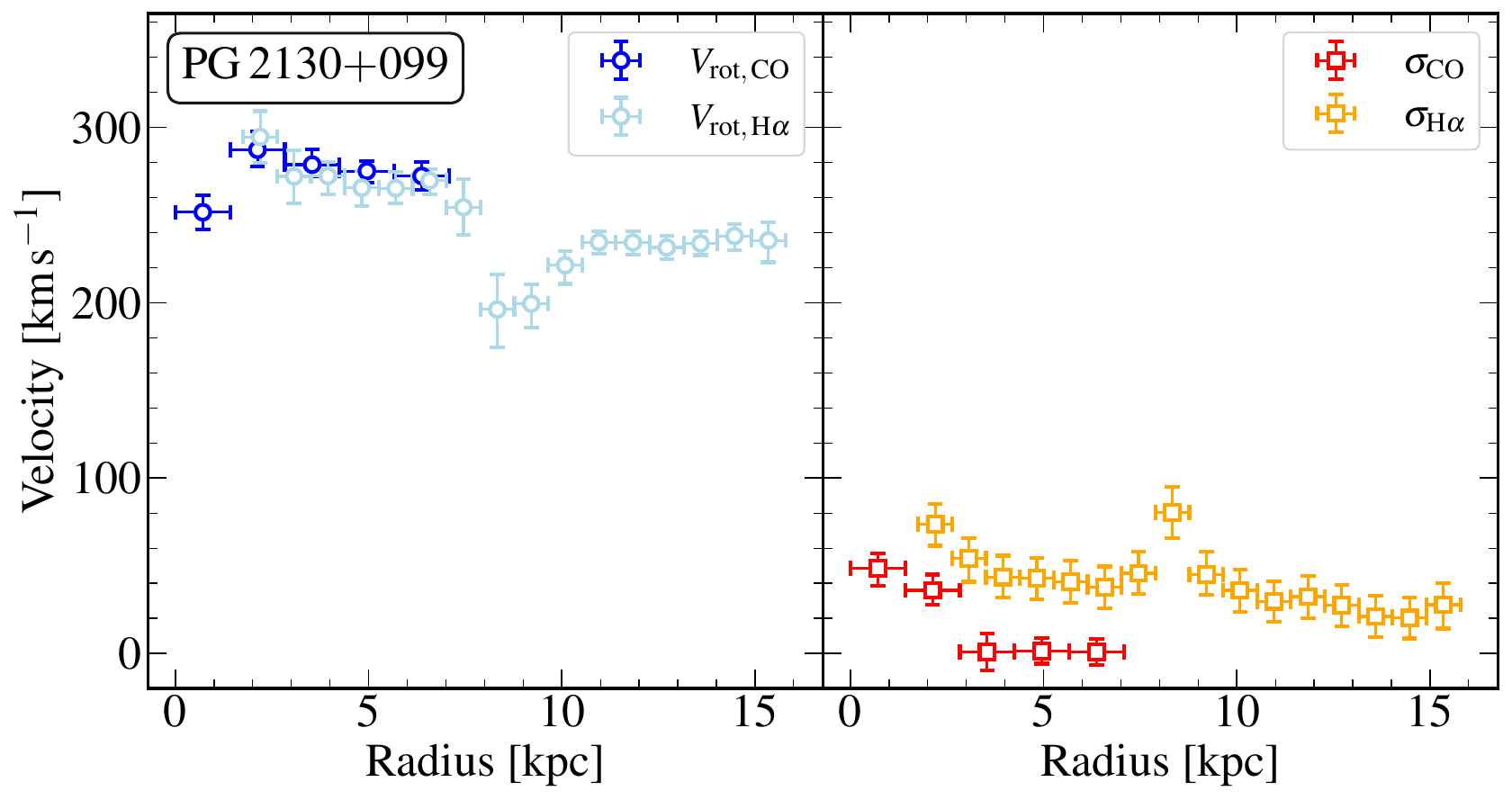}
    \includegraphics[width=\linewidth]{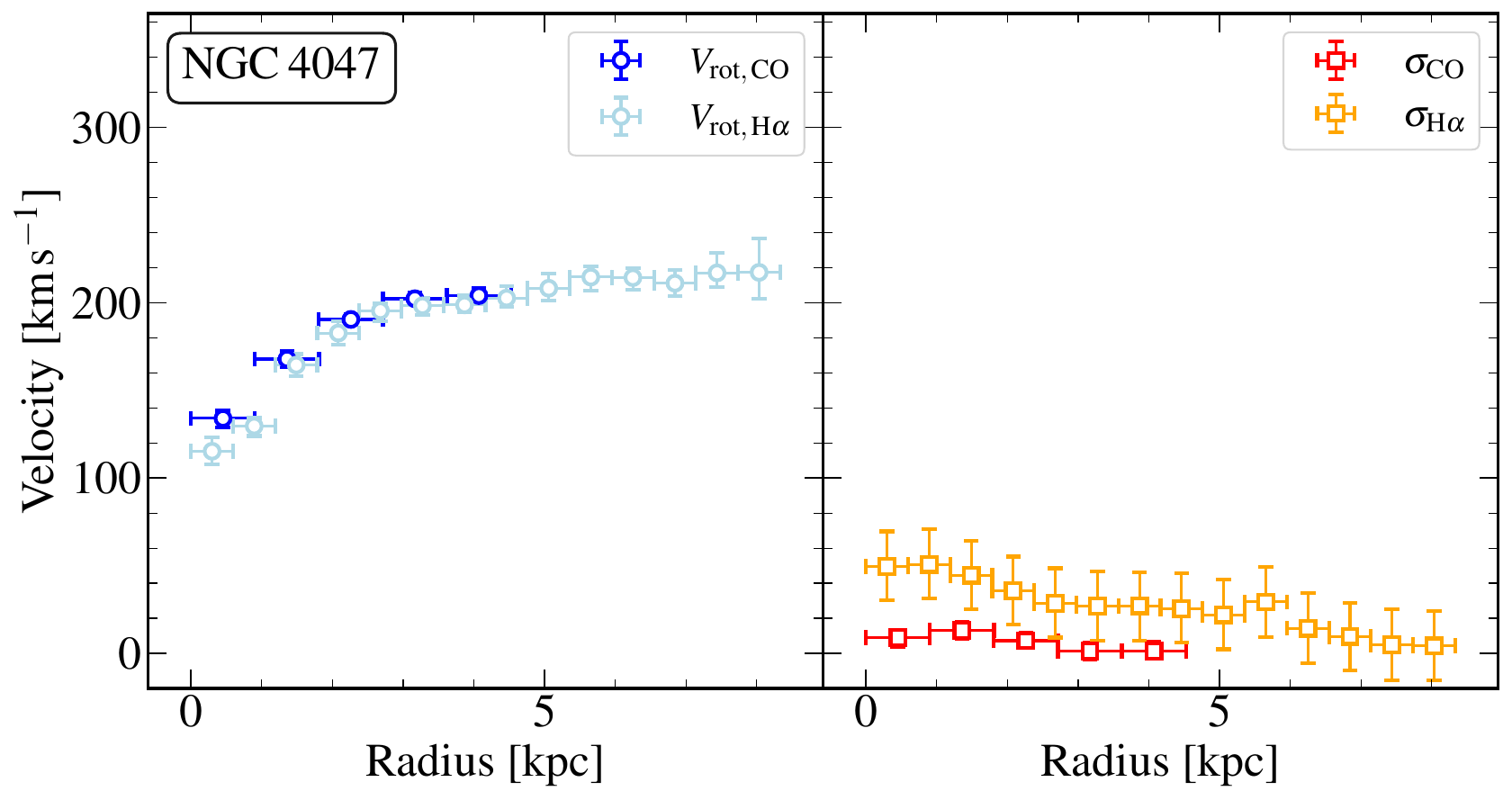}
    \caption{The rotation velocity and velocity dispersion obtained through the \Bbarolo\  analysis of the CO and H$\alpha$ data cubes for three quasar host galaxies and one of the inactive star-forming galaxies. The rotation velocities derived from the CO and H$\alpha$ are represented by blue and lightblue circles. The velocity dispersions for CO and H$\alpha$ are denoted by red and orange squares, respectively. The horizontal errorbar represents the width of the ring used in kinematic modeling.}
    \label{Fig5: velocity profile}
\end{figure}

To derive the intrinsic kinematics of the cold molecular gas and hot ionized gas in each galaxy, we employed the 3D-Based Analysis of Rotating Objects from Line Observations \citep[\Bbarolo][]{Di-Tepdoro&Fraternali2015}, which regards rotating disk as tilted rings and fit each ring independently \citep{Rogstad+1974}. We utilized the \Bbarolo\ to fit the gas kinematics in a two-step procedure, as outlined in \cite{Fei+2023}, with certain modifications. The \Bbarolo\ offers a two-stage fitting mode, where parameters are regularized using specific functions during the second stage to achieve smoother rotation curves. In the first step of fitting, we allowed the kinematic center, systematic velocity, rotation velocity, velocity dispersion, position angle, and inclination angle to be free parameters. The ring width was set to the FWHM of the beam size (for optical IFU data, we adopted the seeing limit as the beam size) in the fitting process. The initial estimates for the position angles and inclination angles were adopted from the literature \citep{Bolatto+2017, Shangguan+2020a}. The initial estimates for the systematic velocity and kinematic center were obtained through the \Bbarolo\  \texttt{SEARCH} task. We employed regularization with a constant function for the kinematic center, the systemic velocity, the inclination, and the position angle during the second stage of this step fitting, which results in only two free parameters, the rotation velocity and velocity dispersion, being fitted in this second stage. The best-fit results for the CO and H$\alpha$ data of all three quasar hosts and an example of the inactive galaxies, NGC\,4047, are illustrated in Figure \ref{Fig3: kinematics for CO} and \ref{Fig4: kinematics for Ha}, respectively. The rotation velocity and velocity dispersion profiles are shown in Figure \ref{Fig5: velocity profile}. The best-fit inclination angle and position angle of molecular gas and ionized gas are provided in Table \ref{Tab3: kinematics}. 

We note that the velocities traced by CO and H$\alpha$ are similar to each other, but there are some objects with relatively low H$\alpha$ velocities compared to CO rotation curves. The H$\alpha$ velocity dispersion given by \Bbarolo\ of these objects is higher than that of the CO. This suggests that these objects may have a turbulence-supported ionized gas disk, in which the movement of gas is supported not only by gravity but also by pressure \citep{Burkert+2010, Levy+2018}. Therefore a pressure-supported velocity correction is expected to recover the rotation velocities that are purely contributed by gravity, so-called asymmetric drift correction. We corrected this effect and estimated the circular velocities that are contributed by the gravitational potential following the method described in \cite{Iorio+2017}. After correcting the pressure and calculating the circular velocity, we found that the circular velocities derived from cold and hot gas are the same, taking into account the uncertainties. We note that the velocity dispersion of H$\alpha$ emission is smaller than the spectral resolution of MUSE ($\sim 120\,\rm km\,s^{-1}$) and CALIFA ($\sim 160\,\rm km\,s^{-1}$) data, which leads to the occurrence of low-velocity dispersion in several targets (near 10\,$\rm km\,s^{-1}$). The velocity dispersion of H$\alpha$ of those targets should be underestimated, and the asymmetric drift correction is unreliable. However, we note that all of those targets are rotation-dominated systems. Therefore our following analysis is not sensitive to the underestimation of velocity dispersion.


\figsetgrpstart
\figsetgrpnum{3.1}
\figsetgrptitle{CO kinematics of PG0923+129}
\figsetplot{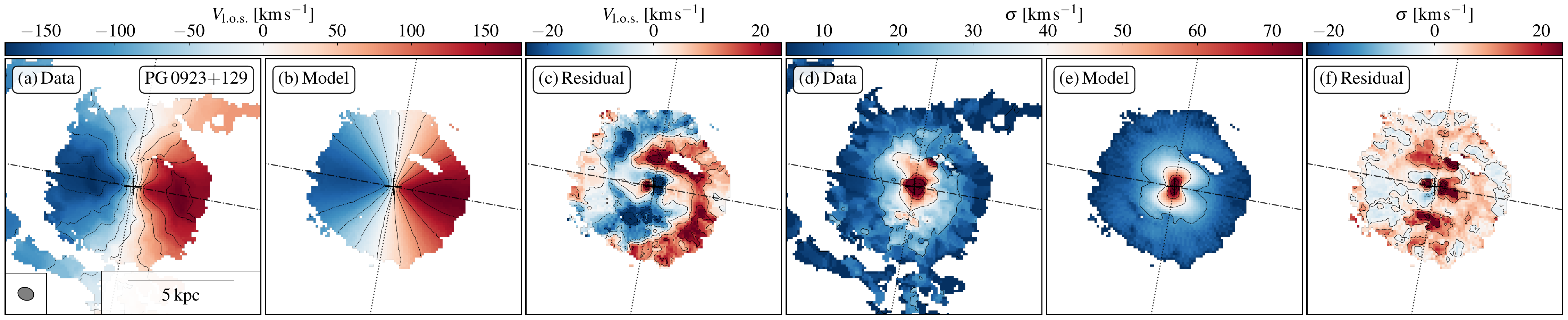}
\figsetgrpnote{The \Bbarolo\  best-fit results for the CO(2--1) emission of a quasar host galaxy, PG\,0923$+$129. Panels (a) and (b) depict the line-of-sight (los) velocity maps and the corresponding best-fit models. Panel (c) illustrates the residual map, showing the differences between the observation and the best-fit model. Panels (d) and (e) exhibit the velocity dispersion maps, respectively. Panel (f) shows the residual map of the velocity dispersion.
The contours in panels (a) and (b) are plotted starting from $-200\,{\rm km\,s^{-1}}$, with steps of $40\,{\rm km\,s^{-1}}$. The contours in panels (d) and (e) start from $0\,{\rm km\,s^{-1}}$, with steps of $20\,{\rm km\,s^{-1}}$. The residual maps' contours range from $-50\,{\rm km\,s^{-1}}$ to $50\,{\rm km\,s^{-1}}$, in steps of $10\,{\rm km\,s^{-1}}$. Each subpanel features a black cross, indicating the kinematic center. The dashed-dotted and dotted lines represent the major and minor kinematic axes, respectively. Synthesized beams are depicted at the bottom-left corner of each panel, and scale bars are provided at the bottom-right corner.}
\figsetgrpend

\figsetgrpstart
\figsetgrpnum{3.2}
\figsetgrptitle{CO kinematics of PG1126-041}
\figsetplot{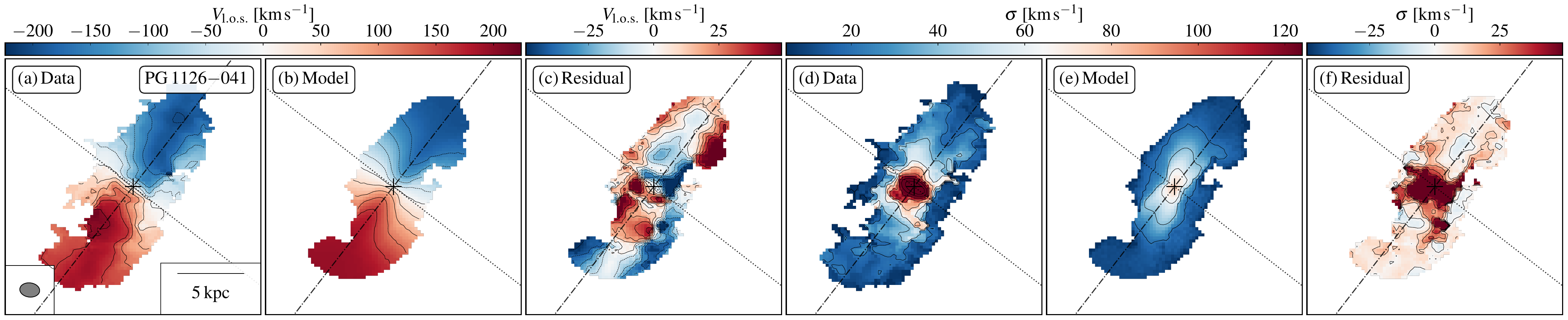}
\figsetgrpnote{The \Bbarolo\  best-fit results for the CO(2--1) emission of a quasar host galaxy, PG\,1126$-$041. Panels (a) and (b) depict the line-of-sight (los) velocity maps and the corresponding best-fit models. Panel (c) illustrates the residual map, showing the differences between the observation and the best-fit model. Panels (d) and (e) exhibit the velocity dispersion maps, respectively. Panel (f) shows the residual map of the velocity dispersion.
The contours in panels (a) and (b) are plotted starting from $-200\,{\rm km\,s^{-1}}$, with steps of $40\,{\rm km\,s^{-1}}$. The contours in panels (d) and (e) start from $0\,{\rm km\,s^{-1}}$, with steps of $20\,{\rm km\,s^{-1}}$. The residual maps' contours range from $-50\,{\rm km\,s^{-1}}$ to $50\,{\rm km\,s^{-1}}$, in steps of $10\,{\rm km\,s^{-1}}$. Each subpanel features a black cross, indicating the kinematic center. The dashed-dotted and dotted lines represent the major and minor kinematic axes, respectively. Synthesized beams are depicted at the bottom-left corner of each panel, and scale bars are provided at the bottom-right corner.}
\figsetgrpend

\figsetgrpstart
\figsetgrpnum{3.3}
\figsetgrptitle{CO kinematics of PG2130+099}
\figsetplot{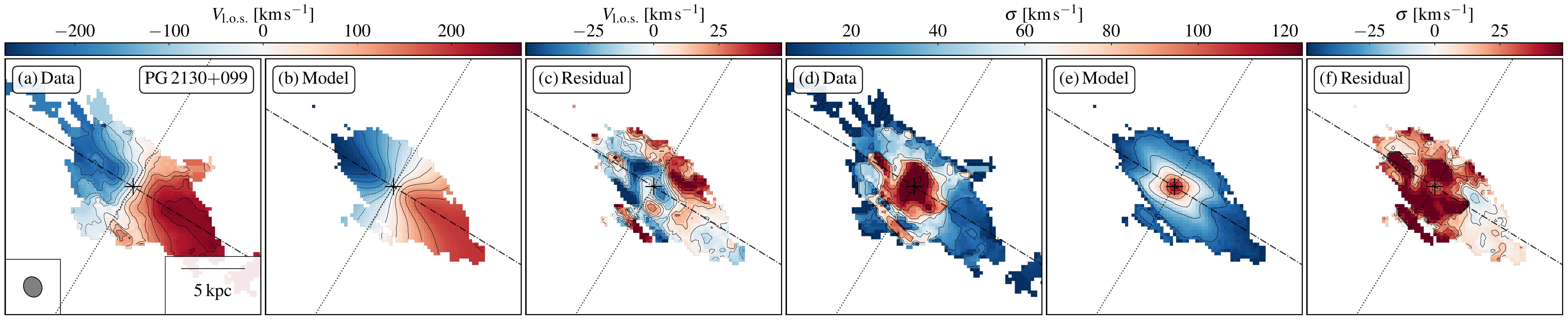}
\figsetgrpnote{The \Bbarolo\  best-fit results for the CO(2--1) emission of a quasar host galaxy, PG\,2130$+$099. Panels (a) and (b) depict the line-of-sight (los) velocity maps and the corresponding best-fit models. Panel (c) illustrates the residual map, showing the differences between the observation and the best-fit model. Panels (d) and (e) exhibit the velocity dispersion maps, respectively. Panel (f) shows the residual map of the velocity dispersion.
The contours in panels (a) and (b) are plotted starting from $-200\,{\rm km\,s^{-1}}$, with steps of $40\,{\rm km\,s^{-1}}$. The contours in panels (d) and (e) start from $0\,{\rm km\,s^{-1}}$, with steps of $20\,{\rm km\,s^{-1}}$. The residual maps' contours range from $-50\,{\rm km\,s^{-1}}$ to $50\,{\rm km\,s^{-1}}$, in steps of $10\,{\rm km\,s^{-1}}$. Each subpanel features a black cross, indicating the kinematic center. The dashed-dotted and dotted lines represent the major and minor kinematic axes, respectively. Synthesized beams are depicted at the bottom-left corner of each panel, and scale bars are provided at the bottom-right corner.}
\figsetgrpend

\figsetgrpstart
\figsetgrpnum{3.4}
\figsetgrptitle{CO kinematics of IC0944}
\figsetplot{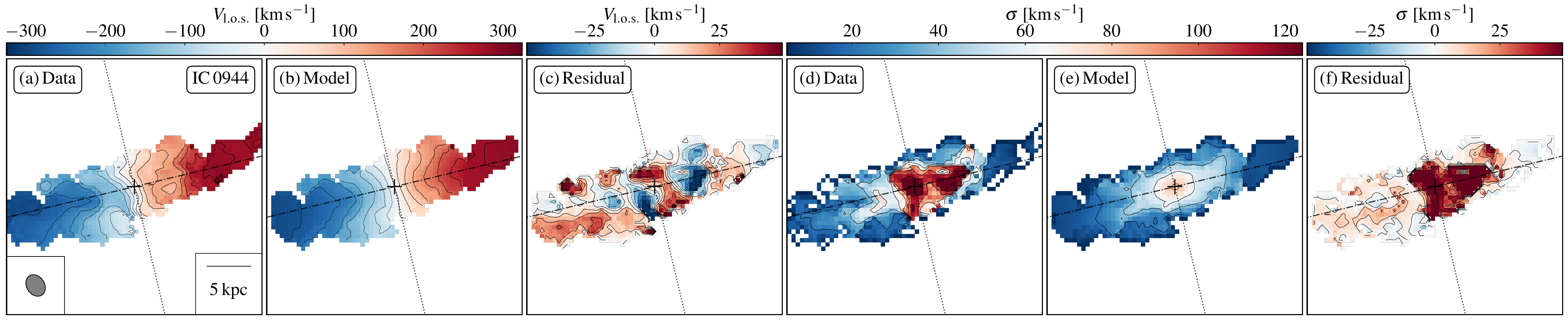}
\figsetgrpnote{The \Bbarolo\  best-fit results for the CO(1--0) emission of an inactive star-forming galaxy, IC0944. Panels (a) and (b) depict the line-of-sight (los) velocity maps and the corresponding best-fit models. Panel (c) illustrates the residual map, showing the differences between the observation and the best-fit model. Panels (d) and (e) exhibit the velocity dispersion maps, respectively. Panel (f) shows the residual map of the velocity dispersion.
The contours in panels (a) and (b) are plotted starting from $-200\,{\rm km\,s^{-1}}$, with steps of $40\,{\rm km\,s^{-1}}$. The contours in panels (d) and (e) start from $0\,{\rm km\,s^{-1}}$, with steps of $20\,{\rm km\,s^{-1}}$. The residual maps' contours range from $-50\,{\rm km\,s^{-1}}$ to $50\,{\rm km\,s^{-1}}$, in steps of $10\,{\rm km\,s^{-1}}$. Each subpanel features a black cross, indicating the kinematic center. The dashed-dotted and dotted lines represent the major and minor kinematic axes, respectively. Synthesized beams are depicted at the bottom-left corner of each panel, and scale bars are provided at the bottom-right corner.}
\figsetgrpend

\figsetgrpstart
\figsetgrpnum{3.5}
\figsetgrptitle{CO kinematics of IC1199}
\figsetplot{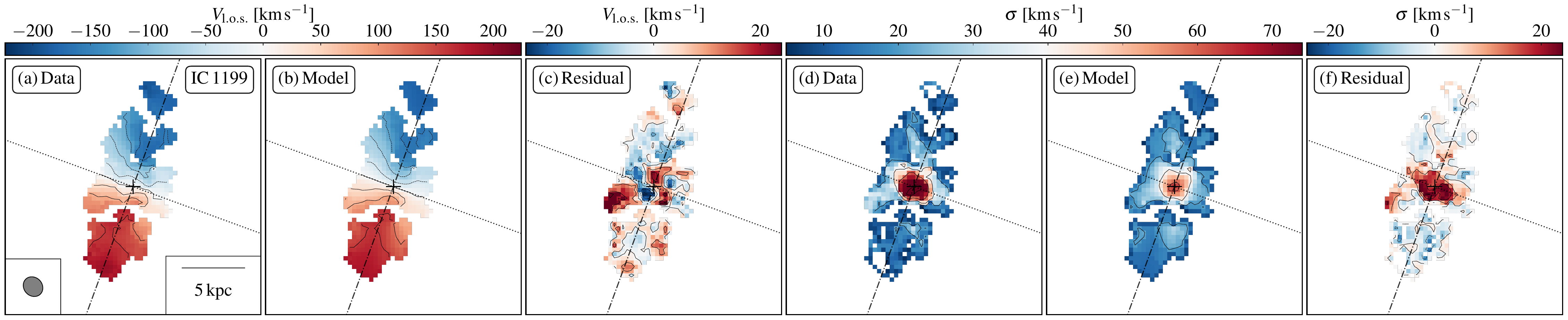}
\figsetgrpnote{The \Bbarolo\  best-fit results for the CO(1--0) emission of an inactive star-forming galaxy, IC1199. Panels (a) and (b) depict the line-of-sight (los) velocity maps and the corresponding best-fit models. Panel (c) illustrates the residual map, showing the differences between the observation and the best-fit model. Panels (d) and (e) exhibit the velocity dispersion maps, respectively. Panel (f) shows the residual map of the velocity dispersion.
The contours in panels (a) and (b) are plotted starting from $-200\,{\rm km\,s^{-1}}$, with steps of $40\,{\rm km\,s^{-1}}$. The contours in panels (d) and (e) start from $0\,{\rm km\,s^{-1}}$, with steps of $20\,{\rm km\,s^{-1}}$. The residual maps' contours range from $-50\,{\rm km\,s^{-1}}$ to $50\,{\rm km\,s^{-1}}$, in steps of $10\,{\rm km\,s^{-1}}$. Each subpanel features a black cross, indicating the kinematic center. The dashed-dotted and dotted lines represent the major and minor kinematic axes, respectively. Synthesized beams are depicted at the bottom-left corner of each panel, and scale bars are provided at the bottom-right corner.}
\figsetgrpend

\figsetgrpstart
\figsetgrpnum{3.6}
\figsetgrptitle{CO kinematics of IC1683}
\figsetplot{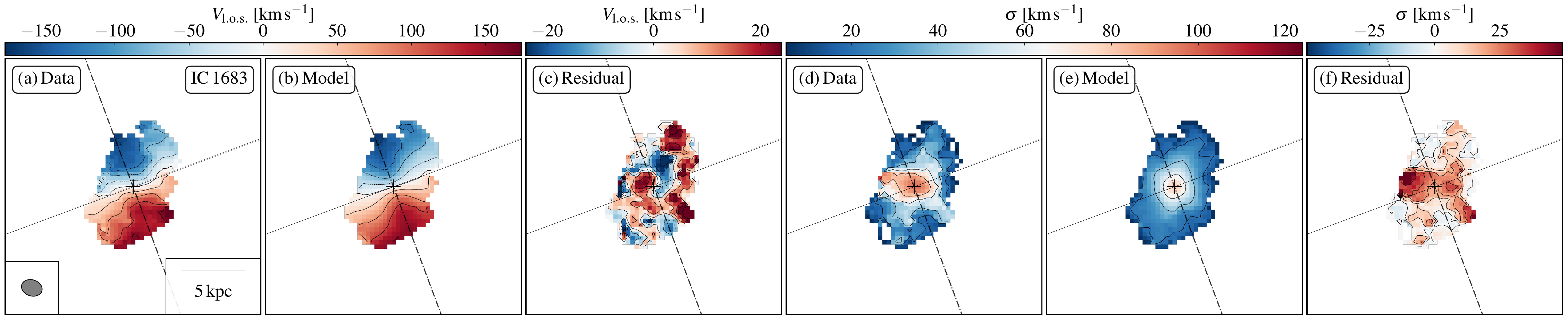}
\figsetgrpnote{The \Bbarolo\  best-fit results for the CO(1--0) emission of an inactive star-forming galaxy, IC1683. Panels (a) and (b) depict the line-of-sight (los) velocity maps and the corresponding best-fit models. Panel (c) illustrates the residual map, showing the differences between the observation and the best-fit model. Panels (d) and (e) exhibit the velocity dispersion maps, respectively. Panel (f) shows the residual map of the velocity dispersion.
The contours in panels (a) and (b) are plotted starting from $-200\,{\rm km\,s^{-1}}$, with steps of $40\,{\rm km\,s^{-1}}$. The contours in panels (d) and (e) start from $0\,{\rm km\,s^{-1}}$, with steps of $20\,{\rm km\,s^{-1}}$. The residual maps' contours range from $-50\,{\rm km\,s^{-1}}$ to $50\,{\rm km\,s^{-1}}$, in steps of $10\,{\rm km\,s^{-1}}$. Each subpanel features a black cross, indicating the kinematic center. The dashed-dotted and dotted lines represent the major and minor kinematic axes, respectively. Synthesized beams are depicted at the bottom-left corner of each panel, and scale bars are provided at the bottom-right corner.}
\figsetgrpend

\figsetgrpstart
\figsetgrpnum{3.7}
\figsetgrptitle{CO kinematics of NGC0496}
\figsetplot{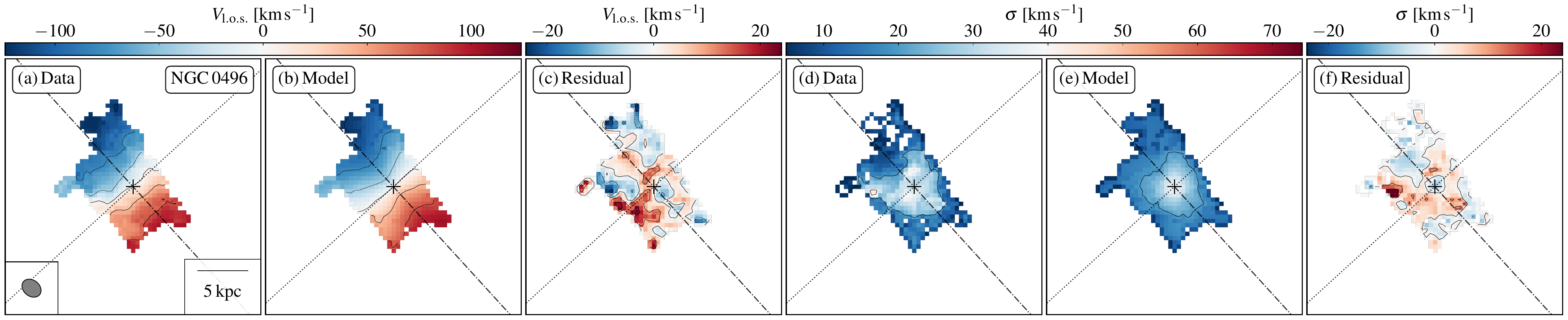}
\figsetgrpnote{The \Bbarolo\  best-fit results for the CO(1--0) emission of an inactive star-forming galaxy, NGC0496. Panels (a) and (b) depict the line-of-sight (los) velocity maps and the corresponding best-fit models. Panel (c) illustrates the residual map, showing the differences between the observation and the best-fit model. Panels (d) and (e) exhibit the velocity dispersion maps, respectively. Panel (f) shows the residual map of the velocity dispersion.
The contours in panels (a) and (b) are plotted starting from $-200\,{\rm km\,s^{-1}}$, with steps of $40\,{\rm km\,s^{-1}}$. The contours in panels (d) and (e) start from $0\,{\rm km\,s^{-1}}$, with steps of $20\,{\rm km\,s^{-1}}$. The residual maps' contours range from $-50\,{\rm km\,s^{-1}}$ to $50\,{\rm km\,s^{-1}}$, in steps of $10\,{\rm km\,s^{-1}}$. Each subpanel features a black cross, indicating the kinematic center. The dashed-dotted and dotted lines represent the major and minor kinematic axes, respectively. Synthesized beams are depicted at the bottom-left corner of each panel, and scale bars are provided at the bottom-right corner.}
\figsetgrpend

\figsetgrpstart
\figsetgrpnum{3.8}
\figsetgrptitle{CO kinematics of NGC2906}
\figsetplot{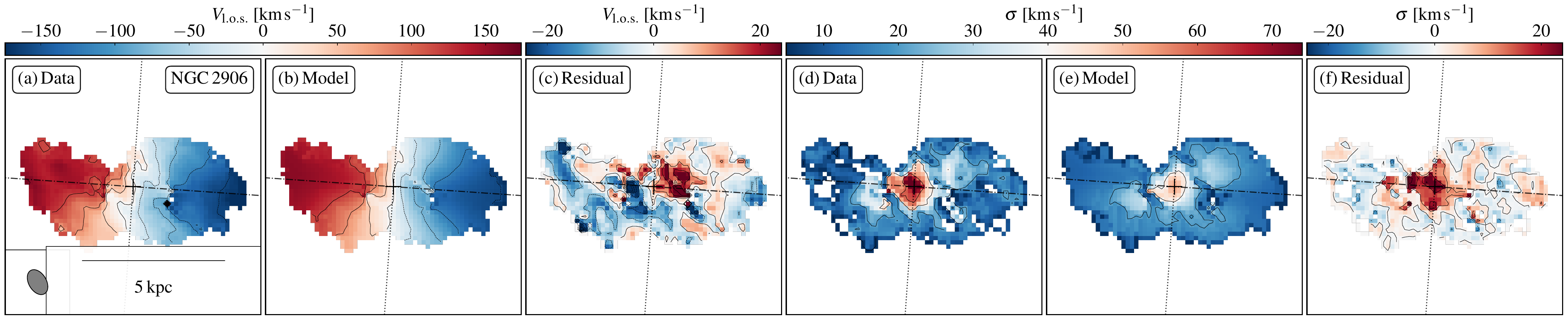}
\figsetgrpnote{The \Bbarolo\  best-fit results for the CO(1--0) emission of an inactive star-forming galaxy, NGC2906. Panels (a) and (b) depict the line-of-sight (los) velocity maps and the corresponding best-fit models. Panel (c) illustrates the residual map, showing the differences between the observation and the best-fit model. Panels (d) and (e) exhibit the velocity dispersion maps, respectively. Panel (f) shows the residual map of the velocity dispersion.
The contours in panels (a) and (b) are plotted starting from $-200\,{\rm km\,s^{-1}}$, with steps of $40\,{\rm km\,s^{-1}}$. The contours in panels (d) and (e) start from $0\,{\rm km\,s^{-1}}$, with steps of $20\,{\rm km\,s^{-1}}$. The residual maps' contours range from $-50\,{\rm km\,s^{-1}}$ to $50\,{\rm km\,s^{-1}}$, in steps of $10\,{\rm km\,s^{-1}}$. Each subpanel features a black cross, indicating the kinematic center. The dashed-dotted and dotted lines represent the major and minor kinematic axes, respectively. Synthesized beams are depicted at the bottom-left corner of each panel, and scale bars are provided at the bottom-right corner.}
\figsetgrpend

\figsetgrpstart
\figsetgrpnum{3.9}
\figsetgrptitle{CO kinematics of NGC3994}
\figsetplot{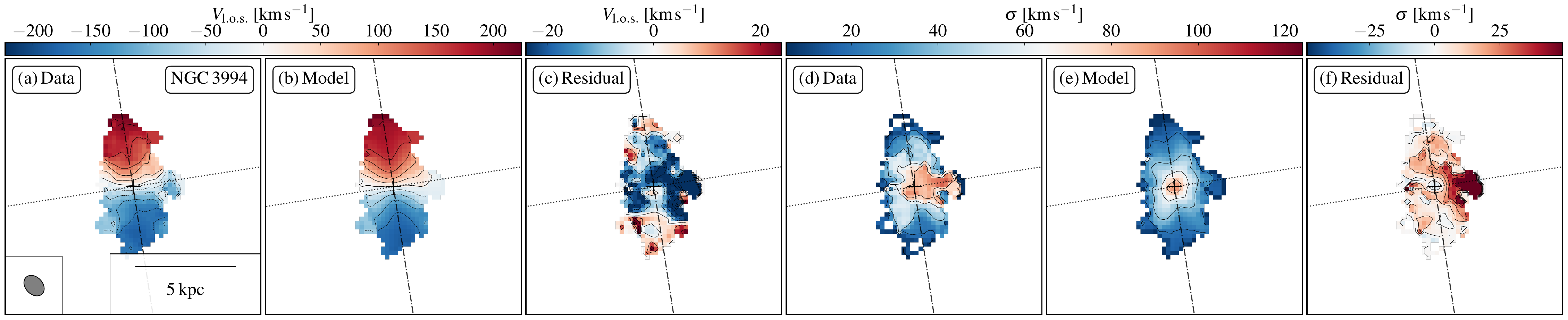}
\figsetgrpnote{The \Bbarolo\  best-fit results for the CO(1--0) emission of an inactive star-forming galaxy, NGC3994. Panels (a) and (b) depict the line-of-sight (los) velocity maps and the corresponding best-fit models. Panel (c) illustrates the residual map, showing the differences between the observation and the best-fit model. Panels (d) and (e) exhibit the velocity dispersion maps, respectively. Panel (f) shows the residual map of the velocity dispersion.
The contours in panels (a) and (b) are plotted starting from $-200\,{\rm km\,s^{-1}}$, with steps of $40\,{\rm km\,s^{-1}}$. The contours in panels (d) and (e) start from $0\,{\rm km\,s^{-1}}$, with steps of $20\,{\rm km\,s^{-1}}$. The residual maps' contours range from $-50\,{\rm km\,s^{-1}}$ to $50\,{\rm km\,s^{-1}}$, in steps of $10\,{\rm km\,s^{-1}}$. Each subpanel features a black cross, indicating the kinematic center. The dashed-dotted and dotted lines represent the major and minor kinematic axes, respectively. Synthesized beams are depicted at the bottom-left corner of each panel, and scale bars are provided at the bottom-right corner.}
\figsetgrpend

\figsetgrpstart
\figsetgrpnum{3.10}
\figsetgrptitle{CO kinematics of NGC4047}
\figsetplot{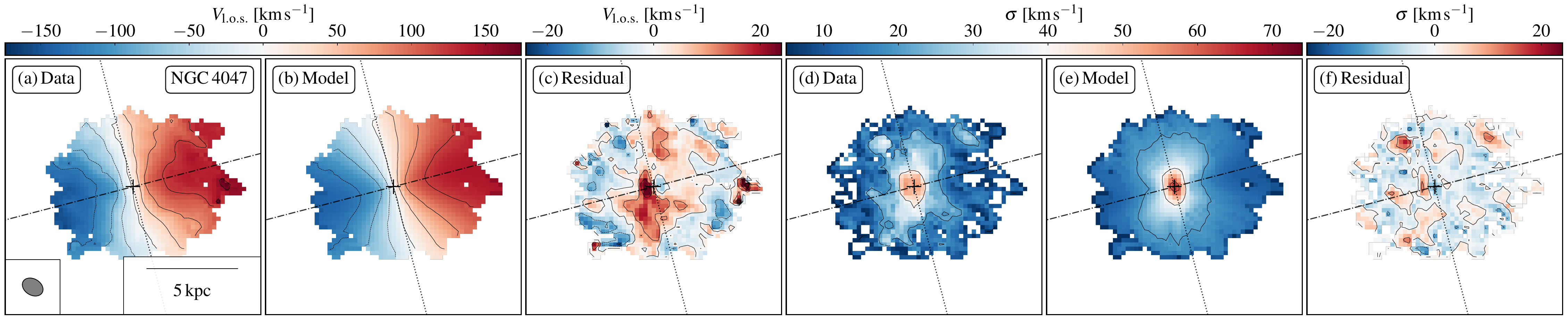}
\figsetgrpnote{The \Bbarolo\  best-fit results for the CO(1--0) emission of an inactive star-forming galaxy, NGC4047. Panels (a) and (b) depict the line-of-sight (los) velocity maps and the corresponding best-fit models. Panel (c) illustrates the residual map, showing the differences between the observation and the best-fit model. Panels (d) and (e) exhibit the velocity dispersion maps, respectively. Panel (f) shows the residual map of the velocity dispersion.
The contours in panels (a) and (b) are plotted starting from $-200\,{\rm km\,s^{-1}}$, with steps of $40\,{\rm km\,s^{-1}}$. The contours in panels (d) and (e) start from $0\,{\rm km\,s^{-1}}$, with steps of $20\,{\rm km\,s^{-1}}$. The residual maps' contours range from $-50\,{\rm km\,s^{-1}}$ to $50\,{\rm km\,s^{-1}}$, in steps of $10\,{\rm km\,s^{-1}}$. Each subpanel features a black cross, indicating the kinematic center. The dashed-dotted and dotted lines represent the major and minor kinematic axes, respectively. Synthesized beams are depicted at the bottom-left corner of each panel, and scale bars are provided at the bottom-right corner.}
\figsetgrpend

\figsetgrpstart
\figsetgrpnum{3.11}
\figsetgrptitle{CO kinematics of NGC4644}
\figsetplot{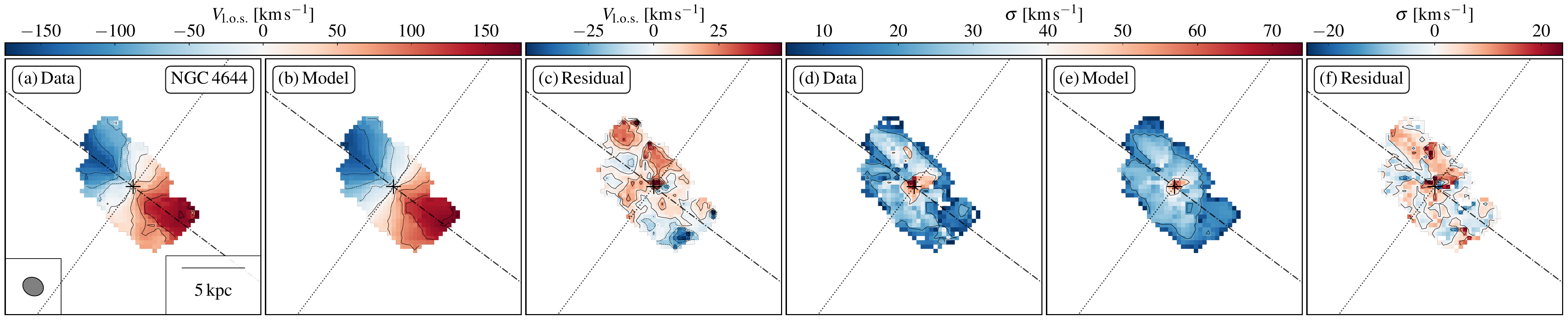}
\figsetgrpnote{The \Bbarolo\  best-fit results for the CO(2--1) emission of quasar host galaxies and for the CO(1--0) emission of one inactive star-forming galaxy NGC\,4047. Panels (a) and (b) depict the line-of-sight (los) velocity maps and the corresponding best-fit models. Panel (c) illustrates the residual map, showing the differences between the observation and the best-fit model. Panels (d) and (e) exhibit the velocity dispersion maps, respectively. Panel (f) shows the residual map of the velocity dispersion.
The contours in panels (a) and (b) are plotted starting from $-200\,{\rm km\,s^{-1}}$, with steps of $40\,{\rm km\,s^{-1}}$. The contours in panels (d) and (e) start from $0\,{\rm km\,s^{-1}}$, with steps of $20\,{\rm km\,s^{-1}}$. The residual maps' contours range from $-50\,{\rm km\,s^{-1}}$ to $50\,{\rm km\,s^{-1}}$, in steps of $10\,{\rm km\,s^{-1}}$. Each subpanel features a black cross, indicating the kinematic center. The dashed-dotted and dotted lines represent the major and minor kinematic axes, respectively. Synthesized beams are depicted at the bottom-left corner of each panel, and scale bars are provided at the bottom-right corner.}
\figsetgrpend

\figsetgrpstart
\figsetgrpnum{3.12}
\figsetgrptitle{CO kinematics of NGC4711}
\figsetplot{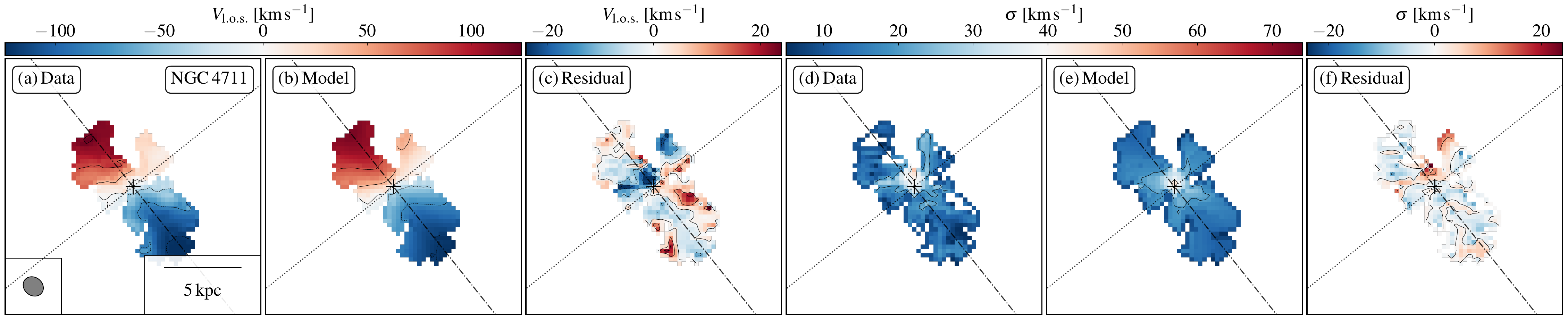}
\figsetgrpnote{The \Bbarolo\  best-fit results for the CO(1--0) emission of an inactive star-forming galaxy, NGC4711. Panels (a) and (b) depict the line-of-sight (los) velocity maps and the corresponding best-fit models. Panel (c) illustrates the residual map, showing the differences between the observation and the best-fit model. Panels (d) and (e) exhibit the velocity dispersion maps, respectively. Panel (f) shows the residual map of the velocity dispersion.
The contours in panels (a) and (b) are plotted starting from $-200\,{\rm km\,s^{-1}}$, with steps of $40\,{\rm km\,s^{-1}}$. The contours in panels (d) and (e) start from $0\,{\rm km\,s^{-1}}$, with steps of $20\,{\rm km\,s^{-1}}$. The residual maps' contours range from $-50\,{\rm km\,s^{-1}}$ to $50\,{\rm km\,s^{-1}}$, in steps of $10\,{\rm km\,s^{-1}}$. Each subpanel features a black cross, indicating the kinematic center. The dashed-dotted and dotted lines represent the major and minor kinematic axes, respectively. Synthesized beams are depicted at the bottom-left corner of each panel, and scale bars are provided at the bottom-right corner.}
\figsetgrpend

\figsetgrpstart
\figsetgrpnum{3.13}
\figsetgrptitle{CO kinematics of NGC5480}
\figsetplot{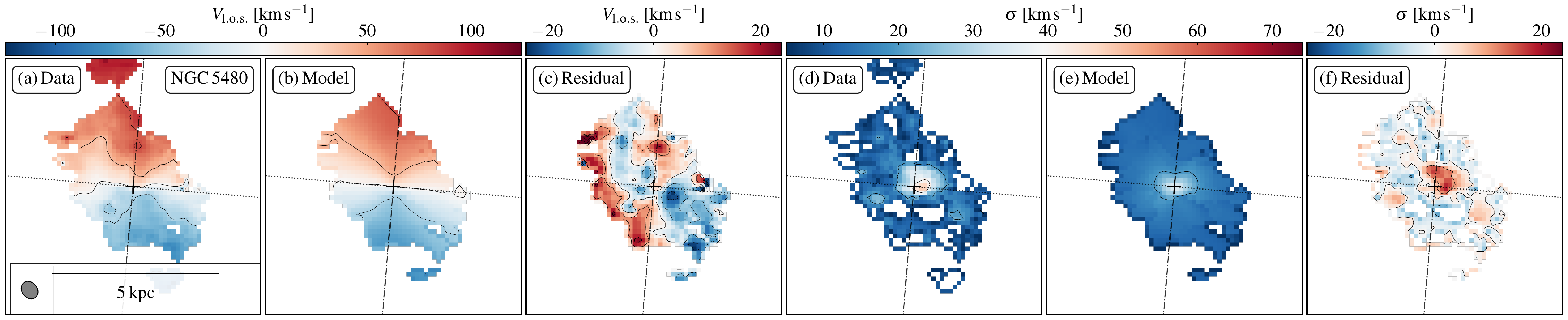}
\figsetgrpnote{The \Bbarolo\  best-fit results for the CO(1--0) emission of an inactive star-forming galaxy, NGC5480. Panels (a) and (b) depict the line-of-sight (los) velocity maps and the corresponding best-fit models. Panel (c) illustrates the residual map, showing the differences between the observation and the best-fit model. Panels (d) and (e) exhibit the velocity dispersion maps, respectively. Panel (f) shows the residual map of the velocity dispersion.
The contours in panels (a) and (b) are plotted starting from $-200\,{\rm km\,s^{-1}}$, with steps of $40\,{\rm km\,s^{-1}}$. The contours in panels (d) and (e) start from $0\,{\rm km\,s^{-1}}$, with steps of $20\,{\rm km\,s^{-1}}$. The residual maps' contours range from $-50\,{\rm km\,s^{-1}}$ to $50\,{\rm km\,s^{-1}}$, in steps of $10\,{\rm km\,s^{-1}}$. Each subpanel features a black cross, indicating the kinematic center. The dashed-dotted and dotted lines represent the major and minor kinematic axes, respectively. Synthesized beams are depicted at the bottom-left corner of each panel, and scale bars are provided at the bottom-right corner.}
\figsetgrpend

\figsetgrpstart
\figsetgrpnum{3.14}
\figsetgrptitle{CO kinematics of NGC5980}
\figsetplot{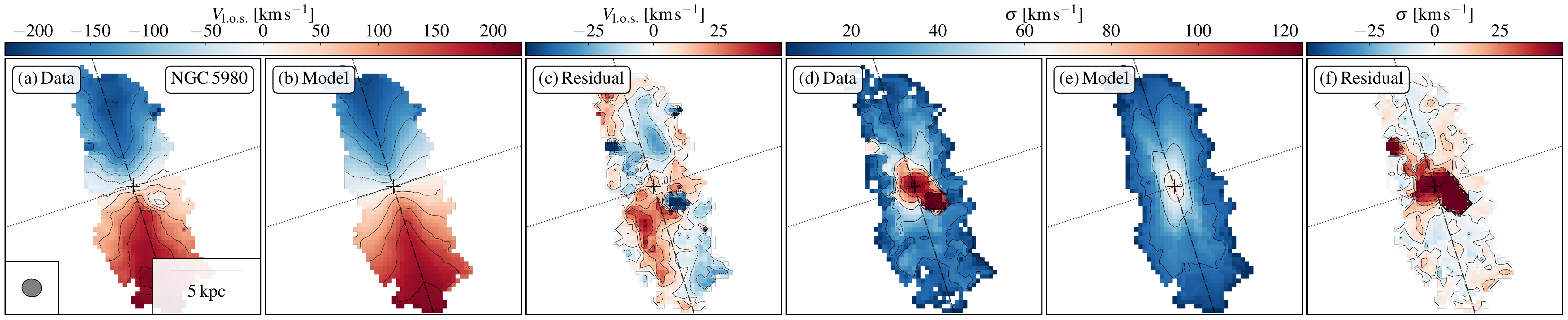}
\figsetgrpnote{The \Bbarolo\  best-fit results for the CO(1--0) emission of an inactive star-forming galaxy, NGC5980. Panels (a) and (b) depict the line-of-sight (los) velocity maps and the corresponding best-fit models. Panel (c) illustrates the residual map, showing the differences between the observation and the best-fit model. Panels (d) and (e) exhibit the velocity dispersion maps, respectively. Panel (f) shows the residual map of the velocity dispersion.
The contours in panels (a) and (b) are plotted starting from $-200\,{\rm km\,s^{-1}}$, with steps of $40\,{\rm km\,s^{-1}}$. The contours in panels (d) and (e) start from $0\,{\rm km\,s^{-1}}$, with steps of $20\,{\rm km\,s^{-1}}$. The residual maps' contours range from $-50\,{\rm km\,s^{-1}}$ to $50\,{\rm km\,s^{-1}}$, in steps of $10\,{\rm km\,s^{-1}}$. Each subpanel features a black cross, indicating the kinematic center. The dashed-dotted and dotted lines represent the major and minor kinematic axes, respectively. Synthesized beams are depicted at the bottom-left corner of each panel, and scale bars are provided at the bottom-right corner.}
\figsetgrpend

\figsetgrpstart
\figsetgrpnum{3.15}
\figsetgrptitle{CO kinematics of NGC6060}
\figsetplot{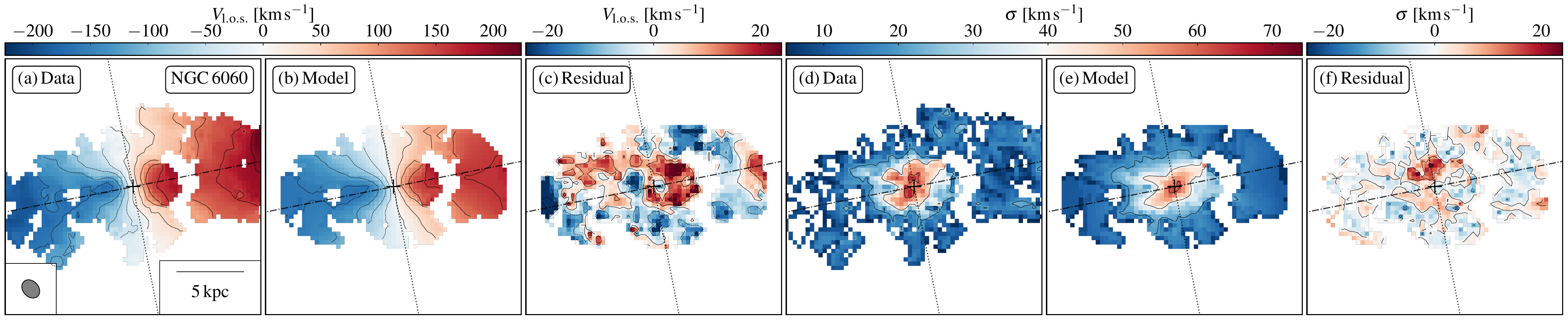}
\figsetgrpnote{The \Bbarolo\  best-fit results for the CO(1--0) emission of an inactive star-forming galaxy, NGC6060. Panels (a) and (b) depict the line-of-sight (los) velocity maps and the corresponding best-fit models. Panel (c) illustrates the residual map, showing the differences between the observation and the best-fit model. Panels (d) and (e) exhibit the velocity dispersion maps, respectively. Panel (f) shows the residual map of the velocity dispersion.
The contours in panels (a) and (b) are plotted starting from $-200\,{\rm km\,s^{-1}}$, with steps of $40\,{\rm km\,s^{-1}}$. The contours in panels (d) and (e) start from $0\,{\rm km\,s^{-1}}$, with steps of $20\,{\rm km\,s^{-1}}$. The residual maps' contours range from $-50\,{\rm km\,s^{-1}}$ to $50\,{\rm km\,s^{-1}}$, in steps of $10\,{\rm km\,s^{-1}}$. Each subpanel features a black cross, indicating the kinematic center. The dashed-dotted and dotted lines represent the major and minor kinematic axes, respectively. Synthesized beams are depicted at the bottom-left corner of each panel, and scale bars are provided at the bottom-right corner.}
\figsetgrpend

\figsetgrpstart
\figsetgrpnum{3.16}
\figsetgrptitle{CO kinematics of NGC6301}
\figsetplot{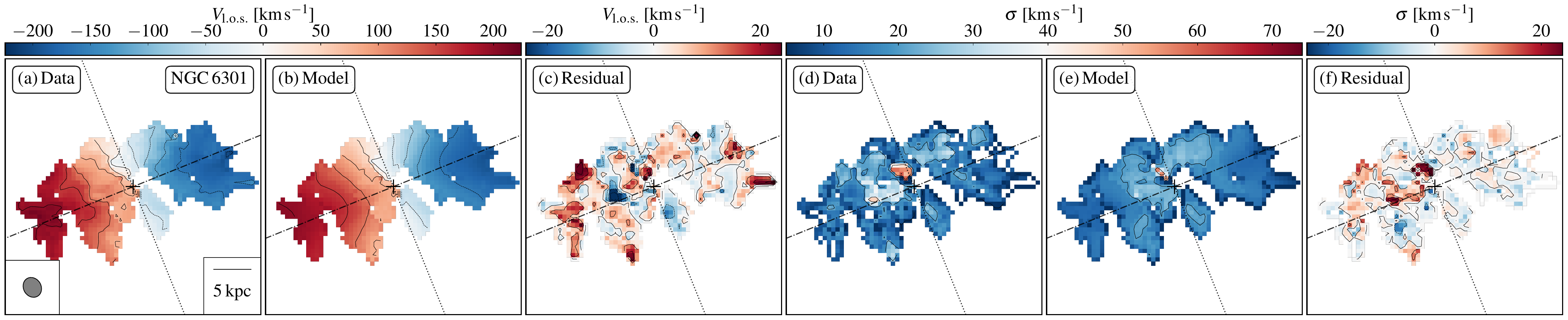}
\figsetgrpnote{The \Bbarolo\  best-fit results for the CO(1--0) emission of an inactive star-forming galaxy, NGC6301. Panels (a) and (b) depict the line-of-sight (los) velocity maps and the corresponding best-fit models. Panel (c) illustrates the residual map, showing the differences between the observation and the best-fit model. Panels (d) and (e) exhibit the velocity dispersion maps, respectively. Panel (f) shows the residual map of the velocity dispersion.
The contours in panels (a) and (b) are plotted starting from $-200\,{\rm km\,s^{-1}}$, with steps of $40\,{\rm km\,s^{-1}}$. The contours in panels (d) and (e) start from $0\,{\rm km\,s^{-1}}$, with steps of $20\,{\rm km\,s^{-1}}$. The residual maps' contours range from $-50\,{\rm km\,s^{-1}}$ to $50\,{\rm km\,s^{-1}}$, in steps of $10\,{\rm km\,s^{-1}}$. Each subpanel features a black cross, indicating the kinematic center. The dashed-dotted and dotted lines represent the major and minor kinematic axes, respectively. Synthesized beams are depicted at the bottom-left corner of each panel, and scale bars are provided at the bottom-right corner.}
\figsetgrpend

\figsetgrpstart
\figsetgrpnum{3.17}
\figsetgrptitle{CO kinematics of NGC6478}
\figsetplot{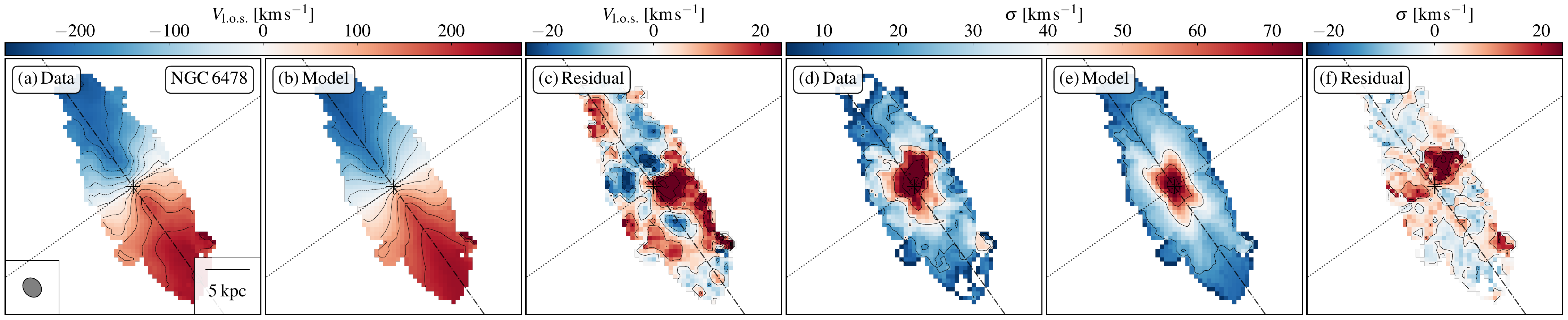}
\figsetgrpnote{The \Bbarolo\  best-fit results for the CO(1--0) emission of an inactive star-forming galaxy, NGC6478. Panels (a) and (b) depict the line-of-sight (los) velocity maps and the corresponding best-fit models. Panel (c) illustrates the residual map, showing the differences between the observation and the best-fit model. Panels (d) and (e) exhibit the velocity dispersion maps, respectively. Panel (f) shows the residual map of the velocity dispersion.
The contours in panels (a) and (b) are plotted starting from $-200\,{\rm km\,s^{-1}}$, with steps of $40\,{\rm km\,s^{-1}}$. The contours in panels (d) and (e) start from $0\,{\rm km\,s^{-1}}$, with steps of $20\,{\rm km\,s^{-1}}$. The residual maps' contours range from $-50\,{\rm km\,s^{-1}}$ to $50\,{\rm km\,s^{-1}}$, in steps of $10\,{\rm km\,s^{-1}}$. Each subpanel features a black cross, indicating the kinematic center. The dashed-dotted and dotted lines represent the major and minor kinematic axes, respectively. Synthesized beams are depicted at the bottom-left corner of each panel, and scale bars are provided at the bottom-right corner.}
\figsetgrpend

\figsetgrpstart
\figsetgrpnum{3.18}
\figsetgrptitle{CO kinematics of UGC09067}
\figsetplot{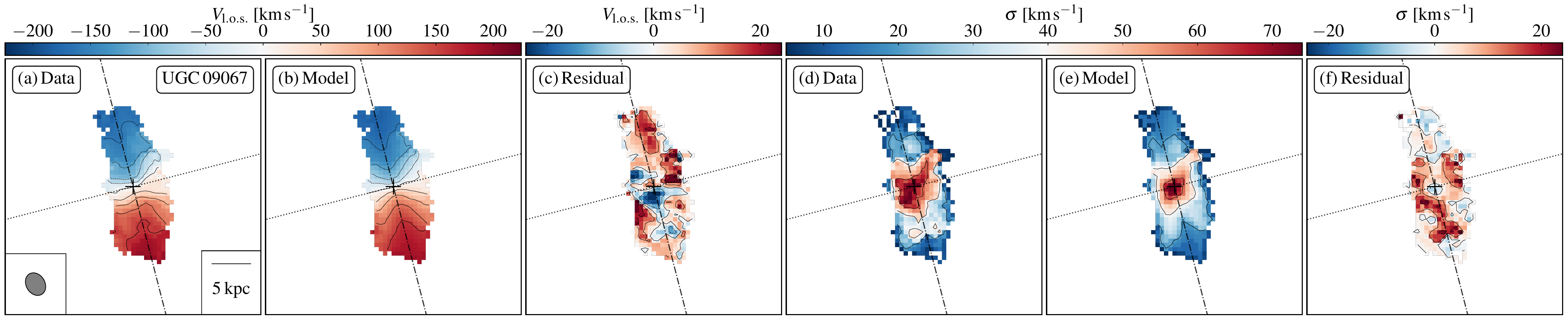}
\figsetgrpnote{The \Bbarolo\  best-fit results for the CO(1--0) emission of an inactive star-forming galaxy, UGC09067. Panels (a) and (b) depict the line-of-sight (los) velocity maps and the corresponding best-fit models. Panel (c) illustrates the residual map, showing the differences between the observation and the best-fit model. Panels (d) and (e) exhibit the velocity dispersion maps, respectively. Panel (f) shows the residual map of the velocity dispersion.
The contours in panels (a) and (b) are plotted starting from $-200\,{\rm km\,s^{-1}}$, with steps of $40\,{\rm km\,s^{-1}}$. The contours in panels (d) and (e) start from $0\,{\rm km\,s^{-1}}$, with steps of $20\,{\rm km\,s^{-1}}$. The residual maps' contours range from $-50\,{\rm km\,s^{-1}}$ to $50\,{\rm km\,s^{-1}}$, in steps of $10\,{\rm km\,s^{-1}}$. Each subpanel features a black cross, indicating the kinematic center. The dashed-dotted and dotted lines represent the major and minor kinematic axes, respectively. Synthesized beams are depicted at the bottom-left corner of each panel, and scale bars are provided at the bottom-right corner.}
\figsetgrpend

\figsetgrpstart
\figsetgrpnum{4.1}
\figsetgrptitle{Ha kinematics of PG0923+129}
\figsetplot{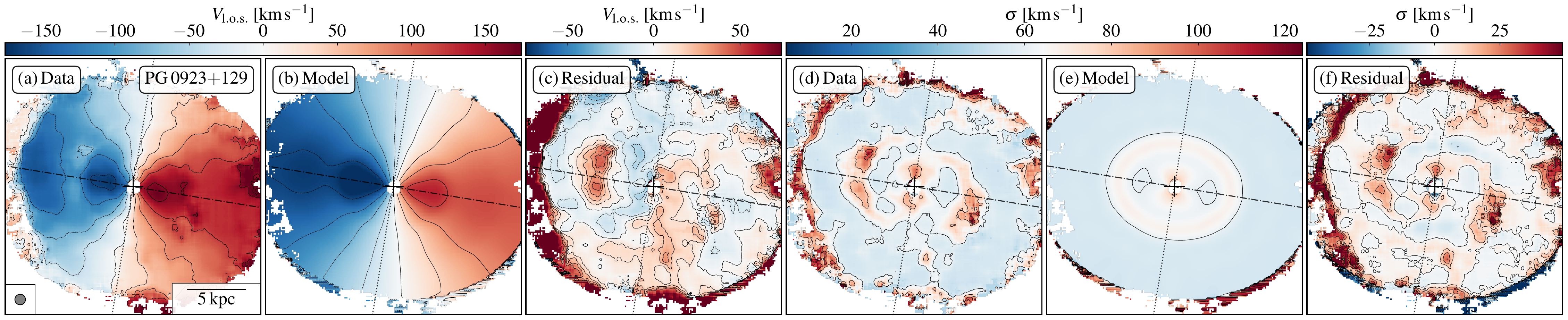}
\figsetgrpnote{The \Bbarolo\ best-fit results for the H$\alpha$ emission of a quasar host galaxy PG\,0923$+$129.}
\figsetgrpend

\figsetgrpstart
\figsetgrpnum{4.2}
\figsetgrptitle{Ha kinematics of PG1126-041}
\figsetplot{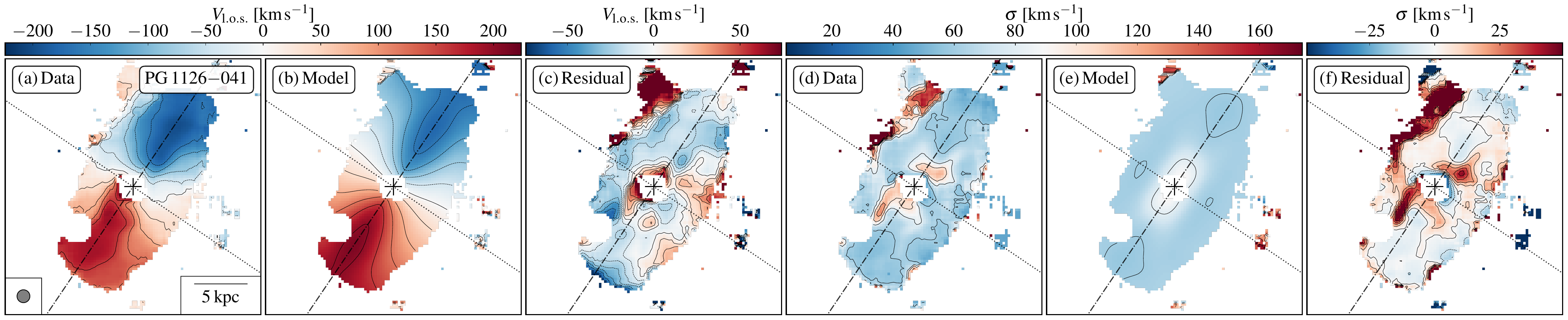}
\figsetgrpnote{The \Bbarolo\ best-fit results for the H$\alpha$ emission of a quasar host galaxy PG\,1126$-$041.}
\figsetgrpend

\figsetgrpstart
\figsetgrpnum{4.3}
\figsetgrptitle{Ha kinematics of PG2130+099}
\figsetplot{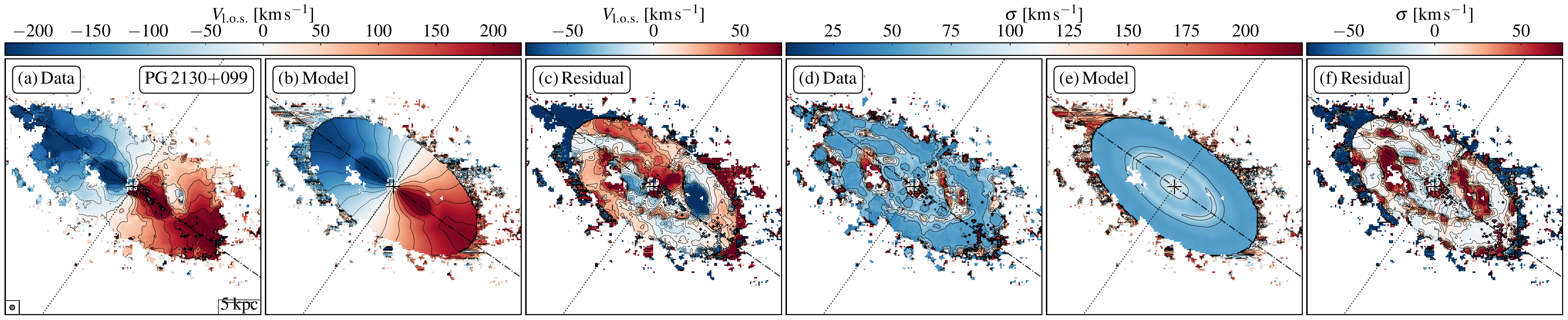}
\figsetgrpnote{The \Bbarolo\ best-fit results for the H$\alpha$ emission of a quasar host galaxy PG\,2130$+$099.}
\figsetgrpend

\figsetgrpstart
\figsetgrpnum{4.4}
\figsetgrptitle{Ha kinematics of IC0944}
\figsetplot{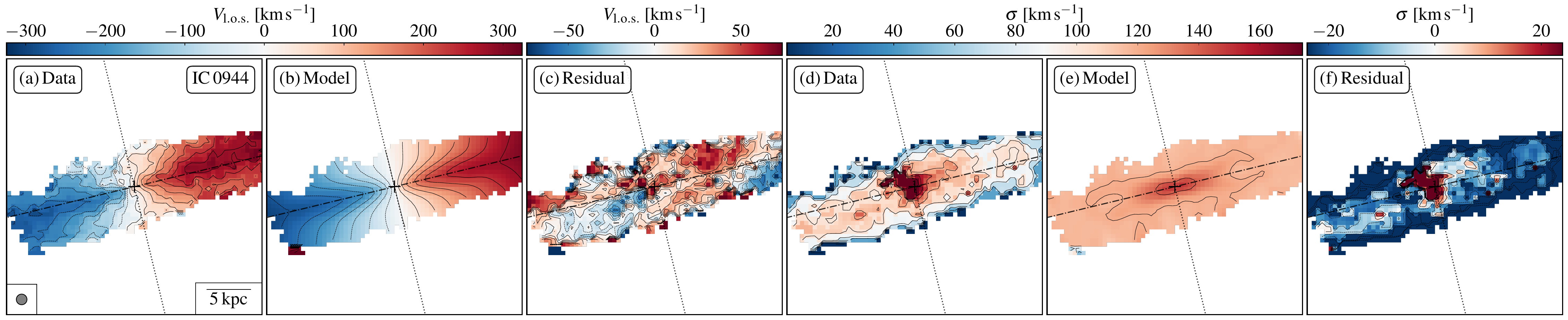}
\figsetgrpnote{The \Bbarolo\ best-fit results for the H$\alpha$ emission of an inactive star-forming galaxy, IC0944.}
\figsetgrpend

\figsetgrpstart
\figsetgrpnum{4.5}
\figsetgrptitle{Ha kinematics of IC1199}
\figsetplot{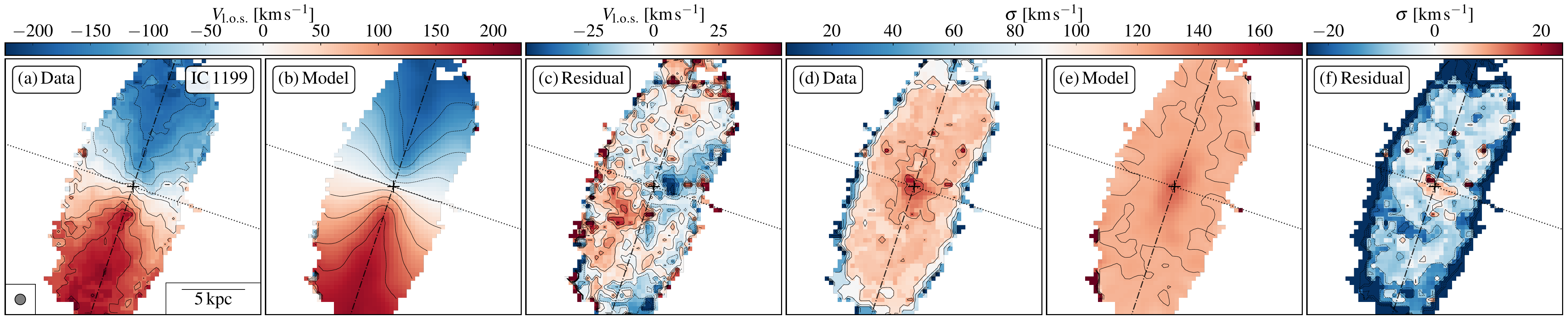}
\figsetgrpnote{The \Bbarolo\ best-fit results for the H$\alpha$ emission of an inactive star-forming galaxy, IC1199.}
\figsetgrpend

\figsetgrpstart
\figsetgrpnum{4.6}
\figsetgrptitle{Ha kinematics of IC1683}
\figsetplot{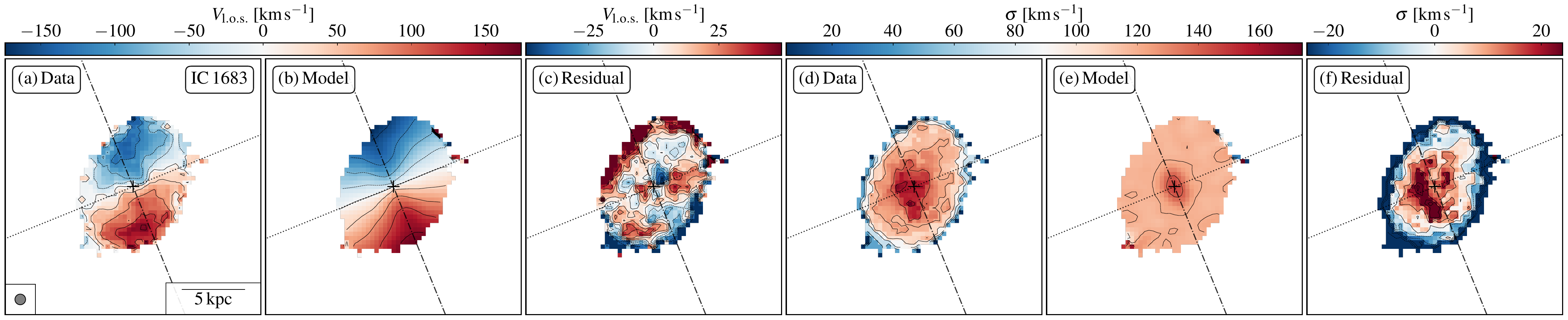}
\figsetgrpnote{The \Bbarolo\ best-fit results for the H$\alpha$ emission of an inactive star-forming galaxy, IC1683.}
\figsetgrpend

\figsetgrpstart
\figsetgrpnum{4.7}
\figsetgrptitle{Ha kinematics of NGC0496}
\figsetplot{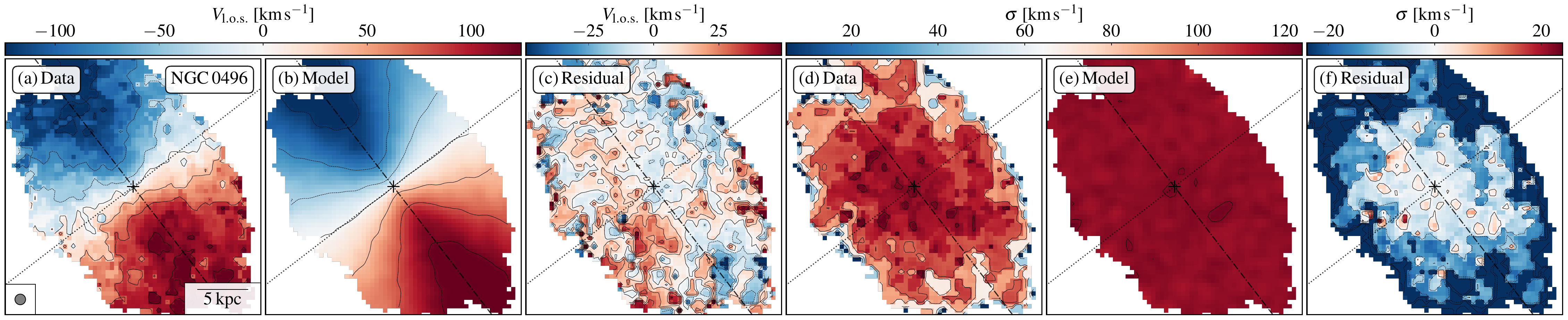}
\figsetgrpnote{The \Bbarolo\ best-fit results for the H$\alpha$ emission of an inactive star-forming galaxy, NGC0496.}
\figsetgrpend

\figsetgrpstart
\figsetgrpnum{4.8}
\figsetgrptitle{Ha kinematics of NGC2906}
\figsetplot{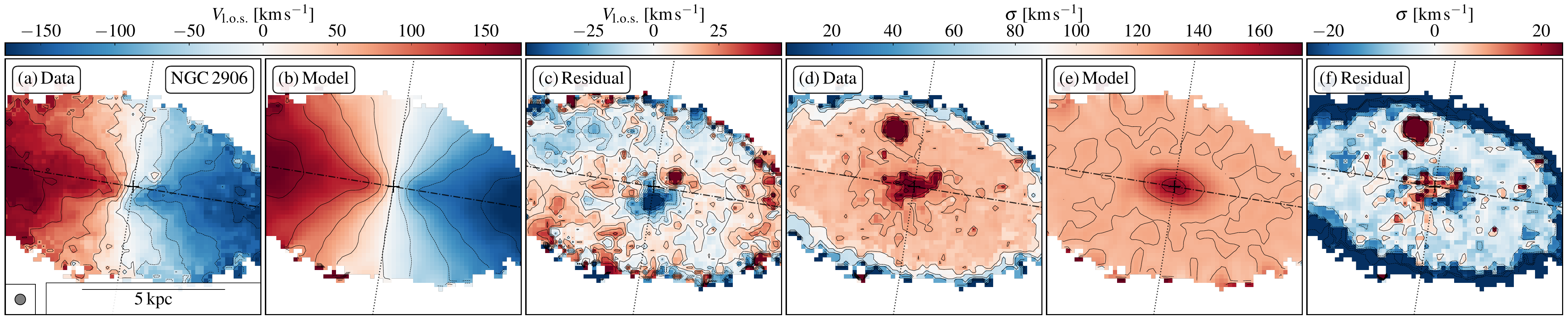}
\figsetgrpnote{The \Bbarolo\ best-fit results for the H$\alpha$ emission of an inactive star-forming galaxy, NGC2906.}
\figsetgrpend

\figsetgrpstart
\figsetgrpnum{4.9}
\figsetgrptitle{Ha kinematics of NGC3994}
\figsetplot{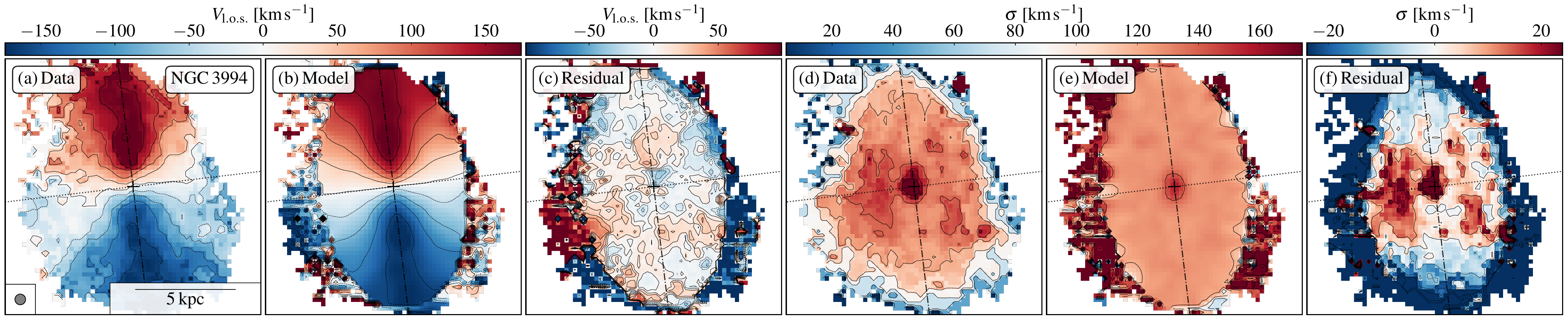}
\figsetgrpnote{The \Bbarolo\ best-fit results for the H$\alpha$ emission of an inactive star-forming galaxy, NGC3994.}
\figsetgrpend

\figsetgrpstart
\figsetgrpnum{4.10}
\figsetgrptitle{Ha kinematics of NGC4047}
\figsetplot{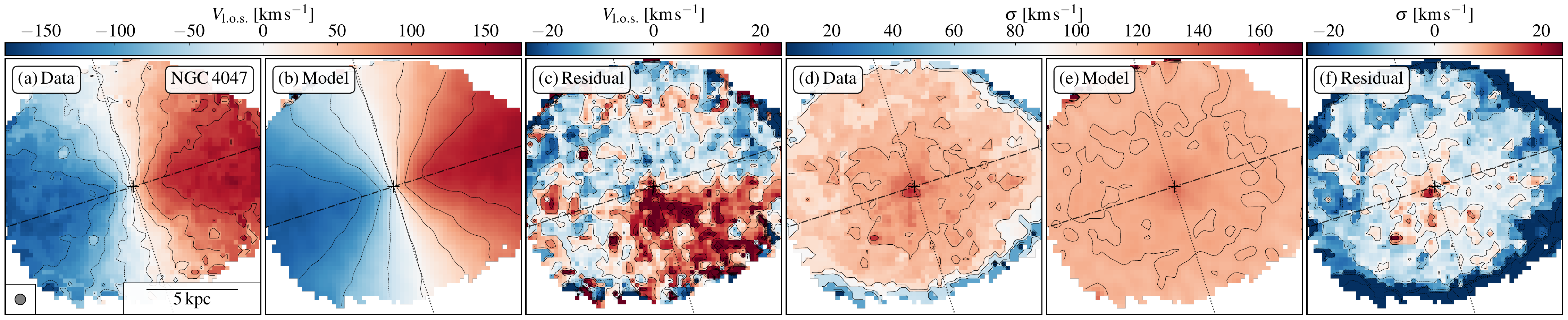}
\figsetgrpnote{The \Bbarolo\ best-fit results for the H$\alpha$ emission of an inactive star-forming galaxy, NGC4047.}
\figsetgrpend

\figsetgrpstart
\figsetgrpnum{4.11}
\figsetgrptitle{Ha kinematics of NGC4644}
\figsetplot{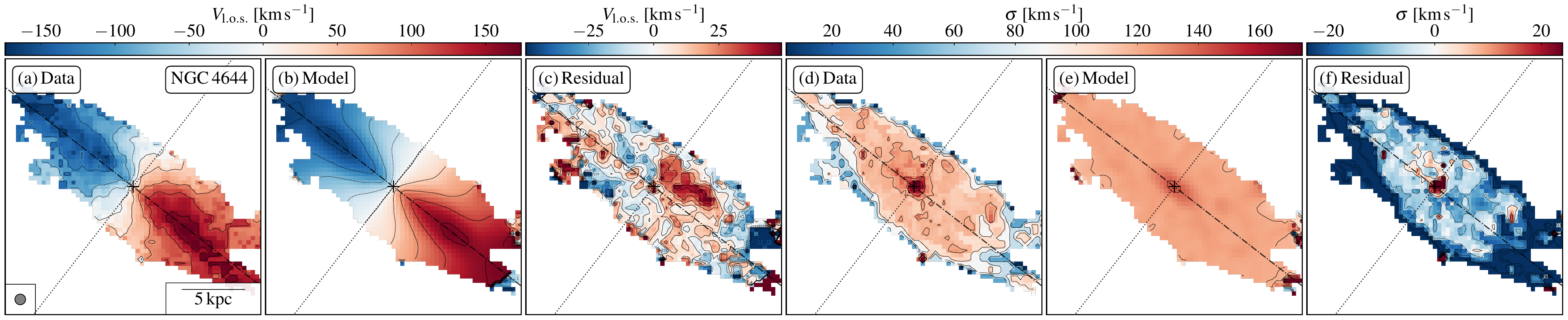}
\figsetgrpnote{The \Bbarolo\ best-fit results for the H$\alpha$ emission of an inactive star-forming galaxy, NGC4644.}
\figsetgrpend

\figsetgrpstart
\figsetgrpnum{4.12}
\figsetgrptitle{Ha kinematics of NGC4711}
\figsetplot{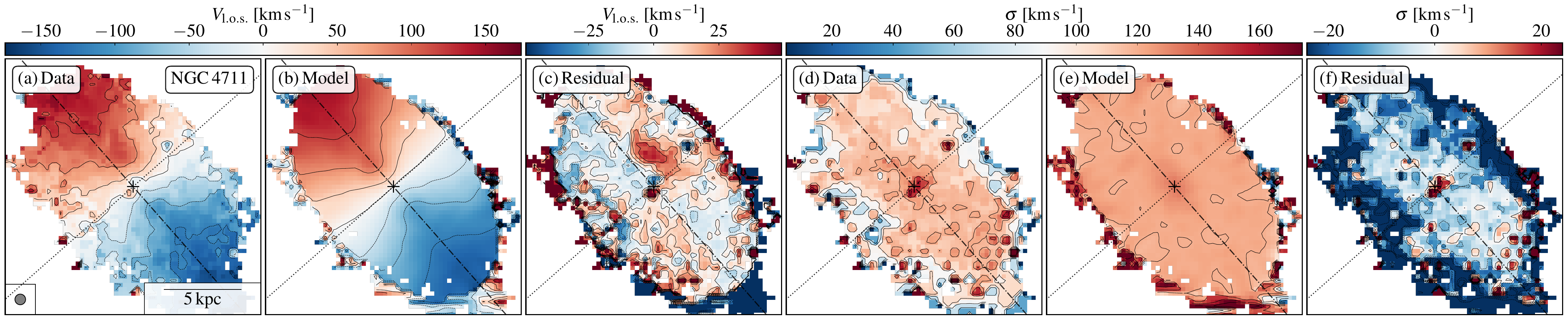}
\figsetgrpnote{The \Bbarolo\ best-fit results for the H$\alpha$ emission of an inactive star-forming galaxy, NGC4711.}
\figsetgrpend

\figsetgrpstart
\figsetgrpnum{4.13}
\figsetgrptitle{Ha kinematics of NGC5480}
\figsetplot{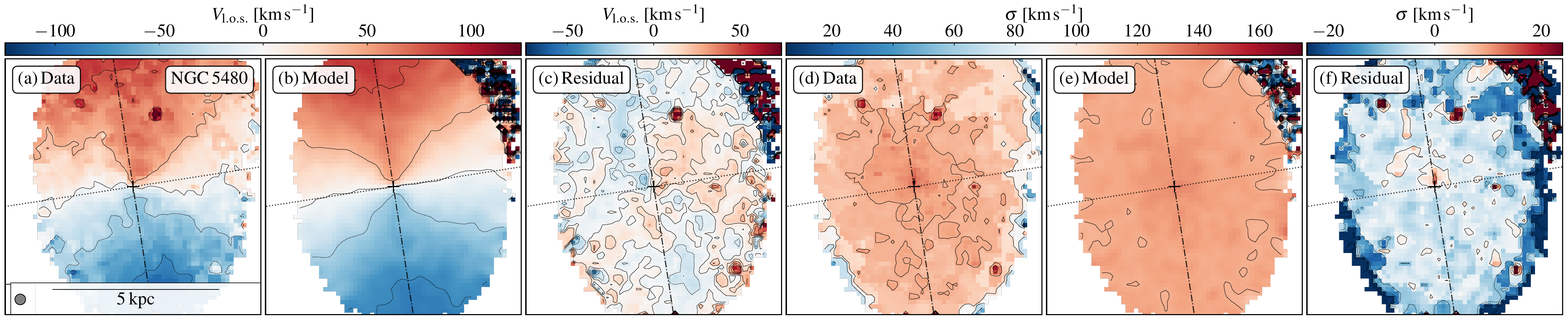}
\figsetgrpnote{The \Bbarolo\ best-fit results for the H$\alpha$ emission of an inactive star-forming galaxy, NGC5480.}
\figsetgrpend

\figsetgrpstart
\figsetgrpnum{4.14}
\figsetgrptitle{Ha kinematics of NGC5980}
\figsetplot{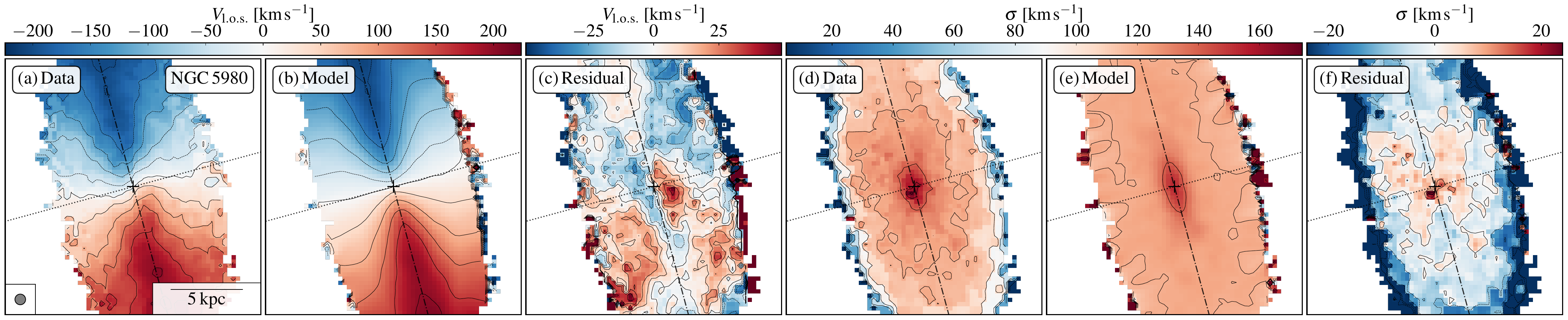}
\figsetgrpnote{The \Bbarolo\ best-fit results for the H$\alpha$ emission of an inactive star-forming galaxy, NGC5980.}
\figsetgrpend

\figsetgrpstart
\figsetgrpnum{4.15}
\figsetgrptitle{Ha kinematics of NGC6060}
\figsetplot{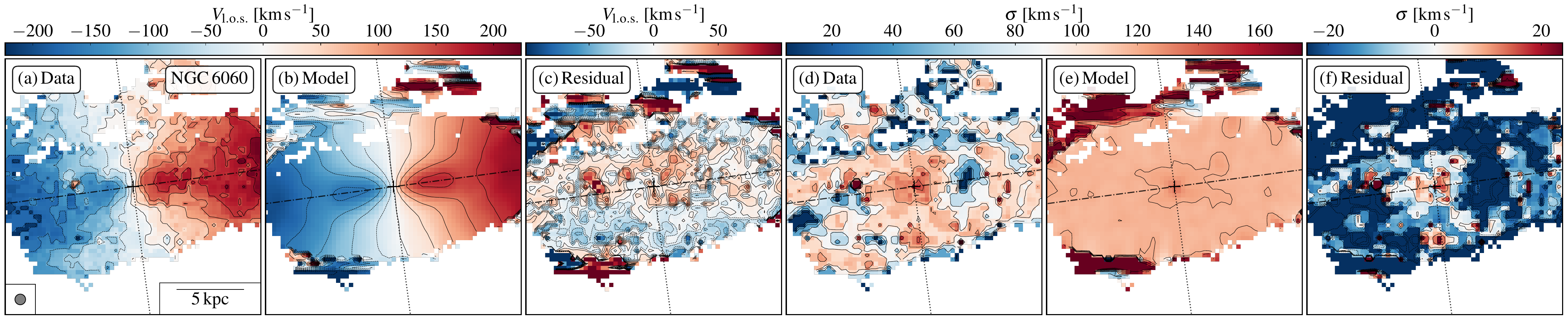}
\figsetgrpnote{The \Bbarolo\ best-fit results for the H$\alpha$ emission of an inactive star-forming galaxy, NGC6060.}
\figsetgrpend

\figsetgrpstart
\figsetgrpnum{4.16}
\figsetgrptitle{Ha kinematics of NGC6301}
\figsetplot{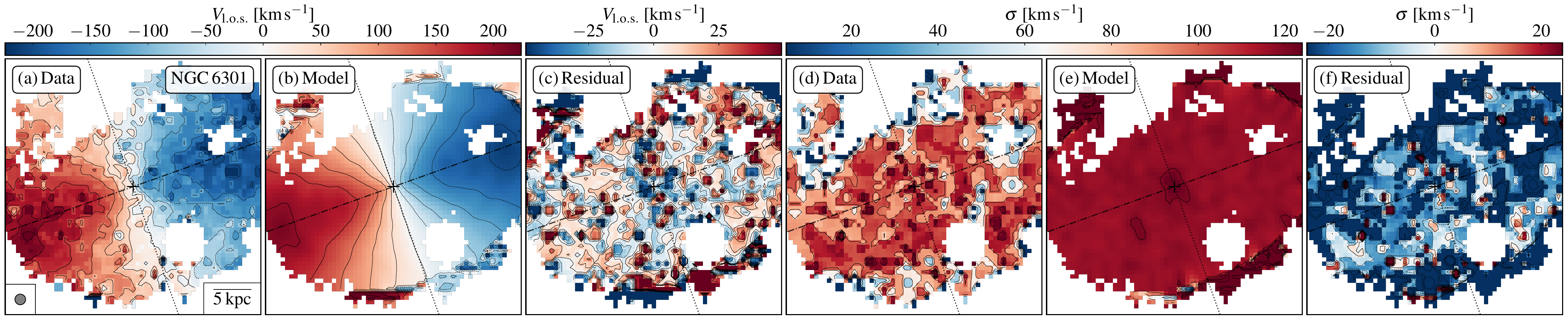}
\figsetgrpnote{The \Bbarolo\ best-fit results for the H$\alpha$ emission of an inactive star-forming galaxy, NGC6301.}
\figsetgrpend

\figsetgrpstart
\figsetgrpnum{4.17}
\figsetgrptitle{Ha kinematics of NGC6478}
\figsetplot{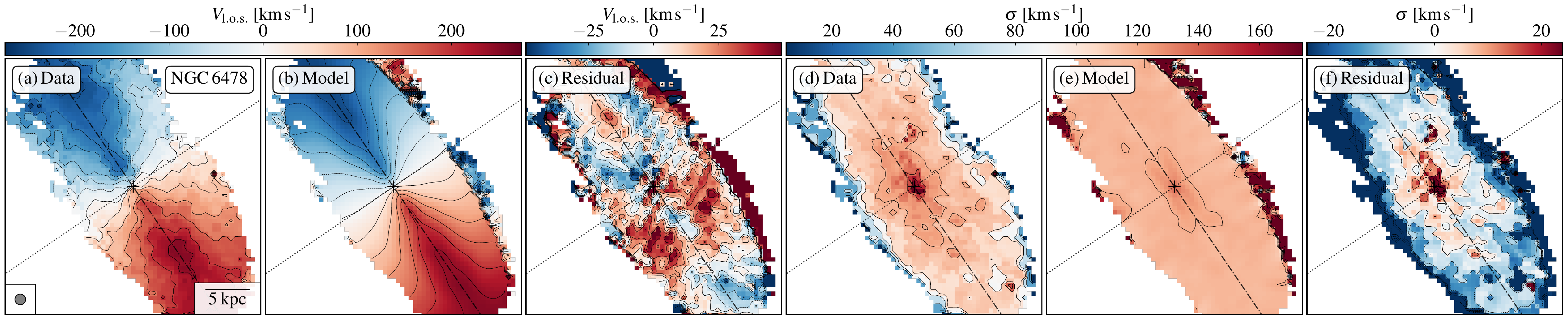}
\figsetgrpnote{The \Bbarolo\ best-fit results for the H$\alpha$ emission of an inactive star-forming galaxy, NGC6478.}
\figsetgrpend

\figsetgrpstart
\figsetgrpnum{4.18}
\figsetgrptitle{Ha kinematics of UGC09067}
\figsetplot{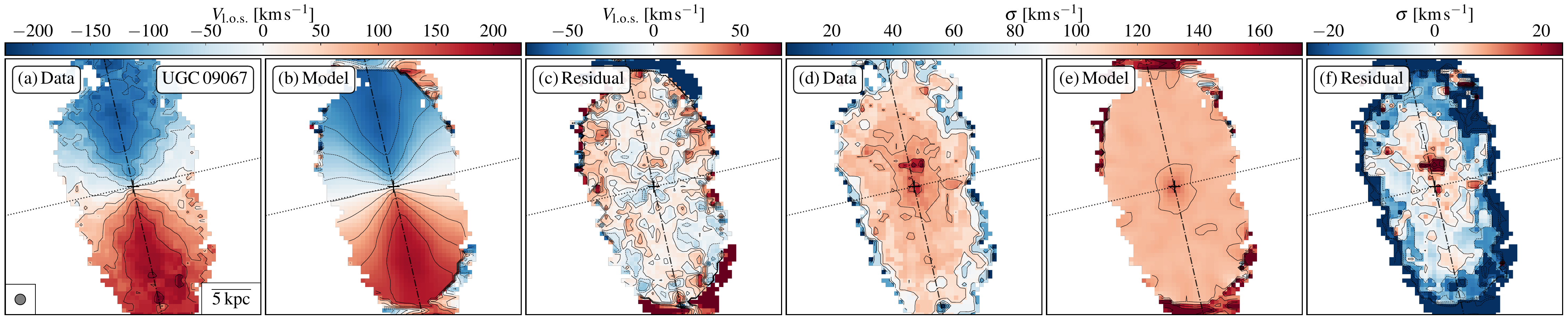}
\figsetgrpnote{The \Bbarolo\ best-fit results for the H$\alpha$ emission of an inactive star-forming galaxy, UGC09067.}
\figsetgrpend

\figsetgrpstart
\figsetgrpnum{5.1}
\figsetgrptitle{Radial profile of PG0923+129}
\figsetplot{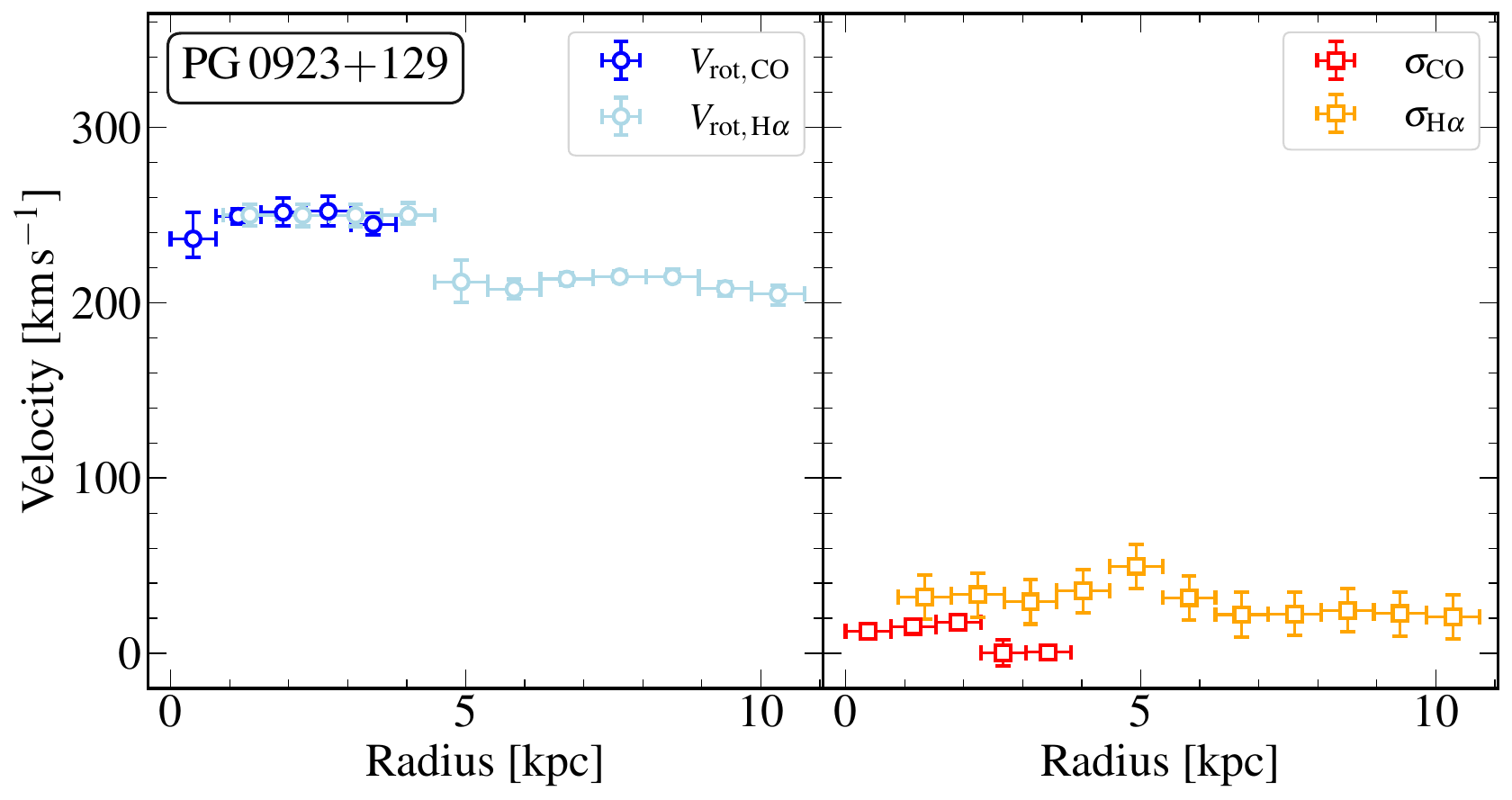}
\figsetgrpnote{The rotation velocity and velocity dispersion obtained through the \Bbarolo\  analysis of the CO and H$\alpha$ data cubes for a quasar host galaxy, PG\,0923$+$129. The rotation velocities derived from the CO and H$\alpha$ are represented by blue and lightblue circles. The velocity dispersions for CO and H$\alpha$ are denoted by red and orange squares, respectively. The horizontal errorbar represents the width of the ring used in kinematic modeling.}
\figsetgrpend

\figsetgrpstart
\figsetgrpnum{5.2}
\figsetgrptitle{Radial profile of PG1126-041}
\figsetplot{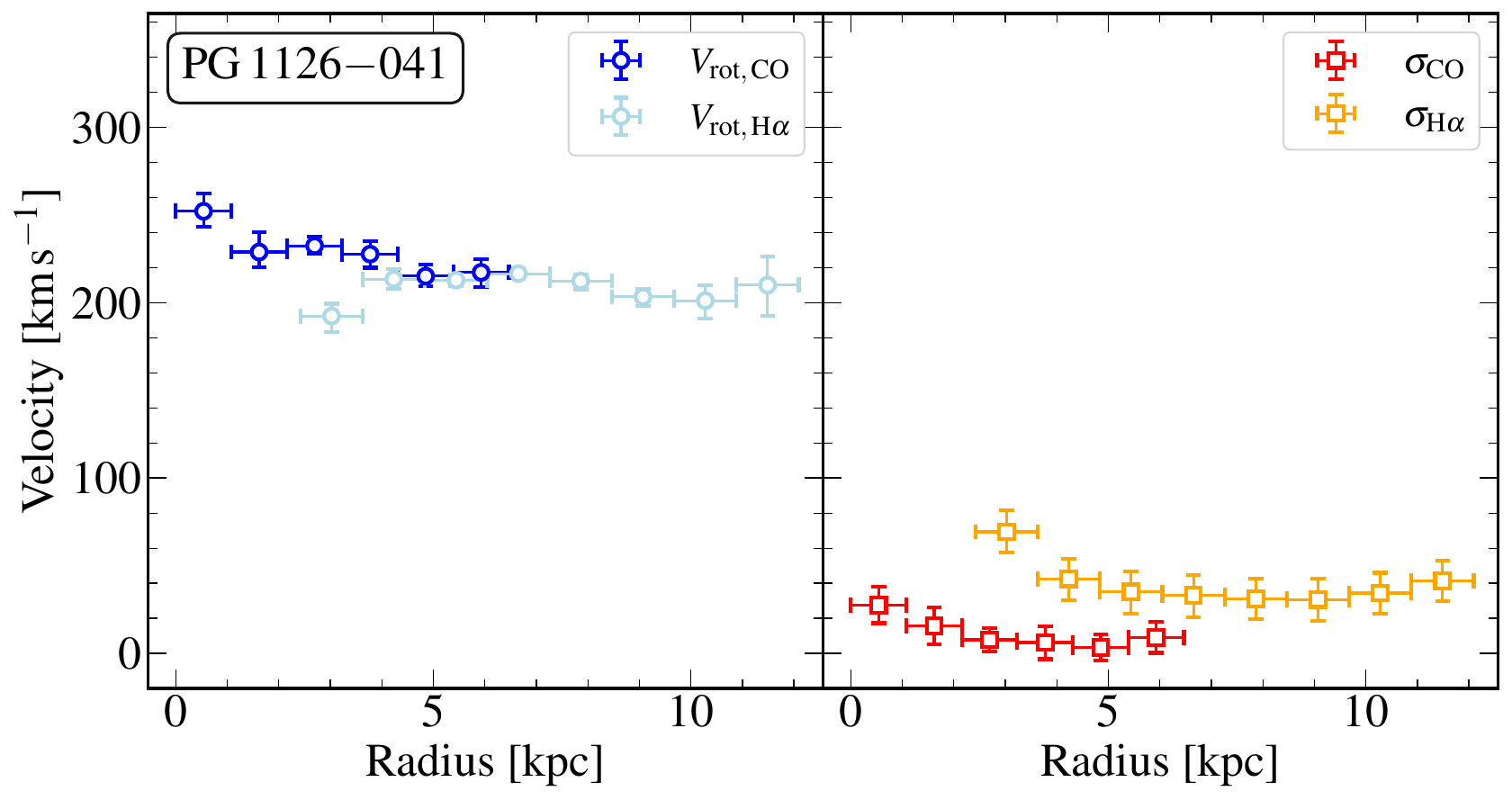}
\figsetgrpnote{The rotation velocity and velocity dispersion obtained through the \Bbarolo\  analysis of the CO and H$\alpha$ data cubes for a quasar host galaxy, PG\,1126$-$041. The rotation velocities derived from the CO and H$\alpha$ are represented by blue and lightblue circles. The velocity dispersions for CO and H$\alpha$ are denoted by red and orange squares, respectively. The horizontal errorbar represents the width of the ring used in kinematic modeling.}
\figsetgrpend

\figsetgrpstart
\figsetgrpnum{5.3}
\figsetgrptitle{Radial profile of PG2130+099}
\figsetplot{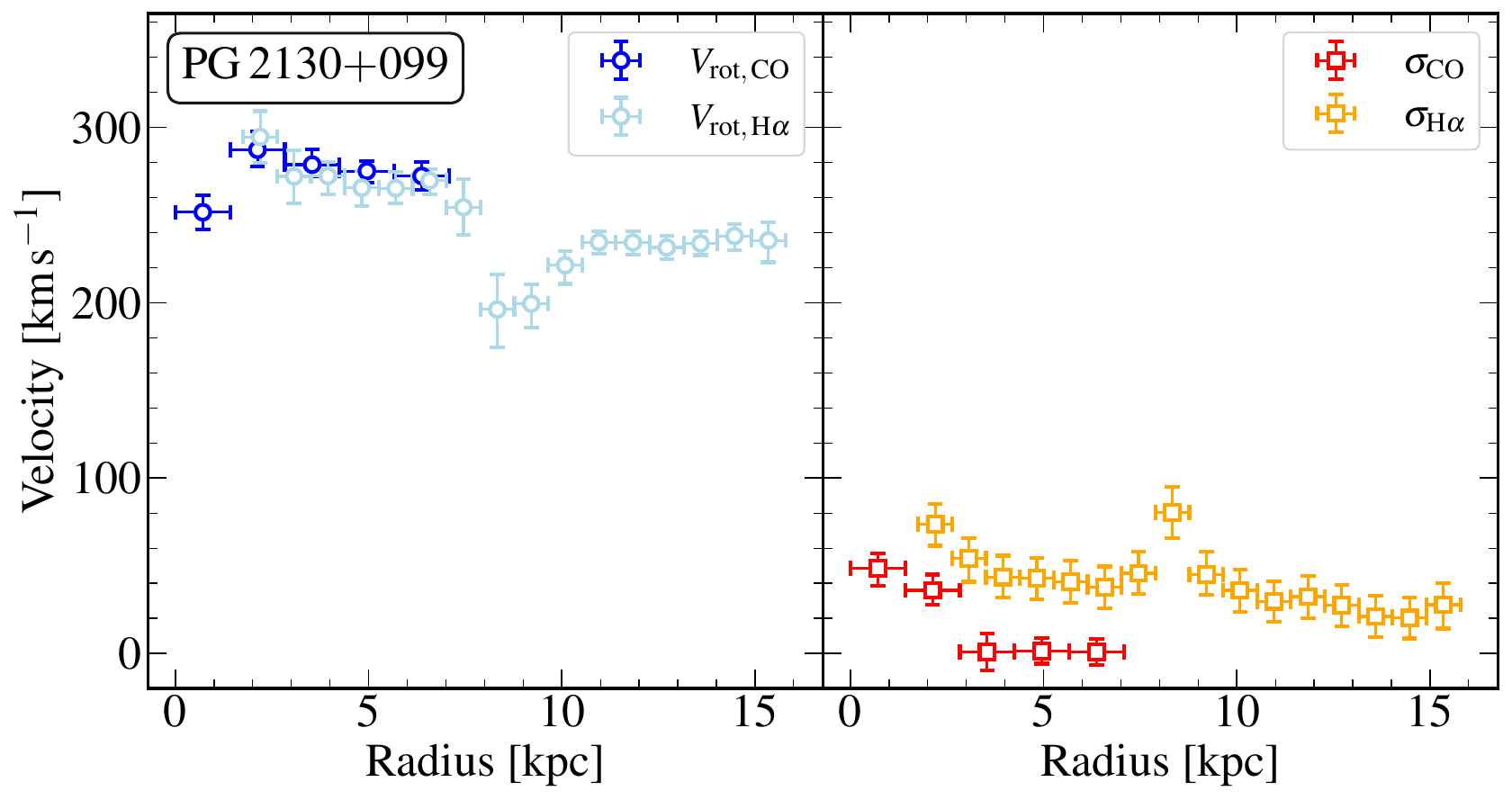}
\figsetgrpnote{The rotation velocity and velocity dispersion obtained through the \Bbarolo\  analysis of the CO and H$\alpha$ data cubes for a quasar host galaxy, PG\,2130$+$099. The rotation velocities derived from the CO and H$\alpha$ are represented by blue and lightblue circles. The velocity dispersions for CO and H$\alpha$ are denoted by red and orange squares, respectively. The horizontal errorbar represents the width of the ring used in kinematic modeling.}
\figsetgrpend

\figsetgrpstart
\figsetgrpnum{5.4}
\figsetgrptitle{Radial profile of IC0944}
\figsetplot{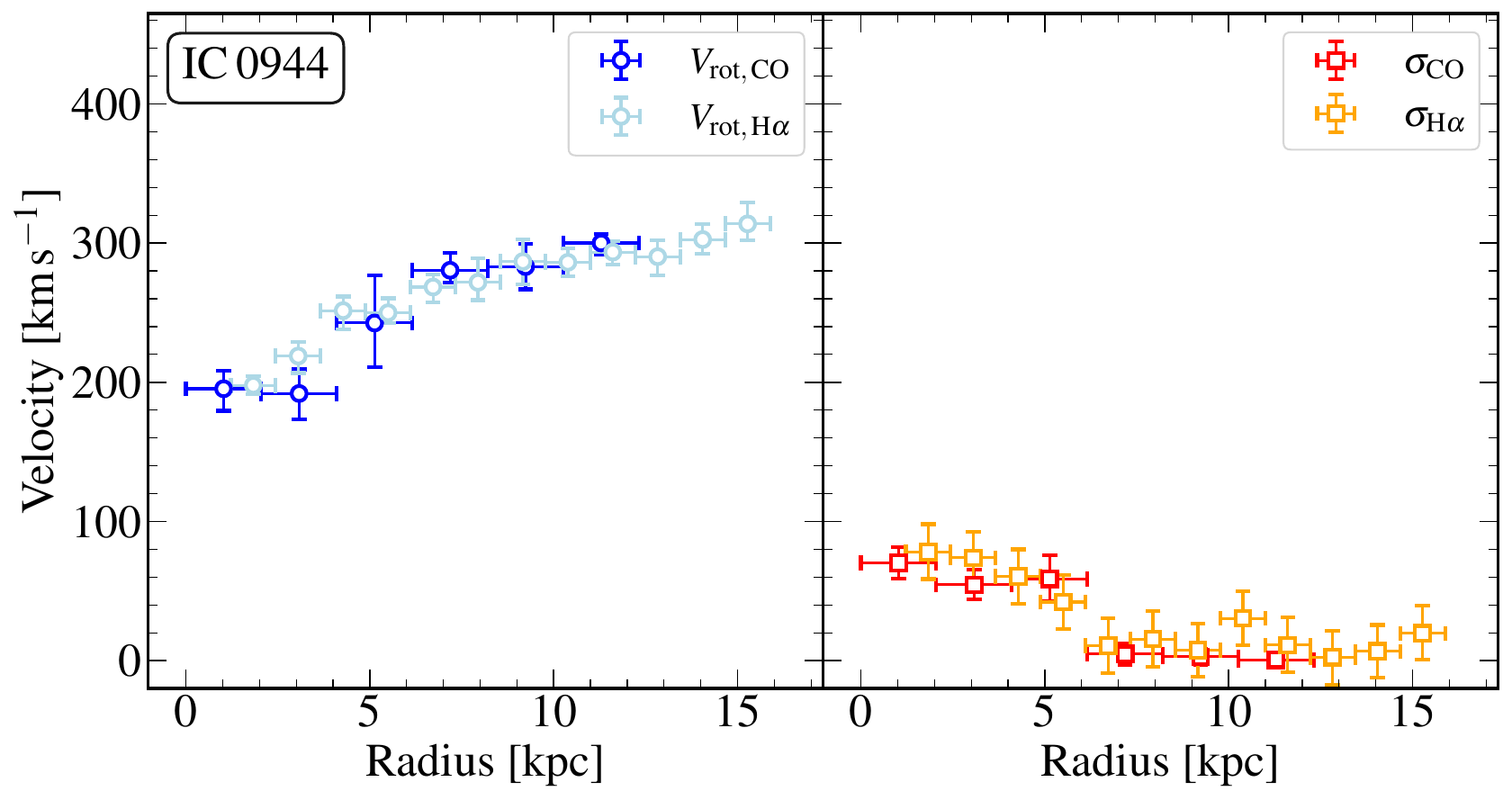}
\figsetgrpnote{The rotation velocity and velocity dispersion obtained through the \Bbarolo\  analysis of the CO and H$\alpha$ data cubes for an inactive star-forming galaxy, IC0944. The rotation velocities derived from the CO and H$\alpha$ are represented by blue and lightblue circles. The velocity dispersions for CO and H$\alpha$ are denoted by red and orange squares, respectively. The horizontal errorbar represents the width of the ring used in kinematic modeling.}
\figsetgrpend

\figsetgrpstart
\figsetgrpnum{5.5}
\figsetgrptitle{Radial profile of IC1199}
\figsetplot{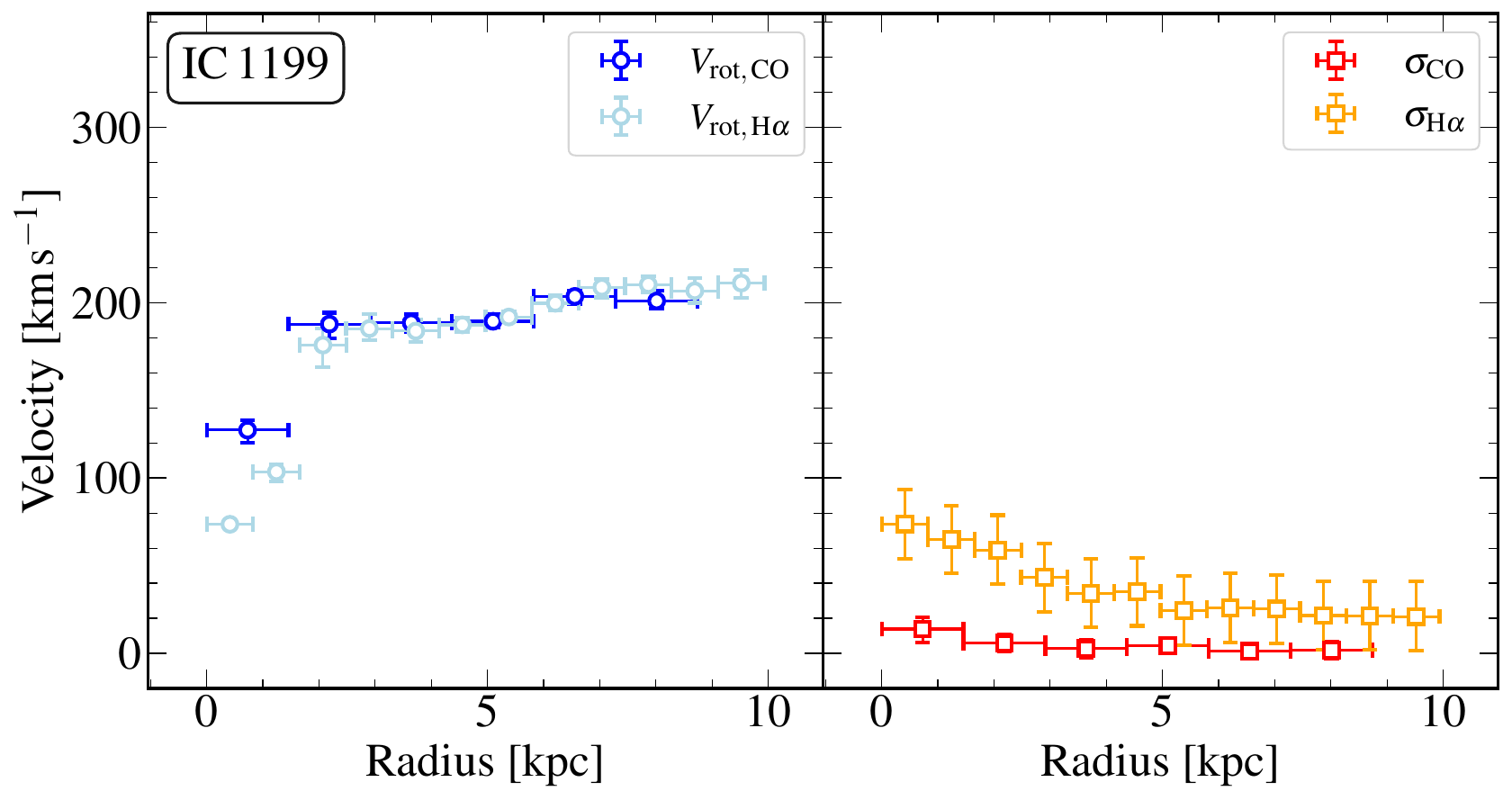}
\figsetgrpnote{The rotation velocity and velocity dispersion obtained through the \Bbarolo\  analysis of the CO and H$\alpha$ data cubes for an inactive star-forming galaxy, IC1199. The rotation velocities derived from the CO and H$\alpha$ are represented by blue and lightblue circles. The velocity dispersions for CO and H$\alpha$ are denoted by red and orange squares, respectively. The horizontal errorbar represents the width of the ring used in kinematic modeling.}
\figsetgrpend

\figsetgrpstart
\figsetgrpnum{5.6}
\figsetgrptitle{Radial profile of IC1683}
\figsetplot{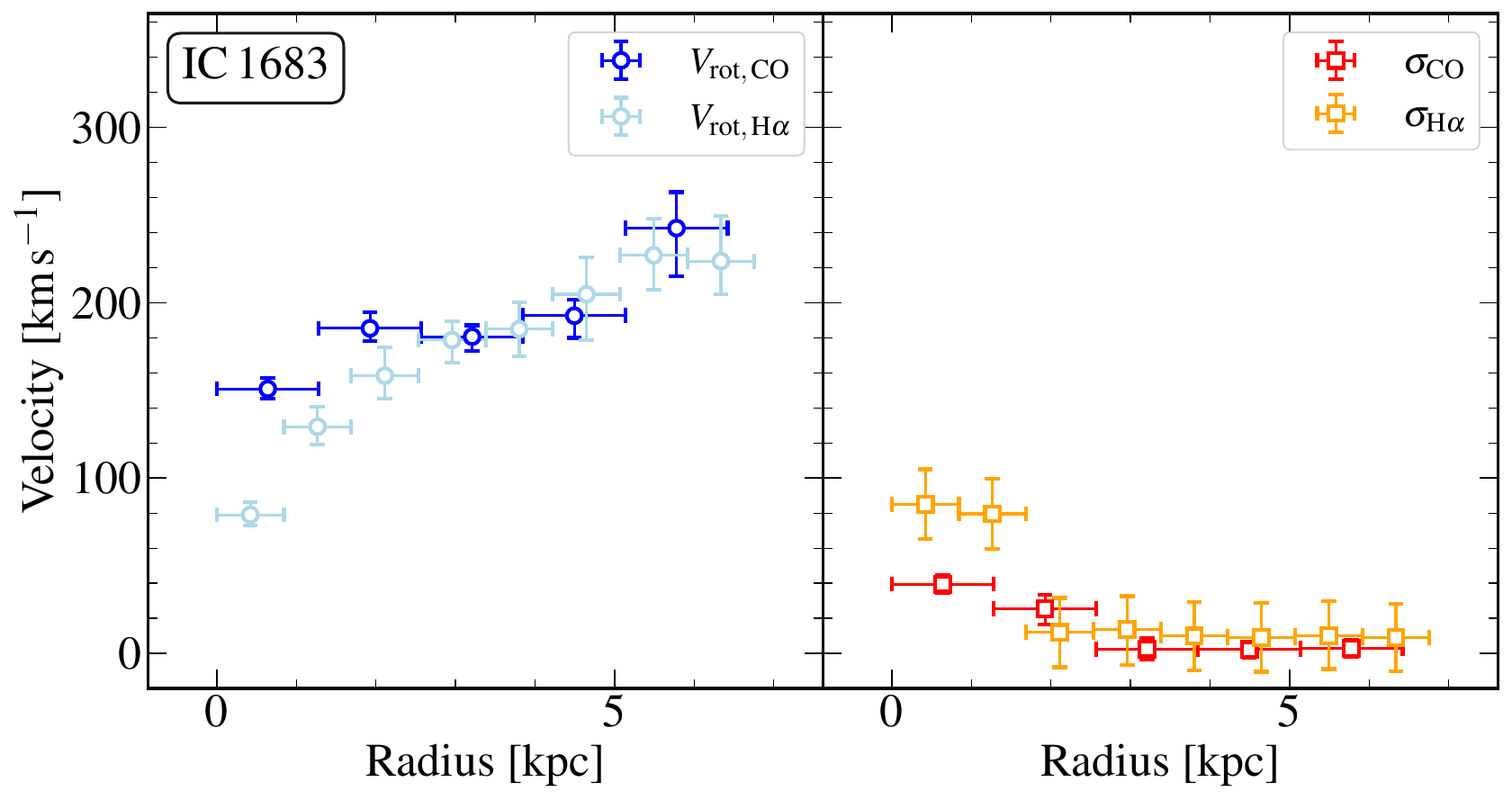}
\figsetgrpnote{The rotation velocity and velocity dispersion obtained through the \Bbarolo\  analysis of the CO and H$\alpha$ data cubes for an inactive star-forming galaxy, IC1683. The rotation velocities derived from the CO and H$\alpha$ are represented by blue and lightblue circles. The velocity dispersions for CO and H$\alpha$ are denoted by red and orange squares, respectively. The horizontal errorbar represents the width of the ring used in kinematic modeling.}
\figsetgrpend

\figsetgrpstart
\figsetgrpnum{5.7}
\figsetgrptitle{Radial profile of NGC0496}
\figsetplot{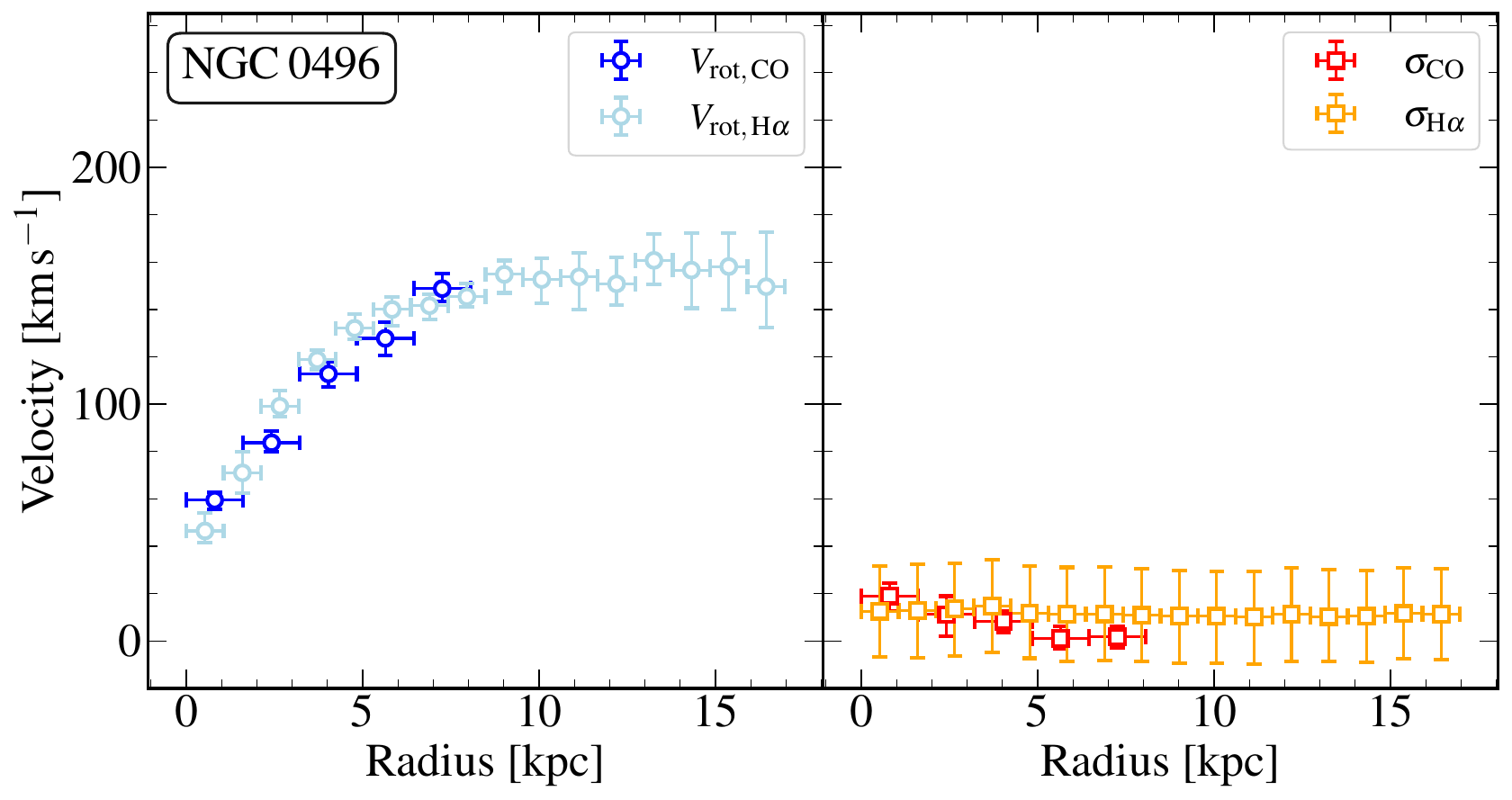}
\figsetgrpnote{The rotation velocity and velocity dispersion obtained through the \Bbarolo\  analysis of the CO and H$\alpha$ data cubes for an inactive star-forming galaxy, NGC0496. The rotation velocities derived from the CO and H$\alpha$ are represented by blue and lightblue circles. The velocity dispersions for CO and H$\alpha$ are denoted by red and orange squares, respectively. The horizontal errorbar represents the width of the ring used in kinematic modeling.}
\figsetgrpend

\figsetgrpstart
\figsetgrpnum{5.8}
\figsetgrptitle{Radial profile of NGC2906}
\figsetplot{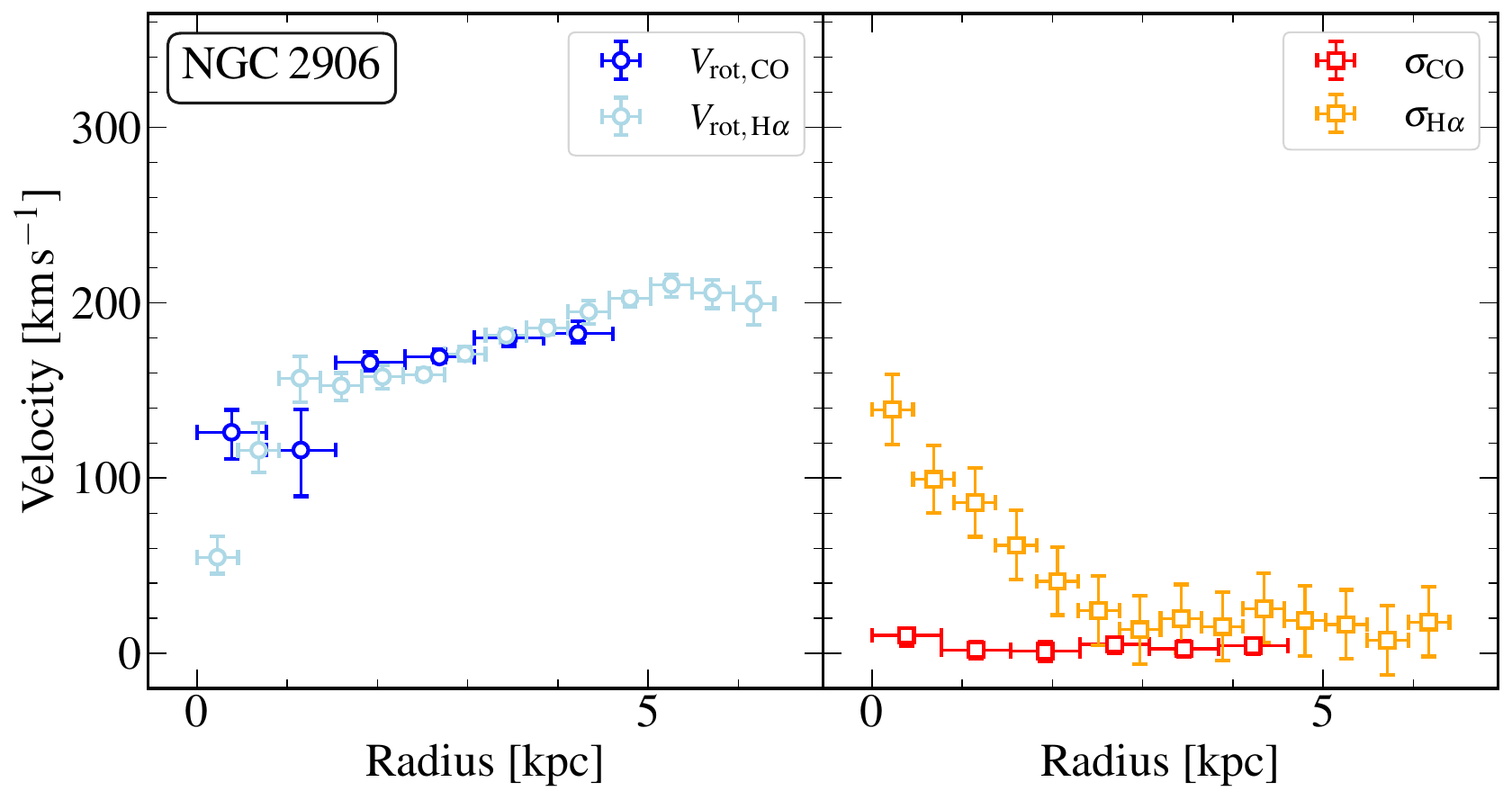}
\figsetgrpnote{The rotation velocity and velocity dispersion obtained through the \Bbarolo\  analysis of the CO and H$\alpha$ data cubes for an inactive star-forming galaxy, NGC2906. The rotation velocities derived from the CO and H$\alpha$ are represented by blue and lightblue circles. The velocity dispersions for CO and H$\alpha$ are denoted by red and orange squares, respectively. The horizontal errorbar represents the width of the ring used in kinematic modeling.}
\figsetgrpend

\figsetgrpstart
\figsetgrpnum{5.9}
\figsetgrptitle{Radial profile of NGC3994}
\figsetplot{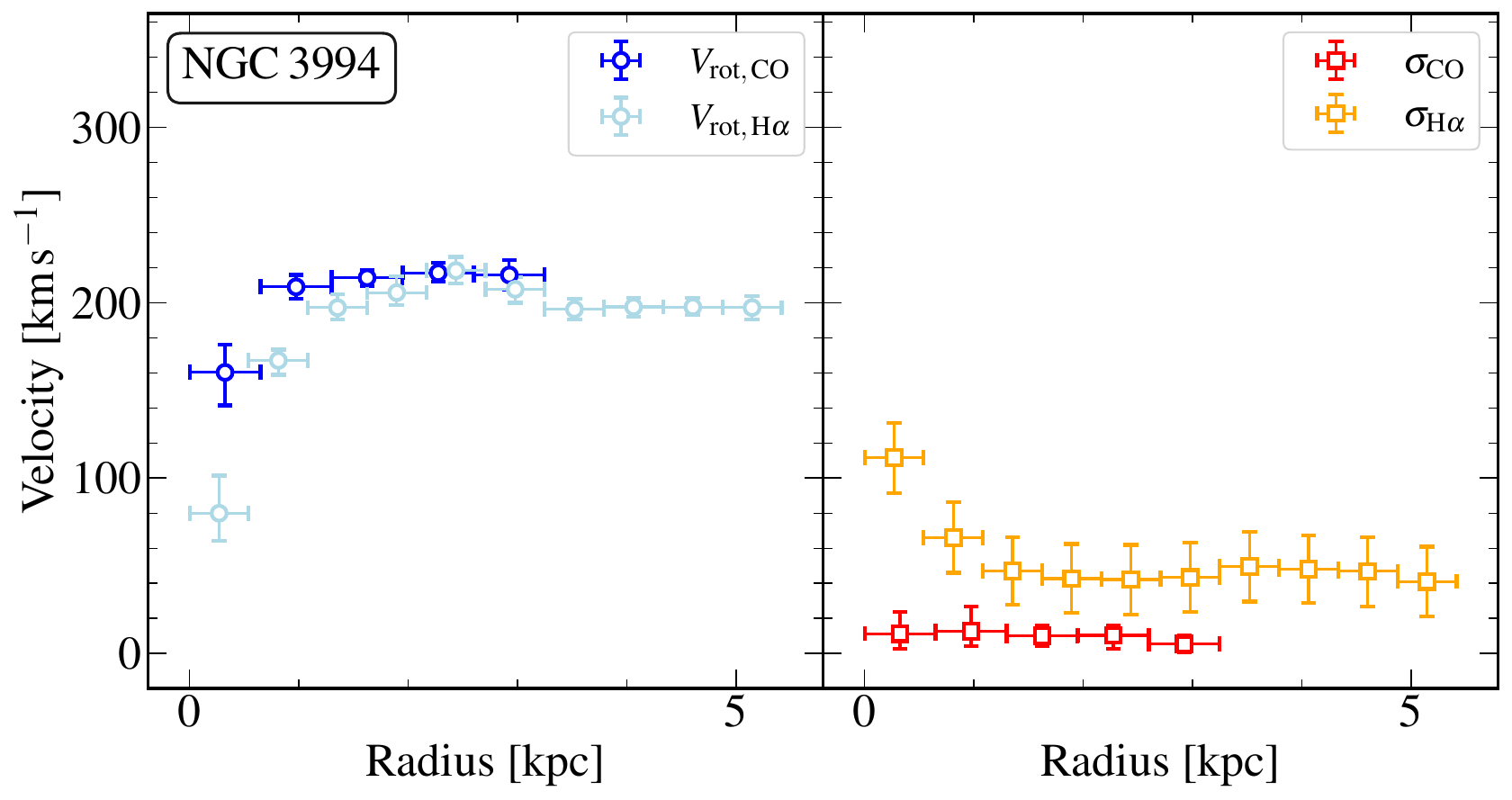}
\figsetgrpnote{The rotation velocity and velocity dispersion obtained through the \Bbarolo\  analysis of the CO and H$\alpha$ data cubes for an inactive star-forming galaxy, NGC3994. The rotation velocities derived from the CO and H$\alpha$ are represented by blue and lightblue circles. The velocity dispersions for CO and H$\alpha$ are denoted by red and orange squares, respectively. The horizontal errorbar represents the width of the ring used in kinematic modeling.}
\figsetgrpend

\figsetgrpstart
\figsetgrpnum{5.10}
\figsetgrptitle{Radial profile of NGC4047}
\figsetplot{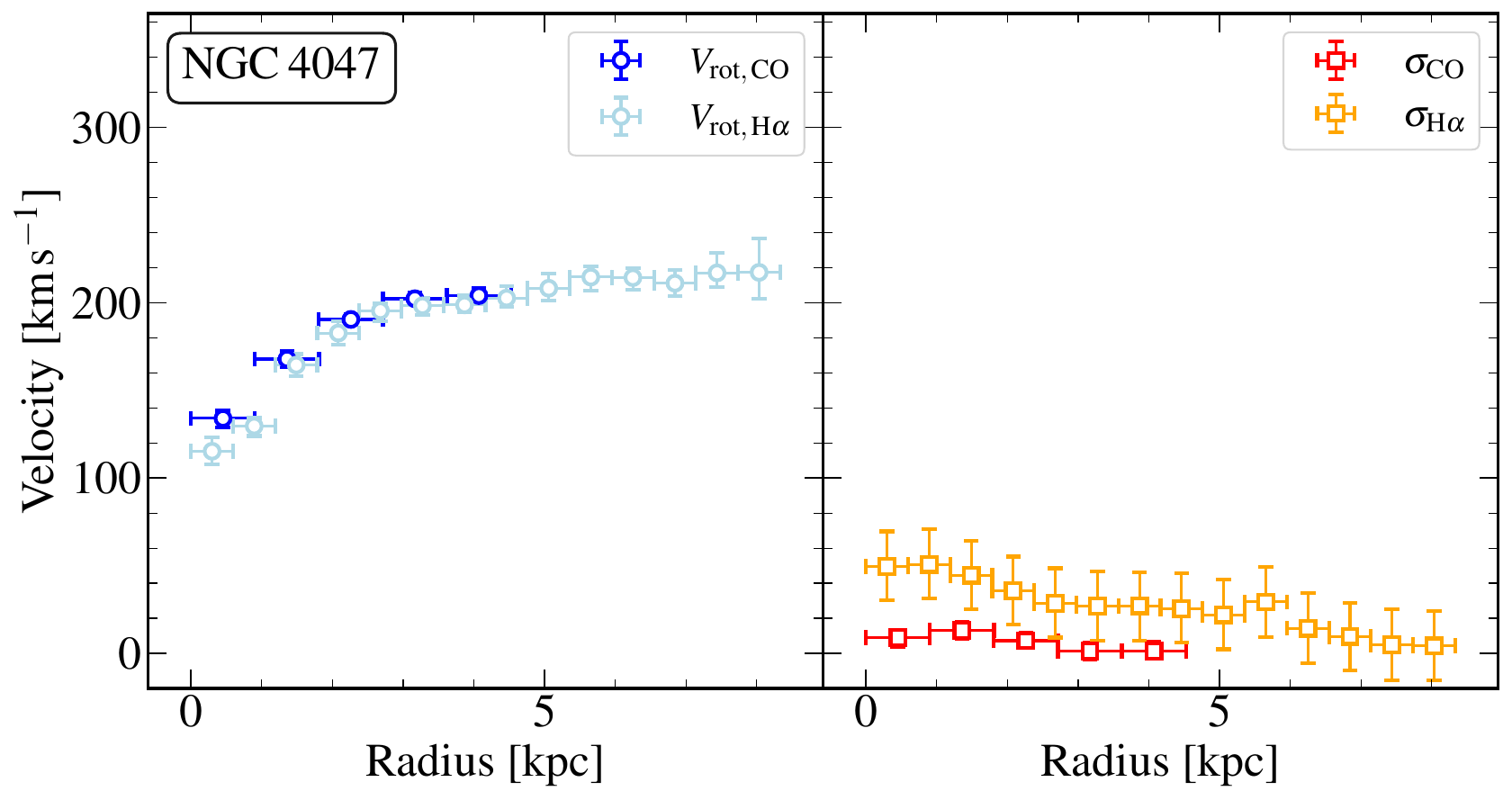}
\figsetgrpnote{The rotation velocity and velocity dispersion obtained through the \Bbarolo\  analysis of the CO and H$\alpha$ data cubes for an inactive star-forming galaxy, NGC4047. The rotation velocities derived from the CO and H$\alpha$ are represented by blue and lightblue circles. The velocity dispersions for CO and H$\alpha$ are denoted by red and orange squares, respectively. The horizontal errorbar represents the width of the ring used in kinematic modeling.}
\figsetgrpend

\figsetgrpstart
\figsetgrpnum{5.11}
\figsetgrptitle{Radial profile of NGC4644}
\figsetplot{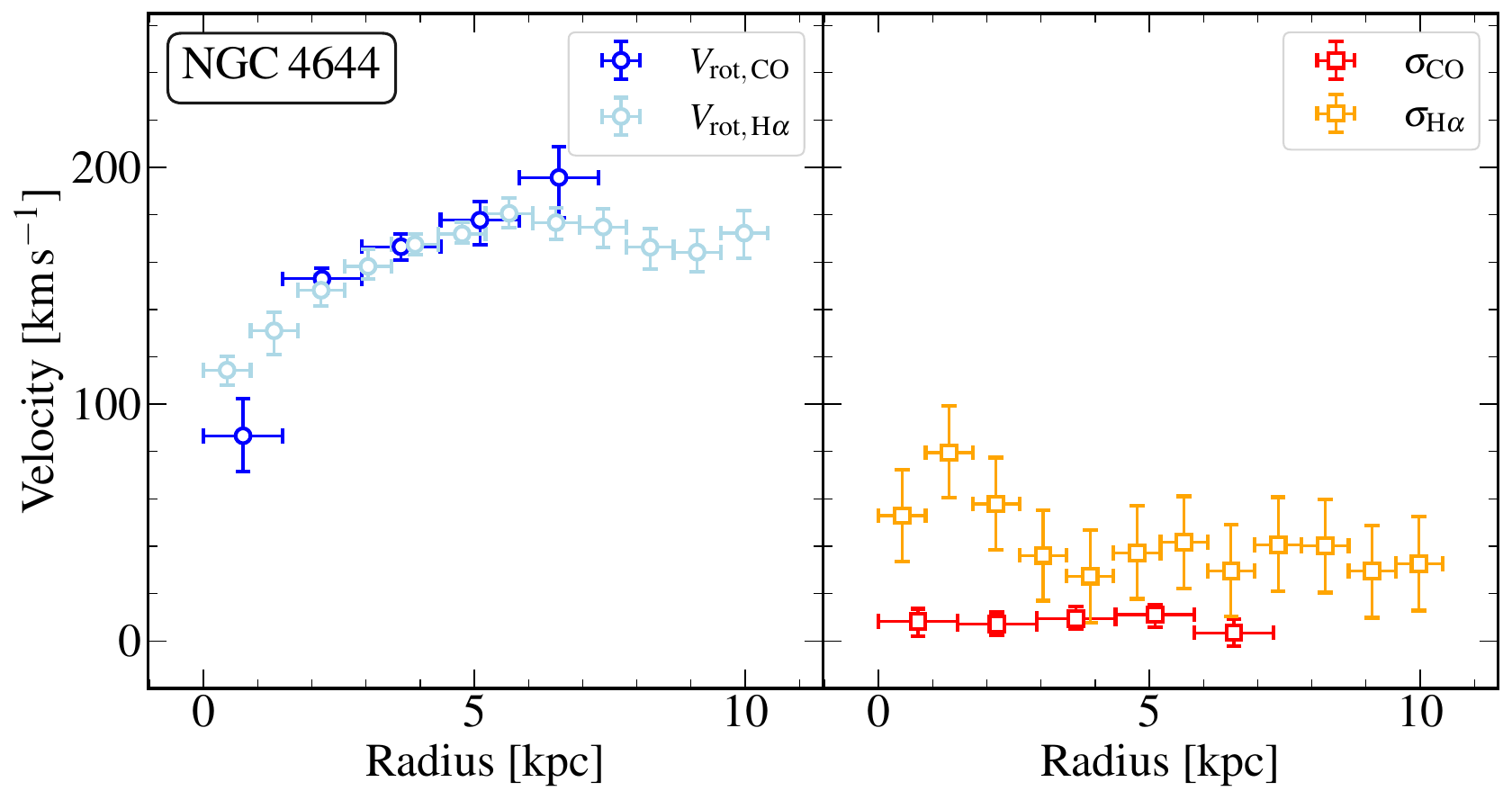}
\figsetgrpnote{The rotation velocity and velocity dispersion obtained through the \Bbarolo\  analysis of the CO and H$\alpha$ data cubes for an inactive star-forming galaxy, NGC4644. The rotation velocities derived from the CO and H$\alpha$ are represented by blue and lightblue circles. The velocity dispersions for CO and H$\alpha$ are denoted by red and orange squares, respectively. The horizontal errorbar represents the width of the ring used in kinematic modeling.}
\figsetgrpend

\figsetgrpstart
\figsetgrpnum{5.12}
\figsetgrptitle{Radial profile of NGC4711}
\figsetplot{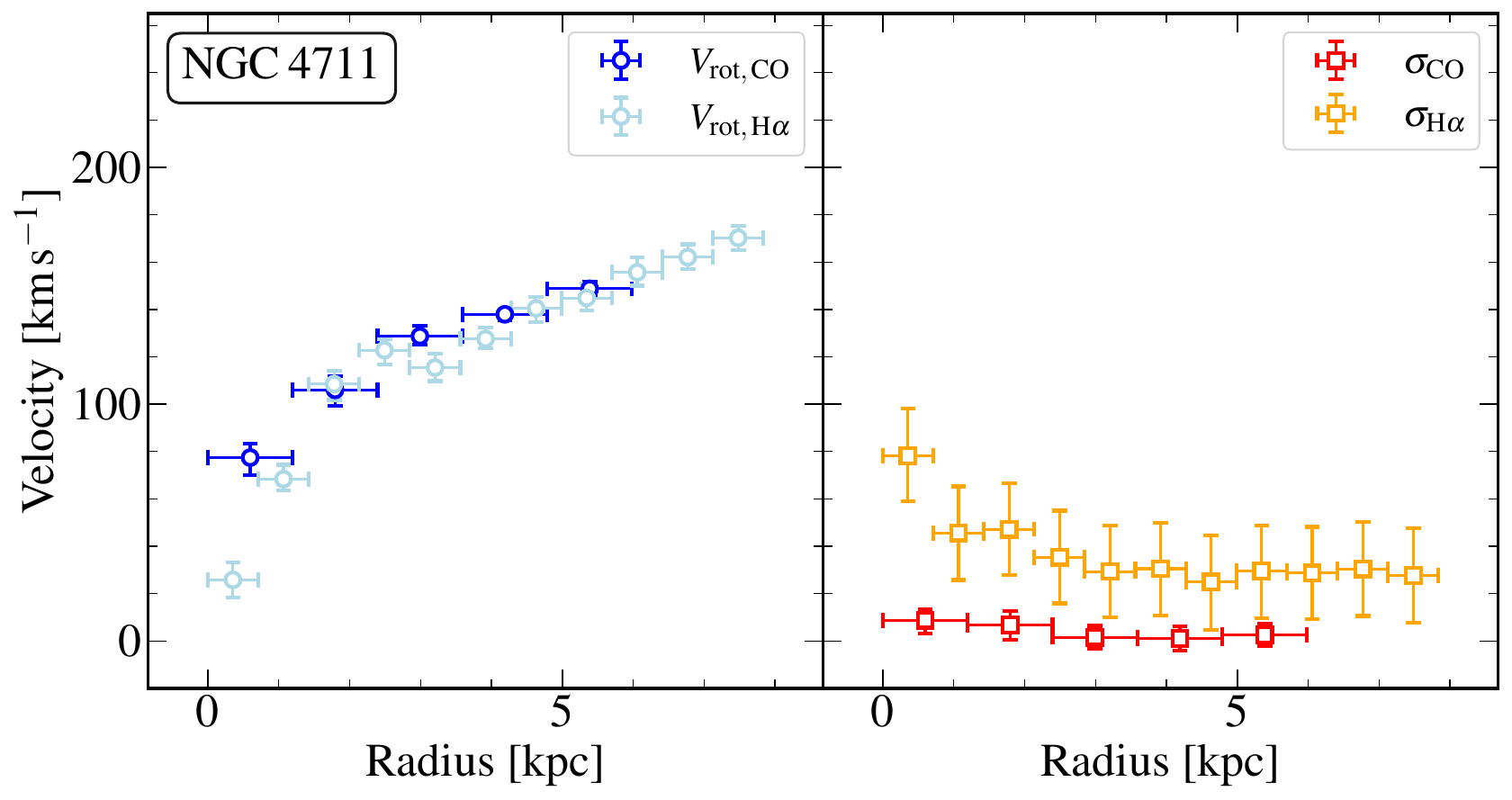}
\figsetgrpnote{The rotation velocity and velocity dispersion obtained through the \Bbarolo\  analysis of the CO and H$\alpha$ data cubes for an inactive star-forming galaxy, NGC4711. The rotation velocities derived from the CO and H$\alpha$ are represented by blue and lightblue circles. The velocity dispersions for CO and H$\alpha$ are denoted by red and orange squares, respectively. The horizontal errorbar represents the width of the ring used in kinematic modeling.}
\figsetgrpend

\figsetgrpstart
\figsetgrpnum{5.13}
\figsetgrptitle{Radial profile of NGC5480}
\figsetplot{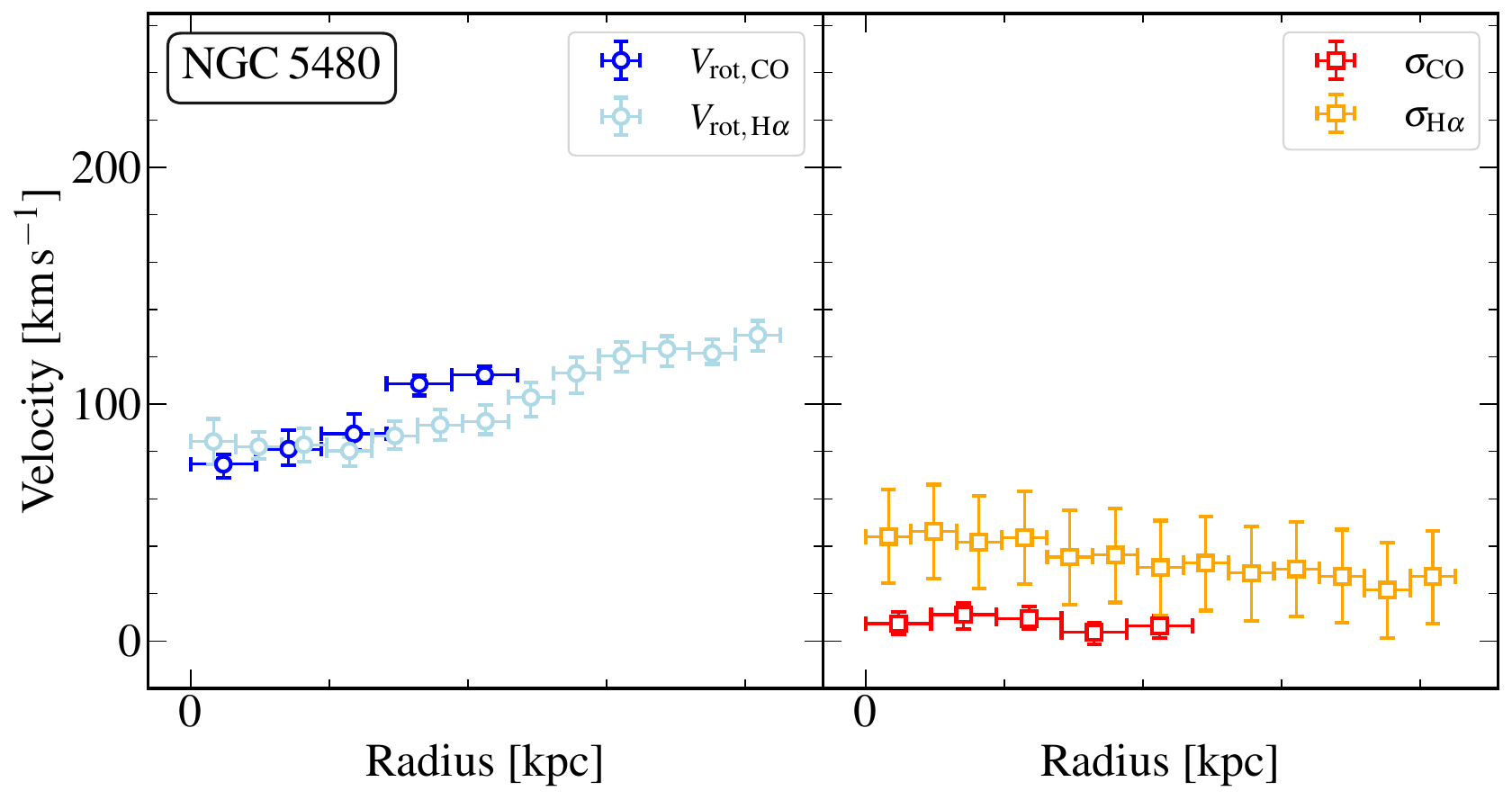}
\figsetgrpnote{The rotation velocity and velocity dispersion obtained through the \Bbarolo\  analysis of the CO and H$\alpha$ data cubes for an inactive star-forming galaxy, NGC5480. The rotation velocities derived from the CO and H$\alpha$ are represented by blue and lightblue circles. The velocity dispersions for CO and H$\alpha$ are denoted by red and orange squares, respectively. The horizontal errorbar represents the width of the ring used in kinematic modeling.}
\figsetgrpend

\figsetgrpstart
\figsetgrpnum{5.14}
\figsetgrptitle{Radial profile of NGC5980}
\figsetplot{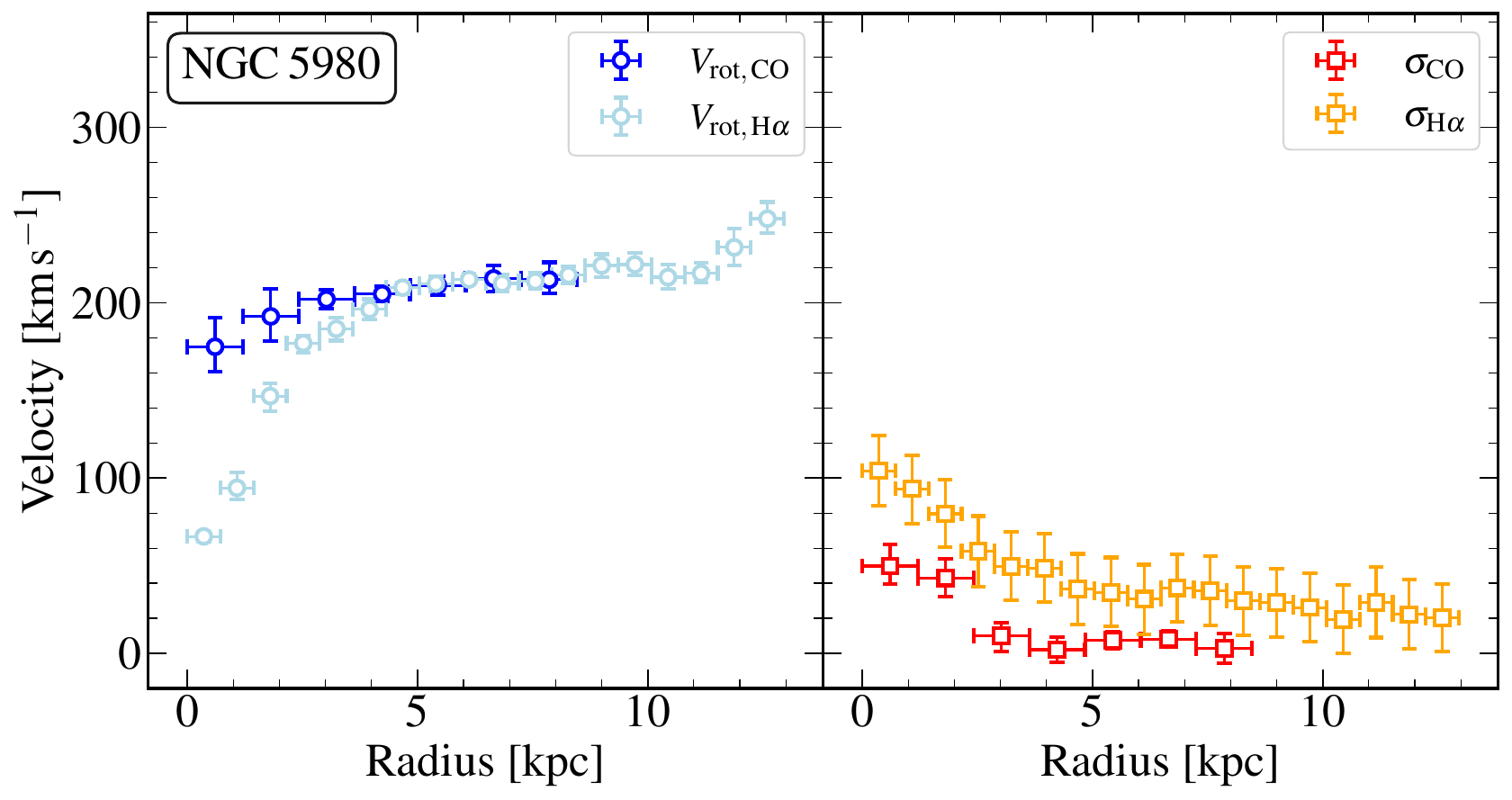}
\figsetgrpnote{The rotation velocity and velocity dispersion obtained through the \Bbarolo\  analysis of the CO and H$\alpha$ data cubes for an inactive star-forming galaxy, Radial profile of NGC5980. The rotation velocities derived from the CO and H$\alpha$ are represented by blue and lightblue circles. The velocity dispersions for CO and H$\alpha$ are denoted by red and orange squares, respectively. The horizontal errorbar represents the width of the ring used in kinematic modeling.}
\figsetgrpend

\figsetgrpstart
\figsetgrpnum{5.15}
\figsetgrptitle{Radial profile of NGC6060}
\figsetplot{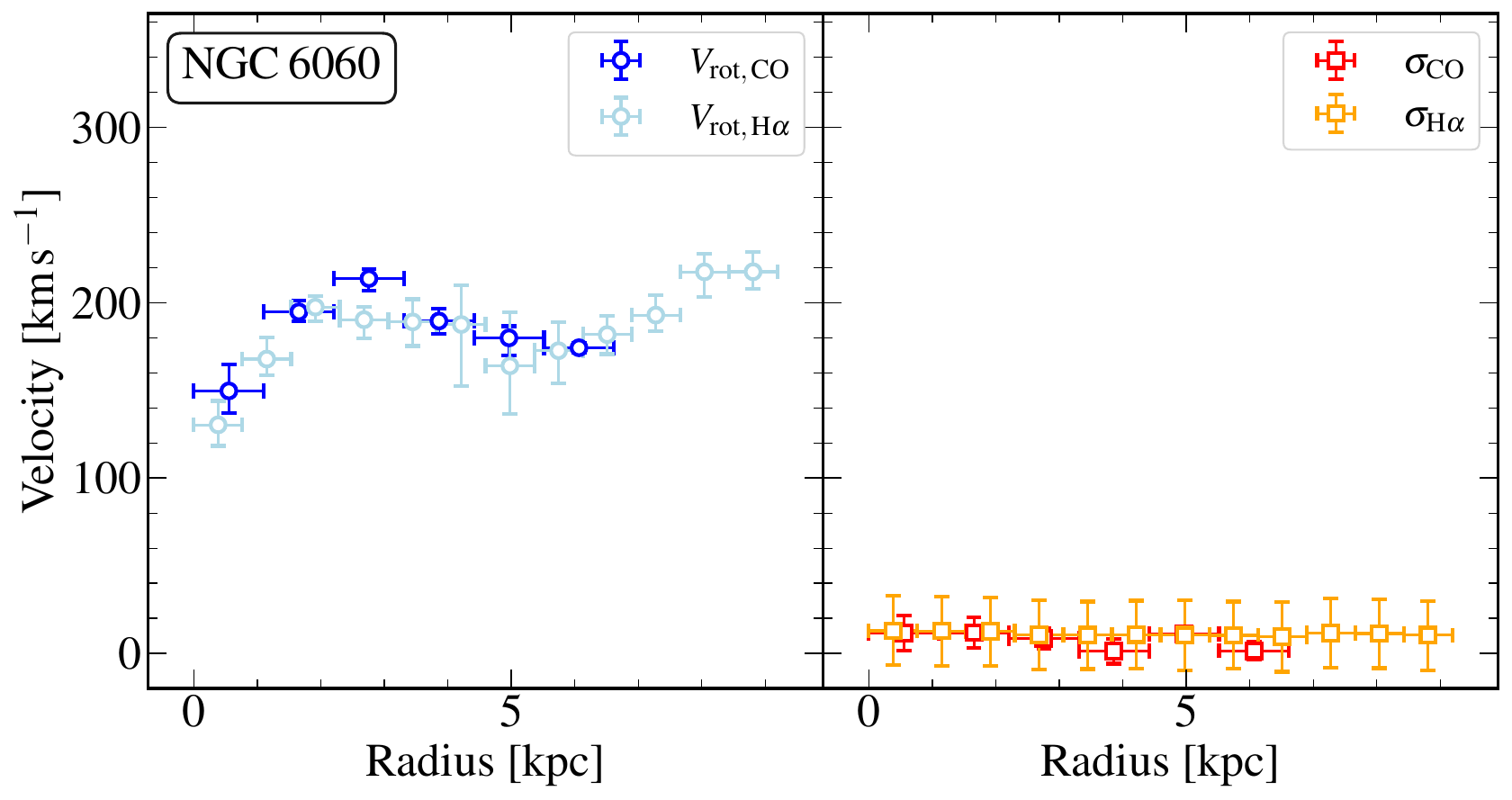}
\figsetgrpnote{The rotation velocity and velocity dispersion obtained through the \Bbarolo\  analysis of the CO and H$\alpha$ data cubes for an inactive star-forming galaxy, NGC6060. The rotation velocities derived from the CO and H$\alpha$ are represented by blue and lightblue circles. The velocity dispersions for CO and H$\alpha$ are denoted by red and orange squares, respectively. The horizontal errorbar represents the width of the ring used in kinematic modeling.}
\figsetgrpend

\figsetgrpstart
\figsetgrpnum{5.16}
\figsetgrptitle{Radial profile of NGC6301}
\figsetplot{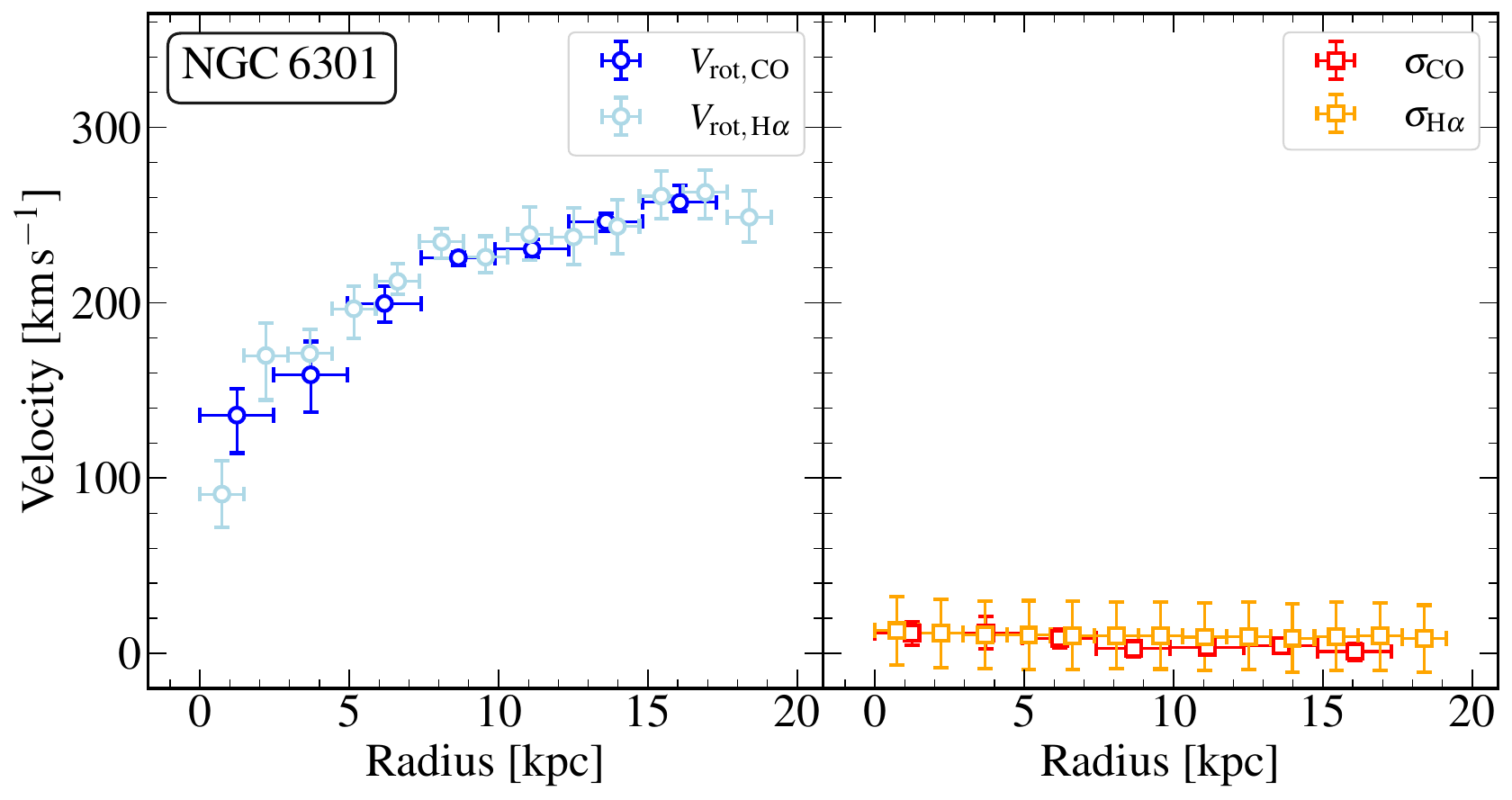}
\figsetgrpnote{The rotation velocity and velocity dispersion obtained through the \Bbarolo\  analysis of the CO and H$\alpha$ data cubes for an inactive star-forming galaxy, NGC6301. The rotation velocities derived from the CO and H$\alpha$ are represented by blue and lightblue circles. The velocity dispersions for CO and H$\alpha$ are denoted by red and orange squares, respectively. The horizontal errorbar represents the width of the ring used in kinematic modeling.}
\figsetgrpend

\figsetgrpstart
\figsetgrpnum{5.17}
\figsetgrptitle{Radial profile of NGC6478}
\figsetplot{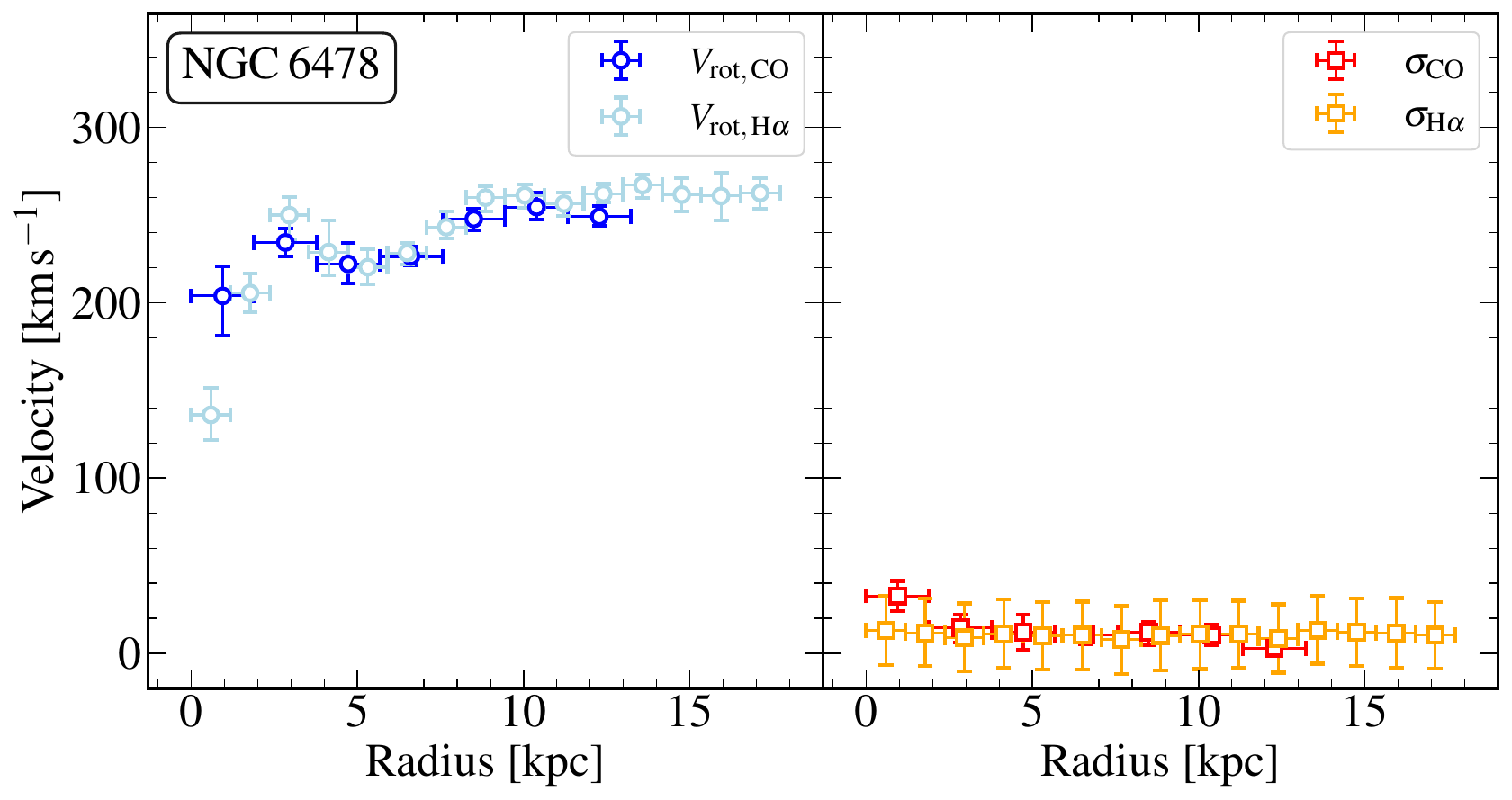}
\figsetgrpnote{The rotation velocity and velocity dispersion obtained through the \Bbarolo\  analysis of the CO and H$\alpha$ data cubes for an inactive star-forming galaxy, NGC6478. The rotation velocities derived from the CO and H$\alpha$ are represented by blue and lightblue circles. The velocity dispersions for CO and H$\alpha$ are denoted by red and orange squares, respectively. The horizontal errorbar represents the width of the ring used in kinematic modeling.}
\figsetgrpend

\figsetgrpstart
\figsetgrpnum{5.18}
\figsetgrptitle{Radial profile of UGC09067}
\figsetplot{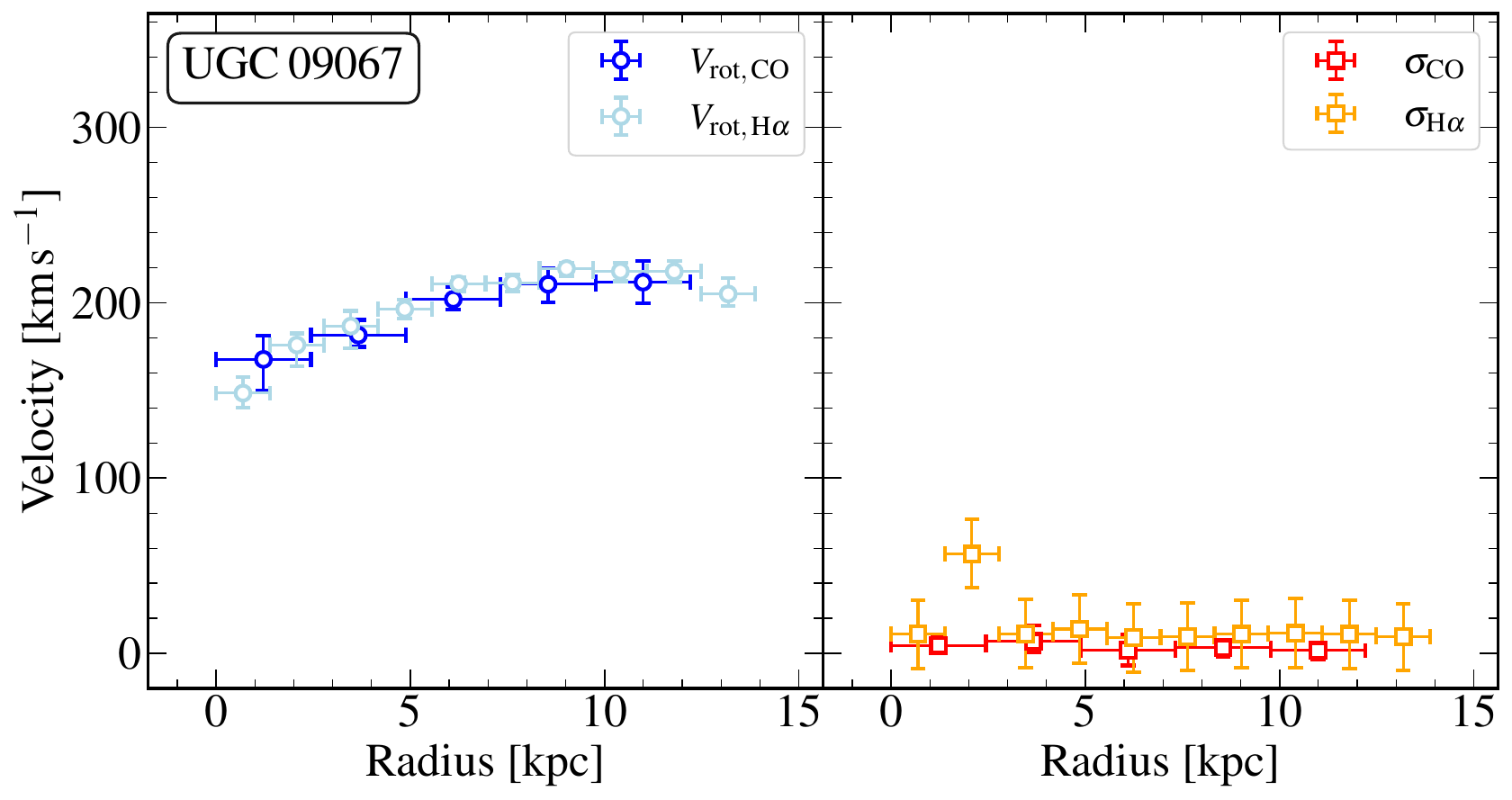}
\figsetgrpnote{The rotation velocity and velocity dispersion obtained through the \Bbarolo\  analysis of the CO and H$\alpha$ data cubes for an inactive star-forming galaxy, UGC09067. The rotation velocities derived from the CO and H$\alpha$ are represented by blue and lightblue circles. The velocity dispersions for CO and H$\alpha$ are denoted by red and orange squares, respectively. The horizontal errorbar represents the width of the ring used in kinematic modeling.}
\figsetgrpend

\subsection{Non-circular motions}
\label{subsec3.3: kinematics results}

The \Bbarolo\ best-fit model and the observations are shown in Figure \ref{Fig3: kinematics for CO} and Figure \ref{Fig4: kinematics for Ha}, respectively. The agreement between the model and the observations, as indicated by the residual velocity map, demonstrates that these systems exhibit coherent and well-organized rotating disks \citep{Levy+2018, Molina+2021}. The residuals are small compared to the total observed velocities (on average about 10\%). However, some features are still visible in the residual map, suggesting the presence of non-circular gas motion.

The CO residual velocity map of PG\,0923+129 conspicuously exhibits a well-defined spiral pattern characterized by discernible regions of both red- and blue-shifted velocities. This pronounced spiral pattern is strongly indicative of a substantial gas inflow phenomenon, which is primarily driven by the inner spiral arms within the system, as expounded upon in prior work \citep{Molina+2021}. A closely analogous spiral-like structure is also notably shown in the residual velocity-dispersion map. Considering its regular, spiral-like morphology, this non-circular motion is probably induced by gravitational perturbation driven by inner spiral arms. In contrast, we did not identify similar velocity perturbations in the H$\alpha$ map, which intrinsically traces a much larger scale of the disk where the perturbation is minor. Nevertheless, a conspicuous elevation in velocity dispersion is prominently evident within the H$\alpha$ velocity-dispersion residual map. Notably, these excesses in velocity dispersion, which are not accounted for by the \Bbarolo\ model, have been attributed to AGN-driven outflows in previous investigations \citep{Molina+2022}.

Regarding the host galaxy of the quasar PG\,1126$-$041, we remove the regular rotating component in the gas velocity field with \Bbarolo\ model, which reveals pronounced red- and blue-shifted gas components in the south-east and north-west directions relative to the kinematic center. Remarkably, these shifted gas constituents spatially coincide with the region where high-velocity dispersion gas has been previously identified \citep{Molina+2022}, an observation suggestive of a plausible influence from the central AGN. Additionally, it is noteworthy that the velocity dispersion of CO emission exhibits an exceptionally high magnitude ($\sim 50\,\rm km\,s^{-1}$) when compared to the primary disk component ($\sim 10\,\rm km\,s^{-1}$). This enhanced velocity dispersion is further corroborated by its manifestation in the residual CO velocity dispersion map, indicative of an unresolved nuclear disk structure \citep{Shangguan+2020a, Izumi+2020}.

In the case of the host galaxy of the quasar PG\,2130$+$099, the residual CO velocity map reveals a relatively substantial residual component located at the periphery of the galaxy. The residual velocity dispersion with an amplitude of $\sim 50\,\rm km\,s^{-1}$ is observed in the in residual velocity dispersion exhibits, suggesting apparent non-circular motions. The residuals are associated with the spiral arms suggesting the secular evolution of gas in the disk. However, an AGN-driven outflow cannot be fully ruled out. Through the same analysis procedure, we found that the H$\alpha$ map only presents inflow gas. The residual CO velocity dispersion map reveals significant residuals, suggesting that the cold molecular gas is highly perturbated. We also note that there are large velocity residuals along the inter-arm regions in the H$\alpha$ velocity map, accompanied also by very high-velocity dispersion. This is mainly due to the low surface brightness and poor SNR in these regions.

\section{Dynamical modeling}
\label{sec4: gas dynamics}
The rotation curve derived through \Bbarolo\ modeling offers a valuable independent constraint on the mass distribution within the line-emitting region. In this section, we employ a multicomponent dynamical model to fit the rotation curve, aiming to explore the mass contributions from various components, including the stellar, molecular gas, and dark matter halo. 

We focus on modeling the gas dynamics and performing a fit to the rotation curve obtained through \Bbarolo\, using data derived from CO and H$\alpha$ line emission within the region of interest. To ensure robust and reliable results, we exercise caution in the fitting process by excluding certain data points. Specifically, we omit the innermost data points, where complex gas kinematics may be present and not adequately captured by \Bbarolo, leading to potentially unreliable estimates of circular velocities. Similarly, we discard the outermost data points due to their relatively lower SNR at larger radii, which can introduce greater uncertainties into the analysis.

During the kinematic modeling process, we consider four main components contributing to the circular motion of gas within the galaxies: the stellar bulge, the stellar disk, the molecular gas disk, and the dark matter halo. We disregard the \textsc{Hi} gas component due to its typically much larger spatial extent compared to stars and molecular gas, making it dominant only on larger scales \citep{de-Blok+2008}. Additionally, we neglect the gravitational potential contributed by the ionized gas, as its mass is negligible compared to the other components. Similarly, the contribution from the SMBH is not taken into account. Given the typical SMBH mass in our sample, which is on the order of $\sim10^7-10^8\,M_\odot$, its impact on the circular velocities at kiloparsec scales is negligible. Hence, the total circular velocity can be expressed as follows:
\begin{eqnarray}
    V_{\rm circ}^2 = V_{\rm bulge}^2 + V_{\rm disk}^2 + V_{\rm halo}^2 + V_{\rm gas}^2,
    \label{equ5: vcirc}
\end{eqnarray}
where $V_{\rm bulge},\,V_{\rm disk},\, V_{\rm halo},$ and $V_{\rm gas}$ are the circular velocities contributed by the stellar bulge, stellar disk, dark matter halo, and molecular gas, respectively.

\subsection{Circular velocity derivations}
\label{subsec4.1: circular velocity}
We derive the circular velocities by employing the action-based galaxy modelling architecture (\textsc{Agama}; \citealt{Vasiliev2019}). This software framework offers versatile tools for computing the gravitational potential of various analytic density profiles, making it suitable for analyzing the dynamics of stars in galaxies. By utilizing \textsc{Agama}, we are able to accurately model the mass distribution and subsequently obtain the circular velocities associated with the galaxies under study. In the following, we provide a concise overview of the density profiles employed by \textsc{Agama} for the computation of gravitational potentials and circular velocities. 

Regarding the stellar bulge component, we adopted the deprojection of the S{\'e}rsic profile provided by \textsc{Agama} for calculating the gravitational potential. The deprojection of the S{\'e}rsic function incorporates five parameters, in addition to the parameters described in Section \ref{subsec3.1: gas distribution}. Furthermore, \textsc{Agama} provides the axis ratio to account for a triaxial distribution of the bulge component. However, in this study, we treat the stellar bulge component as spheroidal, thus neglecting its asymmetry. This assumption might introduce a smaller bulge mass than the real value.

Concerning the stellar disk component, we adopted the disk profile form provided by \textsc{Agama} as follows:
\begin{eqnarray}
    \rho =\Sigma_0\exp\left[-\left(\frac{R}{R_d}\right)^{1/n}\right]\times \frac{1}{2h}\exp\left(-\left|\frac{z}{h}\right|\right),
\end{eqnarray}
Here, $\Sigma_0$ represents the surface density at galaxy center, $R_d$ is the scale radius, $n$ denotes the S{\'e}rsic index, and $h$ represents the scale height of the disk. This form is also employed for calculating the gravitational potential contributed by the molecular gas disk.

For the dark matter component, we employ a simulation-motivated Navarro-Frank-White (NFW) model \citep{Navarro+1996}, which is characterized by its density profile:
\begin{eqnarray}
\frac{\rho(r)}{\rho_{\rm crit}} = \frac{\delta_c}{(r/r_s)(1+r/r_s)^2},
\end{eqnarray}
Here, $r_s=r_{200}/c$ represents the characteristic radius, $\rho_{\rm crit}=3H^2/8\pi G$ denotes the critical density, and $\delta_c$ and $c$ are two dimensionless parameters \citep[further details, refer to ][Section 4.1]{Navarro+1996}.

\subsection{Asymmetric Drift Correction}
\label{subsec4.2: pressure correction}
The pressure gradients within the interstellar medium (ISM) can provide additional support to the gas, counteracting the galaxy's self-gravity. To account for this and adjust the rotation curve accordingly, an asymmetric drift correction is applied, as outlined by \cite{Burkert+2010}, \cite{Lang+2017}, and \cite{Iorio+2017}. The corrected rotation velocity, $V_{\rm rot}$, is related to the circular velocity, $V_{\rm circ}$—which traces the gravitational potential—through the following equation:
\begin{eqnarray}
    V_{\rm rot}^2 = V_{\rm circ}^2 + \frac{1}{\rho}\frac{d\left(\rho \sigma_{\rm g}^2\right)}{d \ln r},
\end{eqnarray}
where the rightmost term accounts for the effects of asymmetric drift, typically leading to $V_{\rm rot} < V_{\rm circ}$. In this equation, $\sigma_{\rm g}$ denotes the velocity dispersion, assumed to be isotropic, which can be derived from moment 2 maps or using \Bbarolo. The symbol $\rho$ represents the gas density, and $r$ is the galactocentric radius. The asymmetric drift correction is crucial for accurately incorporating the ISM pressure support and obtaining reliable rotation velocity estimates.

To precisely calculate the asymmetric drift correction, we utilized the methodology proposed by \cite{Iorio+2017}, which involves simultaneously fitting the gas surface density and velocity dispersion profiles to estimate the gas pressure. This approach allows for potential variations in velocity dispersion across different radii, ensuring a more accurate correction \citep[for further details, see ][]{Iorio+2017}.

\subsection{Best-fit Dynamical modeling and Results}
\label{subsec4.3: dynamical fitting}

We used the Python package \texttt{emcee} to fit the rotation velocities of \Bbarolo\ by implementing Equ. \ref{equ5: vcirc}. This allowed us to obtain the best-fit values for our model parameters.  During the fitting process, we employ the following free parameters, the stellar bulge mass $\left(\log M_{\rm b}\right)$, the effective radius of stellar bulge $\left(R_{\rm e,b}\right)$, the S{\'e}rsic index of stellar bulge $\left(n\right)$, the stellar disk mass $\left(\log M_{\rm d}\right)$, the effective radius of stellar disk $\left(R_{\rm e,d}\right)$, the stellar-to-dark matter mass ratio $\left(\log f_\star\right)$, the deviation from the ``expected concentration'' following the mass-concentration relation \citep[$\left(\Delta c\right)$;][]{Dutton+2014}, and the CO-to-H$_2$ conversion factor $\left(\alpha_{\rm CO}\right)$. For the host galaxy of PG quasars, the HST image analysis provides a non-unity S{\'e}rsic index for the stellar disk, therefore we also set a free stellar disk S{\'e}rsic index during the fitting. For the host galaxy of PG\,0923$+$129, the HST image analysis indicates two stellar disks, which are considered during the fitting. At least 8 parameters are used to fit the rotation curves.

In order to effectively constrain the model and mitigate parameter degeneracy, we implemented prior constraints based on established stellar galaxy morphology models for the host galaxies, as well as the well-known relation between dark matter halo and stellar mass content. For the host galaxies of PG quasars, we applied Gaussian priors to the effective radius of the stellar bulges and stellar disks, as well as to the S{\'e}rsic index of the bulge and disks. These priors were informed by the fitting results and uncertainties obtained through HST image modeling \citep{Veilleux+2009, Kim+2017, Zhao+2021}. Regarding inactive galaxies, the prior constraints for their stellar components were adopted from the 2-dimensional analysis provided by \cite{Mendez-Abreu+2017}.

Regarding the dark matter halo component, we assigned a prior constraint on the stellar-to-dark matter mass ratio, setting it to range from $-2$ to $-1.5$ uniformly, which is appropriate for galaxies at low-$z$ \citep{Behroozi+2010}. Additionally, we set the halo concentration to follow the mass-concentration relation with a scatter of 0.2 dex. For $\alpha_{\rm CO}$, we implemented a Gaussian prior centered at $3.1\,M_\odot\,\left(\rm K\,km\,s^{-1}\,pc^{-2}\right)^{-1}$ with an uncertainty of 0.3 dex, as suggested by \cite{Sandstrom+2013}. Those prior constraints are so strong that effectively help to reduce the degeneracy between parameters. In this context, those parameters are somehow ``fixed'' during the fitting. Only several parameters, e.g., the stellar bulge mass, and the stellar disk mass, are free parameters. 

\begin{figure*}
    \centering
    \includegraphics[width=0.49\linewidth]{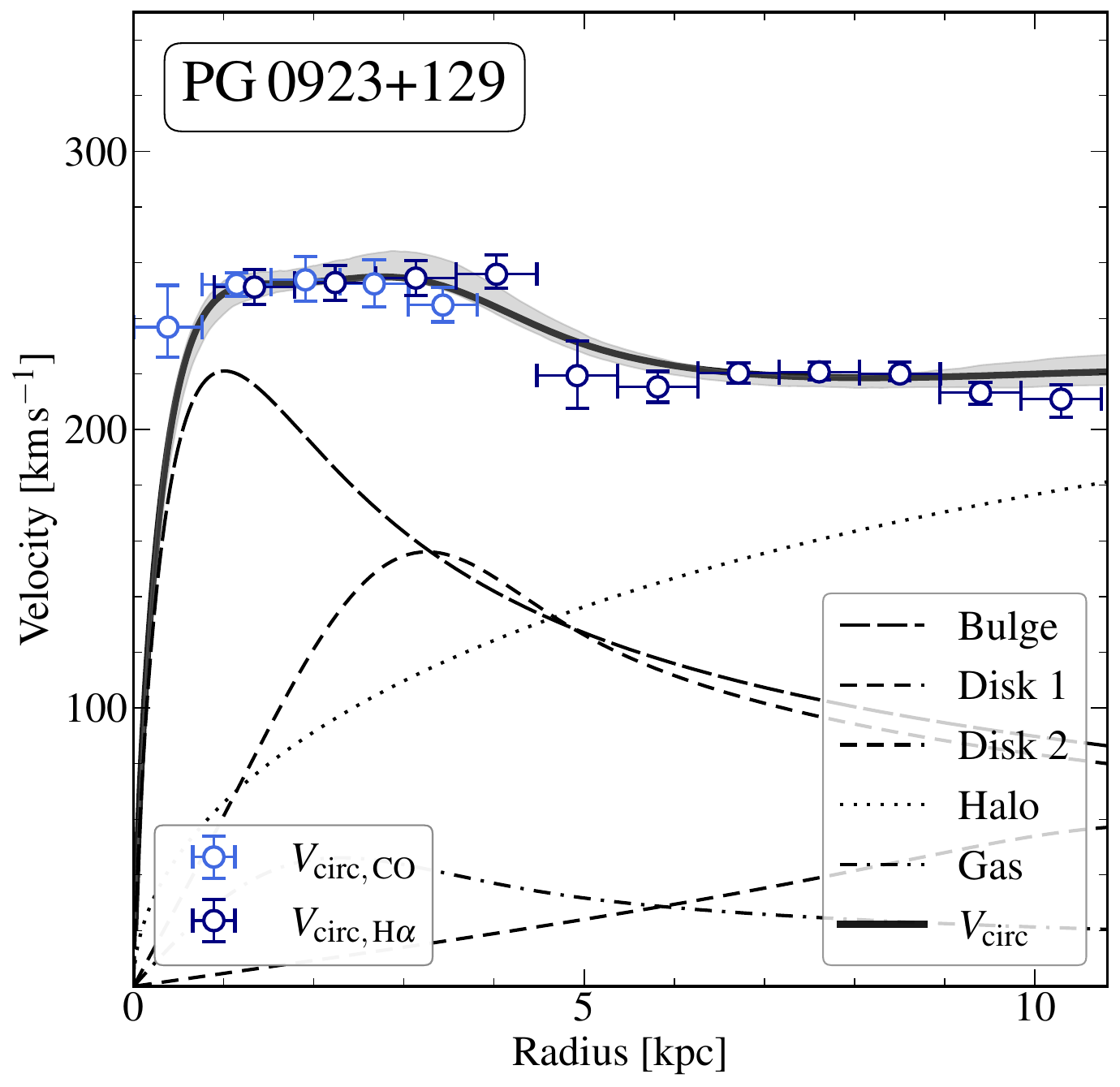}
    \includegraphics[width=0.49\linewidth]{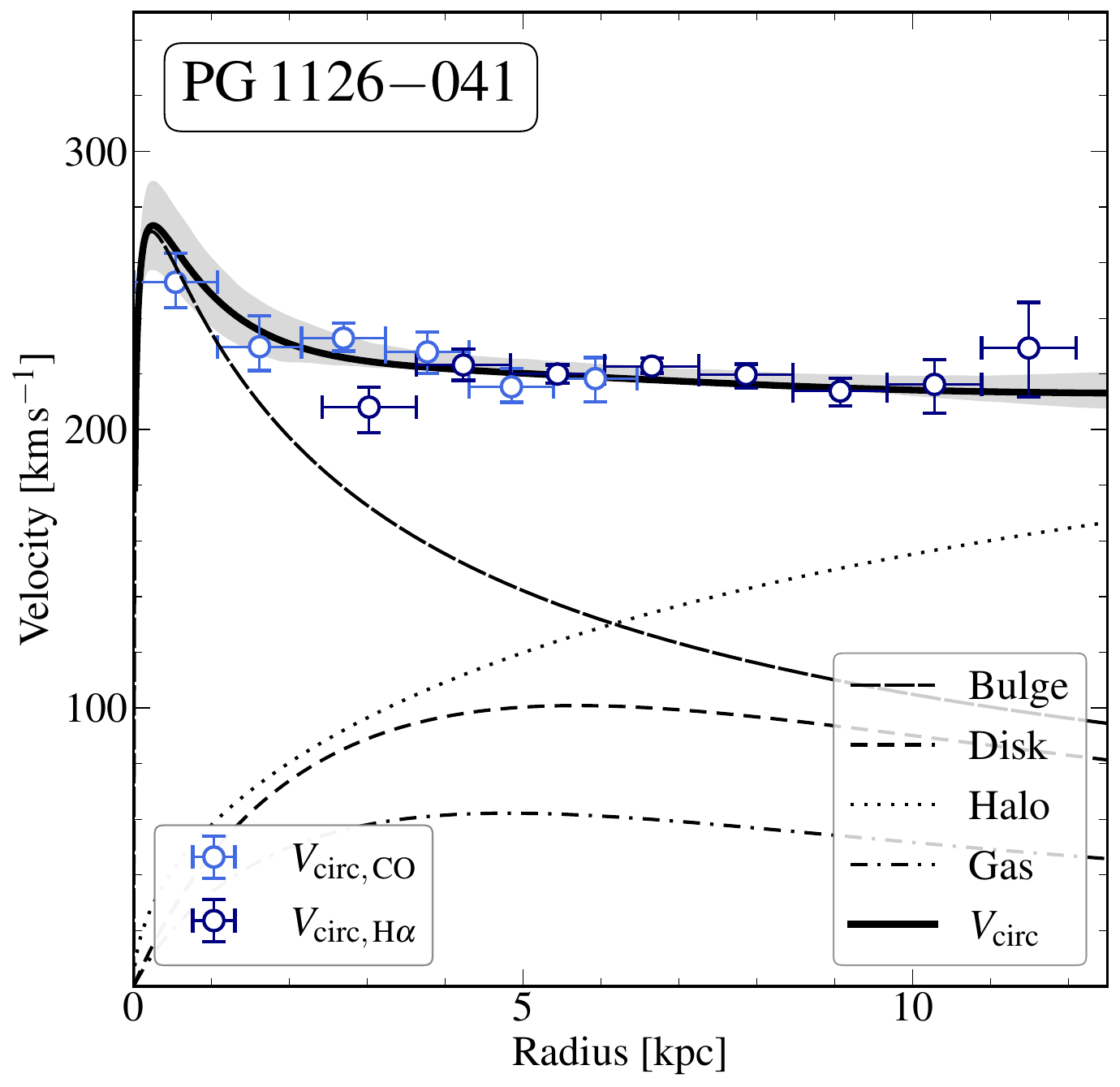}
    \includegraphics[width=0.49\linewidth]{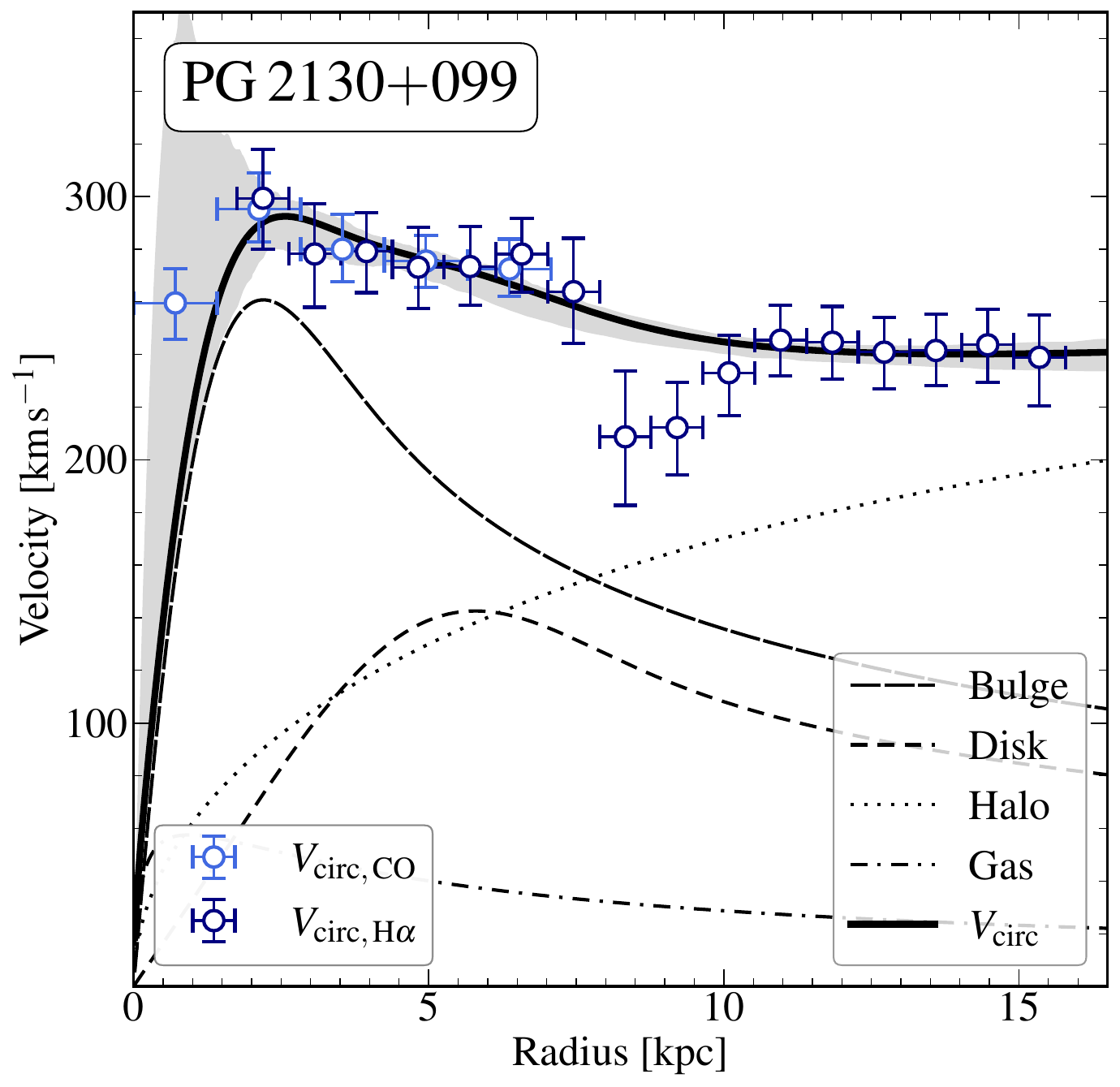}
    \includegraphics[width=0.49\linewidth]{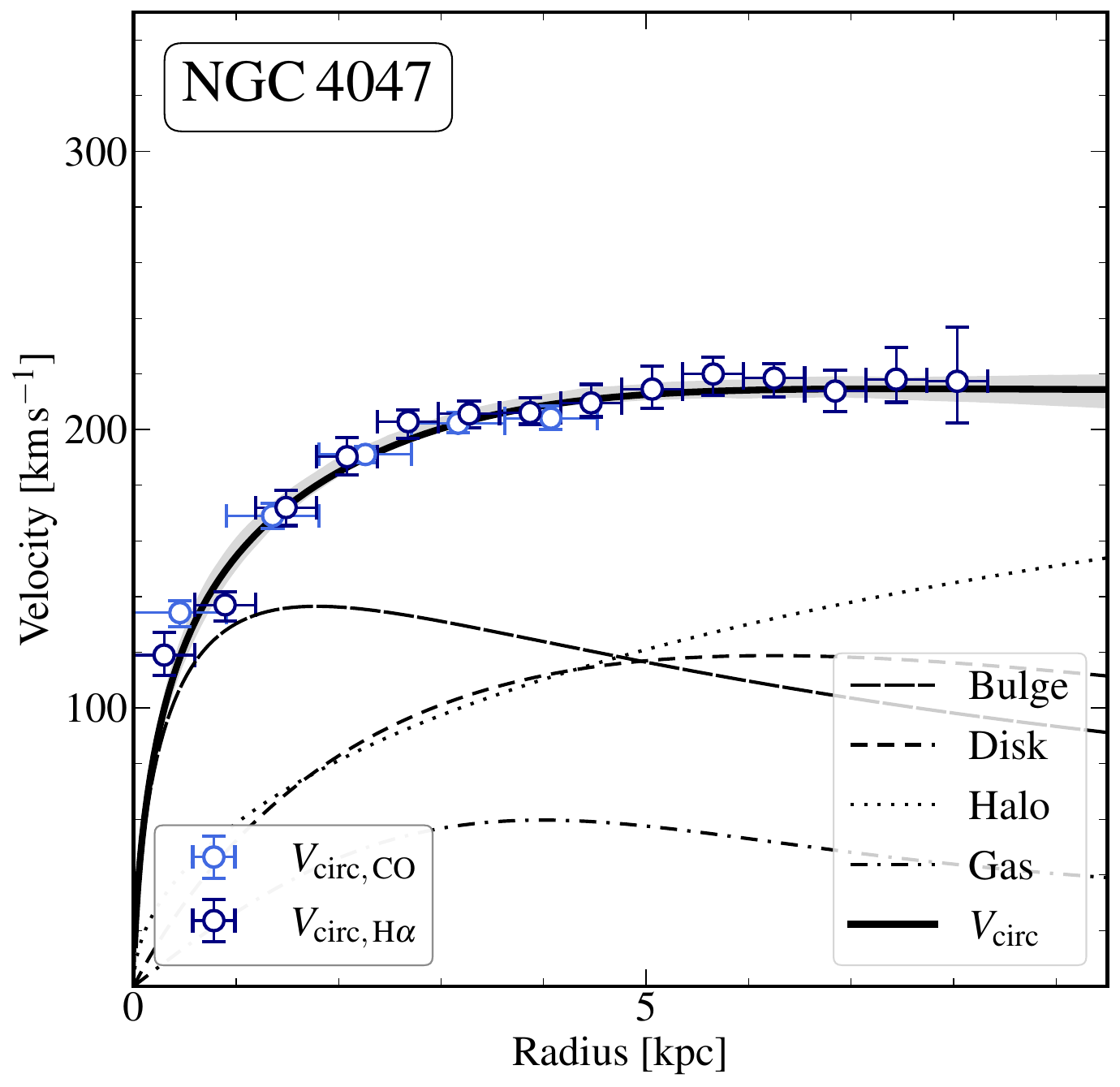}
    \caption{The circular velocities derived from the \Bbarolo\ analysis and the best-fit dynamical model for three quasar host galaxies and one inactive star-forming galaxy. The names of targets are plotted on the bottom right corner of each sub-panel. The circular velocities obtained from \Bbarolo\ are represented by blue and cyan circles, accompanied by the shaded region indicating the associated uncertainties. The long dashed line corresponds to the circular velocities contributed by the stellar bulge, the short dashed line represents the stellar disk, the dotted line represents the dark matter component, the dash-dotted line represents the cold molecular gas, and the thick solid line represents the overall circular velocity. The gray-shaded region represents the uncertainties of the circular velocities derived from the dynamical method.}
    \label{Fig6: dynamical fitting}
\end{figure*}

We initiate the MCMC sampling process with 400 walkers, each taking 1000 steps, following a burn-in phase of 500 steps. After the burn-in phase, we identify the ``maximum a posteriori'' (MAP) values for the parameters. These MAP values are then used as the initial parameter values to commence the second round of MCMC sampling, utilizing the same number of walkers and steps. The first 500 steps of all walkers are discarded, and we retain the subsequent chains, resulting in $400 \times 500$ points, to construct the full posterior probability distribution (PDF). We utilize the 50th percentile of the MCMC samples as the best-fit results and assess the uncertainties of the parameters by calculating the 16th and 84th percentiles of the MCMC samples. The best-fit results for quasar host galaxies and one example of inactive star-forming galaxies, NGC\,4047, are illustrated in Figure \ref{Fig6: dynamical fitting}, and the best-fit parameter values are listed in Table \ref{Tab4: dyn paras}.



\figsetgrpstart
\figsetgrpnum{6.1}
\figsetgrptitle{Circular velocities for PG0923}
\figsetplot{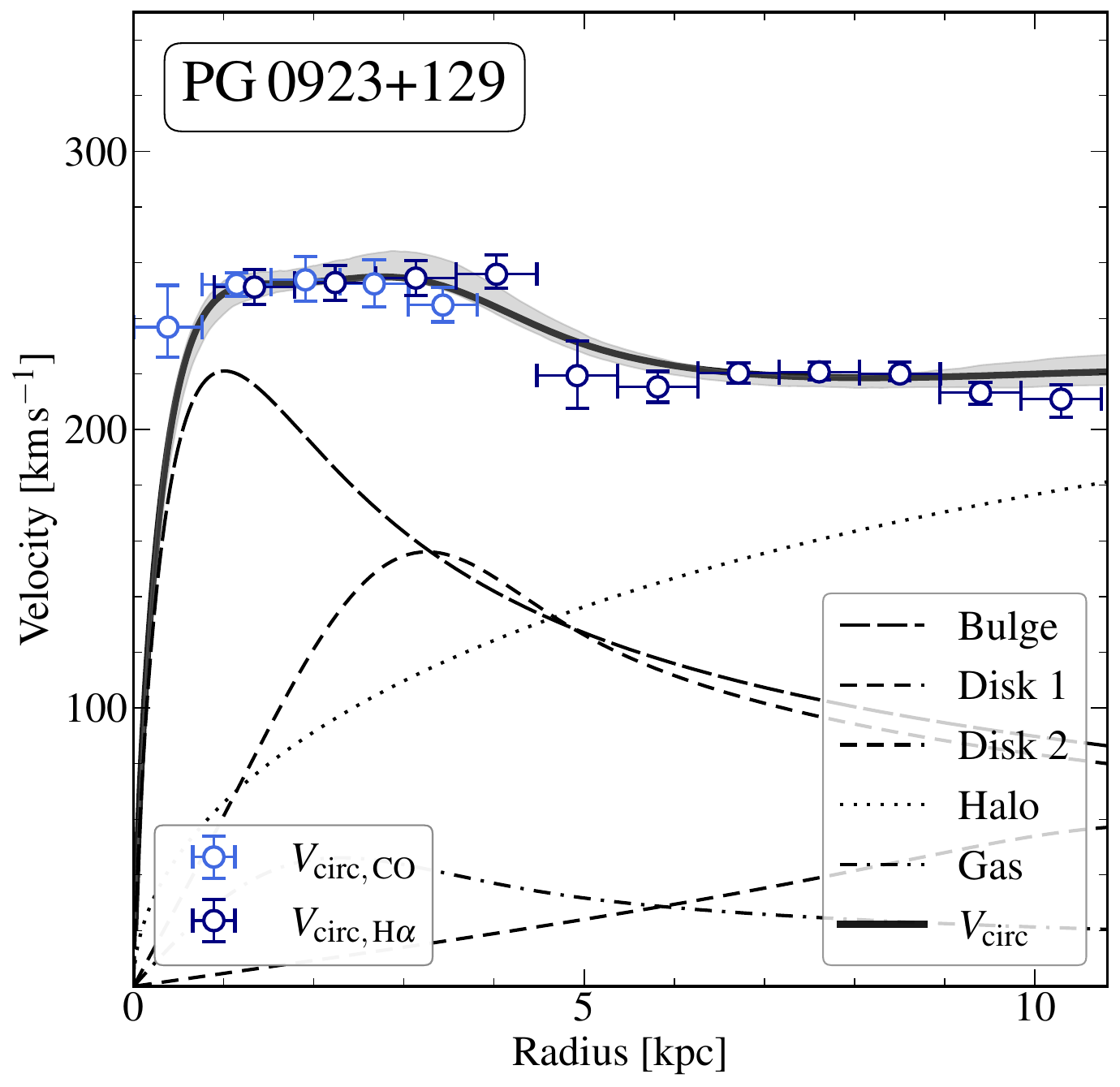}
\figsetgrpnote{The circular velocities derived from the \Bbarolo\ analysis and the best-fit dynamical model for a quasar host galaxy, PG\,0923$+$129. The names of targets are plotted on the bottom right corner of each sub-panel. The circular velocities obtained from \Bbarolo\ are represented by blue and cyan circles, accompanied by the shaded region indicating the associated uncertainties. The long dashed line corresponds to the circular velocities contributed by the stellar bulge, the short dashed line represents the stellar disk, the dotted line represents the dark matter component, the dash-dotted line represents the cold molecular gas, and the thick solid line represents the overall circular velocity. The gray-shaded region represents the uncertainties of the circular velocities derived from the dynamical method.}
\figsetgrpend

\figsetgrpstart
\figsetgrpnum{6.2}
\figsetgrptitle{Circular velocities for PG1126}
\figsetplot{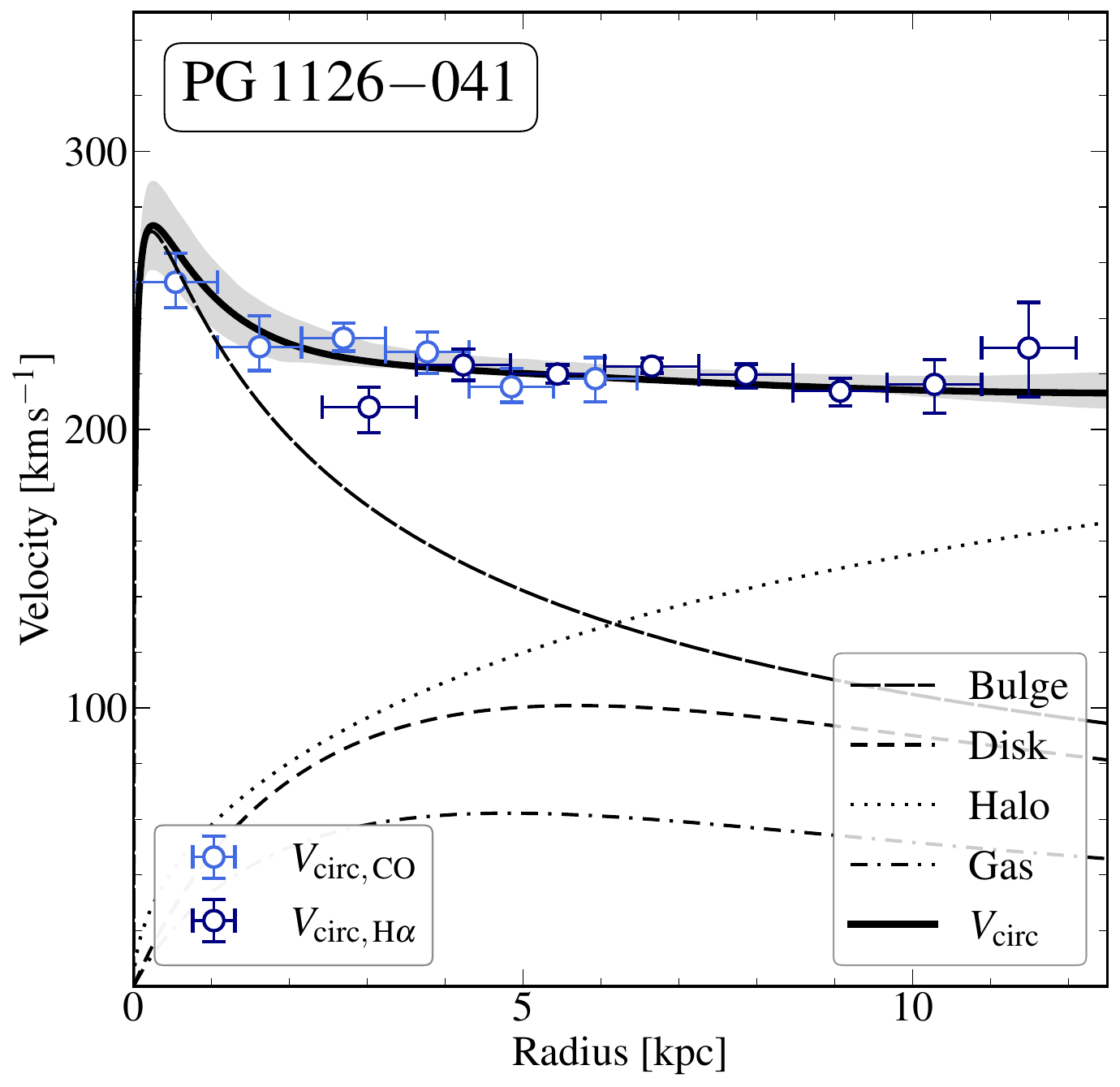}
\figsetgrpnote{The circular velocities derived from the \Bbarolo\ analysis and the best-fit dynamical model for a quasar host galaxy, PG\,1126$-$041. The names of targets are plotted on the bottom right corner of each sub-panel. The circular velocities obtained from \Bbarolo\ are represented by blue and cyan circles, accompanied by the shaded region indicating the associated uncertainties. The long dashed line corresponds to the circular velocities contributed by the stellar bulge, the short dashed line represents the stellar disk, the dotted line represents the dark matter component, the dash-dotted line represents the cold molecular gas, and the thick solid line represents the overall circular velocity. The gray-shaded region represents the uncertainties of the circular velocities derived from the dynamical method.}
\figsetgrpend

\figsetgrpstart
\figsetgrpnum{6.3}
\figsetgrptitle{Circular velocities for PG2130}
\figsetplot{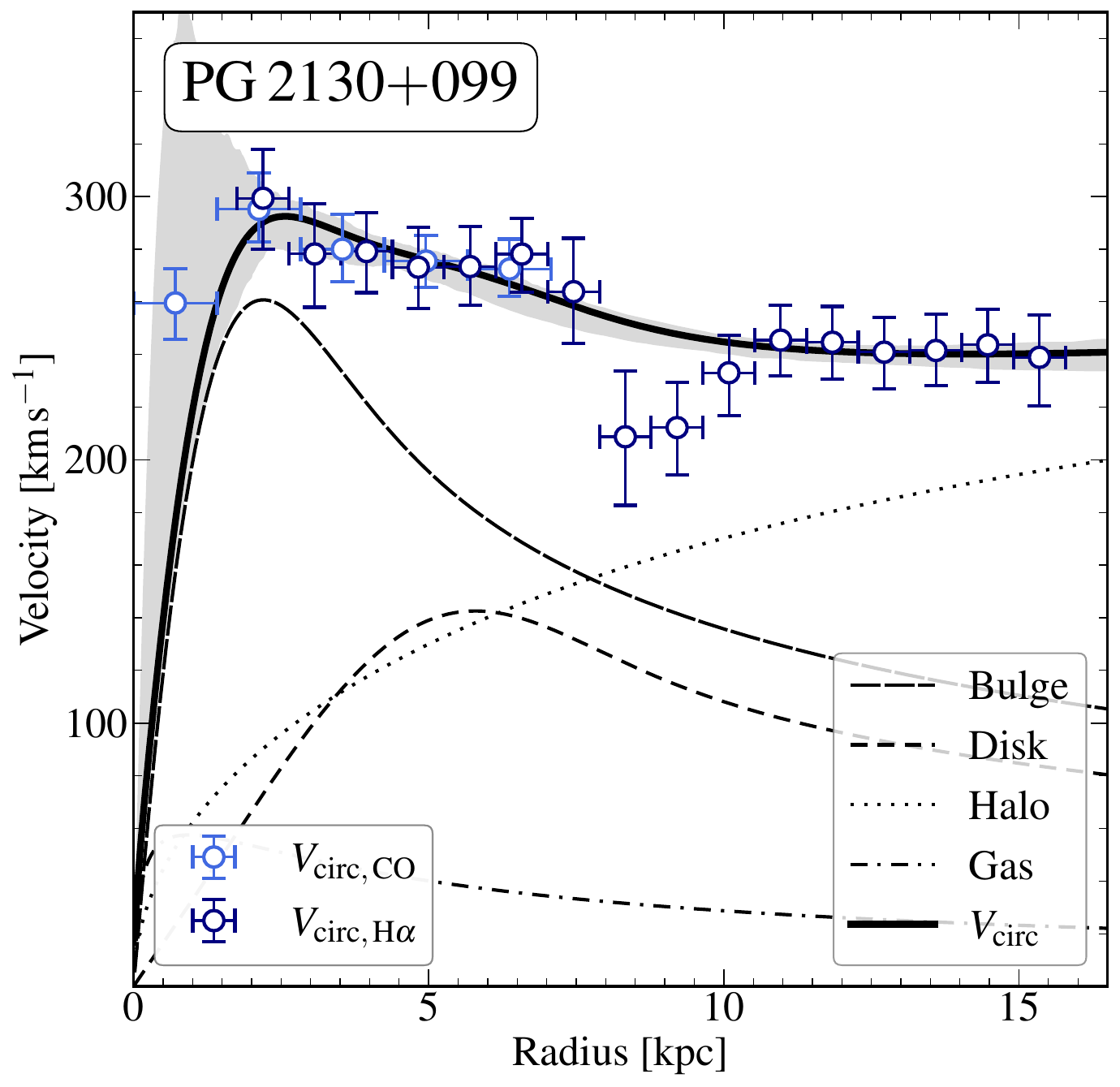}
\figsetgrpnote{The circular velocities derived from the \Bbarolo\ analysis and the best-fit dynamical model for a quasar host galaxy, PG\,2130$+$099. The names of targets are plotted on the bottom right corner of each sub-panel. The circular velocities obtained from \Bbarolo\ are represented by blue and cyan circles, accompanied by the shaded region indicating the associated uncertainties. The long dashed line corresponds to the circular velocities contributed by the stellar bulge, the short dashed line represents the stellar disk, the dotted line represents the dark matter component, the dash-dotted line represents the cold molecular gas, and the thick solid line represents the overall circular velocity. The gray-shaded region represents the uncertainties of the circular velocities derived from the dynamical method.}
\figsetgrpend

\figsetgrpstart
\figsetgrpnum{6.4}
\figsetgrptitle{Circular velocities for IC0944}
\figsetplot{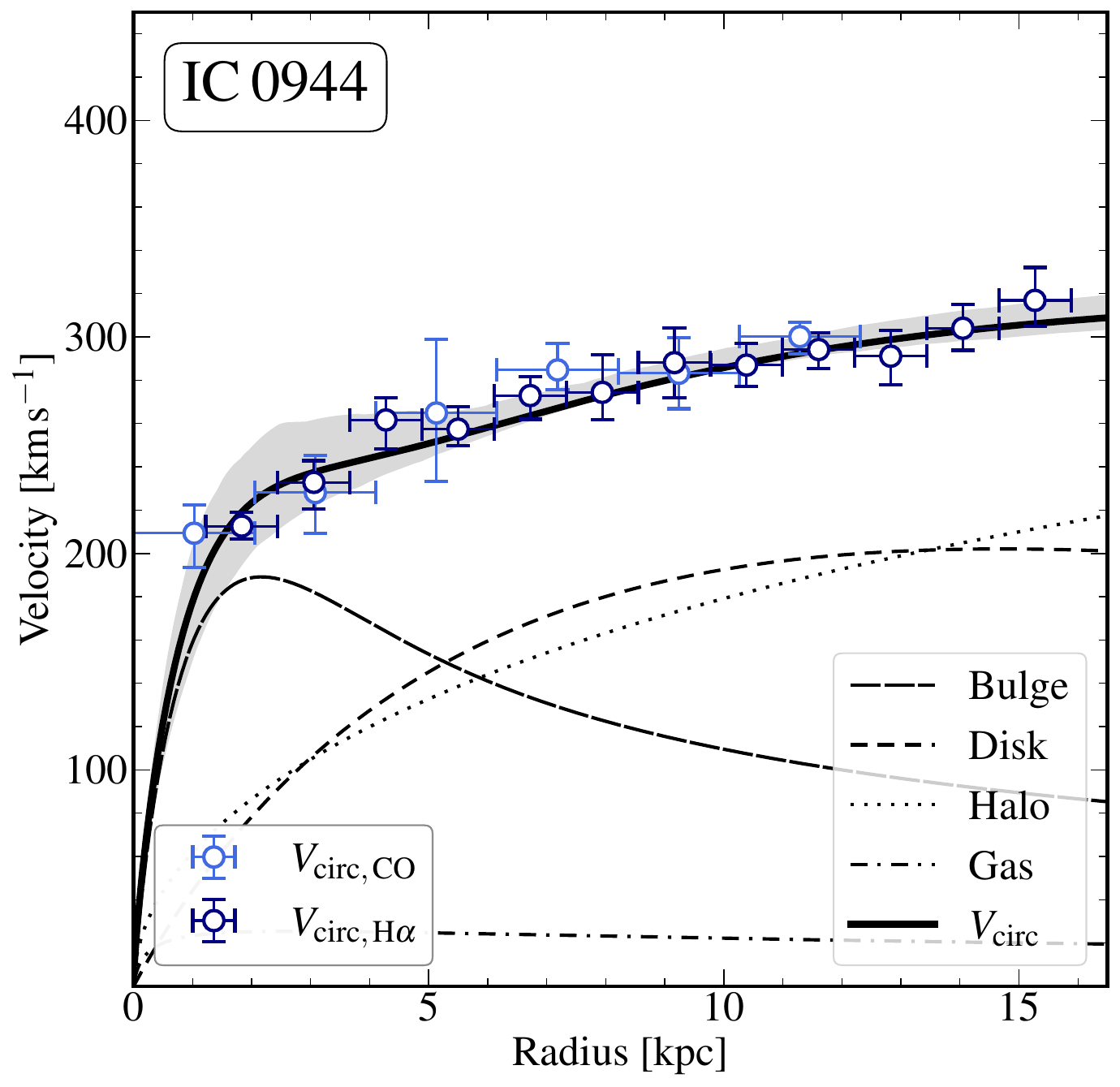}
\figsetgrpnote{The circular velocities derived from the \Bbarolo\ analysis and the best-fit dynamical model for an inactive star-forming galaxy, IC0944. The names of targets are plotted on the bottom right corner of each sub-panel. The circular velocities obtained from \Bbarolo\ are represented by blue and cyan circles, accompanied by the shaded region indicating the associated uncertainties. The long dashed line corresponds to the circular velocities contributed by the stellar bulge, the short dashed line represents the stellar disk, the dotted line represents the dark matter component, the dash-dotted line represents the cold molecular gas, and the thick solid line represents the overall circular velocity. The gray-shaded region represents the uncertainties of the circular velocities derived from the dynamical method.}
\figsetgrpend

\figsetgrpstart
\figsetgrpnum{6.5}
\figsetgrptitle{Circular velocities for IC1199}
\figsetplot{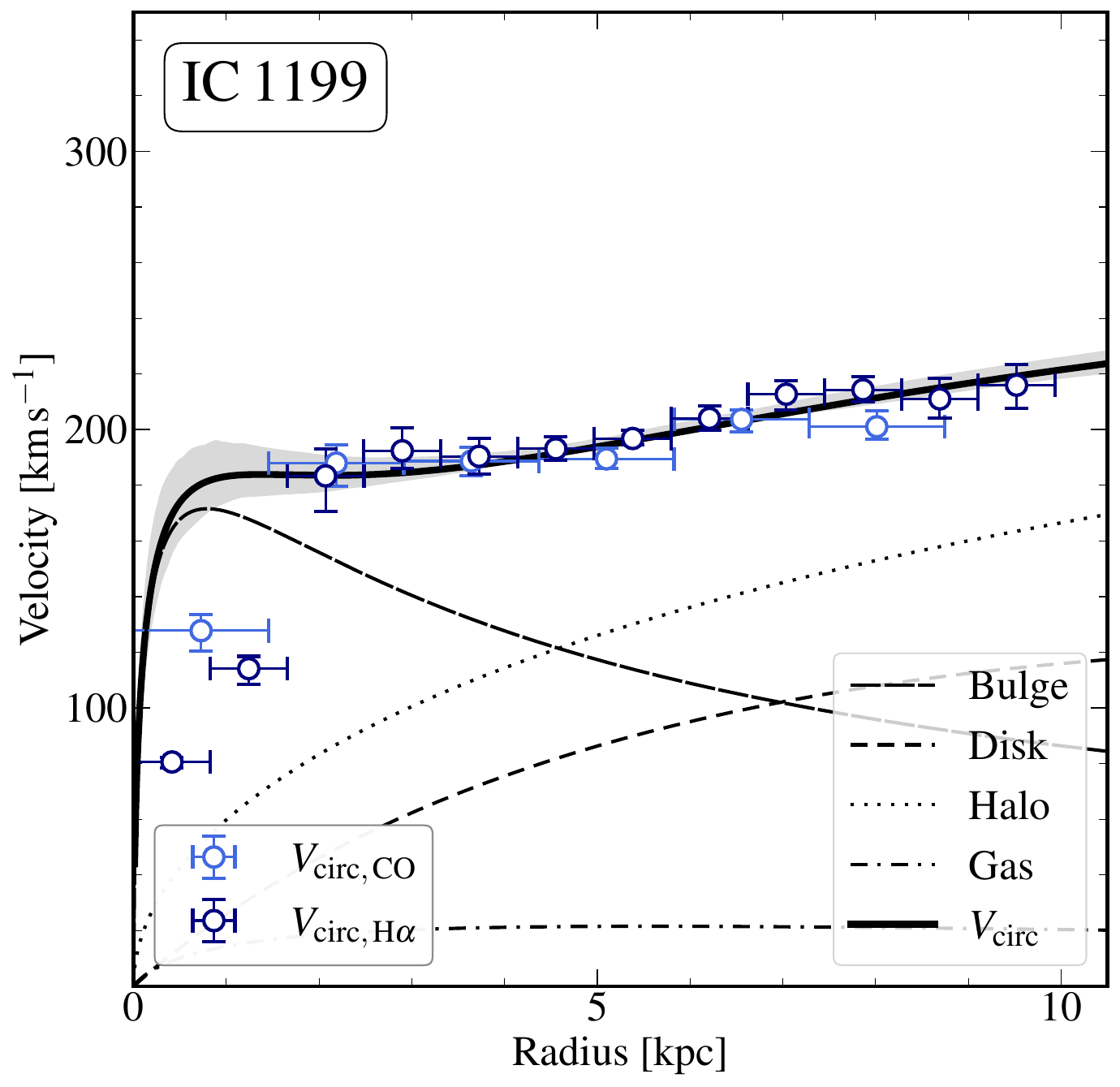}
\figsetgrpnote{The circular velocities derived from the \Bbarolo\ analysis and the best-fit dynamical model for an inactive star-forming galaxy, IC1199. The names of targets are plotted on the bottom right corner of each sub-panel. The circular velocities obtained from \Bbarolo\ are represented by blue and cyan circles, accompanied by the shaded region indicating the associated uncertainties. The long dashed line corresponds to the circular velocities contributed by the stellar bulge, the short dashed line represents the stellar disk, the dotted line represents the dark matter component, the dash-dotted line represents the cold molecular gas, and the thick solid line represents the overall circular velocity. The gray-shaded region represents the uncertainties of the circular velocities derived from the dynamical method.}
\figsetgrpend

\figsetgrpstart
\figsetgrpnum{6.6}
\figsetgrptitle{Circular velocities for IC1683}
\figsetplot{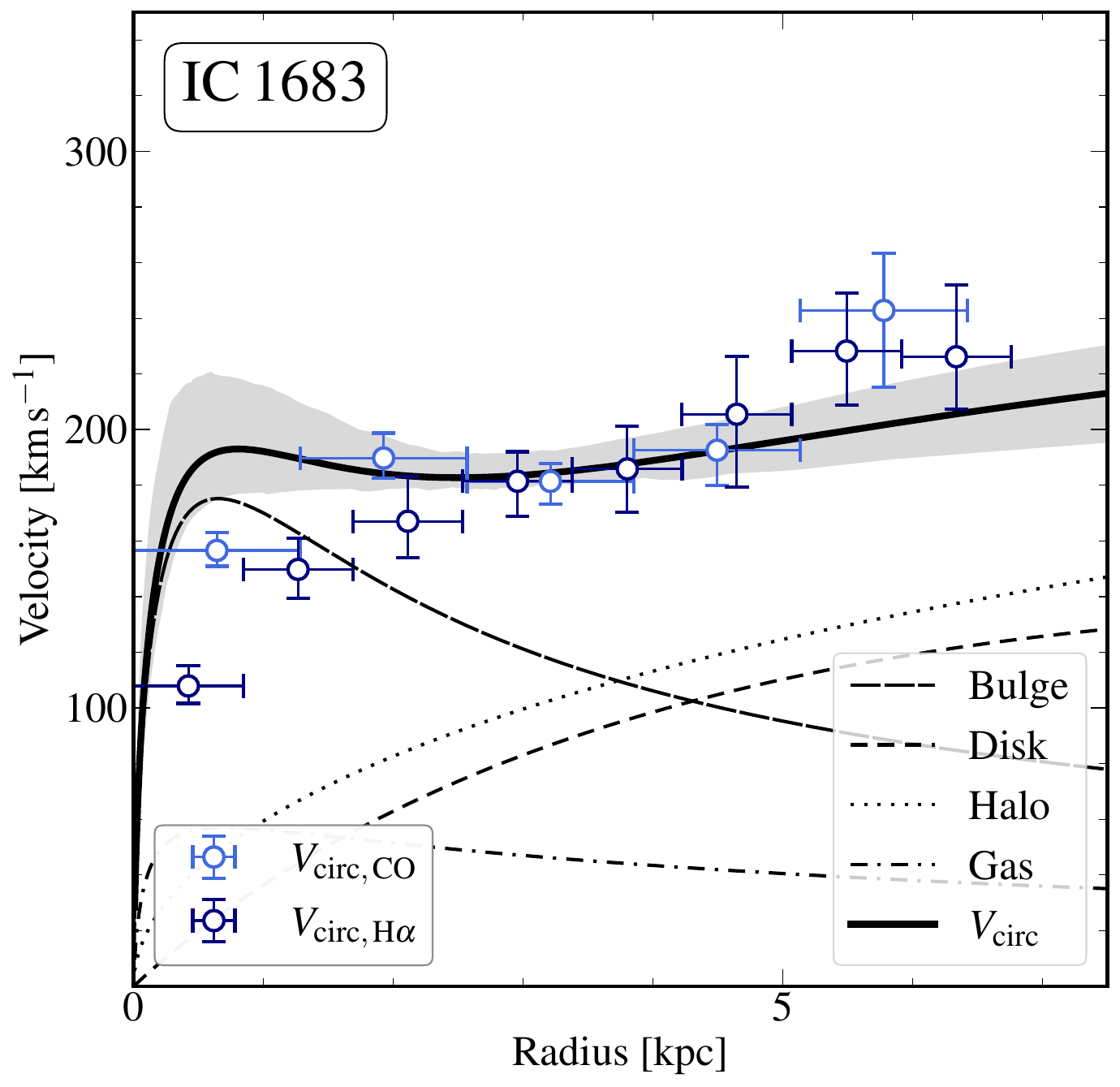}
\figsetgrpnote{The circular velocities derived from the \Bbarolo\ analysis and the best-fit dynamical model for an inactive star-forming galaxy, IC1683. The names of targets are plotted on the bottom right corner of each sub-panel. The circular velocities obtained from \Bbarolo\ are represented by blue and cyan circles, accompanied by the shaded region indicating the associated uncertainties. The long dashed line corresponds to the circular velocities contributed by the stellar bulge, the short dashed line represents the stellar disk, the dotted line represents the dark matter component, the dash-dotted line represents the cold molecular gas, and the thick solid line represents the overall circular velocity. The gray-shaded region represents the uncertainties of the circular velocities derived from the dynamical method.}
\figsetgrpend

\figsetgrpstart
\figsetgrpnum{6.7}
\figsetgrptitle{Circular velocities for NGC0496}
\figsetplot{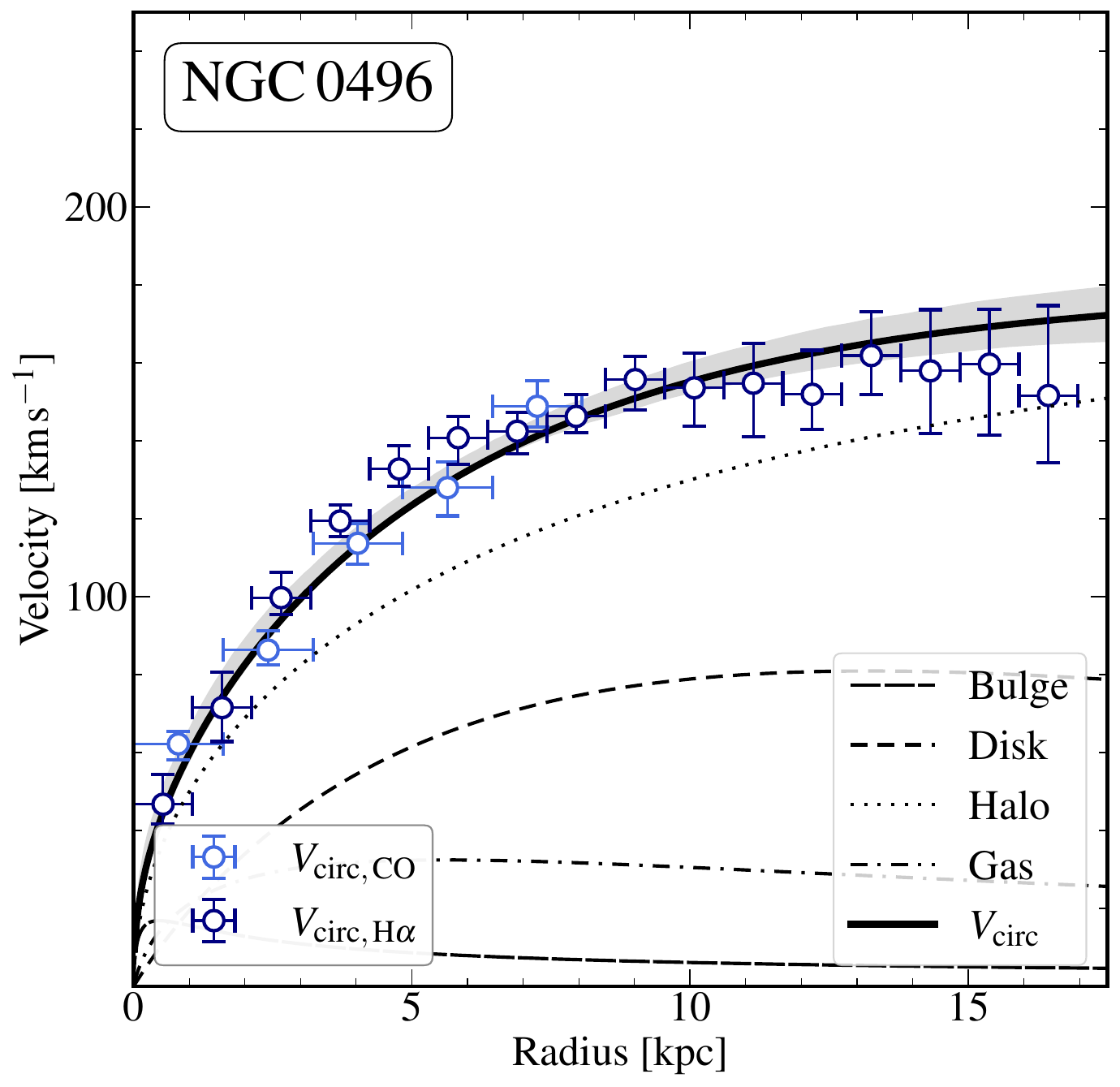}
\figsetgrpnote{The circular velocities derived from the \Bbarolo\ analysis and the best-fit dynamical model for an inactive star-forming galaxy, NGC0496. The names of targets are plotted on the bottom right corner of each sub-panel. The circular velocities obtained from \Bbarolo\ are represented by blue and cyan circles, accompanied by the shaded region indicating the associated uncertainties. The long dashed line corresponds to the circular velocities contributed by the stellar bulge, the short dashed line represents the stellar disk, the dotted line represents the dark matter component, the dash-dotted line represents the cold molecular gas, and the thick solid line represents the overall circular velocity. The gray-shaded region represents the uncertainties of the circular velocities derived from the dynamical method.}
\figsetgrpend

\figsetgrpstart
\figsetgrpnum{6.8}
\figsetgrptitle{Circular velocities for NGC2906}
\figsetplot{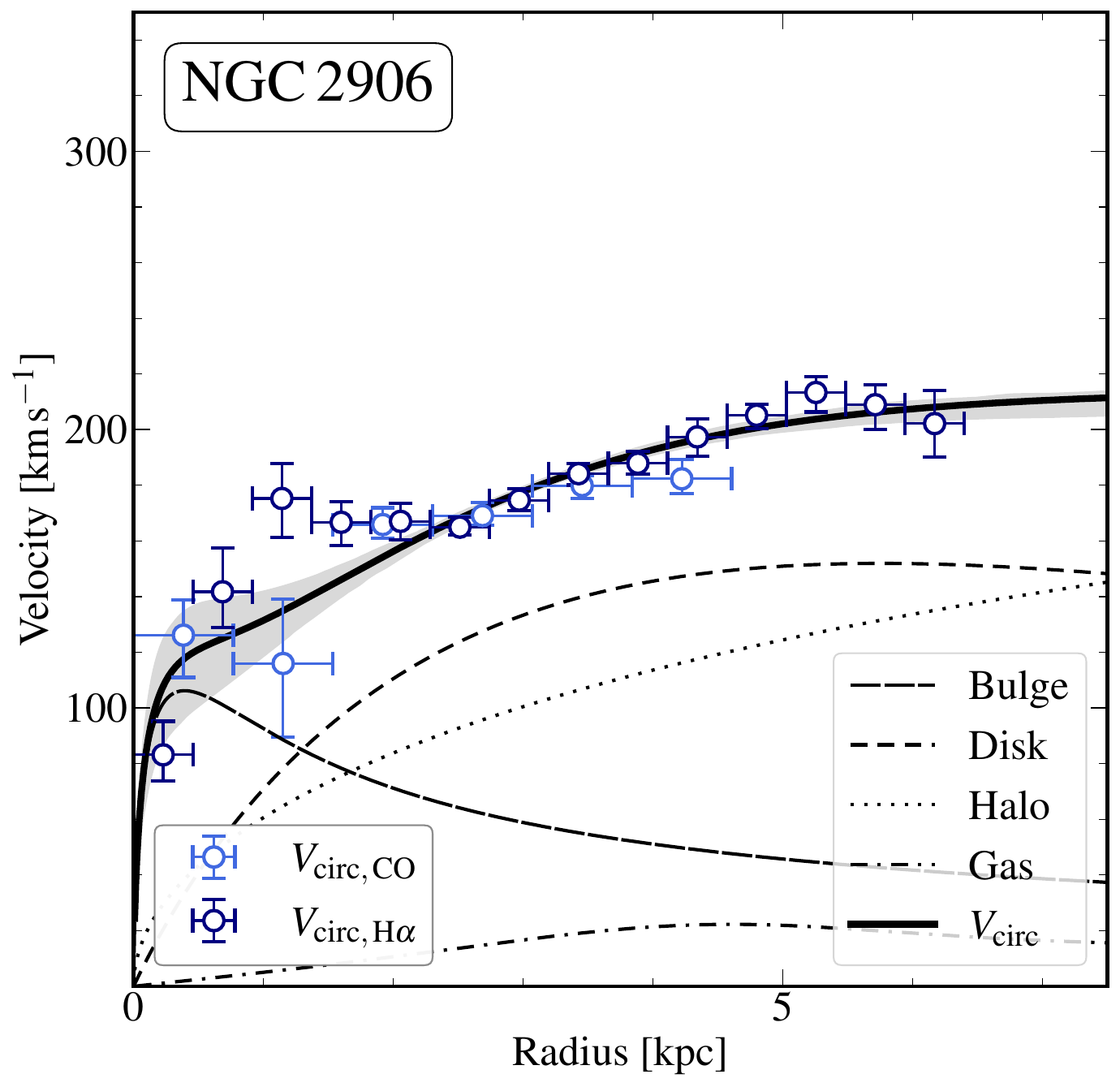}
\figsetgrpnote{The circular velocities derived from the \Bbarolo\ analysis and the best-fit dynamical model for an inactive star-forming galaxy, NGC2906. The names of targets are plotted on the bottom right corner of each sub-panel. The circular velocities obtained from \Bbarolo\ are represented by blue and cyan circles, accompanied by the shaded region indicating the associated uncertainties. The long dashed line corresponds to the circular velocities contributed by the stellar bulge, the short dashed line represents the stellar disk, the dotted line represents the dark matter component, the dash-dotted line represents the cold molecular gas, and the thick solid line represents the overall circular velocity. The gray-shaded region represents the uncertainties of the circular velocities derived from the dynamical method.}
\figsetgrpend

\figsetgrpstart
\figsetgrpnum{6.9}
\figsetgrptitle{Circular velocities for NGC3994}
\figsetplot{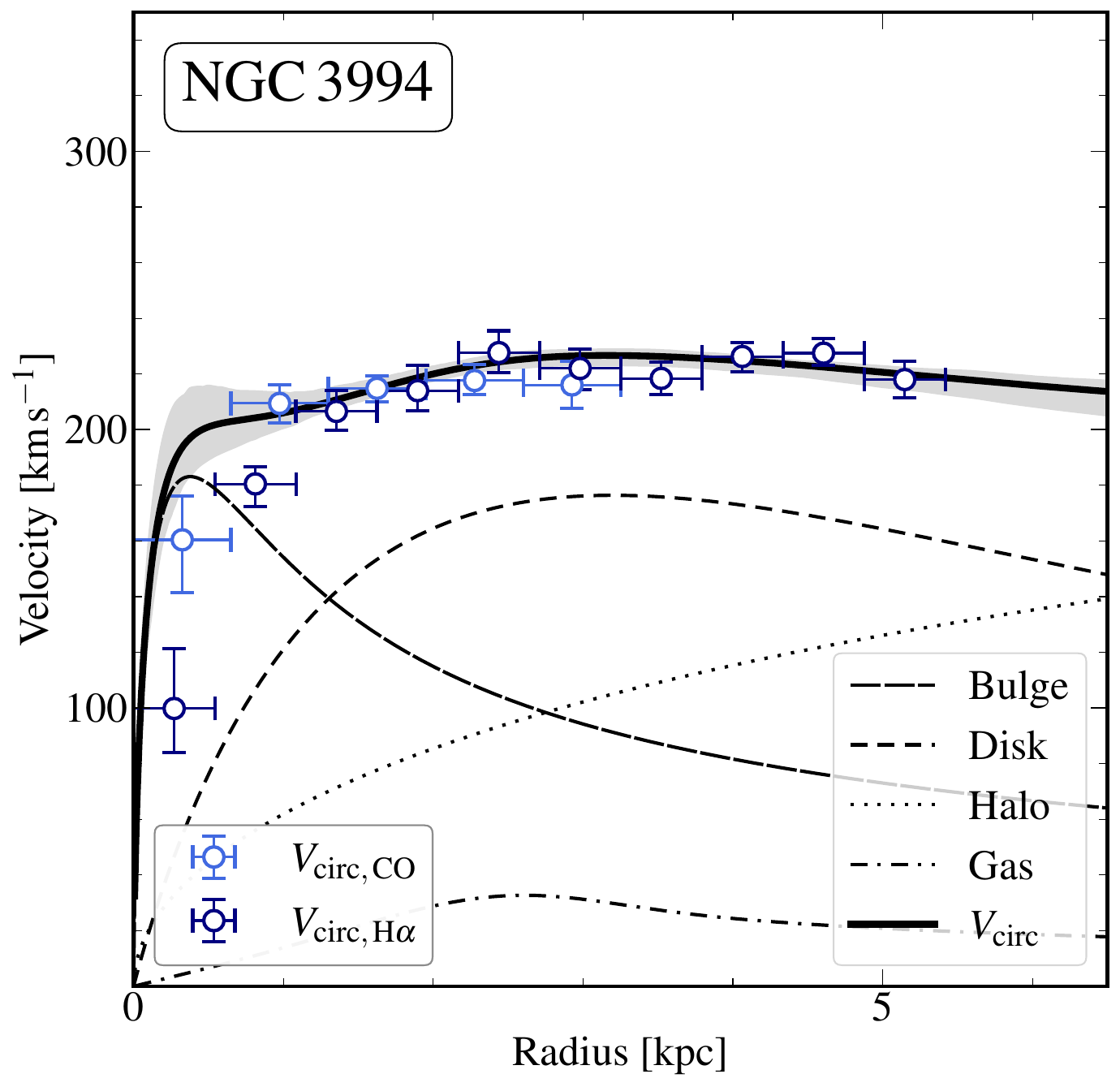}
\figsetgrpnote{The circular velocities derived from the \Bbarolo\ analysis and the best-fit dynamical model for an inactive star-forming galaxy, NGC3994. The names of targets are plotted on the bottom right corner of each sub-panel. The circular velocities obtained from \Bbarolo\ are represented by blue and cyan circles, accompanied by the shaded region indicating the associated uncertainties. The long dashed line corresponds to the circular velocities contributed by the stellar bulge, the short dashed line represents the stellar disk, the dotted line represents the dark matter component, the dash-dotted line represents the cold molecular gas, and the thick solid line represents the overall circular velocity. The gray-shaded region represents the uncertainties of the circular velocities derived from the dynamical method.}
\figsetgrpend

\figsetgrpstart
\figsetgrpnum{6.10}
\figsetgrptitle{Circular velocities for NGC4047}
\figsetplot{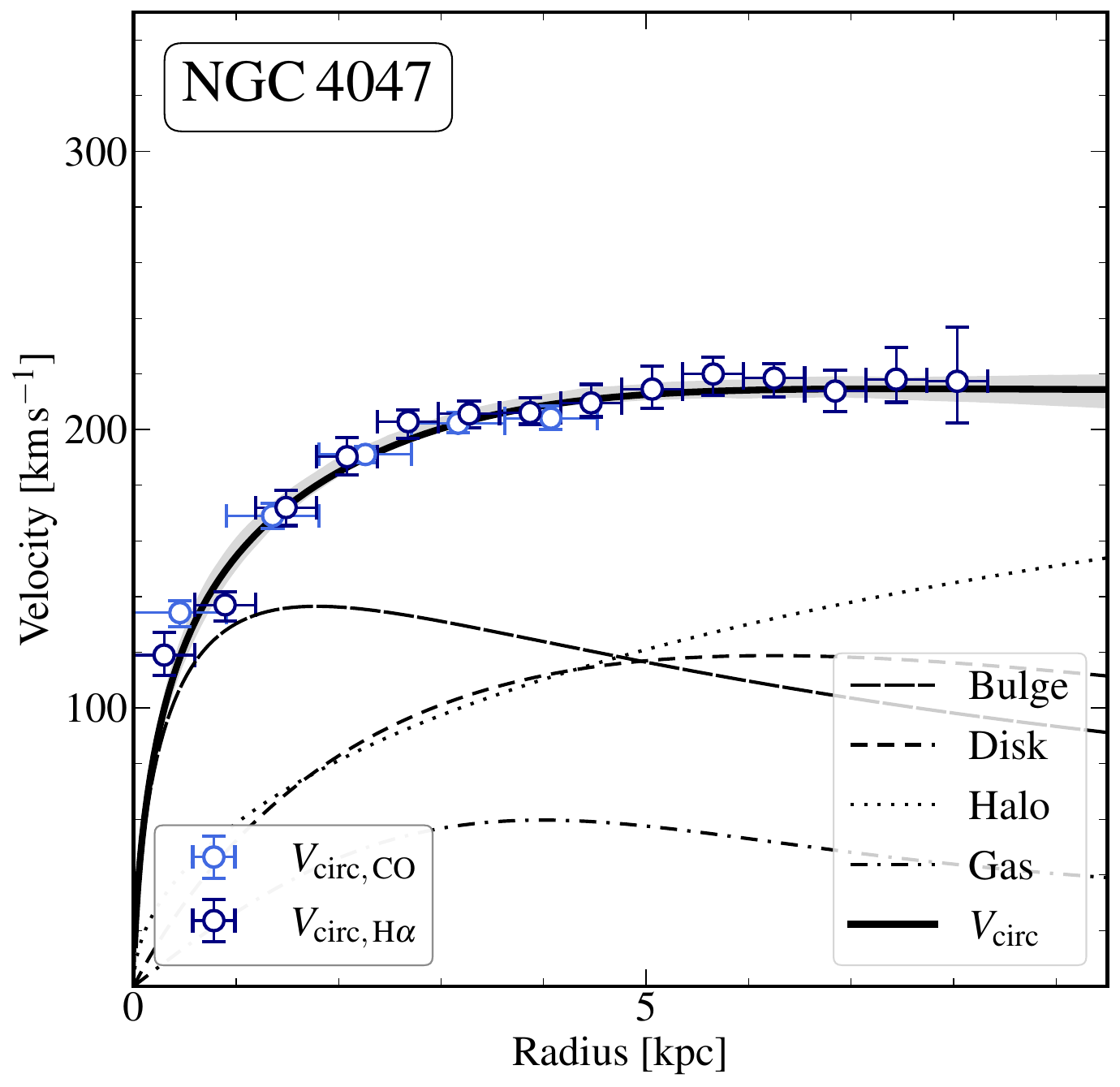}
\figsetgrpnote{The circular velocities derived from the \Bbarolo\ analysis and the best-fit dynamical model for an inactive star-forming galaxy, NGC4047. The names of targets are plotted on the bottom right corner of each sub-panel. The circular velocities obtained from \Bbarolo\ are represented by blue and cyan circles, accompanied by the shaded region indicating the associated uncertainties. The long dashed line corresponds to the circular velocities contributed by the stellar bulge, the short dashed line represents the stellar disk, the dotted line represents the dark matter component, the dash-dotted line represents the cold molecular gas, and the thick solid line represents the overall circular velocity. The gray-shaded region represents the uncertainties of the circular velocities derived from the dynamical method.}
\figsetgrpend

\figsetgrpstart
\figsetgrpnum{6.11}
\figsetgrptitle{Circular velocities for NGC4644}
\figsetplot{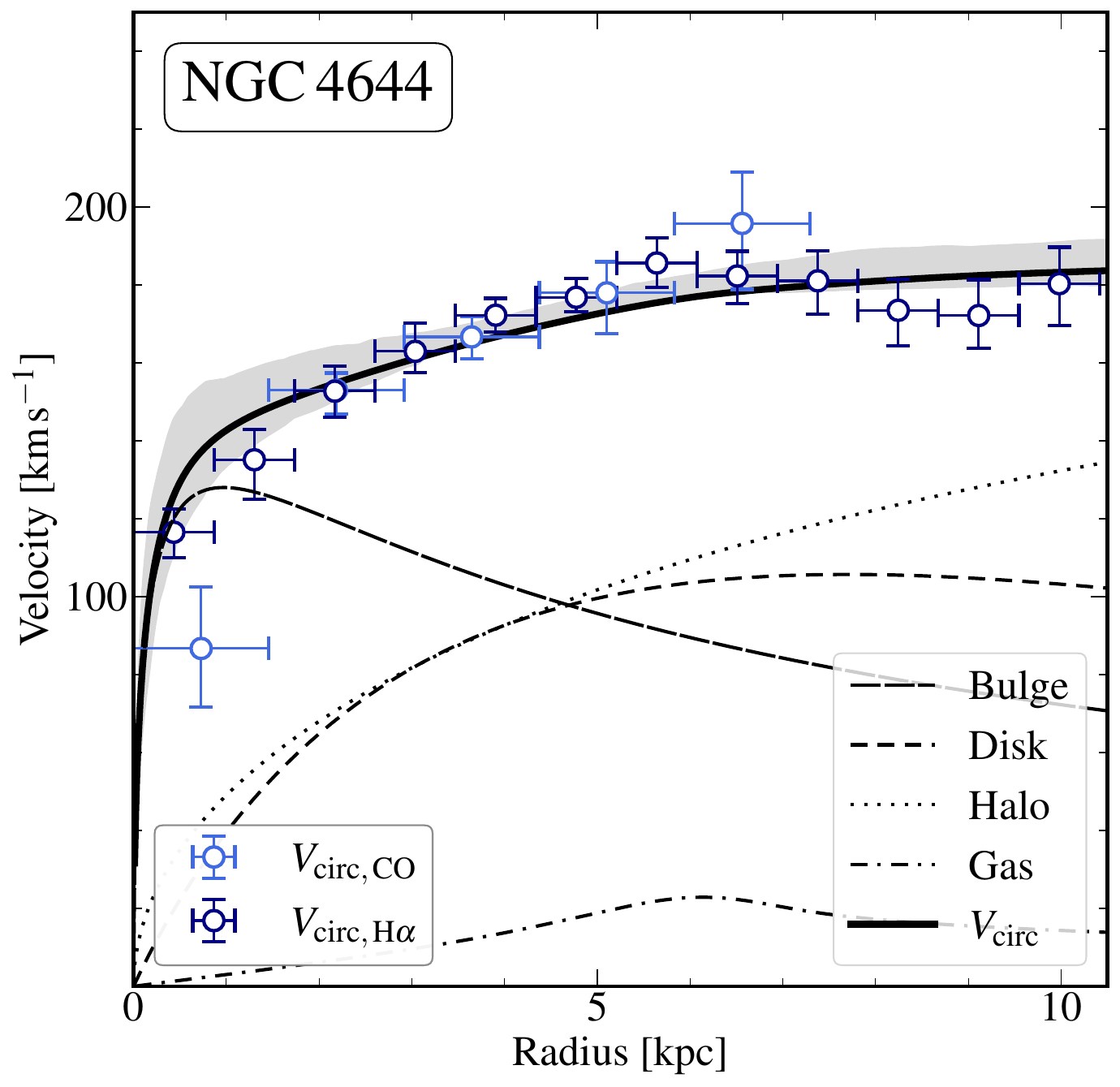}
\figsetgrpnote{The circular velocities derived from the \Bbarolo\ analysis and the best-fit dynamical model for an inactive star-forming galaxy, NGC4644. The names of targets are plotted on the bottom right corner of each sub-panel. The circular velocities obtained from \Bbarolo\ are represented by blue and cyan circles, accompanied by the shaded region indicating the associated uncertainties. The long dashed line corresponds to the circular velocities contributed by the stellar bulge, the short dashed line represents the stellar disk, the dotted line represents the dark matter component, the dash-dotted line represents the cold molecular gas, and the thick solid line represents the overall circular velocity. The gray-shaded region represents the uncertainties of the circular velocities derived from the dynamical method.}
\figsetgrpend

\figsetgrpstart
\figsetgrpnum{6.12}
\figsetgrptitle{Circular velocities for NGC4711}
\figsetplot{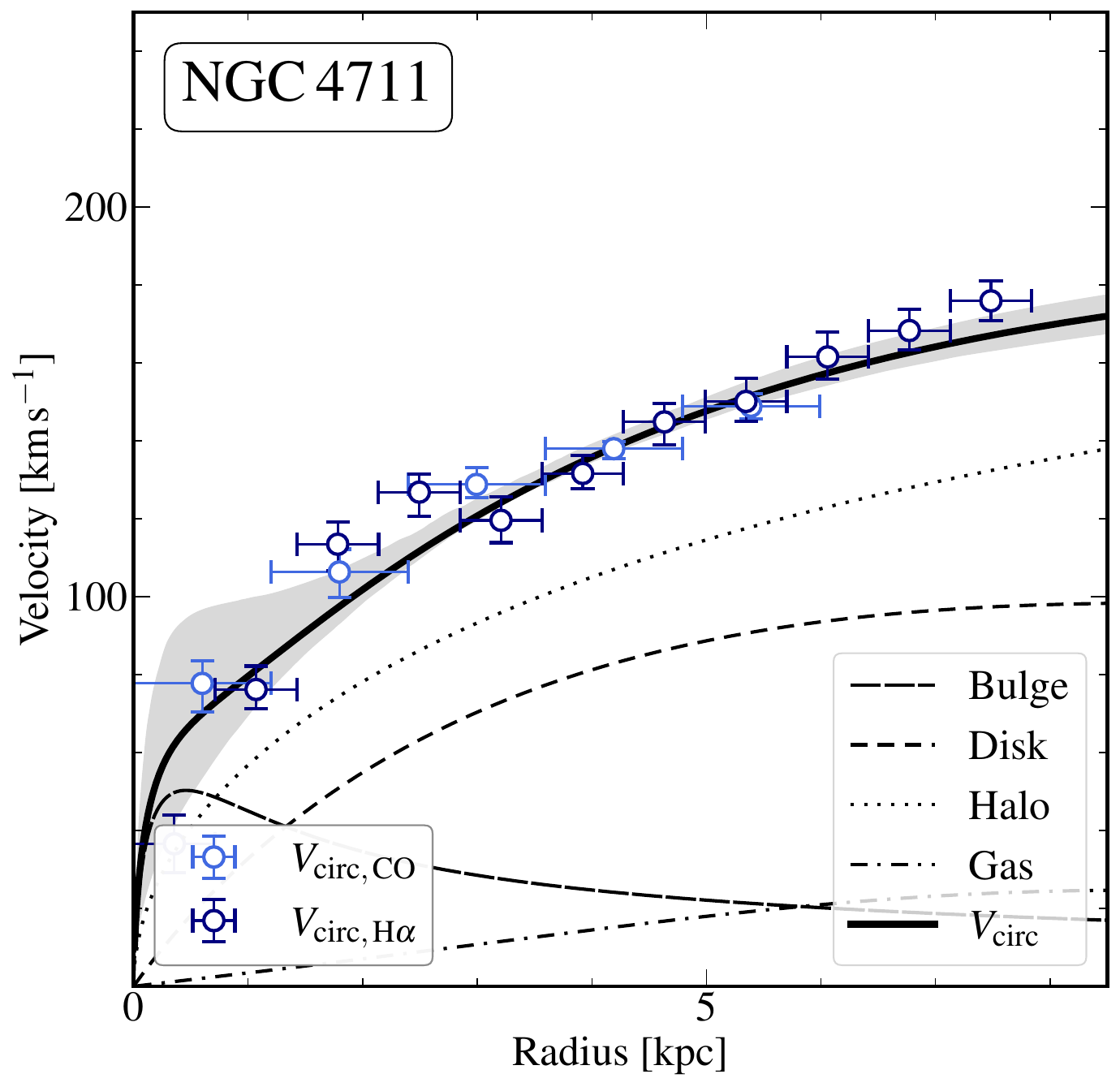}
\figsetgrpnote{The circular velocities derived from the \Bbarolo\ analysis and the best-fit dynamical model for an inactive star-forming galaxy, NGC4711. The names of targets are plotted on the bottom right corner of each sub-panel. The circular velocities obtained from \Bbarolo\ are represented by blue and cyan circles, accompanied by the shaded region indicating the associated uncertainties. The long dashed line corresponds to the circular velocities contributed by the stellar bulge, the short dashed line represents the stellar disk, the dotted line represents the dark matter component, the dash-dotted line represents the cold molecular gas, and the thick solid line represents the overall circular velocity. The gray-shaded region represents the uncertainties of the circular velocities derived from the dynamical method.}
\figsetgrpend

\figsetgrpstart
\figsetgrpnum{6.13}
\figsetgrptitle{Circular velocities for NGC5480}
\figsetplot{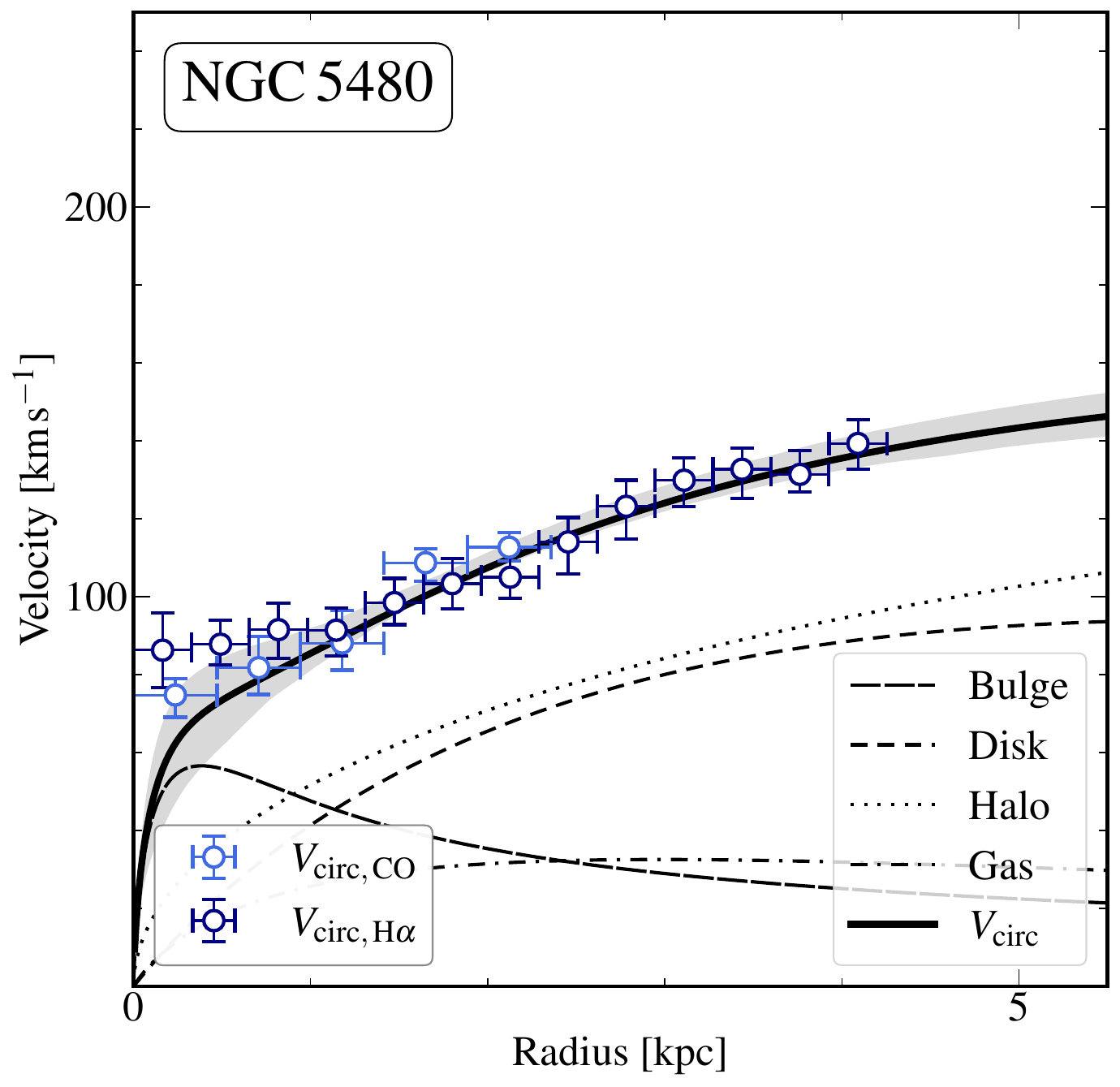}
\figsetgrpnote{The circular velocities derived from the \Bbarolo\ analysis and the best-fit dynamical model for an inactive star-forming galaxy, NGC5480. The names of targets are plotted on the bottom right corner of each sub-panel. The circular velocities obtained from \Bbarolo\ are represented by blue and cyan circles, accompanied by the shaded region indicating the associated uncertainties. The long dashed line corresponds to the circular velocities contributed by the stellar bulge, the short dashed line represents the stellar disk, the dotted line represents the dark matter component, the dash-dotted line represents the cold molecular gas, and the thick solid line represents the overall circular velocity. The gray-shaded region represents the uncertainties of the circular velocities derived from the dynamical method.}
\figsetgrpend

\figsetgrpstart
\figsetgrpnum{6.14}
\figsetgrptitle{Circular velocities for NGC5980}
\figsetplot{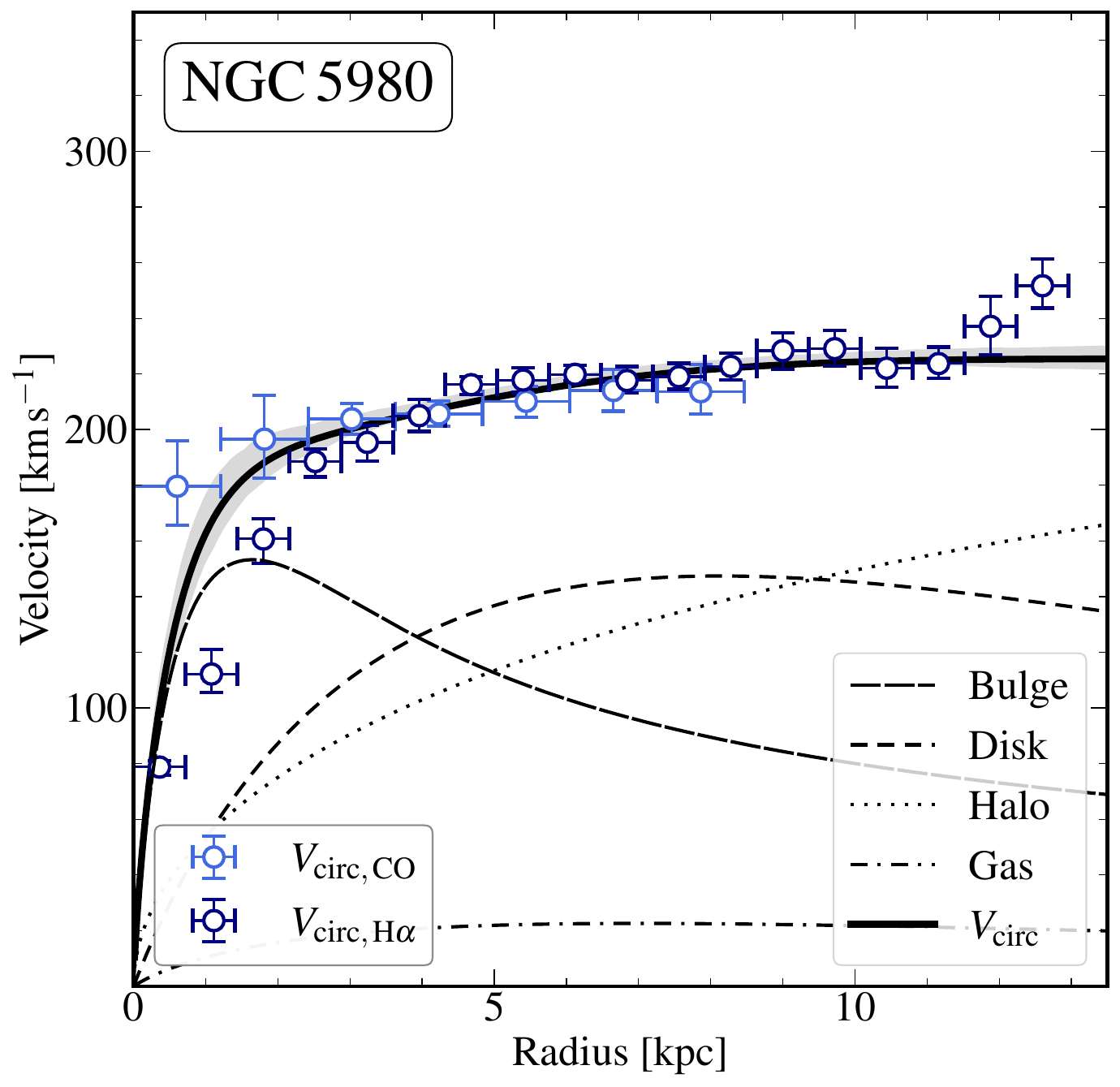}
\figsetgrpnote{The circular velocities derived from the \Bbarolo\ analysis and the best-fit dynamical model for an inactive star-forming galaxy, NGC5980. The names of targets are plotted on the bottom right corner of each sub-panel. The circular velocities obtained from \Bbarolo\ are represented by blue and cyan circles, accompanied by the shaded region indicating the associated uncertainties. The long dashed line corresponds to the circular velocities contributed by the stellar bulge, the short dashed line represents the stellar disk, the dotted line represents the dark matter component, the dash-dotted line represents the cold molecular gas, and the thick solid line represents the overall circular velocity. The gray-shaded region represents the uncertainties of the circular velocities derived from the dynamical method.}
\figsetgrpend

\figsetgrpstart
\figsetgrpnum{6.15}
\figsetgrptitle{Circular velocities for NGC6060}
\figsetplot{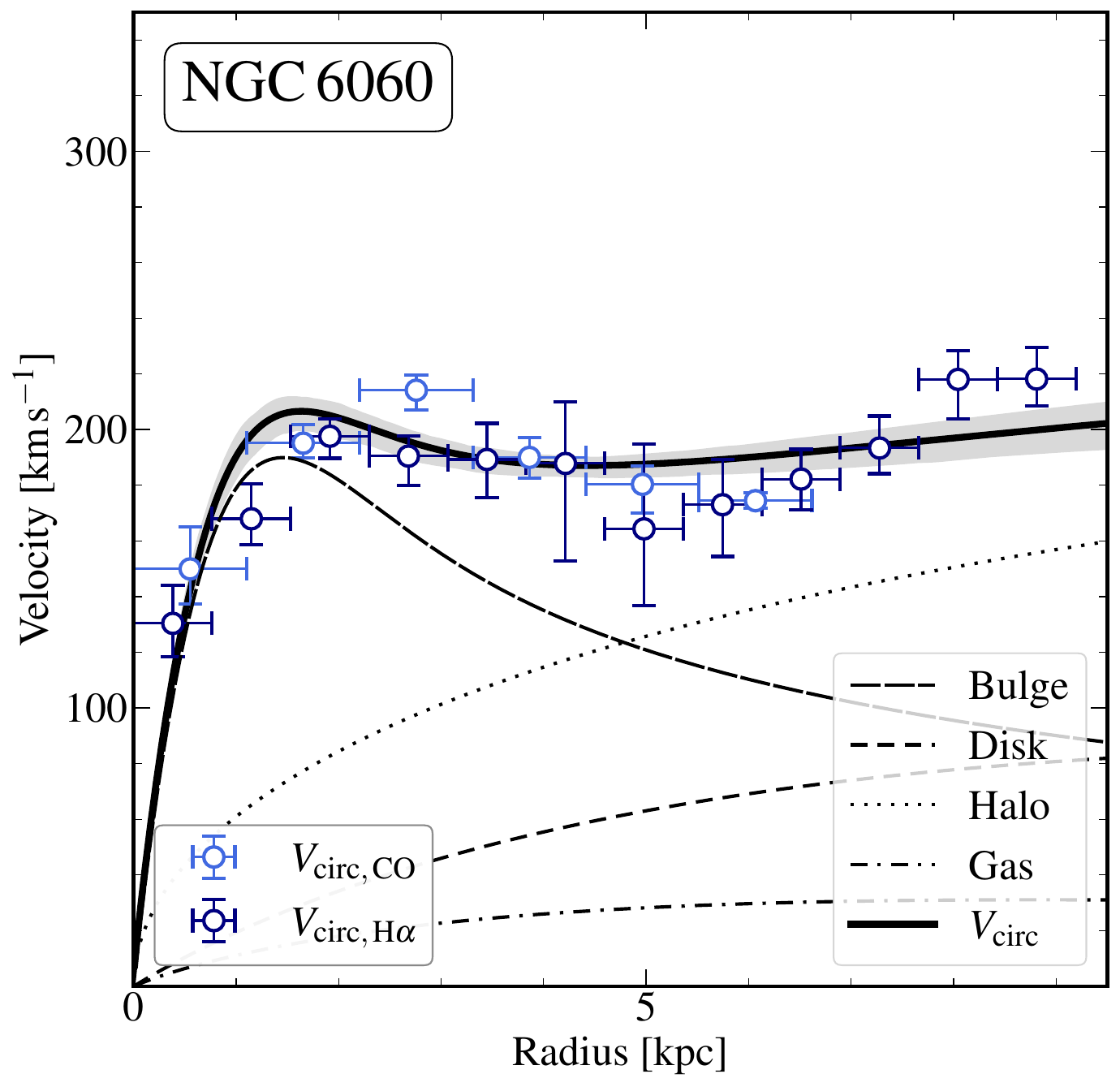}
\figsetgrpnote{The circular velocities derived from the \Bbarolo\ analysis and the best-fit dynamical model for an inactive star-forming galaxy, NGC6060. The names of targets are plotted on the bottom right corner of each sub-panel. The circular velocities obtained from \Bbarolo\ are represented by blue and cyan circles, accompanied by the shaded region indicating the associated uncertainties. The long dashed line corresponds to the circular velocities contributed by the stellar bulge, the short dashed line represents the stellar disk, the dotted line represents the dark matter component, the dash-dotted line represents the cold molecular gas, and the thick solid line represents the overall circular velocity. The gray-shaded region represents the uncertainties of the circular velocities derived from the dynamical method.}
\figsetgrpend

\figsetgrpstart
\figsetgrpnum{6.16}
\figsetgrptitle{Circular velocities for NGC6301}
\figsetplot{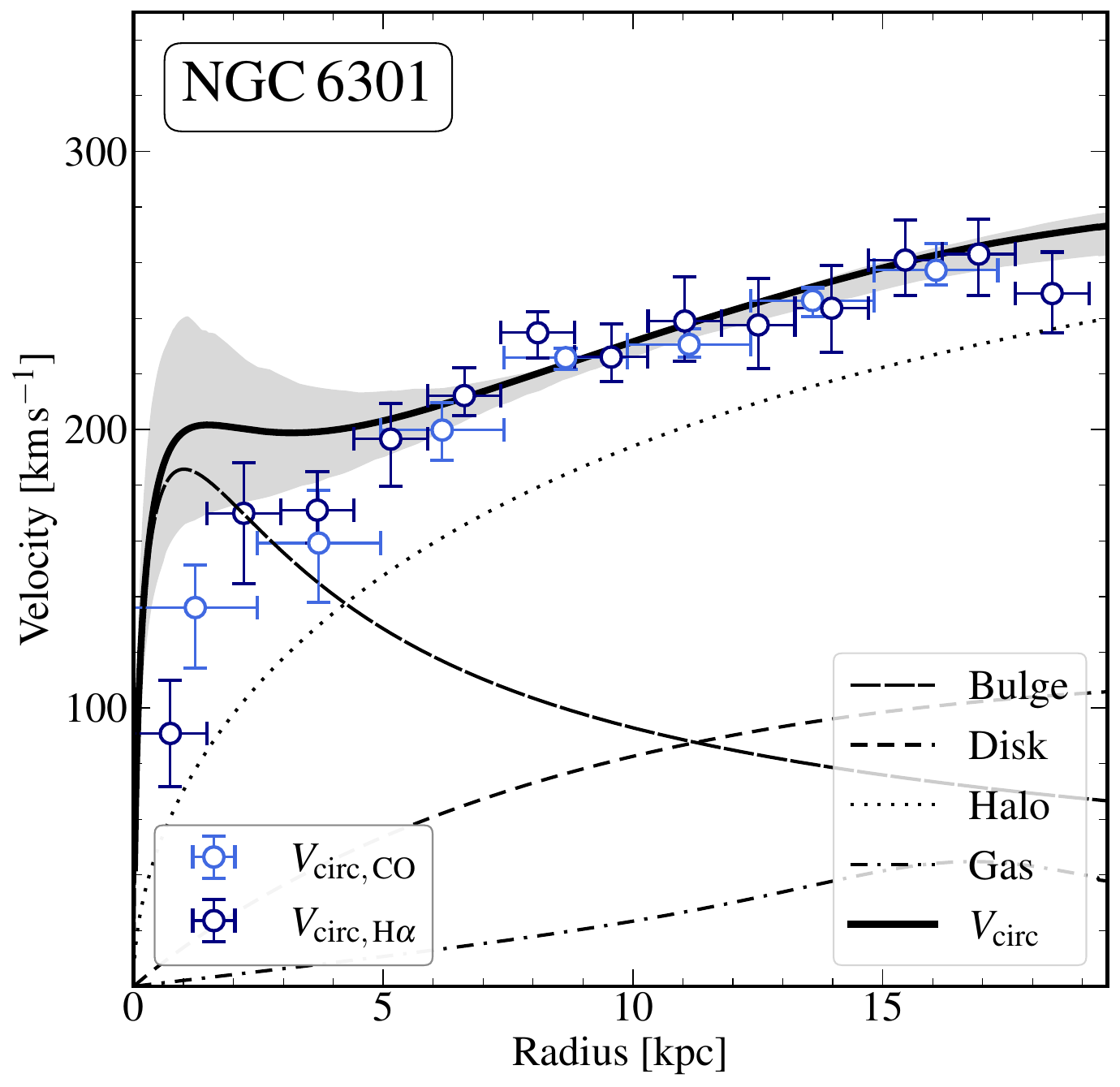}
\figsetgrpnote{The circular velocities derived from the \Bbarolo\ analysis and the best-fit dynamical model for an inactive star-forming galaxy, NGC6301. The names of targets are plotted on the bottom right corner of each sub-panel. The circular velocities obtained from \Bbarolo\ are represented by blue and cyan circles, accompanied by the shaded region indicating the associated uncertainties. The long dashed line corresponds to the circular velocities contributed by the stellar bulge, the short dashed line represents the stellar disk, the dotted line represents the dark matter component, the dash-dotted line represents the cold molecular gas, and the thick solid line represents the overall circular velocity. The gray-shaded region represents the uncertainties of the circular velocities derived from the dynamical method.}
\figsetgrpend

\figsetgrpstart
\figsetgrpnum{6.17}
\figsetgrptitle{Circular velocities for NGC6478}
\figsetplot{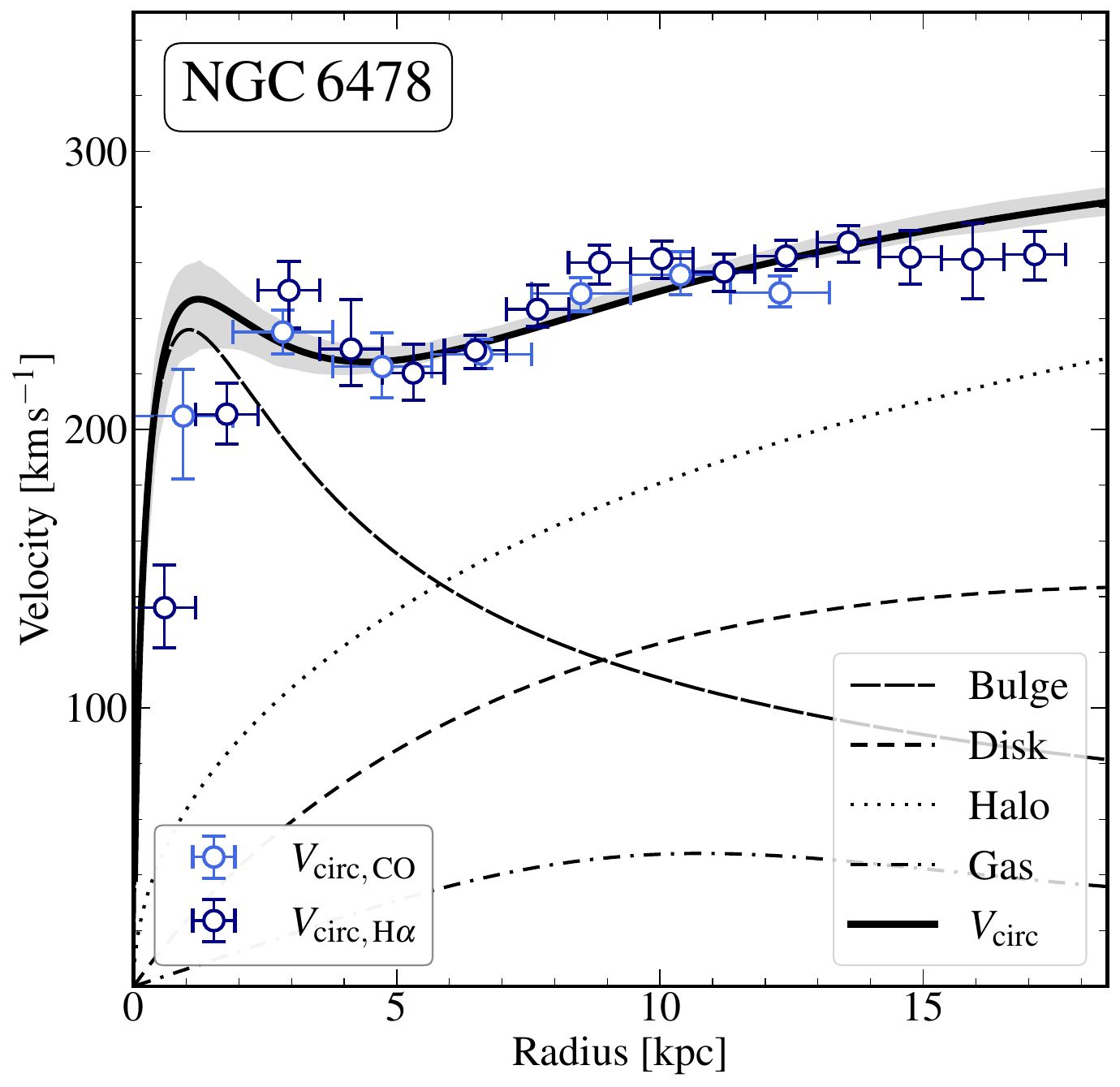}
\figsetgrpnote{The circular velocities derived from the \Bbarolo\ analysis and the best-fit dynamical model for an inactive star-forming galaxy, NGC6478. The names of targets are plotted on the bottom right corner of each sub-panel. The circular velocities obtained from \Bbarolo\ are represented by blue and cyan circles, accompanied by the shaded region indicating the associated uncertainties. The long dashed line corresponds to the circular velocities contributed by the stellar bulge, the short dashed line represents the stellar disk, the dotted line represents the dark matter component, the dash-dotted line represents the cold molecular gas, and the thick solid line represents the overall circular velocity. The gray-shaded region represents the uncertainties of the circular velocities derived from the dynamical method.}
\figsetgrpend

\figsetgrpstart
\figsetgrpnum{6.18}
\figsetgrptitle{Circular velocities for UGC09067}
\figsetplot{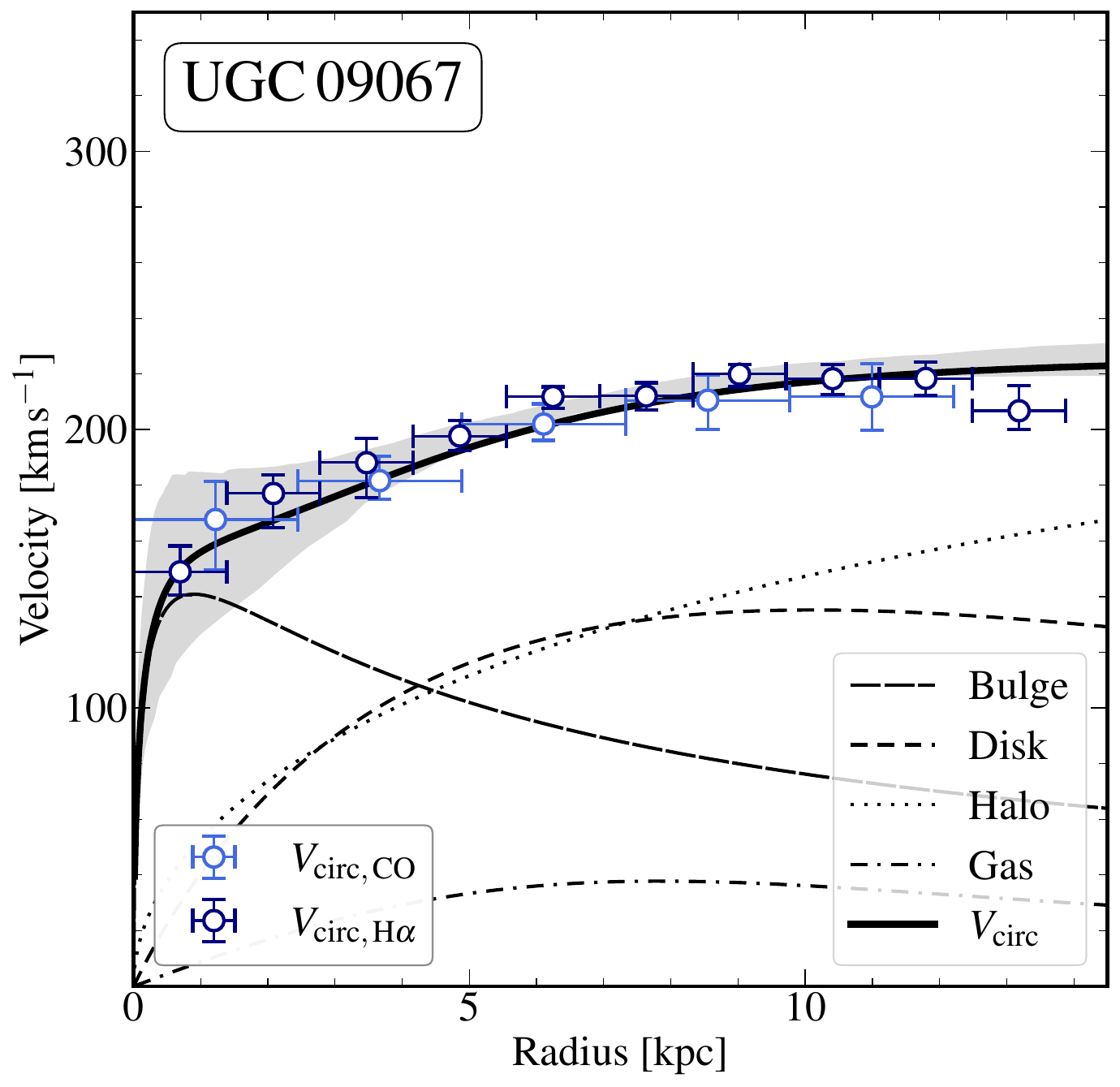}
\figsetgrpnote{The circular velocities derived from the \Bbarolo\ analysis and the best-fit dynamical model for an inactive star-forming galaxy, UGC09067. The names of targets are plotted on the bottom right corner of each sub-panel. The circular velocities obtained from \Bbarolo\ are represented by blue and cyan circles, accompanied by the shaded region indicating the associated uncertainties. The long dashed line corresponds to the circular velocities contributed by the stellar bulge, the short dashed line represents the stellar disk, the dotted line represents the dark matter component, the dash-dotted line represents the cold molecular gas, and the thick solid line represents the overall circular velocity. The gray-shaded region represents the uncertainties of the circular velocities derived from the dynamical method.}
\figsetgrpend



\subsection{Mass comparison}
\label{subsec4.4: mass comparison}
\begin{figure}
    \centering
    \includegraphics[width=\linewidth]{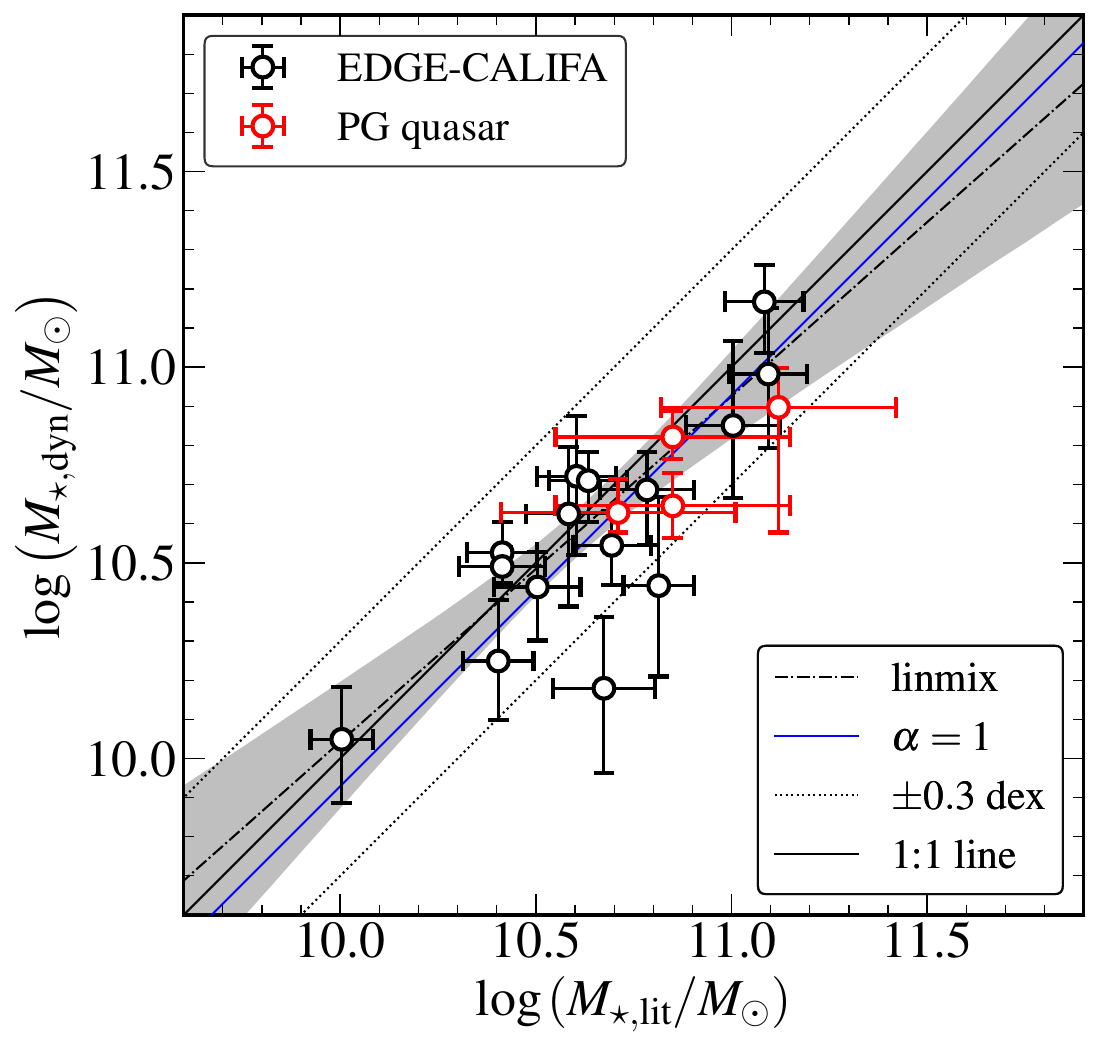}
    \caption{Comparison between the stellar masses obtained through our dynamical methodology and those documented in the literature. The black solid line corresponds to the equality line, while the dotted lines represent the extent of the associated 0.3 dex uncertainties. The best-fit outcome of the \texttt{linmix} fitting procedure is represented by the dash-dotted line, while the associated uncertainties inherent in the \texttt{linmix} results are shown as the shaded gray region. The mass of I\,Zw\,1 is also included in this figure \citep{Fei+2023}.} 
    \label{Fig7: mass comparison}
\end{figure}

We compare the total stellar masses obtained from our dynamical procedure described above to the stellar masses given by \citet{Bolatto+2017, Shangguan+2018}. The comparison result is presented in Figure \ref{Fig7: mass comparison}. Our approach capitalizes on a robust and widely accepted model of galactic dynamics, thus engendering independent estimates of stellar mass. This methodology obviates the need for making assumptions about fundamental factors such as the initial mass function (IMF) or the star formation history (SFH), both of which possess the potential to exert systematic influences upon the inference of stellar mass \citep{Mitchell+2013, Lower+2020}.

To study the inherent systematic difference arising from the adoption of different IMF in the literature, we undertake an adjustment aimed at equating stellar masses. Specifically, we apply a scaling factor of 1.5 to recalibrate the stellar mass determinations for inactive galaxies, effecting a transition from results predicated upon the Salpeter IMF to the Kroupa IMF \citep{Kennicutt+2012}. Our derived stellar masses are consistent with those provided by the literature after applying the IMF calibration, within the confines of a 0.3\,dex uncertainty. To investigate the systemic offset between the stellar masses estimated from the two methods, we fit the linear relationship between the stellar masses derived from the two methods of the EDGE-CALIFA sample while fixing the slope to 1. The fitting result is shown as the blue line shown in Figure \ref{Fig7: mass comparison}, which gives $\beta=-0.07\pm0.05$, which suggests a 0.07\,dex offset between two methods. While considering the uncertainties, the stellar mass derived from two methods are consistent. 

Furthermore, we assess the comparability between the stellar masses derived from the two methodologies by employing the \texttt{linmix} Python package to model their relationship. The optimal fit outcome using \texttt{linmix} suggests a notable congruence near the 1:1 alignment ($\log M_{\star,\rm dyn}=0.88_{-0.24}^{+0.26}\log M_{\star,\rm lit}+1.20_{-2.76}^{+2.54}$; dash-dotted line in Fig. \ref{Fig7: mass comparison}), yet displays deviations at the lower mass end. It should be emphasized that our analytical scrutiny is focused upon star-forming galaxies characterized by moderate stellar masses spanning the range of $10^{10}$--$10^{11.5}\,M_\odot$. To investigate a more comprehensive and robust comparison of stellar masses derived via disparate methodologies, it is necessary to study a more extensive sample size that spans a wider parameter space. 

We then assess the suitability of the input prior and the probability distribution function (PDF) of $\alpha_{\rm CO}$ for both quasar hosts and star-forming galaxies. Utilizing our mass model, we estimate a mid-plane density ranging from $10^{1.4}$ to $10^{2.2}\,M_\odot\, \rm pc^{-2}$, which, according to the empirical relation for star-forming galaxies \citep{Bolatto+2017}, corresponds to an $\alpha_{\rm CO}$ value of $3.1\, M_\odot\, \rm (K\,km\,s^{-1}\,pc^{-2})^{-1}$. Additionally, $\alpha_{\rm CO}$ can be related to metallicity and the star-formation rate, as described by \cite{Accurso+2017}. By estimating metallicity using the $M_\star - Z$ relation \citep{Kewley+2008}, and noting that our targets are situated on the star-formation main sequence, we found that our $\alpha_{\rm CO}$ values are in good agreement with the empirical relation.

\begin{deluxetable*}{cccccccccc}
    \tablenum{4}
    \tablecaption{Best-fit parameters\label{Tab4: dyn paras}}
    \tablewidth{0pt}
    \tablehead{
        \colhead{Name} & \colhead{$\log M_{\rm b}$} & \colhead{$R_{\rm e,\,b}$} & \colhead{$n_{\rm b}$} & \colhead{$\log M_{\rm d}$} & \colhead{$R_{\rm e,\,d}$} & \colhead{$n_{\rm d}$} & \colhead{$\log M_{\rm halo}$} & \colhead{$c$} & \colhead{$\alpha_{\rm CO}$} \\
        \colhead{} & \colhead{$\left(M_\odot\right)$} & \colhead{(kpc)} & \colhead{} & \colhead{$\left(M_\odot\right)$} & \colhead{(kpc)} & \colhead{} & \colhead{$\left(M_\odot\right)$} & \colhead{} & \colhead{$\left[M_\odot\,\rm (K\,km\,^{-1}\,pc^{-2})^{-1}\right]$}
    }
    \decimalcolnumbers
    \startdata
        PG\,0923$+$129$^a$ & $10.28_{-0.04}^{+0.03}$ & $0.60_{-0.03}^{+0.03}$ & $1.04_{-0.05}^{+0.05}$ & \makecell[c]{$10.22_{-0.09}^{+0.07}$ \\ $9.82_{-0.35}^{+0.35}$} & \makecell[c]{$2.08_{-0.03}^{+0.03}$ \\ $7.45_{-0.22}^{+0.22}$} & \makecell[c]{$0.30_{-0.03}^{+0.03}$ \\ $0.08_{-0.03}^{+0.03}$} & $12.42_{-0.15}^{+0.15}$ & $12.93_{-2.33}^{+2.10}$ & $2.99_{-1.40}^{+2.66}$ \\
        PG\,1126$-$041 & $10.43_{-0.06}^{+0.05}$ & $0.79_{-0.05}^{+0.05}$ & $4.00_{-0.05}^{+0.05}$ & $10.24_{-0.35}^{+0.22}$ & $4.14_{-0.05}^{+0.05}$ & $1^{b}$ & $12.41_{-0.16}^{+0.15}$ & $11.13_{-3.07}^{+3.03}$ & $3.54_{-1.81}^{+3.35}$ \\
        PG\,2130$+$099 & $10.62_{-0.14}^{+0.15}$ & $1.36_{-0.76}^{+0.67}$ & $0.61_{-0.32}^{+0.36}$ & $10.40_{-0.37}^{+0.18}$ & $3.67_{-0.98}^{+1.33}$ & $0.33_{-0.18}^{+0.22}$ & $12.61_{-0.16}^{+0.16}$ & $11.06_{-2.34}^{+2.55}$ & $3.46_{-1.82}^{+3.48}$ \\
        IC\,0944 & $10.45_{-0.18}^{+0.13}$ & $1.31_{-0.14}^{+0.14}$ & $0.91_{-0.05}^{+0.05}$ & $11.23_{-0.16}^{+0.11}$ & $10.71_{-0.69}^{+0.69}$ & 1 & $13.02_{-0.15}^{+0.18}$ & $8.78_{-2.53}^{+3.54}$ & $3.21_{-1.59}^{+3.38}$ \\ 
        IC\,1199 & $10.25_{-0.07}^{+0.06}$ & $1.03_{-0.12}^{+0.11}$ & $2.43_{-0.23}^{+0.23}$ & $10.82_{-0.26}^{+0.18}$ & $11.17_{-0.47}^{+0.48}$ & 1 & $12.66_{-0.17}^{+0.20}$ & $10.23_{-3.22}^{+3.69}$ & $3.28_{-1.66}^{+2.99}$ \\ 
        IC\,1683 & $10.03_{-0.16}^{+0.11}$ & $0.56_{-0.11}^{+0.11}$ & $1.69_{-0.24}^{+0.24}$ & $10.79_{-0.31}^{+0.20}$ & $8.53_{-0.35}^{+0.36}$ & 1 & $12.60_{-0.23}^{+0.22}$ & $10.50_{-3.38}^{+3.63}$ & $3.56_{-1.80}^{+3.71}$ \\ 
        NGC\,0496 & $7.94_{-0.64}^{+0.68}$ & $0.51_{-0.08}^{+0.07}$ & $2.16_{-0.31}^{+0.31}$ & $10.39_{-0.22}^{+0.18}$ & $9.64_{-0.41}^{+0.43}$ & 1 & $12.11_{-0.18}^{+0.19}$ & $10.80_{-3.14}^{+3.51}$ & $3.42_{-1.71}^{+3.06}$ \\ 
        NGC\,2906 & $9.39_{-0.27}^{+0.16}$ & $0.35_{-0.02}^{+0.03}$ & $1.77_{-0.06}^{+0.06}$ & $10.58_{-0.09}^{+0.09}$ & $4.19_{-0.12}^{+0.11}$ & 1 & $12.39_{-0.16}^{+0.14}$ & $12.16_{-3.63}^{+2.93}$ & $3.29_{-1.67}^{+3.14}$ \\ 
        NGC\,3994 & $9.79_{-0.09}^{+0.08}$ & $0.30_{-0.03}^{+0.03}$ & $1.53_{-0.14}^{+0.14}$ & $10.45_{-0.07}^{+0.07}$ & $2.33_{-0.09}^{+0.10}$ & 1 & $12.33_{-0.16}^{+0.13}$ & $12.88_{-3.74}^{+2.48}$ & $2.98_{-1.47}^{+2.82}$ \\ 
        NGC\,4047 & $10.28_{-0.07}^{+0.07}$ & $1.66_{-0.13}^{+0.13}$ & $1.85_{-0.10}^{+0.09}$ & $10.40_{-0.22}^{+0.15}$ & $4.53_{-0.08}^{+0.08}$ & 1 & $12.41_{-0.15}^{+0.16}$ & $11.46_{-3.46}^{+3.34}$ & $3.11_{-1.51}^{+2.68}$ \\ 
        NGC\,4644 & $10.10_{-0.14}^{+0.12}$ & $1.34_{-0.33}^{+0.35}$ & $2.54_{-0.35}^{+0.36}$ & $10.39_{-0.27}^{+0.15}$ & $5.61_{-0.52}^{+0.52}$ & 1 & $12.27_{-0.15}^{+0.18}$ & $9.64_{-2.57}^{+4.44}$ & $3.45_{-1.70}^{+3.61}$ \\ 
        NGC\,4711 & $8.75_{-1.08}^{+0.46}$ & $0.36_{-0.04}^{+0.04}$ & $1.54_{-0.15}^{+0.15}$ & $10.39_{-0.16}^{+0.16}$ & $6.48_{-0.27}^{+0.27}$ & 1 & $12.16_{-0.15}^{+0.19}$ & $12.62_{-3.82}^{+3.30}$ & $3.30_{-1.61}^{+3.23}$ \\ 
        NGC\,5480 & $8.77_{-0.41}^{+0.20}$ & $0.29_{-0.02}^{+0.02}$ & $1.48_{-0.08}^{+0.08}$ & $10.19_{-0.17}^{+0.14}$ & $4.54_{-0.08}^{+0.08}$ & 1 & $11.94_{-0.17}^{+0.19}$ & $12.48_{-4.07}^{+4.05}$ & $3.91_{-2.04}^{+4.12}$ \\ 
        NGC\,5980 & $10.17_{-0.07}^{+0.06}$ & $1.04_{-0.12}^{+0.11}$ & $1.05_{-0.10}^{+0.10}$ & $10.70_{-0.15}^{+0.10}$ & $5.92_{-0.26}^{+0.25}$ & 1 & $12.52_{-0.15}^{+0.18}$ & $9.54_{-2.49}^{+3.69}$ & $3.30_{-1.66}^{+3.34}$ \\ 
        NGC\,6060 & $10.23_{-0.05}^{+0.04}$ & $0.83_{-0.06}^{+0.05}$ & $0.69_{-0.02}^{+0.02}$ & $10.48_{-0.46}^{+0.31}$ & $10.43_{-0.30}^{+0.30}$ & 1 & $12.44_{-0.19}^{+0.24}$ & $11.92_{-4.01}^{+3.24}$ & $3.33_{-1.69}^{+3.39}$ \\ 
        NGC\,6301 & $10.30_{-0.22}^{+0.14}$ & $0.95_{-0.10}^{+0.10}$ & $1.87_{-0.18}^{+0.17}$ & $10.96_{-0.25}^{+0.24}$ & $19.49_{-0.85}^{+0.85}$ & 1 & $12.77_{-0.13}^{+0.23}$ & $12.07_{-3.66}^{+2.12}$ & $2.95_{-1.41}^{+2.57}$ \\ 
        NGC\,6478 & $10.46_{-0.06}^{+0.05}$ & $0.83_{-0.09}^{+0.09}$ & $1.52_{-0.15}^{+0.15}$ & $11.08_{-0.24}^{+0.20}$ & $14.86_{-0.62}^{+0.66}$ & 1 & $12.90_{-0.18}^{+0.20}$ & $9.74_{-3.26}^{+3.51}$ & $4.04_{-2.10}^{+4.61}$ \\ 
        UGC\,09067 & $10.14_{-0.27}^{+0.17}$ & $1.21_{-0.33}^{+0.34}$ & $2.50_{-0.34}^{+0.35}$ & $10.72_{-0.20}^{+0.14}$ & $7.37_{-0.67}^{+0.67}$ & 1 & $12.52_{-0.16}^{+0.18}$ & $9.31_{-2.54}^{+4.08}$ & $3.63_{-1.84}^{+3.74}$ \\ 
        \enddata
    \tablecomments{(1) Galaxy name; (2) Stellar bulge mass; (3) Effective radius of the stellar bulge; (4) S{\'e}rsic index of the stellar bulge; (5) Mass of the stellar disk; (6) Effective radius of the stellar disk; (7) S{\'e}rsic index of the stellar disk$^c$; (8) The mass of the dark matter halo; (9) The concentration parameter of the dark matter halo; (10) The CO-to-H$_2$ conversion factor. \\
    $^a$ Two stellar disks were identified in PG\,0923$+$129 from the HST image \citep{Zhao+2021}, therefore both disk components are included during the dynamical fitting. \\ 
    $^b$ Parameters that are fixed during the dynamical fitting. \\
    $^c$ The S\'ersic indices of inactive star-forming galaxies are all fixed to 1 according to \cite{Mendez-Abreu+2017}.}
\end{deluxetable*}

\section{Discussion}
\label{sec5: discussions}

\subsection{The kinetic pressure, scale height, and the weight of the cold ISM}
\label{subsec5.1: gas disk scale height}
\begin{figure}
    \centering
    \includegraphics[width=\linewidth]{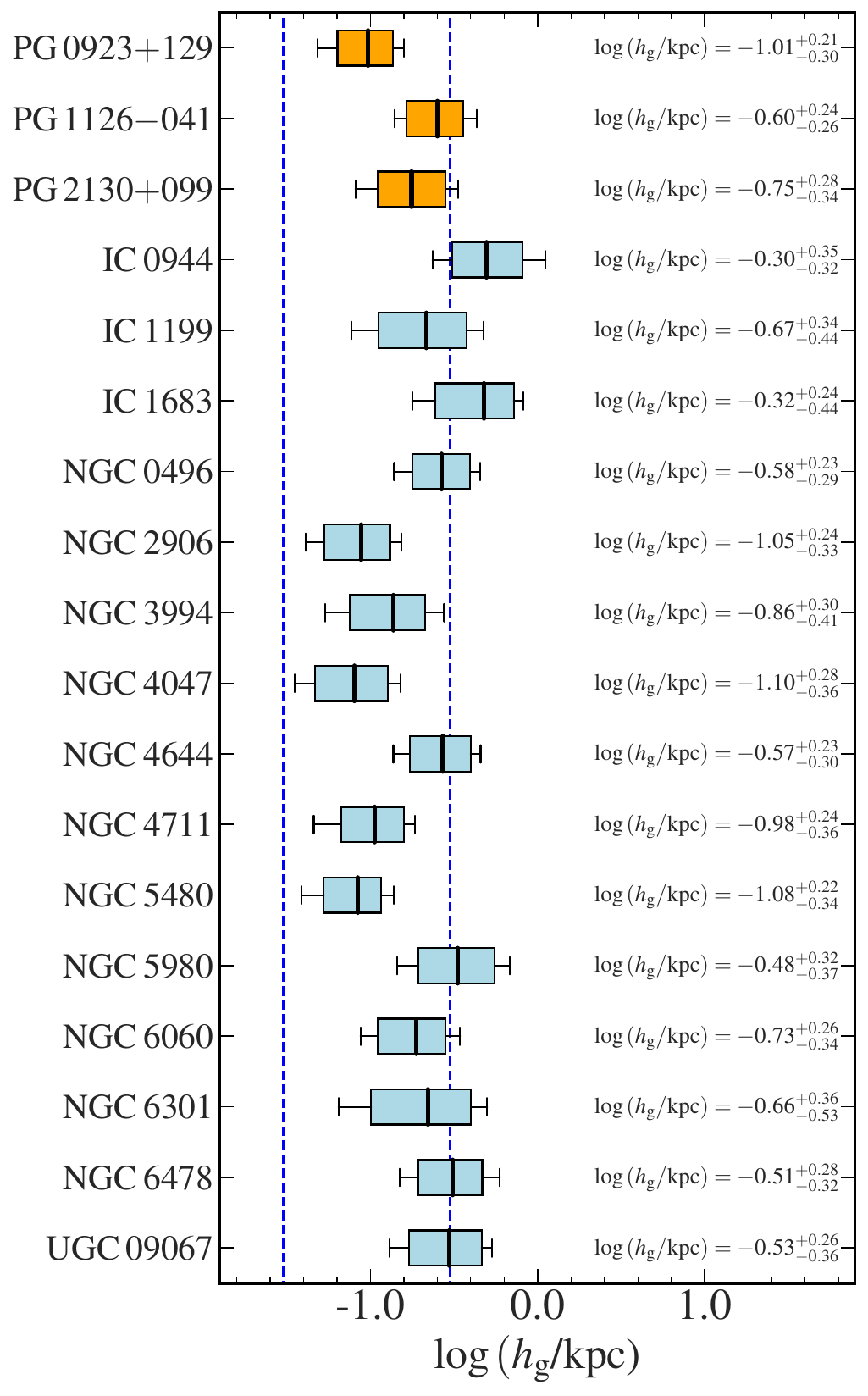}
    \caption{The scale height of the molecular gas disk in quasar host galaxies (orange boxes) and in the control sample (light blue). Each box represents the range between the 25th and 75th percentile of the scale height, with the median value indicated by a thick vertical line. The extended caps on the boxes represent the range between the 16th and 84th percentile of the scale height, which are used to estimate the uncertainty. The values and uncertainties of the scale height for each galaxy are labeled on the right side of the figure. The vertical dashed lines represent the range of gas disk scale height given by previous studies about normal star-forming galaxies and starburst galaxies \citep{Bacchini+2019, Wilson+2019}.}
    \label{Fig8: gas disk height}
\end{figure}
By delving into the intricacies of the vertical momentum equation that governs the behavior of molecular gas within galactic disks, \cite{Boulares&Cos1990} introduced the notion that a state of equilibrium is attained when the kinetic pressure, encompassing magnetic and cosmic ray pressures at the midplane, effectively counterbalances the gravitational force exerted by the galaxies themselves. This equilibrium pivots on various factors, including the gas surface density, as well as the potentials attributed to the stellar and dark matter constituents, as comprehensively described by \cite{Ostriker&Kim2022}. Nevertheless, the persistence of this delicate equilibrium within AGN host galaxies remains an enigma, primarily due to the potential influence of AGN feedback mechanisms (e.g., \citealt{Fabian2012}). This feedback is widely recognized for its pivotal role in modulating the trajectory of galactic evolution by infusing energy and momentum into the ISM \citep{Silk&Rees1998, Yuan+2018}.

To assess the impact of AGN feedback in quasar host galaxies, we undertake an assessment involving the quantification of the midplane's kinetic pressure ($P_{\rm ISM}$) and the gravitational pressure on the interstellar medium (weight; $\mathcal{W}$). The estimation of both pressures is based on the mass model acquired through our dynamical methodology. By investigating the relationship between these pressures in quasar host galaxies, we aim to substantiate the substantive role of AGN feedback. We also compute the corresponding $P_{\rm ISM}$ and $\mathcal{W}$ values for inactive galaxies, thereby facilitating a comparative analysis.

The weight of the ISM can be expressed as follows \citep{Ostriker&Kim2022}:
\begin{eqnarray}
    \mathcal{W} = \pi G \Sigma_{\rm g}^2 + 4\pi \zeta_{\rm d} G \Sigma_{\rm g} \rho_{\rm sd} h_{\rm g},
\end{eqnarray}
where $G$ is the gravitational constant; $\Sigma_{\rm g}$ represents the surface density of the gas; $\zeta_{\rm d}$ is a dimensionless number depending on, but not sensitive to, the geometry of gas disk which is close to one-third \citep{Ostriker+2010}; $\rho_{\rm sd}$ is the external density attained to other galaxy components; $h_{\rm g}$ is the half-thickness of the gas disk. The first term on the right side of this equation encapsulates the weight generated by the self-gravitational pull of the cold molecular gas disk \citep{Spitzer1942, Elmegreen1989}. It is pertinent to note that the actual gas surface density should encompass both molecular and atomic gas constituents; however, for our analysis, we assume the atomic gas contribution negligible, permitting us to replace $\Sigma_{\rm g}$ with $\Sigma_{\rm mol}$. The second term on the right side of the equation delineates the gravitational weight of the interstellar medium arising from external sources, encompassing the stellar component and the dark matter halo.

The kinetic pressure of the ISM at the midplane is defined by the difference in the total vertical momentum flux across the gas layer \citep{Boulares&Cos1990, Ostriker&Kim2022}, and thus can be expressed as
\begin{eqnarray}
    P_{\rm ISM} = \rho_{\rm mid} \sigma_{\rm eff}^2=\frac{\Sigma_{\rm g}}{2h_{\rm g}}\sigma_{\rm g}^2(1+\alpha),
\end{eqnarray}
where $\sigma_{\rm g}$ represents the velocity dispersion of molecular gas, and the $\alpha\sim 0.3$ represents the contribution of magnetic field and the cosmic ray  \citep{Wilson+2019,Ostriker&Kim2022}. It is noteworthy that both the kinetic pressure and the weight of the interstellar medium are reliant on the scale height of the molecular gas disk, $h_{\rm g}$, which could demonstrate variations across different galaxies \citep{Wilson+2019, Patra2020}, and is hard to determine. The $P_{\rm ISM}$ strongly depends on how to estimate the gas velocity dispersion. Here we consider two kinds of velocity dispersions, one is given by \Bbarolo\ model ($\sigma_{\rm mod}$), which only considers the gas in the regular rotating disk, and the other is the intrinsic gas velocity dispersion ($\sigma_{\rm intr}$; please see more details in Section \ref{subsec5.2: equilibrium}). We calculate the kinetic pressure using these two velocity dispersions and estimated $h_{\rm g}$ for each pixel by assuming the equality between $\mathcal{W}$ and $P_{\rm ISM,\,mod}$ using $\sigma_{\rm mod}$ using the Equation (5) in \cite{Ostriker&Kim2022}.

The distribution of $h_{\rm g}$ is depicted in Figure \ref{Fig8: gas disk height}. Notably, the scale height of the gas disk spans a range from 50\,pc to 500\,pc across the sample and exhibits considerable variation even within individual galaxies. Furthermore, our findings align with the established literature, wherein the gas disk scale height is reported to be within the range of $h_{\rm g}\sim 30-300\,$pc \citep{Bacchini+2019, Wilson+2019, Molina+2020, Jeffreson+2022}, except for IC\,0944, whose high inclination angle may cause highly uncertain beam-smearing effect correction. Notably, these studies derived $h_{\rm g}$ using a similar assumption of hydrostatic equilibrium while adjusting the velocity dispersion, which differs from our methodology. Nevertheless, it is easy to observe that the distribution of $h_{\rm g}$ in quasar host galaxies ($h_{\rm g}=0.17\pm0.13\,\rm kpc$) is similar to that of the control sample ($h_{\rm g}=0.30\pm0.30\,\rm kpc$). The molecular gas in these AGN host galaxies does not show a tendency to move to higher latitudes. This result suggests that either there is no noticeable external energy budget present in the quasar host galaxies or the influence of AGN feedback is so weak that it has little effect on the dynamic states of cold molecular gas.

\subsection{The cold gas equilibrium in quasar hosts and inactive galaxies}
\label{subsec5.2: equilibrium}

\begin{figure*}[ht]
    \centering
    \includegraphics[width=\linewidth]{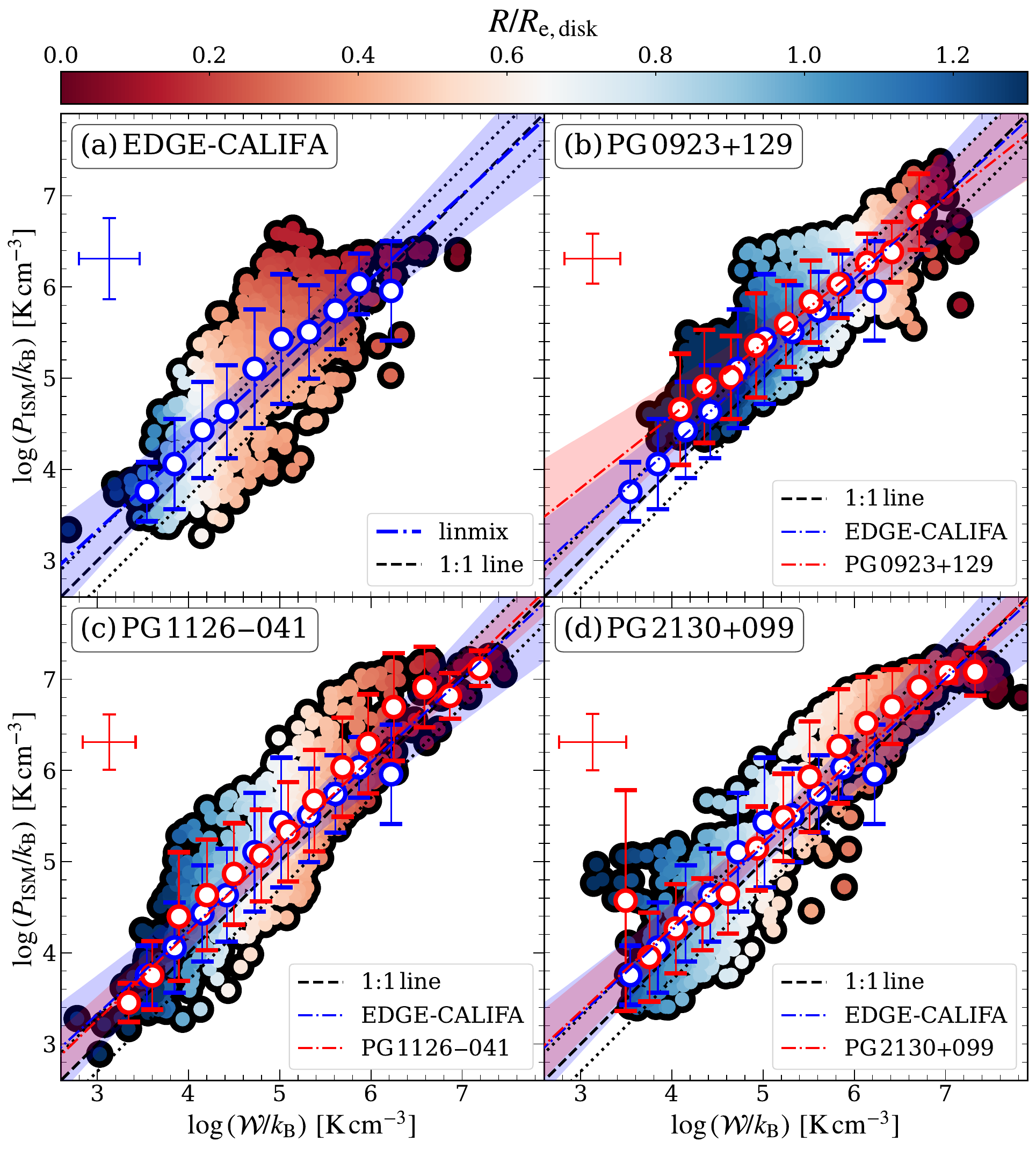}
    \caption{The relationship between the kinetic pressure of the ISM ($P_{\rm ISM}$) and the weight of the ISM ($\mathcal{W}$) for inactive galaxies (\textit{panel (a)}) and for quasar host galaxies ({\it panel (b,c,d)}). Each data point represents a specific pixel. The blue circles illustrate the mean values of $P_{\rm ISM}$ for normal star-forming galaxies within 0.3\,dex $\mathcal{W}$ bins, while red circles denote that for PG quasars. Error bars indicate the standard deviation of $P_{\rm ISM}$ within each $\mathcal{W}$ bin. The black dashed line represents the line of equality, with an associated scatter of 0.3\,dex by the dotted line. The blue dash-dotted line represents the fitting result obtained through the usage of {\tt linmix}, with uncertainties represented by the blue shaded region.}
    \label{Fig9: pressure-weight}
\end{figure*}

Now, we estimate the kinetic pressure and the ISM weight of each pixel in every inactive galaxy and quasar host with the real velocity dispersion and the $h_{\rm g}$ derived from Section \ref{subsec5.1: gas disk scale height}. Although \Bbarolo\ can correct the beam-smearing effect, it is still hard to directly obtain the pixel-wise velocity dispersion due to the projection effect. Due to the inclination angle of galaxies, the velocity dispersion of each pixel is always enhanced since more than one velocity components are included along the line of sight. This enhanced velocity dispersion is not caused by the limited spatial resolution but just arises from the overlapped gaseous clouds. According to this reason, we eliminate the spurious contribution arising from the beam-smearing effect to obtain a pixel-wise velocity dispersion, following the procedure \citep{Levy+2018, Adamczyk+2023}. The procedure involves generating moment 2 maps through the same mask method outlined in Section \ref{subsec3.1: gas distribution}. We utilize \Bbarolo\ to simulate the velocity dispersion solely induced by the beam-smearing effect. This simulation involves endowing each galaxy model with equivalent CO brightness and rotation velocities derived from the best-fit kinematics, albeit with a velocity dispersion that is only contributed by the thermal broadening, $\sigma_{\rm th}\sim 0.3\,\rm km\,s^{-1}$ . Following this, the model is convolved with the synthesized beam and used to calculate the simulated velocity dispersion ($\sigma_{\rm sim}$). The simulated velocity dispersion is then subtracted in quadrature from the observed moment 2 map to obtain the beam-smearing-corrected (bsc) velocity dispersion ($\sigma_{\rm bsc}^2=\sigma_{\rm obs}^2-\sigma_{\rm sim}^2$). Finally, we manually add the thermal broadening to the bsc velocity dispersion and estimate the intrinsic gas velocity dispersion ($\sigma_{\rm intr}^2 = \sigma_{\rm bsc}^2 + \sigma_{\rm th}^2$).

In order to determine the uncertainties of $P_{\rm ISM}$ and $\mathcal{W}$, we have created $N$ realizations of each pixel by randomly selecting parameters from the posterior distribution of parameters obtained in Section \ref{subsec4.3: dynamical fitting}. We have also considered the uncertainties of velocity dispersion at each pixel by generating a normal distribution with the center from the velocity dispersion adopted from the above-mentioned and the uncertainty given by the \texttt{maskmoment} in Section \ref{subsec3.2: gas kinematics}. After generating the realizations, we calculate the final $P_{\rm ISM}$ and $\mathcal{W}$ of each pixel from the mean value of those realizations. The uncertainties are estimated from the standard deviation of the sample. We have found that $N=1000$ is sufficient to obtain a good description of the error.

Based on the line ratio of each galaxy given by \citet{Shangguan+2020a}, we estimate the CO(1--0) line surface brightness for PG quasar host galaxies. We adopted $R_{21}=0.53$ for PG\,1126$-$041 and $R_{21}=0.90$ for PG\,2130$+$099, which are measured from the global CO(1--0) and CO(2--1) observation. We note that the PG\,0923$+$129 has no literature CO(1--0) observation, therefore we adopted the mean value of the low-$z$ quasar host galaxies, $R_{21}=0.62$ \citep{Shangguan+2020a}. We then estimate the molecular gas surface density by adopting the $\alpha_{\rm CO}$ values from the dynamical model as shown in Section \ref{sec4: gas dynamics}. The $\rho_{\rm sd}$ is also estimated from the dynamical model. We removed those pixels whose beam-smearing corrected velocity dispersion was less than the $\sigma$ of the channel width (for ALMA data, $\sigma_{\rm cw}\sim14\,\rm km\,s^{-1}$, for EDGE data, $\sigma_{\rm cw}\sim 8.5\,\rm km\,s^{-1}$) or those pixels with SNR$<3$.


The relationships between $P_{\rm ISM}$ and $\mathcal{W}$ for inactive galaxies and quasar host galaxies are shown in Figure \ref{Fig9: pressure-weight}. Since the EDGE data are not as deep as our ALMA observation, it is hard to investigate the relationship for individual targets, here we present the result of stacking all of the targets. The color of each data point represents the distance between each pixel and the galaxy center. The white circles with error bars represent the intensity-weighted mean and standard deviation error of $P_{\rm ISM}$ within each 0.3\,dex $\mathcal{W}$ bins. 

For PG\,0923$+$129, we found that $P_{\rm ISM}$ is enhanced at large radii and reduced at small radii, while the global trend follows the 1:1 relation, suggesting an equilibrium between kinetic pressure and the gravitational pressure across the majority of the gas disk. However, several data points are located above the hydrodynamic equilibrium line, suggesting a perturbation of gas within those regions. We regard this perturbation as a result of spiral arms, mentioned in Section \ref{subsec3.3: kinematics results}. Similar results are also found in PG\,1126$-$041 and PG\,2130$+$099. However, note that these two quasar hosts have relatively larger inclination angles, this effect might be caused by the residual beam-smearing effect that is not fully removed in the previous procedure \citep{Adamczyk+2023}. For those data points with small radii, we found a deviation from the 1:1 relation in PG\,1126$-$041 and PG\,2130$+$099, potentially suggesting the effect of AGN feedback. Moreover, we find that $P_{\rm ISM}$ in all three targets decreases in the inner region (dark red pixels). Since those pixels are close to the galaxy center, it is hard to provide any constraints on the physical scale of those pixels, except for providing an upper limit. The spatial resolution of our PG quasars corresponds to $\sim 500\,\rm pc$ for PG\,0923$+$129, and corresponds to $\sim 1.2\,\rm kpc$ for the other two targets. The possibility of the deviation can be caused by several reasons. The molecular gas could be a complex structure at this scale that cannot be fully described with the current single disk model \citep{Garcia-Burillo+2014, Combes+2019}. The AGN feedback can also contribute to breaking the equilibrium. Due to the limited resolution of ALMA data, it is hard to investigate which reason dominates the phenomenon. Although the AGN feedback might be more influential in those regimes, we note that for the majority of the galaxy disk, the molecular gas satisfies the hydrostatic equilibrium and the feedback impact is negligible.

To avoid the effect of those specific data points, we investigated the relationship between $P_{\rm ISM}$ and $\mathcal{W}$ using the binned $P_{\rm ISM}$ and $\mathcal{W}$, we fit the relation between these two parameters in logarithmic space using the Python package \texttt{linmix} \citep{Kelly2007}. This yields the best-fitting power-law relations for inactive star-forming galaxies (blue dashed-dotted line in Figure \ref{Fig9: pressure-weight}):
\begin{eqnarray}
    \log &\left(\frac{P_{\rm ISM}}{k_{\rm B}\,\rm K\,cm^{-3}}\right) = 0.57_{-0.97}^{+0.98} \\
    &+ \left(0.92_{-0.20}^{+0.20}\right)\log \left(\frac{\mathcal{W}}{k_{\rm B}\,\rm K\,cm^{-3}}\right),
\end{eqnarray}
where $k_{\rm B}$ is the Boltzmann constant. The best-fit result suggests a linear relationship between the $P_{\rm ISM}$ and the $\mathcal{W}$ in inactive star-forming galaxies, which is similar to the theoretical 1:1 relation proposed by \cite{Ostriker&Kim2022}. 

The relation between $P_{\rm ISM}$ and $\mathcal{W}$ for quasar host galaxies are illustrated as red dash-dotted lines in Figure \ref{Fig9: pressure-weight}. We note that the relationships within the host galaxies of quasars are consistent with those in actively star-forming galaxies and the theoretical 1:1 relation, accounting for the uncertainties. This finding implies that the dynamical states of cold molecular gas in quasar host galaxies bear resemblance to those in inactive star-forming galaxies, suggesting that the impact of AGN feedback on the cold interstellar medium is likely to be negligible. 

\subsection{The molecular gas outflow}
\label{subsec5.3: gas outflow}
\begin{figure*}
    \centering
    \includegraphics[width=0.45\linewidth]{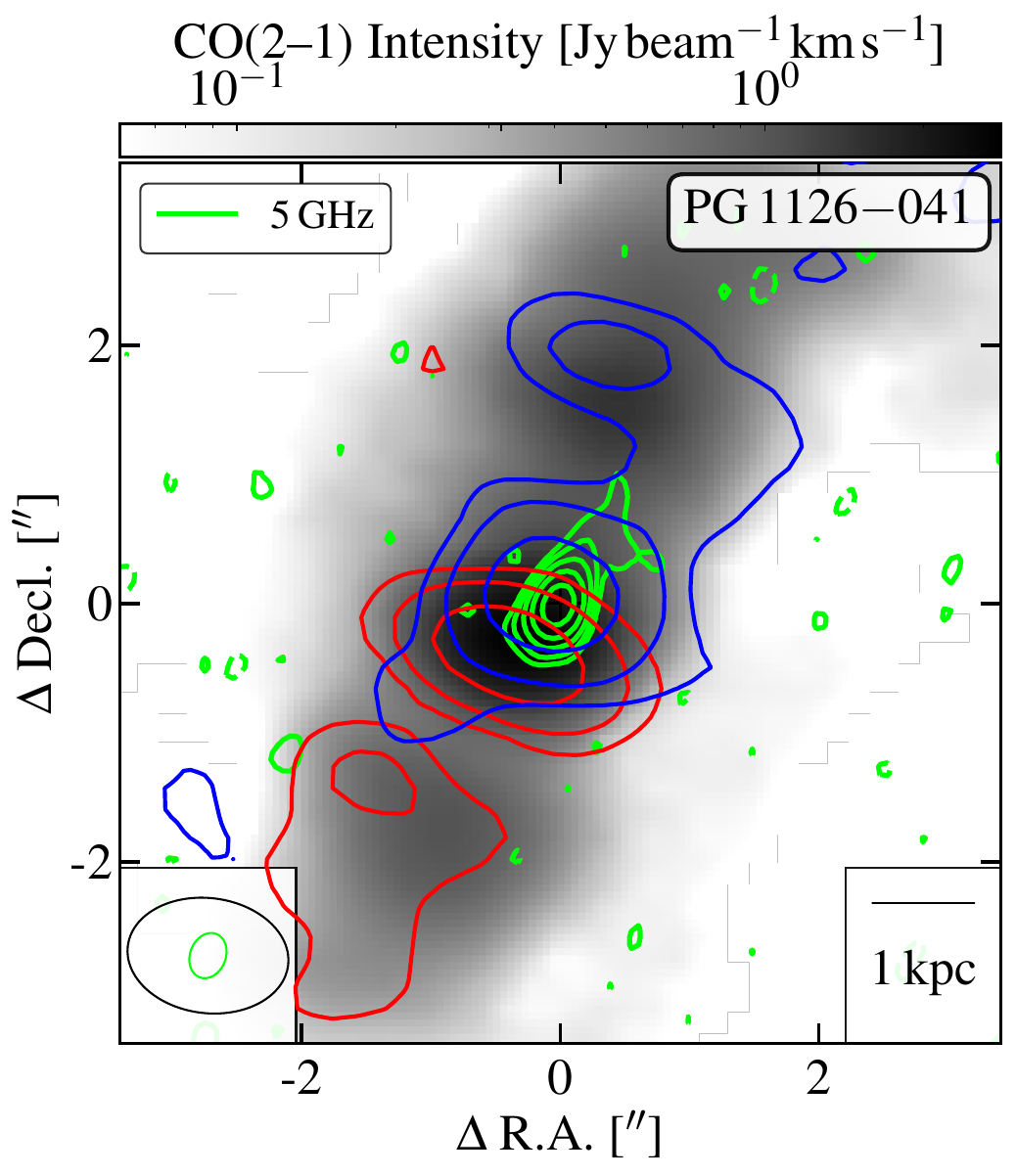}
    \includegraphics[width=0.45\linewidth]{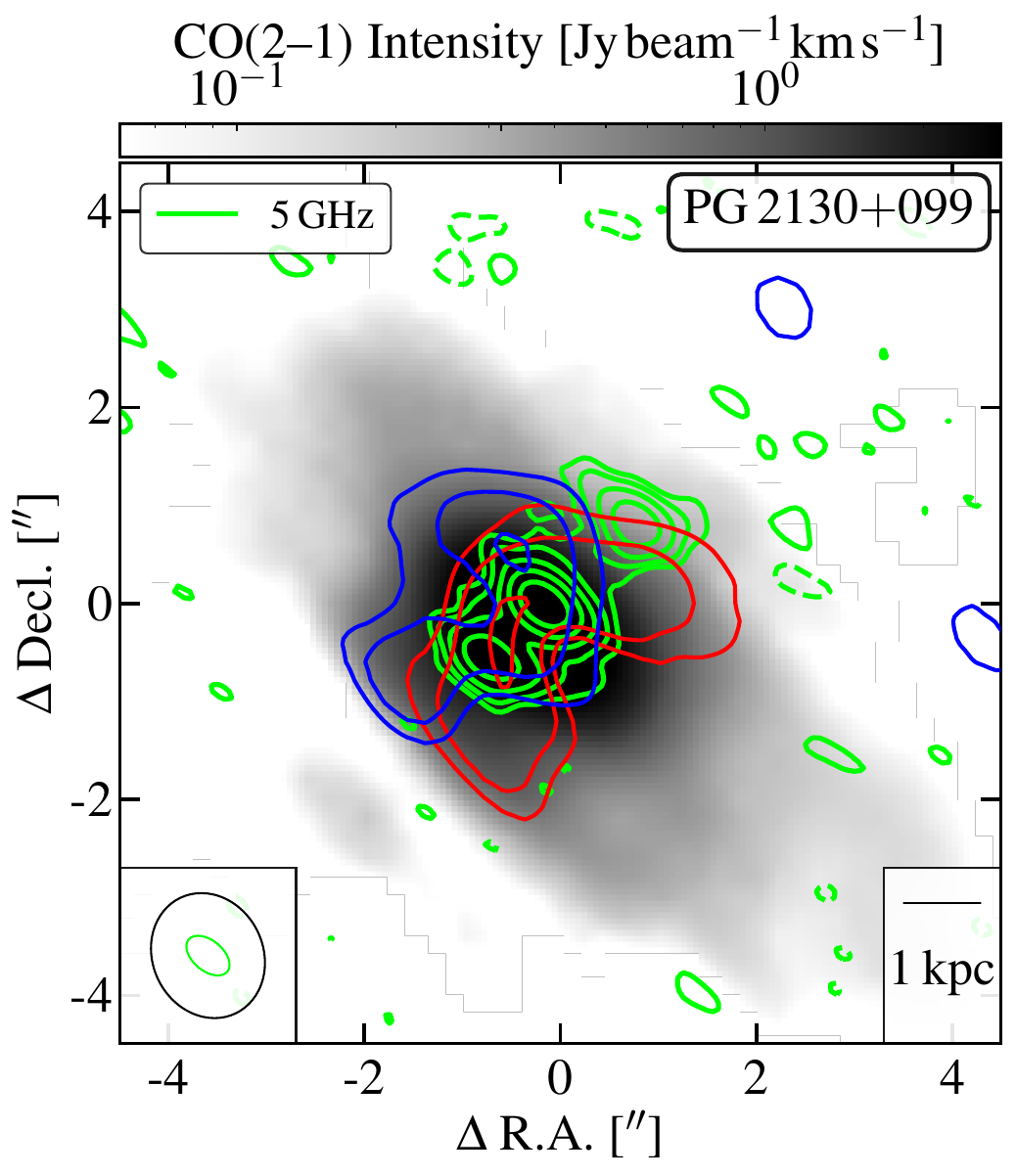}
    \caption{The map of molecular gas non-circular motions and the radio emission in two PG quasars is presented. The grey maps illustrate the CO(2--1) intensity map. The blue and red contours represent the blue- and red-shifted velocity integrated intensity map. The magenta contours represent the flux-weighted velocity dispersion map, starting from 100\,$\rm km\,s^{-1}$ and in steps of 30\,$\rm km\,s^{-1}$. Green contours portray the 5\,GHz radio emission, derived from observations using the Jansky Very Large Array (VLA). These contours span a range from -3 to 48 times the rms noise; $\sigma$ is 3.9\,$\mu \rm Jy\,beam^{-1}$ for PG\,1126$-$041 and 5.4\,$\mu \rm Jy\,beam^{-1}$ for PG\,2130$+$099. The ellipses in black at the lower bottom corner of each panel indicate the synthesized beam of the CO map, while the green ellipses represent the beam of the radio image. The scale bar is shown in the bottom right of each panel.}
    \label{Fig10: outflow map}
\end{figure*}

\begin{figure}
    \centering
    \includegraphics[width=\linewidth]{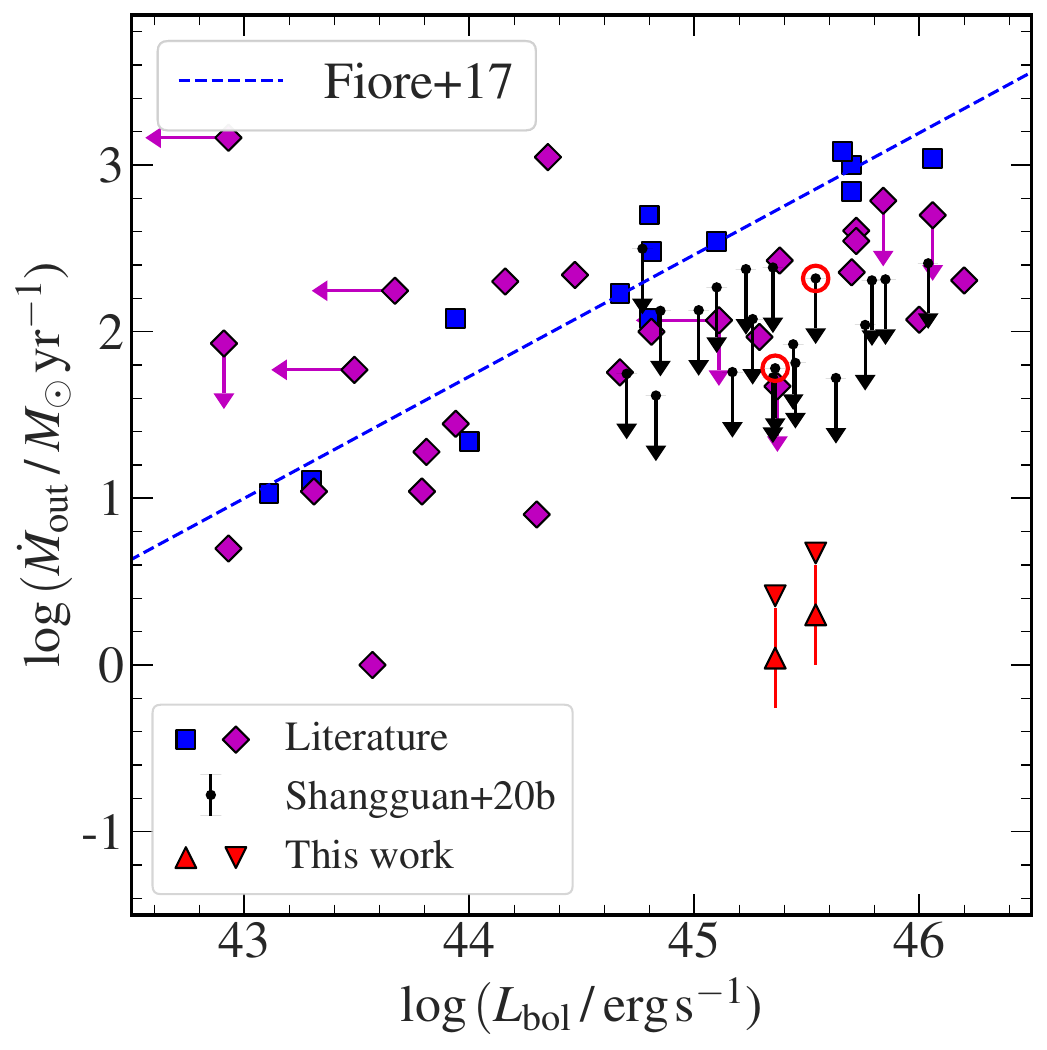}
    \caption{The molecular gas mass outflow vs. the AGN bolometric luminosity for our quasars. The red triangles represent the resulting outflow rates of an expanding shell, and the inverted triangles represent the result of a homogeneously filled cone. Quasars from previous studies are marked as blue and purple squares \citep{Fiore+2017, Fluetsch+2019}. Additionally, black arrows indicate the upper limits of outflow rates taken from ACA data \citep{Shangguan+2020b}, with two targets analyzed in this research highlighted with red circles. The blue dashed line represents an empirical relationship that shows molecular gas outflow rates relative to bolometric luminosities, located 2 dex above our measurements.}
    \label{Fig11: outflow rate}
\end{figure}

Section \ref{subsec3.3: kinematics results} shows that the cold molecular gas motion in PG\,1126$-$041 and in PG\,2130$+$099 can not be fully described by a regular rotating disk, where non-circular motions are considerably presented in the residual map and comparable to the 0th order. Although we noticed residual velocities for PG\,0923$+$129, we found the regular pattern of the velocity residual follows the spiral arms, suggesting that the non-circular motion is caused by the gravitational perturbation rather than the AGN feedback. Therefore we remove this target in the following analysis. For the remaining two targets, we tried to identify whether the non-circular motions found in Section \ref{subsec3.3: kinematics results} are inflow or outflow following the methods introduced by \cite{Genzel+2023}. Our analysis reveals an outflow nature of non-circular motion in PG\,1126$-$041, while a combination of inflow and outflow in PG\,2130$+$099. For simplicity, we regard the non-circular motion as an outflow in PG\,2130$+$099 in the following analysis to constrain the maximum contribution from possible AGN feedback. We caution that the divergency of inflow and outflow depends on the inclination angle of galaxies, which is hard to determine.

We now estimate the mass-outflow rates ($\dot{M}_{\rm out}$) following the methodology outlined by \cite{Fiore+2017, Feruglio+2020}. Taking into consideration that the kinetic model generated by \Bbarolo\ characterizes the orderly rotating aspect of the molecular gas disk (as detailed in Section \ref{subsec3.2: gas kinematics}), we distinguish the non-circular motion, which is regarded as the outflow component, by subtracting the \Bbarolo\ model from the actual observed data cubes. 

In order to characterize the attributes of the outflow component, we separated the red and blue-shifted outflow gas from the residual cube. We create a map of velocity-integrated intensity, a map of los velocities, and a map of velocity dispersion for the red- and blue-shifted components, which represent the flux, the los velocity, and the velocity dispersion of the outflow component. In Figure \ref{Fig10: outflow map}, the blue and red contours show the intensity map of outflow components in two quasars. To better understand the distribution of these outflow components, we overlayed the CO intensity map, the velocity dispersion map of CO emission, and the 5\,GHz radio emission. We obtained the radio data from the Jansky Very Large Array (JVLA) data archive, with data for PG\,1126$-$041 sourced from 16B-084 (PI: M. Perez-Torres) and data for PG\,2130$+$099 sourced from 22A-042 (PI: J. Gelfand). We calibrated the data using CASA pipeline and reduced radio frequency interference (RFI) with CASA task \texttt{rflag}. Next, we imaged and cleaned the continuum data with Briggs weighting (robustness parameter = 0.5) and a stop threshold 2.5 times the rms noise of the source-free region. The green contours on top of Figure \ref{Fig10: outflow map} show the continuum. It is worth noting that although the beam size of the radio image is three times smaller than that of CO(2--1) data, in PG\,1126$-$041, the outflow gas and small-scale radio jet have similar orientations and extents, suggesting that the radio emission might contribute to the production of gas outflow. For PG\,2130$+$099, we note that the radio emission and the outflow alignment of this target is similar to another low-$z$ quasar \citep{Girdhar+2022}. We also roughly estimate the extension of those gas outflows, which are about 1\,kpc after deconvolving the synthesized beam. The outflow extension is similar to the size of the radio jet, supporting the scenario that the radio jet contributes to such outflow. This spatial alignment between the molecular gas outflow and the extended radio emissions implies an inherent correlation, potentially indicative of jet-driven outflows. Similar phenomena have been proposed and observed in numerical simulations (e.g., \citealt{Mukherjee+2016, Mukherjee+2018}) and in several nearby low-luminosity AGN host galaxies \citep{Garcia-Burillo+2014, Feruglio+2020, Nesvadba+2021, Venturi+2021, Girdhar+2022, Murthy+2022, Rao+2023}.

We compute the outflow velocity as $V_{\rm out} = V_{\rm shift} + 2\sigma_{\rm out}$, where $V_{\rm shift}$ denotes the shifted center of the outflow and $\sigma_{\rm out}$ signifies the line width of the outflow, values obtained from the velocity map and the velocity dispersion map of the residual. The corresponding mass of molecular gas in the outflow is then computed as $M_{\rm out} = \alpha_{\rm CO} \times L_{\rm CO,\,out}$, where $\alpha_{\rm CO}$ is assumed to be $0.8\,M_\odot\,\rm (K\,km\,s^{-1}\,pc^{-2})^{-1}$, a commonly employed value for molecular gas outflows \citep{Fiore+2017, Feruglio+2020}, and $L_{\rm CO,\,of}$ is computed from the intensity map derived from the residual cube.

We estimate the outflow rates of both red- and blue-shifted components by assuming a homogeneously filled cone geometry \citep{Fiore+2017},
\begin{eqnarray}
    \dot{M}_{\rm out}=\frac{3M_{\rm out}V_{\rm out}}{R_{\rm out}}.
\end{eqnarray}
For an expanding shell-like shock front outflow scenario \citep{Husemann+2019}, the outflow rate is then estimated by
\begin{eqnarray}
    \dot{M}_{\rm out}=\frac{M_{\rm out}V_{\rm out}}{\Delta R_{\rm out}}.
\end{eqnarray}
In the following analysis, we estimate the outflow rates in these quasar hosts using a scenario, for a homogeneous spherical geometry, with $R_{\rm out}=1\,$kpc and $\alpha_{\rm CO}=3.1\,M_\odot\,\rm (K\,km\,s^{-1}\,pc^{-2})^{-1}$, and for an expanding shell-like scenario, we adopt a ULIRG-like $\alpha_{\rm CO}$ value and $\Delta R=200\,$pc \citep{Fiore+2017, Husemann+2019}.

We present the molecular gas outflow rates against the AGN bolometric luminosities in Figure \ref{Fig11: outflow rate}, and compare our results with those in the literature. We find that the gas outflow rates are lower than the empirical relation by 2 orders of magnitude, while they are consistent with the outflow upper limits estimated by assuming a filled cone geometry \citep{Shangguan+2020b}. We further investigate whether the energy carried by the radio jet is large enough to trigger the molecular gas outflow, following the procedure shown by \cite{Girdhar+2022}. Assuming an MW-like $\alpha_{\rm CO}$ and a homogeneous geometry of the outflow gas, the kinetic outflow rate of molecular gas in PG\,1126$-$041 and PG\,2130$+$099 are around $1.2\times 10^{42}$ and $1.4\times 10^{42}\,\rm erg\,s^{-1}$, respectively. To estimate the kinetic jet power, we used the empirical relation proposed by \cite{Merloni+2007}:
$\log P_{\rm jet} = (0.81\pm0.11)\times \log L_{\rm 5GHz}+11.9_{-4.4}^{+4.1},$
where $L_{\rm 5GHz}$ denotes the 5\,GHz luminosity estimated from the low-resolution 5\,GHz flux density that captured total radio emission \citep{Kellermann+1994, Kukula+1998}, which resulted in $9.8\times 10^{42}\,\rm erg\,s^{-1}$ and $1.6\times 10^{43}\,\rm erg\,s^{-1}$ for PG\,1126$-$041 and PG\,2130$+$099, respectively. We noted that the coupling efficiency, calculated by dividing the kinetic outflow rate by the bolometric luminosity, is less than one percent, which aligns with recent studies \citep{RamosAlmeida+2019, Girdhar+2022}. However, the ratio of gas outflow rates to kinetic jet power indicates a higher coupling efficiency of approximately 10\%. When adopting a ULIRG-like $\alpha_{\rm CO}$, the coupling efficiency remains around 3\%. These results suggest that despite the relatively low power of the radio jet in our sample (as the targets are radio-quiet AGN), the jet is still capable of efficiently transferring energy into the ISM and driving outflows. This phenomenon has been observed in several low-redshift radio-quiet AGNs \citep[e.g.,][]{Villar-Martin+2017, Venturi+2021, Girdhar+2022}. The radio jet may be the dominant source of feedback on sub-kiloparsec scales, as indicated by cosmological simulations \citep[e.g.,][]{Weinberger+2017}. However, it is important to note that the velocity of these outflows is lower than the escape velocity at the corresponding radii (which is roughly 1.4$\times V_{\rm rot}$, see Section \ref{subsec3.2: gas kinematics}), indicating that the outflowing gas is still influenced by the gravitational potential of the host galaxy, thus limiting the impact of AGN feedback. 



\section{Summary and Conclusion}
\label{sec6: summary}
We utilize observational data obtained from the Atacama Large Millimeter/submillimeter Array (ALMA) and the Multi-Unit Spectroscopic Explorer (MUSE) for an investigation encompassing three PG quasar host galaxies at $z\lesssim 0.06$ \citep{Molina+2021, Molina+2022}. With the careful investigation of the distributions and kinematics of both molecular and ionized gas emissions, we perform a systemic analysis exploring the mass distribution characteristics intrinsic to both quasars hosted by disk galaxies and inactive star-forming disk galaxies by modeling the gas dynamics. Our results are enumerated as follows:
\begin{enumerate}
    \item Through kinematic analysis, we have observed that regular rotation plays a dominant role in both molecular and ionized gas components in our sample. Most of the galaxies exhibit rotational motion with some instances of moderate non-circular motions in select quasar host galaxies. Additionally, we have found that the circular velocities of both gas phases are consistent, indicating a uniform rotational behavior across the molecular and ionized gas components.
    \item The stellar masses derived through our dynamical approach demonstrate a notable alignment with those acquired via broad-band photometry, accounting for the inherent uncertainties. Despite the constrained sample size, this comparative scrutiny serves to validate the consistency evident in stellar mass determinations derived from these divergent methodologies. Consequently, this validation establishes a foundational framework for the potential application of these methods in future investigations.
    \item By conducting a comparative investigation into the ISM kinetic pressure and ISM gravitational pressure utilizing our galaxy mass model, we discern the hydrostatic equilibrium state of molecular gas within both quasar host galaxies and inactive star-forming galaxies. Despite the similarity between the $P_{\rm ISM}-\mathcal{W}$ relationships in inactive star-forming galaxies and quasar host galaxies, both closely align with the theoretical 1:1 relation. This parallel relationship implies a limited role for AGN feedback in governing the molecular gas dynamics on a galactic scale.
    \item We explore the distribution of the scale height of molecular gas disks by assuming a hydrostatic equilibrium between kinetic pressure and the weight of ISM. Our investigation reveals a comparable range of molecular gas disk scale height ($h_{\rm g}$) for both quasar host galaxies and the control sample. The observed similarity in the $h_{\rm g}$ range suggests that the thickness of molecular gas disks can be influenced by the self-gravity inherent to galaxies, obviating the necessity for an external energy budget to expel cold molecular gas.
    \item Within the framework of non-circular motion as an integral component of the outflow phenomenon, we undertake the computation of both outflow velocity and outflow rate. Notably, the spatial distribution of molecular gas outflow rates fortuitously aligns with the pattern of radio emissions, suggesting an origin linked to the activity of the AGN. Our molecular gas outflow rates are much less than the literature prediction, while more close to the most recent results.
\end{enumerate}

\begin{acknowledgements}
    We acknowledge supports from the National Natural Science Foundation of China (NSFC) with grant Nos. 11991052, 12173002. The research grants from Ministry of Science and Technology of the People’s Republic of China (NO. 2022YFA1602902), and the science research grants from the China Manned Space Project with Nos. CMS-CSST-2021-A06. Q.F. gratefully acknowledges financial support from the China Scholarship Council. LCH was supported by the National Science Foundation of China (11991052, 12011540375, 12233001), the National Key R\&D Program of China (2022YFF0503401), and the China Manned Space Project (CMS-CSST-2021-A04, CMS-CSST-2021-A06). We gratefully acknowledge funding from ANID - Millennium Science Initiative Program - ICN12\_009 (FEB), CATA-BASAL - FB210003 (FEB, ET), and FONDECYT Regular - 1200495 and 1241005 (FEB, ET). This paper makes use of the following ALMA data: ADS/JAO.ALMA\#2017.1.00287.S, \#2018.1.0006.S. ALMA is a partnership of ESO (representing its member states), NSF (USA) and NINS(Japan), together with NRC(Canada), MOST and ASIAA(Taiwan), and KASI(Republic of Korea), in cooperation with the Republic of Chile. The Joint ALMA observatory is operated by ESO, AUI/NRAO, and NAOJ. This research has made use of the services of the ESO Science Archive Facility and is based on observations collected at the European Organization for Astronomical Research in the Southern Hemisphere uncer ESO program IDs 094.B$-$0345(A), 095.B$-$0015(A), 097.B$-$0080(A), 0101.B$-$0368(B), 0103.B$-$0.496(B) and 0104.B$-$0151(A). This study makes use of the VLA data from program 16B-084, 22A-042. This study uses data provided by the Calar Alto Legacy Integral Field Area (CALIFA) survey (https://califa.caha.es/). Based on observations collected at the Centro Astronómico Hispano Alemán (CAHA) at Calar Alto, operated jointly by the Max-Planck-Institut fűr Astronomie and the Instituto de Astrofísica de Andalucía (CSIC). 
\end{acknowledgements}

\software{\textsc{Agama} \citep{Vasiliev2019}; \textsc{Astropy} \citep{Astropy2013}, \textsc{CASA} \citep{McMullin+2007}; \textsc{Emcee} \citep{Foreman-Mackey+2013}; \textsc{Linmix} \citep{Kelly2007}; \textsc{Numpy} \citep{van_der_Walt+2011}; \textsc{pPXF} \citep{Cappellari2017}; \textsc{Scipy} \citep{Virtanen+2020}.}

\bibliography{manuscript}{}
\bibliographystyle{aasjournal}

\appendix

\section{Recover the pixel-wise velocity dispersion}

\begin{figure*}
    \centering
    \includegraphics[width=\linewidth]{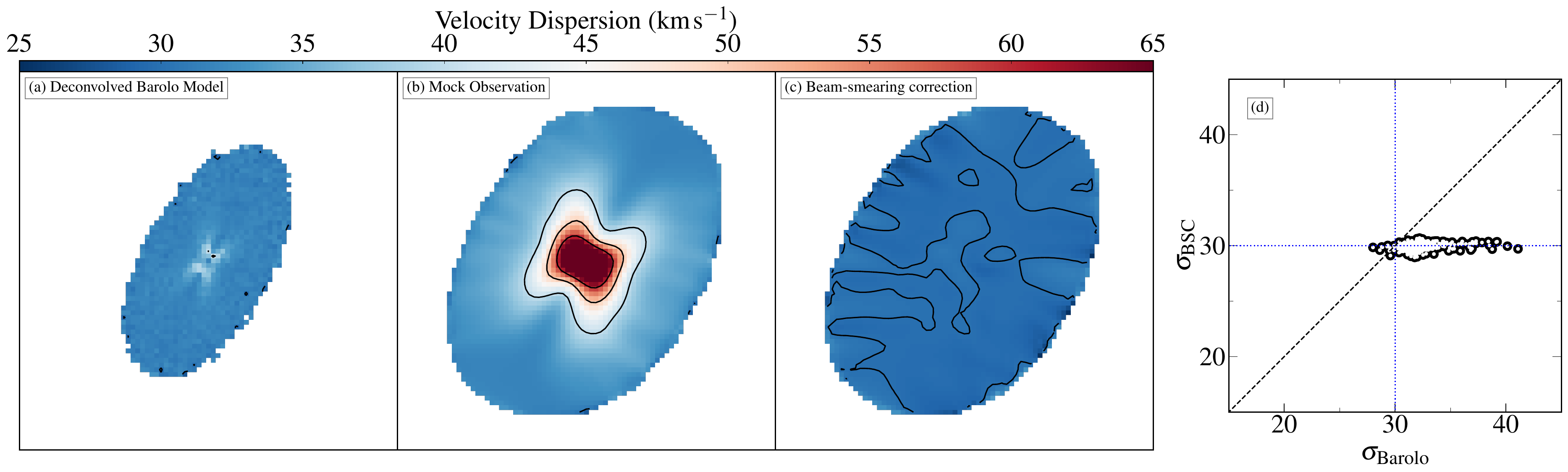}
    \caption{An illustrative example of our analysis. {\it Panel (a) and (b)} show the velocity dispersion map generated from the deconvolved \Bbarolo\ model and the corresponding mock observation, respectively. {\it Panel (c)} displays the $\sigma$ map obtained by applying our correction method to the map in {\it Panel (b)}. {\it Panel (d)} presents a comparison between the pixel-wise velocity dispersions derived from the two methods. The x-axis represents the pixel-wise velocity dispersion directly extracted from the second-moment map generated by the \Bbarolo\ model ({\it Panel a}), while the y-axis represents the velocity dispersion obtained using our beam-smearing correction method. The blue dotted lines indicate the input value of $\sigma$, and the black dashed line denotes the 1:1 relation. In {\it Panels (a)} through {\it (c)}, the black contours start at 30\,$\rm km\,s^{-1}$ and increase in increments of 10\,$\rm km\,s^{-1}$.}
    \label{Fig12: sigma}
\end{figure*}

Correcting for the beam-smearing effect is essential for accurately assessing pixel-wise velocity dispersion in our analysis. We evaluated the recovered pixel-wise $\sigma$ using the method described in Sec.\ref{subsec5.2: equilibrium} and compared it with values directly derived from the ``deconvolved'' \Bbarolo\ model, employing a simulated gas disk model. For simplicity, we generated a data cube of a gaseous disk using the \Bbarolo\ task \texttt{GALMOD}, with a rotational velocity $V_{\rm rot}=200\,\rm km\,s^{-1}$ and a velocity dispersion $\sigma=30\,\rm km\,s^{-1}$ uniformly across all radii. The model's inclination and position angle were set to 60$^\circ$ and 150$^\circ$, respectively. We then convolved the modeled data cube with a synthesized beam adapted from the ALMA observation of PG\,1126$-$041 to simulate observational conditions. Velocity dispersion maps for the ``deconvolved'' model and the mock observation were generated using the CASA task \texttt{immoments}. The beam-smearing correction method detailed in Sec.~\ref{subsec5.2: equilibrium} was subsequently applied to the velocity dispersion map of the mock observation to reproduce the intrinsic pixel-wise $\sigma$. These velocity dispersion maps, along with the beam-smearing corrected $\sigma$ map, are presented and compared in Fig.~\ref{Fig12: sigma}.

A comparison of the $\sigma$ values obtained from the two methods reveals that the $\sigma$ map directly generated from the model can overestimate the velocity dispersion due to projection effects. The velocity dispersion of each pixel is consistently enhanced because multiple velocity components are combined along the line of sight. This enhancement is not due to the mixing of emissions from neighboring pixels but results from the overlapping gaseous clouds. Our correction method effectively addresses this issue and accurately reproduces the intrinsic velocity dispersion. By testing different input inclination angles and velocity dispersion values, we consistently reproduced the input values, demonstrating the reliability of our method.

\end{document}